\documentclass[preprint,aps,showpacs]{revtex4}
\usepackage{epsfig}

\begin{document}

\title{
Electron and Photon Scattering on  Three-Nucleon Bound States
}

\author{J.~Golak, R.~Skibi\'nski, H.~Wita{\l}a}
\affiliation{M. Smoluchowski Institute of Physics, Jagiellonian University,
                    PL-30059 Krak\'ow, Poland}

\author{W.~Gl\"ockle}
\affiliation{Institut f\"ur Theoretische Physik II,
         Ruhr Universit\"at Bochum, D-44780 Bochum, Germany}

\author{A.~Nogga}
\affiliation{Forschungszentrum J\"ulich, IKP 
(Theorie), D-52425 J\"ulich, Germany}

\author{H.~Kamada}
\affiliation{Department of Physics, Faculty of Engineering,
  Kyushu Institute of Technology,
  1-1 Sensuicho, Tobata, Kitakyushu 804-8550, Japan}

\begin{abstract}
A big spectrum of processes induced by real and virtual photons on
the $^3$He and $^3$H nuclei is theoretically investigated through many
examples based on nonrelativistic Faddeev calculations for bound and
continuum states. The modern nucleon-nucleon potential AV18 together with the
three-nucleon force UrbanaIX is used. The single nucleon current is 
augmented by explicit $\pi$- and $\rho$-like two-body currents which
fulfill the current continuity equation together  with the corresponding
 parts of the AV18 potential. We also employ the Siegert theorem,
 which induces many-body contributions to the current  operator.
 The interplay of these different dynamical ingredients in the various
electromagnetic processes is studied and the theory is compared to
 the experimental data. Overall we find fair to good  agreement
but also cases of strong disagreement  between theory and experiment, 
which calls for  improved dynamics. 
In several cases we refer the reader to the work of other groups and 
compare their results with ours.
In addition we list a number of predictions 
for observables in different processes which would challenge this
dynamical scenario even more stringently and systematically.
\end{abstract}
\pacs{21.45+v,21.10-k,25.10+s,25.20-x}
\maketitle

\tableofcontents

\newpage

\section{Introduction}
\label{sec:1}

Real and virtual photon induced processes in the three-nucleon (3N) 
 system have been
studied for a long time and these investigations go on with intensity.
The reason is that beyond the deuteron the 3N system in the form
of $^3$He has been always considered since the very 
beginning of nuclear physics~\cite{Wigner,Schwinger}
as  a challenge to be understood
in terms of the available state of the art forces.
Then the next question followed naturally: what is the 
response of the 3N bound state
to real and
 virtual photon absorption ? Again answers have been 
searched for over the many years to the best of the
available physical insights and technical feasibilities.
Here we point just to a few early studies~\cite{Collard63,Schiff63}
and refer the reader  
to the various reviews given below for the long history
of that research.

While the 3N bound states were 
numerically mastered already in the seventies and early eighties
using nucleon-nucleon (NN) forces with realistic and complex
spin-momentum 
structures~\cite{Laverne, Payne, Hajduk, Gloeckle82, Sasakawa81, Brandenburg}
and later on adding first models for 
3N forces~\cite{Gloeckle82, Sasakawa86, Chen, Stadler, Wu, Nogga2003},
the technical
challenges for the 3N continuum with the complex asymptotic boundary conditions
were much more demanding. But in the  last  10-15 years also
the 3N continuum got more and more under 
control~\cite{Ubad,Friar90a,ourreport,Kievski2001,Gloeckle2002,Nemoto98},
which opened solid  theoretical access to the great diversity of 
inelastic real
and virtual  photon
induced reactions on $^3$He and the nucleon-deuteron (Nd) capture processes. There has been,
and this is going on, an intensive interplay and reciprocal stimulation
of theory and experiment, which justifies, as we think, a review
of the present state of the art.

Elastic electron scattering on $^3$He ($^3$H) has been reviewed many
times over the 
years~\cite{Goldemberg66,Kim74, degliAtti80,Mathiot89,Carlson98}.
The inclusive process $^3{\rm He}(e,e')$ has been reviewed
in~\cite{Carlson98}. 
A very informative monograph on 
electron induced processes on nuclei including the 3N
system is~\cite{boffibook}.
Semiexclusive and above all exclusive electron induced
processes on $^3$He came into the focus only with the high-duty cycle electron
accelerators (NIKHEF, MAMI, Jlab) and reviews about those processes 
 in  the 3N system are not
known to us. 
A good collection of references to
old calculations on the photodisintegration of
$^3$He can be found in~\cite{Klepacki92}.
Recent work on these processes is discussed and cited in~\cite{Carlson98}.

Variational approaches and rudimentary treatments of the 3N continuum
in electromagnetically induced processes
were used before the sixties and still in the early seventies
and we refer the reader to the literature quoted in the above
listed reviews. Then with the Faddeev formulation of the
three-body system~\cite{Faddeev60-61}
or the equivalent  Alt-Grassberger-Sandhas (AGS) equations~\cite{AGS}, 
where the latter ones  are ideally suited for finite
rank forces, a new epoch started. In the following we shall not
distinguish between the two and just call them for short "Faddeev
approach". At that time, due to the lack of sufficiently strong 
computational  resources,  
the nuclear potentials were chosen in a  quite simple form, low 
rank separable ones.
Very first calculations for electrodisintegration 
 of $^3$He and  $^3$H in the Faddeev scheme
were performed in~\cite{Lehman69-71}, where the 3N bound state was treated
correctly but in the final 3N continuum state only the interaction
within the spectator pair was kept
(the two nucleons which have not absorbed the photon under a single
nucleon current assumption). 

Very similar in nature and techniques
is the photodisintegration, where the first Faddeev calculation
for the 3N continuum appeared in ~\cite{Barbour67-70}
and where the
importance of the rescattering with the spectator nucleons was emphasized.
One step further was the work in \cite{Gibson75-76}
where for  the two- and three-body photodisintegration
of $^3$He ($^3$H) both, ground state and 3N continuum,  were treated
consistently as solutions of 
 the Schr\"odinger equation with 
 the same 3N Hamiltonian. This exact treatment, 
though still with simple NN forces, already allowed one to ask detailed
questions~\cite{Lehman79}
like the suppression of the isospin T=1/2 contribution in three-body photodisintegration
of $^3$He. Then the first calculation for two-body electrodisintegration
of $^3$He ($^3$H) came up in~\cite{Heimbach77-77}.
Though also the
formalism for three-body disintegration in the context of separable
forces was formulated, limitations of computer resources prevented
their realization. It then took quite some time that the three-body
electrodisintegration has been treated~\cite{Meijgaard90-90-92} 
 using simple $s$-wave local forces in an unitary pole 
expansion  or only
in the form of the unitary pole approximation. The conclusion was again 
that a proper description has to take into account contributions from
the complete multiple scattering series, or in other words, 
that final state interaction (FSI) are important. Due to the lack of 
kinematically complete
breakup data, the calculation of~\cite{Meijgaard90-90-92} was applied 
to a set of existing
inclusive data, where the two- and three-body electrodisintegration
processes are both involved.

Physically and formally closely related to electron induced processes 
is the proton-deuteron (pd) radiative capture
reaction, where a first configuration space 3N calculation based
on solutions of the Faddeev equation for the 3N bound state and 3N
scattering states appeared in~\cite{Jourdan86}
using the Reid NN force \cite{Reid68}. 
Thereby, as in the following
studies~\cite{Ishikawa92,Fonseca91-92-92,Schmid00,Fonseca00},
the interest was in the sensitivity of tensor analyzing
powers to properties of the 3N bound state and to the NN tensor forces.
The treatment of the initial state interaction in the pd capture
processes turned out to be very crucial as well as  the inclusion
of higher NN force components. In
\cite{Ishikawa92} realistic NN forces and even 3N forces
were used in a consistent 3N Faddeev treatment for both the ground state
and the  continuum states. In~\cite{Fonseca91-92-92} separable 
forces were employed but
also an Ernst-Shakin-Thaler-expansion form of the Paris potential.

At very low energies (neutron-deuteron) nd capture was treated
in~\cite{Friar90} using a
configuration space Faddeev method and realistic NN and 3N forces.
The method of correlated orthogonal states ~\cite{Schiavilla85-87}
represents the continuum to some extent and puts
in short-range correlations. Although the states are not proper
solutions of the 3N Hamiltonian, their use in studying inclusive
response functions clearly showed significant improvements over
plane-wave impulse approximation results and underlined the
importance of treating the correlations between the three nucleons
in the final state as consequently as in the 3N bound state.

Another development was the Euclidean response method~\cite{Carlson92-94}
applied to inclusive responses. By path integral techniques one
calculates the Laplace transform of the response functions and
compares them to the corresponding Laplace transformed data. This
is an exact method and includes the full dynamics of the chosen Hamiltonian.
Related to that are approaches with Stieltjes transforms~\cite{Efros93}
or transformations
by a Lorentz kernel~\cite{Efros94}.

Around that time the first calculations appeared, where realistic
NN forces, with all their complexities and including all the relevant
higher NN force components, were applied to the pd(nd) and three-nucleon
electrodisintegration of $^3$He ($^3$H) 
in the Faddeev scheme~\cite{Ishikawa94,Golak95}. 
In that formulation the pd and ppn breakup 
of $^3$He induced by an external probe can be calculated in "one shot"  
solving a Faddeev like integral equation and avoiding 
the nasty low order rescattering processes occurring in the separate
treatment of the 3N continuum~\cite{Meijgaard90-90-92}.
For inclusive scattering a convenient short cut was found in
\cite{Ishikawa94.2,Golak95.2}
using the closure relation for the eigenstates of the Hamiltonian. In this
manner, one avoids the explicit numerical integrations over  
 all the available 
two-body and three-nucleon disintegration configurations.

In the older investigations mostly only the nonrelativistic single
nucleon current operator has been used. For  real photon induced
processes it was supplemented by 
the Siegert approach, which takes some exchange currents
into account. This is  insufficient and the explicit
use of two-body currents (and possibly three-body contributions when a three nucleon force (3NF) is 
included)  is required.
These dynamical ingredients are as complicated as nuclear forces
and therefore progress is slow. An important practical step was performed
in \cite{Riska1,Riska2} by associating two-body currents to NN forces through
the continuity equation. In the case of the AV18 NN force~\cite{AV18}  
that recipe
has been used quite often~\cite{Carlson98} and is still applied. Closely 
related studies
connecting NN forces and two-body currents appeared in~\cite{MEC2, MEC3}.

Nearly all of the results shown in this review are based on our
own work using the Faddeev scheme in a purely nucleonic Hilbert space.
There are also other groups, which investigate
real and virtual photon induced processes on light systems. For 
the wealth of insight
and achievements in the case of the deuteron we refer 
to~\cite{deuteron,Gilman}.
Here we focus  just on the 3N system.
The group in Pisa uses hyperspherical harmonic expansions of different
types and treats  bound  and continuum states consistently. They use 
 modern  nuclear forces in all their complexities together with related
currents. Their focus is mostly on processes
at very low energies~\cite{Pisa.low}. This includes pd radiative capture, 
inclusive threshold electron scattering on $^3$He, and pd breakup 
electrodisintegration of $^3$He. 
 The Urbana-Argonne group relies beside variational approaches
on the Green-function-Monte-Carlo method~\cite{GFMC}. A good overview
on the theory and their  important
results can be found in ~\cite{Carlson98}. In the 3N system this
comprises work on the elastic form factors, short-range correlations
related to the Coulomb sum rule, Nd capture reactions, and Euclidean
inclusive response functions.
The group in Trento uses the Lorentz integral transform (LIT)
method~\cite{Efros94}
and employs also hyperspherical harmonic expansions. In this method
one avoids the direct treatment of the continuum which requires
the handling of the complex boundary conditions. Instead 
that method  converts the continuum problem into a bound state problem.
The price to be paid is an
inversion of  auxiliary Lorentz transformed amplitudes. The
mathematical properties of that technique are   displayed in ~\cite{Efros99}.
This method is being applied not only to the 3N system but is powerful
enough to go beyond A=3  using ideas of effective
force expansions~\cite{Barnea2000,Orlandini04,Leidemann04}.
More recently the Hanover group also started to thoroughly
investigate the 3N continuum and photon induced reactions
therein~\cite{Deltuva2004,Yuan02.1,Yuan02.2}.
The new feature is the explicit inclusion of the $\Delta$-degree of
freedom. Thus the Hilbert space is the direct sum of NNN and NN$\Delta$ states.
In this  manner a certain subset of 3N forces is taken care of as
well as consistent two-body currents.

Last but not least we would like to point to the very rich list of 
investigations by J.M. Laget who uses a diagrammatic approach. That work 
has stimulated many experimental investigations and sheds light on the 
reaction mechanisms. A recent paper \cite{laget} discusses 
electrodisintegration of few-body systems high in momenta above our 
nonrelativistic domain but also provides many references to earlier studies, 
which are relevant to the work discussed in this review. 

For all technical details used by these other groups we refer the reader
to the cited literature. We shall provide, however, information on
their results at the appropriate places in section \ref{data}.

This review is organized as follows.
In Section~\ref{sec:2} we describe
our approach in the Faddeev scheme for the great diversity of photon
induced processes. A brief review on electromagnetic currents is given
in Section~\ref{currents}.
The observables are defined in Section~\ref{observables}.
Then Section~\ref{performance} describes 
 the way we technically perform
the calculations. 
 Section~\ref{data} is devoted to a comparison of our theoretical results and some
selected results by other groups to the data. Much remains, however,
to be done and we present in Section~\ref{predictions}
an incomplete and subjective list of
theoretical predictions, some of which will hopefully be testified 
in experiments in the near future.
In Section~\ref{addendum} we provide remarks on several issues relevant 
in the 3N system which have not been addressed directly in this review. 
We end up with a summary and outlook in Section~\ref{summary}.

\section{Formalism in the Faddeev scheme}
\label{sec:2}
\begin{figure}[!ht]
\begin{center}
\epsfig{file=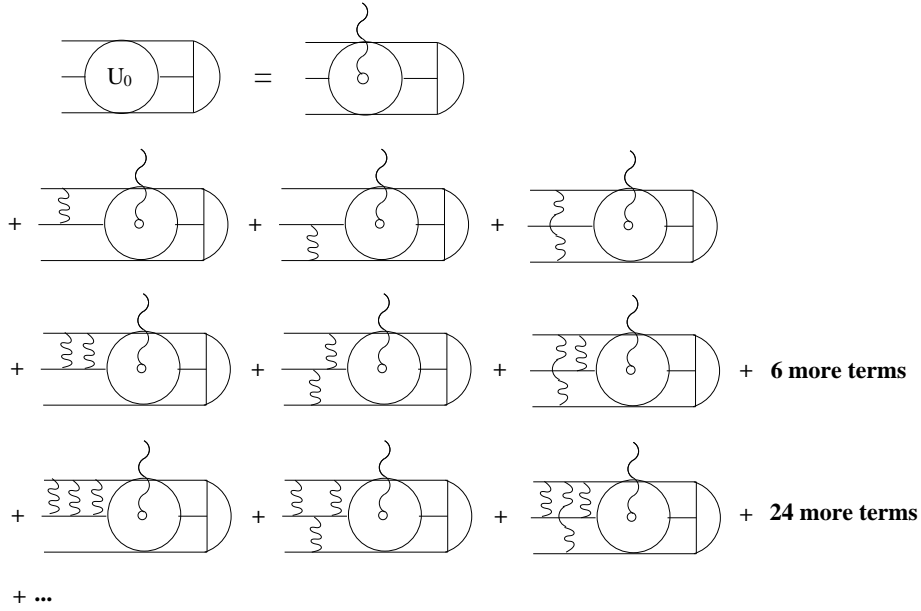,height=8cm}
\caption{
\label{diagram1}
The multiple scattering series for a 3N breakup amplitude due to photon 
absorption. 
 The half moon to the very right stands for the $^3$He state, the
circle with the wiggly line attached to it for the one-photon absorption
process and the wiggly lines for NN forces acting between all pairs to
first order, second order etc. For the sake of notation simplicity
the action of 3N forces has been dropped. The three horizontal lines
between the action of NN forces and between the photon absorption and
the NN forces stand for a free 3N propagation and the
three final horizontal lines to the very left represent the three final
nucleons (their momentum eigenstates).
}
\end{center}
\end{figure}

Let us start with a heuristic approach toward the photon induced complete
breakup of $^3$He. Once the photon has been absorbed inside $^3$He,
the three nucleons
are released but on the way of leaving the space spanned by the $^3$He
state they interact strongly. This is illustrated in Fig.~\ref{diagram1}. 
 Clearly this infinite set of diagrams summarizes all what can happen in
the 3N breakup process under the condition that the  three nucleons are 
interacting by pairwise forces. Because of the strength of nuclear
forces that series is generally diverging for c.m. energies in the 3N
system below the pion production threshold. It has to be summed up to infinite
order. We follow here the Faddeev scheme and perform first a partial
re-summation of the NN forces into NN t-operators. Apparently aside from
the very first term  without 
any interaction after the photon absorption
process ($U_0^{(0)}$) that set of diagrams can be split into 3 subsets
according to the utmost left pair force
\begin{equation}
U_0 = U_0^{(0)} + U_0^{(1)} + U_0^{(2)} + U_0^{(3)}  ,
\label{U0}
\end{equation}
where $U_0^{(i)}$ stands for 
the subset with $V_{jk}$ to the left $(j \ne i \ne  k)$.

\clearpage

Let us regard the first few terms for $U_0^{(1)}$ in Fig.~\ref{diagram2}:
\begin{figure}[!ht]
\begin{center}
\epsfig{file=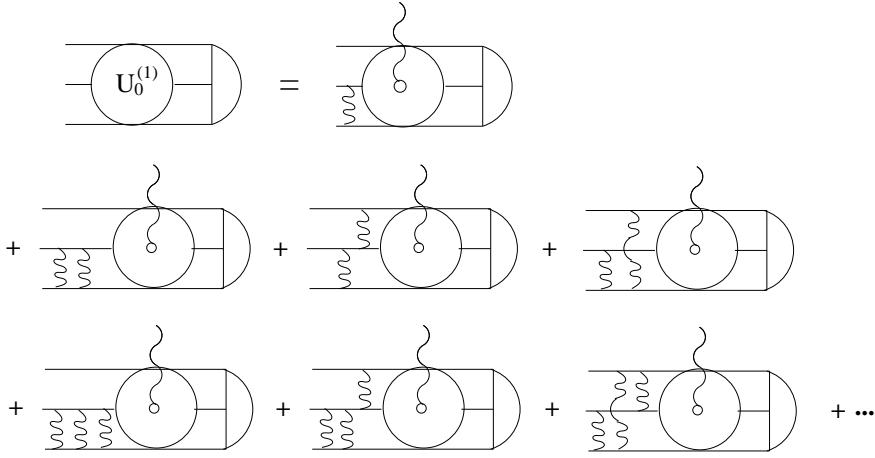,height=6cm}
\caption{
\label{diagram2}
The subset of diagrams ending with $V_{23}$ to the very left. Symbols as in
Fig.~\protect\ref{diagram1}.
}
\end{center}
\end{figure}
\newline
By the very definition of the three subsets this equals
\begin{equation}
U_0^{(1)} =  V_{23} G_0 {\cal O} \mid \Psi \rangle
+ V_{23} G_0 \left( U_0^{(1)} + U_0^{(2)} + U_0^{(3)} \right) ,
\label{U01}
\end{equation}
where ${\cal O}$ is the photon absorption operator, $ \mid \Psi \rangle$
the $^3$He state and $G_0$ the free 3N propagator.
We combine the terms with $U_0^{(1)}$ on the left hand side
\begin{equation}
\left( 1 - V_{23} G_0 \right) U_0^{(1)} =  V_{23} G_0 {\cal O} \mid \Psi \rangle
+ V_{23} G_0 \left( U_0^{(2)} + U_0^{(3)} \right) ,
\label{U01.2}
\end{equation}
invert
\begin{equation}
U_0^{(1)} = \left( 1 - V_{23} G_0 \right)^{-1} V_{23} G_0 {\cal O} \mid \Psi \rangle
+ \left( 1 - V_{23} G_0 \right)^{-1}  V_{23} G_0 \left( U_0^{(2)} + U_0^{(3)} \right) ,
\label{U01.3}
\end{equation}
and introduce the NN t-operator $t_{23}$
\begin{equation}
t_{23} \equiv \left( 1 - V_{23} G_0 \right)^{-1} V_{23}  .
\label{t23}
\end{equation}
Obviously, $t_{23}$ obeys the two-body Lippmann-Schwinger equation
\begin{equation}
t_{23}  = V_{23}  + V_{23} G_0 t_{23} .
\label{t23.2}
\end{equation}
This leads to
\begin{equation}
U_0^{(1)} = t_{23} G_0 {\cal O} \mid \Psi \rangle
+ t_{23} G_0 \left( U_0^{(2)} + U_0^{(3)} \right) .
\label{U01.4}
\end{equation}
Two more equations for $U_0^{(2)}$ and $U_0^{(3)}$ 
arise in exactly the same manner.

\begin{figure}[!ht]
\begin{center}
\epsfig{file=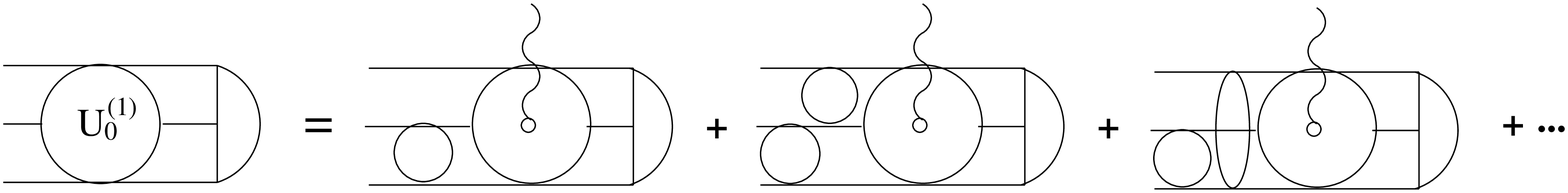,height=1.52cm}
\caption{
\label{diagram3}
In comparison to Fig.~\ref{diagram2} the NN forces are now replaced by NN
$t$-operators represented as circles or as an oval in the case of the pair 13.
In the second and higher orders clearly only consecutive circles acting on different
pairs can appear.
}
\end{center}
\end{figure}

Now we make use of the identity of the three nucleons. Since the photon
absorption operator ${\cal O}$ has to be symmetrical under 
exchange of the three
nucleons and the $^3$He state is antisymmetrical one immediately obtains
\begin{equation}
U_0^{(2)} = P_{12} P_{23} U_0^{(1)}
\label{U02}
\end{equation}
and
\begin{equation}
U_0^{(3)} = P_{13} P_{23} U_0^{(1)} ,
\label{U03}
\end{equation}
where $P_{ij}$ interchanges nucleons $i$ and $j$. It is convenient to define \cite{WGbook}     
\begin{equation}
P \equiv P_{12} P_{23} +  P_{13} P_{23}
\label{P}
\end{equation}
and we obtain
\begin{equation}
U_0^{(1)} = t_{23} G_0 {\cal O} \mid \Psi \rangle
+ t_{23} G_0 P U_0^{(1)} .
\label{U01.6}
\end{equation}
This is already a Faddeev type integral equation, which after iteration
leads to the multiple scattering series, now formulated in terms of NN
t-operators
\begin{equation}
U_0^{(1)} = t G_0 {\cal O} \mid \Psi \rangle
+ t G_0 P t G_0 {\cal O} \mid \Psi \rangle
+ t G_0 P t G_0 P t G_0 {\cal O} \mid \Psi \rangle
+ \dots
\label{U01.7}
\end{equation}
This is graphically depicted in Fig.~\ref{diagram3}.
The whole breakup amplitude is then given as
\begin{equation}
U_0 = U_0^{(0)} + (1 + P) U_0^{(1)} .
\label{U0.2}
\end{equation}
Here $U_0^{(0)}$ is obtained by a simple quadrature and $U_0^{(1)}$ 
arises as solution of the one
Faddeev-like equation (\ref{U01.6}).

Written in a more definite manner as matrix element the breakup amplitude reads
\begin{equation}
U_0 = \langle \phi_0 \mid {\cal O} \mid \Psi \rangle
    + \langle \phi_0 \mid (1 + P) \mid U \rangle
\label{U0.3}
\end{equation}
where the amplitude $\mid U \rangle$ obeys according to (\ref{U01.6})
\begin{equation}
\mid U \rangle = t G_0 {\cal O} \mid \Psi \rangle
    + t G_0 P \mid U \rangle .
\label{U}
\end{equation}
We dropped the index $23$ on t since one can choose any pair and
we introduced the free 3N state $\langle \phi_0 \mid$.
Since $  {\cal O} \mid \Psi \rangle $ and
$(1 + P) \mid U \rangle$ are totally antisymmetrical, we can assume
$\langle \phi_0 \mid$ to be antisymmetrical as well.

Let us now re-derive that result in a more standard algebraic manner
including also 3N forces. 
 The general form of the nuclear matrix element for an electroweak probe
represented by a symmetric operator ${\cal O}$ is given as
\begin{equation}
N = \langle \Psi_f^{(-)} \mid {\cal O} \mid \Psi_i \rangle  .
\label{N}
\end{equation}
Here  $\mid \Psi_i \rangle$ is the initial nucleus state
and $\langle \Psi_f^{(-)} \mid $ the final scattering state
with asymptotic quantum numbers $f$. It is generated as \cite{ME2}
\begin{equation}
\mid \Psi_f^{(-)} \rangle =
\lim_{\epsilon \rightarrow 0^+}
\frac{-i \epsilon}{E - i \epsilon - H } \mid \phi_f \rangle .
\label{Psif}
\end{equation}
In the three-nucleon system and for inelastic processes $f$ stands either
for asymptotic Nd or 3N quantum numbers. In the latter case we already
introduced the fully antisymmetrical state $\mid \phi_0 \rangle $,
which in our notation is given as
\begin{equation}
\mid \phi_0 \rangle = (1 + P) \mid \varphi_0 \rangle
\label{phi0}
\end{equation}
where $\mid \varphi_0 \rangle$ in the nonrelativistic regime is conveniently 
expressed in terms of Jacobi momenta 
\begin{equation}
 \mid \varphi_0 \rangle
\equiv (1 - P_{23}) \mid {\vec p} \,  {\vec q} \rangle
\equiv \mid  {\vec p} \rangle_a \mid  {\vec q} \rangle .
\label{varphi0}
\end{equation}

Depending on
which pair of nucleons is singled out there are three choices for the Jacobi momenta. Let us
choose one of them and define
\begin{equation}
{\vec p} = \frac12 ( {\vec k}_2 - {\vec k}_3 )
\label{jacobip}
\end{equation}
\begin{equation}
{\vec q} = \frac23 \left[ {\vec k}_1 - \frac12 ( {\vec k}_2 + {\vec k}_3 ) \right] ,
\label{jacobiq}
\end{equation}
where the ${\vec k}_i$ are the individual laboratory momenta.
In the notation (\ref{varphi0}) we dropped additional spin 
and isospin quantum numbers.

Let us now firstly stick to the 3N breakup channel,
thus $ \mid \phi_f \rangle = \mid \phi_0 \rangle $.

The Hamiltonian $H$ occurring in (\ref{Psif}) contains 
NN and 3N forces on top of
the kinetic energy $H_0$
\begin{equation}
H = H_0 + \sum\limits_{i<j} V_{ij} + V_{123} .
\label{H}
\end{equation}
One way to handle the 3N force operator $V_{123}$ is to split it into 3
parts
\begin{equation}
V_{123} = V^{(1)} + V^{(2)} + V^{(3)} ,
\label{V123}
\end{equation}
where $V^{(i)}$ is symmetrical under exchange of nucleons $j$ and $k$. 
Such a splitting  is
always possible. Thus it appears natural to combine the
 interactions  as
\begin{equation}
H = H_0
+ (V_{12} + V^{(3)})
+ (V_{23} + V^{(1)})
+ (V_{31} + V^{(2)})
\equiv  H_0 + \sum\limits_{i=1}^3 (V_{i} + V^{(i)}) .
\label{H.2}
\end{equation}
We introduced the standard and convenient notation 
$V_i \equiv V_{jk}, (j \ne i \ne k)$.
Clearly both terms $V_{i}$ and $V^{(i)}$
are symmetrical under exchange of nucleons $j$ and $k$.

Now using the well known  identity between the full resolvent
operator $G^{(-)}$ occurring in (\ref{Psif}) and the free resolvent operator
\begin{equation}
G_0^{(-)} \equiv \frac{1}{E - i \epsilon - H_0 } ,
\label{G0}
\end{equation}
  namely
\begin{equation}
G^{(-)} = G_0^{(-)} + G_0^{(-)} \sum\limits_{i=1}^3 (V_{i} + V^{(i)}) G^{(-)} ,
\label{G}
\end{equation}
one obtains the Lippmann-Schwinger 
equation for $\mid \Psi_0^{(-)} \rangle $   as
\begin{equation}
\mid \Psi_0^{(-)} \rangle = \mid \phi_0 \rangle
+ G_0^{(-)} \sum\limits_{i=1}^3 (V_{i} + V^{(i)}) \mid \Psi_0^{(-)} \rangle .
\label{Psi0}
\end{equation}
This suggests  a decomposition of the total state into three parts
and using again the identity of the three nucleons leads to \cite{ME3}
\begin{equation}
\mid \Psi_0^{(-)} \rangle = (1 + P) \mid \psi^{(-)} \rangle ,
\label{Psi0.2}
\end{equation}
where $\mid \psi^{(-)} \rangle $ obeys the Faddeev-like equation
\begin{equation}
\mid \psi^{(-)} \rangle = \mid \varphi_0^{(-)} \rangle
+ G_0^{(-)} t^{(-)} P \mid \psi^{(-)} \rangle +
\left(
1 + G_0^{(-)} t^{(-)}
\right) G_0^{(-)} V^{(1)} (1 + P) \mid \psi^{(-)} \rangle .
\label{psi}
\end{equation}
The driving term is
\begin{equation}
\mid \varphi_0^{(-)} \rangle
= \left( 1 + G_0^{(-)} t^{(-)} \right) \mid \varphi_0 \rangle
\equiv \mid {\vec p} \rangle_a^{(-)} \, \mid {\vec q} \rangle .
\label{psi.2}
\end{equation}
Thus in the 2N subsystem the antisymmetric free state $\mid {\vec p} \rangle_a$
is replaced by the two-body scattering state $\mid {\vec p} \rangle_a^{(-)} $.

The result for $ \mid \Psi_0^{(-)} \rangle $
can now be inserted into the nuclear matrix element (\ref{N}):
\begin{equation}
N = \langle \psi^{(-)} \mid (1 + P)  {\cal O} \mid \Psi_i \rangle =
\langle \varphi_0 \mid ( 1 + t G_0 ) 
( 1 - K )^{-1} ( 1 + P ) {\cal O} \mid \Psi_i \rangle ,
\label{N.2}
\end{equation}
where
\begin{equation}
K = P t G_0 + (1 + P )  V^{(1)} G_0 ( 1 + t G_0 )
\label{K}
\end{equation}
is the adjoint kernel to the one occurring in (\ref{psi}).

The heuristically derived result (\ref{U0.3}) valid for $ V^{(i)}= 0 $ can now 
be recovered easily.
We use the identity
\begin{equation}
 ( 1 + t G_0 ) ( 1 - P t G_0 )^{-1} = 1 + (1 + P) ( 1 - t G_0 P)^{-1} t G_0
\label{K.2}
\end{equation}
and obtain
\[
N = \langle \varphi_0 \mid ( 1 + P ) {\cal O} \mid \Psi_i \rangle +
    \langle \varphi_0 \mid ( 1 + P ) (1 - t G_0 P)^{-1}  
t G_0 (1 + P) {\cal O} \mid \Psi_i \rangle
\]
\begin{equation}
=
\langle \phi_0 \mid {\cal O} \mid \Psi_i \rangle  +  
\langle \phi_0 \mid U^{\, \prime} \rangle ,
\label{N.3}
\end{equation}
with $  \mid U^{\, \prime} \rangle $     given by the integral equation
\begin{equation}
 \mid U^{\, \prime} \rangle = t G_0 ( 1 + P ) {\cal O} \mid \Psi_i \rangle
+ t G_0 P  \mid U^{\, \prime} \rangle .
\label{Uprime}
\end{equation}

This has  to be compared to the result given
in (\ref{U0.3}) and (\ref{U}). Since 
${\cal O} \mid \Psi_i \rangle $ is antisymmetrical, we obtain
\begin{equation}
t G_0 ( 1 + P ) {\cal O} \mid \Psi_i \rangle 
= 3 t G_0 {\cal O} \mid \Psi_i \rangle  .
\label{symmetry}
\end{equation}
Consequently, $ \mid U^{\, \prime} \rangle = 3  \mid U \rangle $
and the second term in (\ref{N.3}) yields
$ \langle \phi_0 \mid U^{\, \prime} \rangle 
= 3  \langle \phi_0 \mid U \rangle $
which equals the second term in (\ref{U0.3}). This is 
obvious by applying $(1+P)$
to the antisymmetrical state $  \langle \phi_0 \mid $ on 
the left yielding again a
factor of 3. This completes the verification of the 
heuristically derived result.

Including now the 3NF we define according to the expression (\ref{N.2})
\begin{equation}
\mid {\tilde U}^{\, \prime} \rangle \equiv (1 - K)^{-1} 
(1 + P)  {\cal O} \mid \Psi_i \rangle ,
\label{Utildeprime}
\end{equation}
or the equivalent integral equation
\begin{equation}
\mid {\tilde U}^{\, \prime} \rangle =
 (1 + P)  {\cal O} \mid \Psi_i \rangle
+ \left( P t G_0 + (1 + P) V^{(1)} G_0  ( 1 + t G_0 )
\right)  \, \mid {\tilde U}^{\, \prime} \rangle .
\label{Utildeprime.2}
\end{equation}
The breakup matrix element
is determined by means of 
$ \mid {\tilde U}^{\, \prime} \rangle $
according to (\ref{psi}) - (\ref{N.2})
\begin{equation}
N = {}^{(-)}\langle \varphi_0 \mid {\tilde U}^{\, \prime} \rangle .
\label{newU}
\end{equation}
Unfortunately,
the form (\ref{Utildeprime.2}), although suitable for separable forces, 
 is not appropriate  for numerical applications with realistic interactions 
because of the presence of the permutation operator $P$ to 
the very left in the first part of the kernel\cite{Pproblem}. It would ``smear out'' the position of 
the deuteron singularity in the NN t-operator. 
To rewrite (\ref{Utildeprime.2})
into a suitable form we use the following obvious identities
\begin{equation}
 1 + P = \frac12 P (1 + P ) ,
\label{identity1}
\end{equation}
\begin{equation}
\frac12 P (P - 1) = 1 .
\label{identity2}
\end{equation}
Then we obtain from (\ref{Utildeprime.2})
\[
 (P - 1) \mid {\tilde U}^{\, \prime} \rangle  =
(P - 1) (1 + P)  {\cal O} \mid \Psi_i \rangle  \ +
\]
\begin{equation}
\left(
(P - 1) P t G_0  +
(P - 1) (1 + P ) V^{(1)} G_0  ( 1 + t G_0 ) \frac12 P (P - 1)
\right)
\mid {\tilde U}^{\, \prime} \rangle ,
\label{Pminus1}
\end{equation}
or with the definition
\begin{equation}
 (P - 1) \mid {\tilde U}^{\, \prime} \rangle  \equiv \mid {\tilde U} \rangle
\label{Utilde}
\end{equation}
the following equation for ${\tilde U}$ 
\begin{equation}
 \mid {\tilde U} \rangle  =
(1 + P)  {\cal O} \mid \Psi_i \rangle +
\left(
t G_0 P + \frac12 (P + 1) V^{(1)} G_0  ( 1 + t G_0 ) 
P \right) \mid {\tilde U} \rangle .
\label{Utilde.2}
\end{equation}
This integral equation is now suitable for numerical applications
and provides according to (\ref{identity2}) and (\ref{newU})
the nuclear matrix element
\begin{equation}
N = \frac12 \langle \varphi_0 \mid ( 1 + t G_0 ) P \mid {\tilde U} \rangle .
\label{N.4}
\end{equation}

In order to separate the contribution from the
plane wave alone ( $\langle \varphi_0 \mid $ )
and the symmetrized plane wave
($\langle \phi_0 \mid = \langle \varphi_0 \mid (1 + P ) $)
one can modify the driving term in (\ref{Utilde.2})
and solve the following equation for $ \mid \tilde{\tilde U} \rangle$
\[
 \mid \tilde{\tilde U} \rangle =
\left[
t G_0 + \frac12 (P + 1) V^{(1)} G_0  ( 1 + t G_0 )
\right] \, (1 + P)  {\cal O} \mid \Psi_i \rangle
\]
\begin{equation}
+ \
\left(
t G_0 P + \frac12 (P + 1) V^{(1)} G_0  ( 1 + t G_0 ) P \right)
\mid \tilde{\tilde U} \rangle .
\label{Utilde.3}
\end{equation}
With that auxiliary state $\mid \tilde{\tilde U} \rangle $
the amplitude $N$ reads now
\begin{equation}
N =
\langle \varphi_0 \mid ( 1 + t G_0 ) ( 1 + P )  {\cal O} \mid \Psi_i \rangle
+ \langle \varphi_0 \mid ( 1 + t G_0 ) P \mid \tilde{\tilde U} \rangle  .
\label{N.5}
\end{equation}
Dropping the second term and $ t G_0$ in the first term  in (\ref{N.5})
one encounters two plane wave impulse approximations
to the amplitude $N$
\begin{equation}
\label{N.PWIA}
N^{PWIA} \equiv \langle \varphi_0 \mid  {\cal O} \mid \Psi_i \rangle
\end{equation}
and
\begin{equation}
N^{PWIAS} \equiv \langle \varphi_0 \mid  ( 1 + P ) {\cal O} \mid \Psi_i \rangle .
\label{N.PWIAS}
\end{equation}
While in (\ref{N.PWIA}) the final state is antisymmetrized only in one pair, 
in (\ref{N.PWIAS}) it is fully antisymmetrized. 
The verification of (\ref{Utilde.3})
and (\ref{N.5}) requires straightforward algebra.

A completely alternative approach is based on two coupled Faddeev equations,
again starting from (\ref{Utildeprime.2}).
Defining 
\begin{eqnarray}
\mid U' \rangle &\equiv& t G_0 \mid \tilde{U}' \rangle , \label{met09} \\
\mid U^{\, \prime\prime}\rangle  &\equiv& V^{(1)} G_0 (t G_0 + 1) 
\mid \tilde{U}' \rangle , \label{met10}
\end{eqnarray}
and
\begin{equation}
\mid \chi \rangle \equiv (1+P) {\cal O} \mid \Psi_i \rangle ,
\label{newUUU}
\end{equation}
one obviously obtains the set of coupled equations for $U'$ 
and $U^{\, \prime\prime}$ 
\begin{eqnarray}
\mid U' \rangle &=& tG_0 \mid \chi \rangle +
tG_0P \mid U' \rangle + tG_0 (1+P) 
\mid U^{\, \prime\prime} \rangle \nonumber \\
\mid U^{\, \prime\prime} \rangle &=& V^{(1)}G_0(1+tG_0) \mid \chi \rangle
+ V^{(1)}G_0(1+tG_0)P \mid U' \rangle \label{met11a} \nonumber \\
&+& V^{(1)}G_0(1+tG_0) (1+P) \mid U^{\, \prime\prime} \rangle .
\label{met11}
\end{eqnarray}
These three states (\ref{met09}), (\ref{met10}) and (\ref{newUUU})
sum up by definition to
\begin{equation}
\mid \tilde{U}' \rangle = \mid \chi \rangle + P \mid {U}' \rangle
+ ( 1 + P ) \mid {U}'' \rangle ,
\label{newuuuuu}
\end{equation}
which determines according to (\ref{newU}) the breakup matrix element.
Inserting the definition of $ {}^{(-)}\langle \varphi_0 \mid$
and using (\ref{met11}) again, the breakup matrix element is easily turned
into the simpler form
\begin{equation}
N =   \langle \varphi_0 \mid  \left( \mid \chi \rangle
+ (1+P) (\mid U' \rangle + \mid U'' \rangle ) \right). 
\label{newuuuuuuuuuu}
\end{equation}

For the pd breakup of $^3$He, the final channel state regarded up to now
\begin{equation}
{}^{(-)} \langle \varphi_0 \mid = \langle \varphi_0 \mid ( 1 + t G_0 )
\equiv {}^{(-)}_{\ \ a}\langle {\vec p} \, {\vec q} \mid
\label{varphiminus}
\end{equation}
has simply to be replaced by
\begin{equation}
\langle \phi_q \mid \equiv \langle \varphi_d \mid \, \langle {\vec q} \mid .
\label{phiq}
\end{equation}
Thus the two-body scattering state 
 $ {}^{(-)}_{\ \ a}\langle {\vec p}  \mid $ 
 turns into the deuteron state
$ \langle \varphi_d \mid $
and the pd breakup matrix element is given as
\begin{equation}
N^{pd} = \frac12 \langle \phi_q \mid  P \mid {\tilde U} \rangle ,
\label{Npd}
\end{equation}
or
\begin{equation}
N^{pd} =  \langle \phi_q \mid ( 1 + P ) \, {\cal O} \mid \Psi_i \rangle
\ + \
\langle \phi_q \mid  P \mid \tilde{\tilde U} \rangle ,
\label{Npd2}
\end{equation}
if the auxiliary state $\mid \tilde{\tilde U} \rangle$ is employed.

If one uses the coupled set of equations, (\ref{met11}),
 the matrix element $N^{pd}$ will be 
\begin{equation}
N^{pd} = \langle \phi_q \mid {\tilde U}' \rangle
= \langle \phi_q \mid
\left( \mid \chi \rangle
+ P \mid U' \rangle + (1+P) \mid U'' \rangle \right) .
\label{new44444444}
\end{equation}

We refrain  to quote again the simpler
equations given in \cite{Ishikawa94,Ishikawa94.2,Golak95,Golak95.2,Ishikawa98}
valid for NN forces only. The more complex equations 
are necessary since for light nuclei~\cite{Carlson98,Annual}
and few-nucleon scattering processes \cite{ourreport} 3N forces are mandatory.
In the context of effective
field theory constrained  by chiral symmetry NN and three- and more-nucleon
forces are consistently linked to each other~\cite{vkolck94}.
Applications in that
framework to few-nucleon systems~\cite{Evgeni} definitely show that more than 
 pairwise forces are acting and are clearly 
visible in the measured values of the
observables (binding energies and scattering observables). This 
 new approach grounded on effective field theory backs up
the earlier results based on 
phenomenological forces which were 
 constrained only by the one-$\pi$ exchange, that
three-nucleon forces are necessary to describe the data.

The basic equations (\ref{Utilde.2}) or (\ref{met11}) are valid
for electron induced reactions and for real photon
induced processes as well. They only differ in the choice of the  
photon absorption operator ${\cal O}$ (see Section~\ref{currents}).

In the case of nucleon-deuteron capture one can use time reversal invariance
and evaluate the nuclear matrix element via Nd photodisintegration of
the 3N bound state as given in (\ref{Npd}), (\ref{Npd2}) 
or (\ref{new44444444}).
A more direct way is to choose
the matrix element in the form
\begin{equation}
N_{capture} = \langle \Psi \mid  {\cal O} \mid \Psi_i^{(+)} \rangle ,
\label{Ncapture}
\end{equation}
where $ \mid \Psi_i^{(+)} \rangle $ is the Nd scattering state 
with appropriately chosen initial
state quantum numbers $i$, ${\cal O}$ a suitable operator 
depending on the final
photon momentum, and $ \langle \Psi \mid $ the 3N bound state. Here 
we can use directly
the Faddeev equation for the 3N scattering state \cite{ME3}. 
It corresponds 
to (\ref{psi}) and for the initial Nd channel is given as
\begin{equation}
\mid \psi^{(+)} \rangle  = \mid \phi_i \rangle
+ G_0 t P \mid \psi^{(+)} \rangle
+ ( 1 + t G_0 ) G_0 V^{(1)} (1 + P) \mid \psi^{(+)} \rangle .
\label{psiplus}
\end{equation}
Here $ \mid \phi_i \rangle \equiv  \mid \phi_q \rangle $ with 
appropriate initial spin quantum numbers. 
The  total scattering state is then
\begin{equation}
\mid \Psi_i^{(+)} \rangle  = (1 + P) \mid \psi^{(+)} \rangle .
\label{Psiplus2}
\end{equation}

Let us define the amplitude $\mid T \rangle $ by
\begin{equation}
\mid \psi^{(+)} \rangle = \mid \phi_i \rangle + G_0 \mid T \rangle ,
\label{T}
\end{equation}
where $\mid T \rangle $ obeys the Faddeev-like equation
\begin{equation}
\mid T \rangle = t P \mid \phi_i  \rangle  +
( 1 + t G_0 ) V^{(1)} (1 + P) \mid \phi_i \rangle +
 t P G_0 \mid T \rangle + ( 1 + t G_0 ) V^{(1)} (1 + P) \mid T \rangle .
\label{T.2}
\end{equation}

It is this central equation (\ref{T.2}) 
which we solve for 3N scattering \cite{ME3}.
Consequently the nuclear matrix element for Nd capture is obtained in the form
\begin{equation}
N_{capture} = \langle \Psi \mid  {\cal O} (1 + P) \mid \phi_i \rangle +
              \langle \Psi \mid  {\cal O} (1 + P) G_0 \mid T \rangle .
\label{Ncapture.2}
\end{equation}

\section{Current operators}
\label{currents}

While the treatment of the interacting nucleons in the 3N bound and
scattering states is quite well established  
in the framework of the nonrelativistic Schr\"odinger equation,  
for the current operator,
there is still quite some room for improvements. The current operator is a
dynamical object containing in addition to a single nucleon term also 
 the two- and three-body contributions, which are
as complex as nuclear forces themselves. First considerations
can be found in~\cite{first-current.1}
and ~\cite{first-current.2}.
A very nice discussion and review is given in~\cite{Carlson98}. 
Earlier reviews for instance are \cite{riskamec} and \cite{Mathiot89}.
Since our review does not focus on this issue,
we will only briefly describe what underlies our applications.

One approach to include some of the many-body terms in the current, applied 
in the case of photodisintegration (or Nd capture), is based on the
old Siegert idea \cite{Siegert}. The way we use it is described
in section \ref{ourSiegert}. The other
approach used for virtual and real photons is to link a certain subset
of currents via the continuity equation to the NN force AV18, which
has been phrased  ``model independent''  in \cite{Riska1,Riska2}. 
This is briefly reviewed
in section \ref{ourMEC}. 
For additional currents not constrained by the continuity equation we refer 
the reader to \cite{Carlson98}. 
We start with the single nucleon current in section \ref{singlecurrent}. 

\subsection{The single nucleon current}
\label{singlecurrent}

We work in the Hamiltonian formalism and therefore the nucleons
are on the mass shell. The standard single nucleon current at space-time 
point zero $j^\mu_{SN} (0)$ expressed
in terms of the nucleon four momentum 
$ p \equiv (p_0 = \sqrt{ m_N^2 + {\vec p}^{\, 2} }, {\vec p}) $
is
\begin{equation}
j^\mu_{SN} (0) = \bar{u} ( {\vec p}^{~\, \prime} s') ( \gamma^\mu F_1
+ i \sigma^{\mu \nu} ( p' - p)_\nu  F_2 ) u ({\vec p} s )
=
\bar{u} ( {\vec p}^{~\, \prime} s')
( G_M \gamma^\mu - F_2  ( p' + p)^\mu ) u ({\vec p} s ) .
\label{sin.1}
\end{equation}
Here $u$ are Dirac spinors, $F_1( ( p' - p)^2 )$ and $F_2( ( p' - p)^2 )$
 the Dirac and Pauli nucleon form factors, and $G_M \equiv F_1 + 2 m_N F_2 $
the magnetic form factor of the nucleon. That fully relativistic form
can be expressed as a four component $ 2 \times 2 $ matrix operator
$J^\mu ( p' , p) $ acting on Pauli spinors $\xi$:
\begin{equation}
j^\mu_{SN} (0) = \xi^\dagger (s ') J^\mu ( p' , p) \xi (s ) .
\label{jmu0}
\end{equation}
With
\begin{equation}
A=
\sqrt{{ m_N \over p_0}} \sqrt{{ m_N \over {p_0}'}}
\sqrt{{ {{p_0}' + m_N} \over {2 m_N} }} \sqrt{{ {{p_0} + m_N} \over {2 m_N} }}
\label{sin.2}
\end{equation}
the components $J^\mu ( p' , p) $ are written as
\begin{eqnarray}
J^0 =
A \left\{
\left[ G_M - F_2 (p + p')^0 \right] +
\left[ G_M + F_2 (p + p')^0 \right] { {{\vec p}^{\, \prime} \cdot {\vec p}} \over { ({p_0} + m_N) ({p_0}' + m_N)} }
\right\} \cr
+
A \left[ G_M + F_2 (p + p')^0 \right]
{ { i {\vec \sigma} \cdot ({\vec p}^{\, \prime} \times {\vec p}) } \over { ({p_0} + m_N) ({p_0}' + m_N)} }
\label{sin.3}
\end{eqnarray}
and
\begin{eqnarray}
J^k =
 - A F_2 \left(  1 -
{ {{\vec p}^{\, \prime} \cdot {\vec p}} \over { ({p_0} + m_N) ({p_0}' + m_N)} }
\right) \,  (p + p')^k \cr
+ A G_M  \left( { p^k \over {{p_0} + m_N} } + { {p'}^k \over {{p_0}' + m_N} } \right) \cr
+ A F_2 { { (p + p')^k  } \over { ({p_0} + m_N) ({p_0}' + m_N)} } \,
i {\vec \sigma} \cdot ({\vec p}^{\, \prime} \times {\vec p}) \cr
+ A G_M \left[ { 1 \over {({p_0} + m_N)} } \, 
i ( {\vec p} \times {\vec \sigma} )^k
             + { 1 \over {({p_0}' + m_N)} } \, 
i ( {\vec \sigma} \times {\vec p}^{\, \prime} )^k
        \right] .
\label{sin.4}
\end{eqnarray}
This form will be used in section \ref{data}. 

In the bulk of this review we apply only the nonrelativistic limit
leading to the simple forms for
the convection and spin current components:
\begin{equation}
{\vec J} = F_1 \frac { {\vec p} + {\vec p}^{\, \prime} }{2 m_N}
+ \frac{i}{2 m_N} \,
G_M \, {\vec \sigma} \times (  {\vec p}^{\, \prime} - {\vec p} ) .
\label{sin.6}
\end{equation}
The nucleon form factors $F_1$ and $F_2$, and thus $G_M$, 
are normalized for neutron and proton as
\begin{equation}
F_1^n (0) = 0
\label{sin.7}
\end{equation}
\begin{equation}
F_1^p (0) = 1
\label{sin.8}
\end{equation}
\begin{equation}
F_2^n (0) = -1.913 \, \frac1{2 m_N}  = G_M^n (0) \, \frac1{2 m_N}
\label{sin.9}
\end{equation}
\begin{equation}
F_2^p (0) = 1.793 \, \frac1{2 m_N}
= G_M^p (0) \, \frac1{2 m_N} -  \frac1{2 m_N} .
\label{sin.10}
\end{equation}
In the case of the density component 
the leading term in the nonrelativistic limit
\begin{equation}
J^0 = F_1
\label{sin.5}
\end{equation}
is very small for the neutron and therefore one generally adds
the next order relativistic corrections, which are of
the form~\cite{Friar73,deForest66}
\begin{equation}
J^0 = G_E \left( 1 - \frac{{\vec Q}^{\, 2}}{8 m_N^2} \right)
 + i \left( 2 G_M - G_E \right) \, 
\frac { {\vec \sigma} \cdot {\vec p}^{\ \prime} \times {\vec p} }{4 m_N^2} ,
\label{sin.55}
\end{equation}
with the electric form factor
\begin{equation}
G_E \equiv F_1 + \frac {Q^2}{2 m_N} F_2  
\approx F_1 - \frac {{\vec Q}^{\, 2}}{2 m_N} F_2 .
\label{sin.555}
\end{equation}


Due to formal reasons we use that form
also for the proton. Here $Q = (Q_0, \vec Q) \equiv (\omega, \vec Q) $ 
is the real or virtual photon four-momentum and $\vec Q \equiv  \vec p~' - \vec p$. 

In the case of the convection current in (\ref{sin.6}) some 
 authors \cite{Carlson98}
replace $F_1$ by $G_E$ which adds some (not all) relativistic corrections of 
$ O \left( (p/m_N)^2 \right) $ on top of the leading order
going with $F_1$. 
 Once, however, $G_E$ is chosen for the density, then of course $G_E$ should 
also be used for the convection current due to current conservation. 

The choice which underlies our nonrelativistic calculation here is $G_E$ 
for the density in lowest order and $G_E$ in the convection current instead 
of $F_1$ shown in (\ref{sin.6}). The spin current with $G_M$ is used as in 
 (\ref{sin.6}). 

\begin{figure}[!ht]
\begin{center}
\epsfig{file=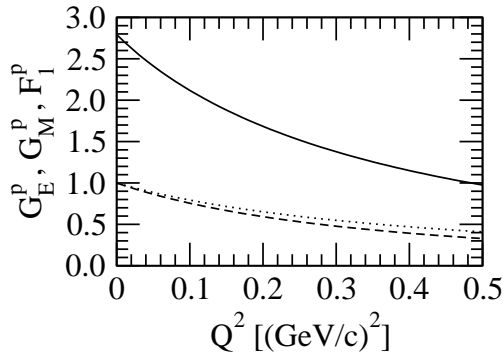,bb=10 515 300 715,clip=true,height=5cm}
\caption{The electromagnetic proton form factors
$G_E^p$ (dashed line) and $G_M^p$ (solid line)
from \cite{HammerMeissner} used in our calculations for this review
as a function of the four-momentum transfer squared $Q^2$.
For the orientation of the reader $F_1^p$ (dotted line) is also shown.}
\label{hmd3_p}
\end{center}
\end{figure}

\begin{figure}[!ht]
\begin{center}
\epsfig{file=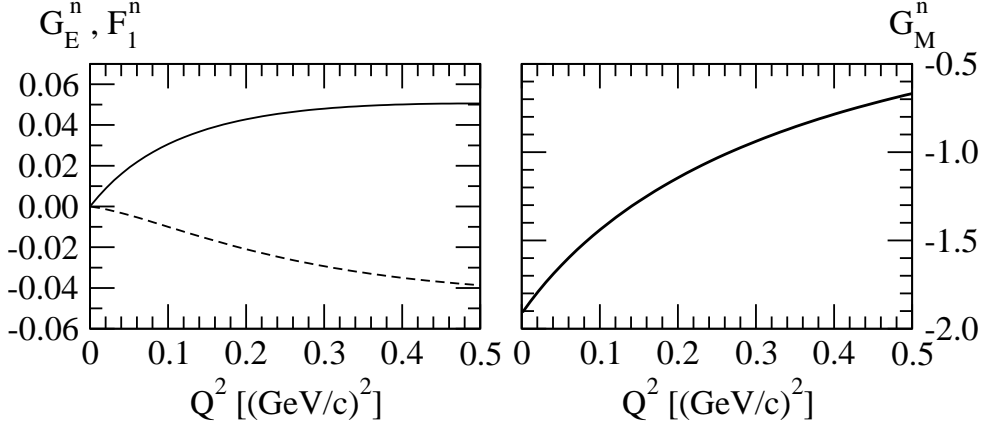,bb=50 515 550 740,clip=true,height=6cm}
\caption{The electromagnetic neutron form factors
$G_E^n$ and $G_M^n$ (solid lines)
from \cite{HammerMeissner} used in our calculations for this review
as functions of the four-momentum transfer squared $Q^2$.
For the orientation of the reader $F_1^n$ (dashed line) is also shown.}
\label{hmd3_n}
\end{center}
\end{figure}

For the convenience of the reader we display the functional
forms of the various nucleon form factors
restricted to our momentum range in Figs.~\ref{hmd3_p}--\ref{hmd3_n}. We 
show theoretical predictions based on a dispersion theoretical analysis constrained by data 
\cite{HammerMeissner}. Recent reviews on nucleon form factors can be found in 
\cite{Jourdan03}. 

In the nonrelativistic regime we choose for virtual photons $\vec Q ^2$ 
to be the argument 
of the nucleon electromagnetic form factors. 
In case of real photons we put 
this argument to be zero since for our momentum range $\vec Q ^2$ is anyway 
very small. 

\subsection{The Siegert approach}
\label{ourSiegert}

Let $ {\vec \epsilon}_\xi ({\vec Q}) $ be the spherical component
of the photon polarization vector
for a photon with three-momentum 
${\vec Q}$ and  ${\vec j} (0)$ the nuclear current
operator at space-time point zero. Then the nuclear matrix element
for photodisintegration is written as
\begin{equation}
N_\xi ({\vec Q}) = \langle {\vec P}^{\, \prime} \, \Psi_f^{(-)} \mid
{\vec \epsilon}_\xi ({\vec Q}) \cdot {\vec j} (0) \mid \Psi_i {\vec P}
\, \rangle
\equiv
{\vec \epsilon}_\xi ({\vec Q}) \cdot {\vec I} ({\vec Q}) .
\label{Nxi}
\end{equation}

As before $ \mid \Psi_i \rangle $
and $ \langle \Psi_f^{(-)} \mid $
are the internal 3N bound and scattering states  and
we added the dependence on the total initial and final 3N  momenta.
Clearly ${\vec P}^{\, \prime} = {\vec P} + {\vec Q}$
as expressed in the overall $\delta$-function of four-momentum
conservation. This  $\delta$-function is taken care of in the evaluation
of the observables.
Each component of the pure nuclear matrix element $ {\vec I} ({\vec Q})$ can be expanded
into spherical harmonics
\begin{equation}
I_k ({\vec Q}) = \sum\limits_{l m}  Y_{l m}^* ({\hat Q})
\int d {\hat Q}^{\, \prime} \, Y_{l m} ({\hat Q}^{\, \prime})
I_k ( \mid {\vec Q} \mid {\hat Q}^{\, \prime})  .
\label{expansion}
\end{equation}
Here and throughout the paper the ``hat'' notation
stands sometimes for a unit vector.
Further the polarization vector $ {\vec \epsilon}_\xi ({\vec Q}) $
for a photon with  momentum in ${\hat Q}$-direction is related
to the photon polarization vector $ {\vec \epsilon}_\xi ({\hat z}) $
for a photon with momentum in z-direction by a rotation.
Since $ {\vec \epsilon}_\xi $    is
a rank one object, one has
\begin{equation}
 {\vec \epsilon}_\xi ({\hat Q}) =
\sum\limits_{\xi^{\prime}} D_{\xi^{\prime} \xi}^1  ({\hat Q})
{\vec \epsilon}_{\xi^{\prime}} ({\hat z})
\label{rank1}
\end{equation}
and using the Clebsch-Gordan coefficients we combine ${\vec \epsilon}_{\xi^{\prime}} ({\hat z}) $
with $Y_{l m} ({\hat Q}^{\, \prime}) $ 
to vector spherical harmonics~\cite{Edmonds}
\begin{equation}
{\vec \epsilon}_{\xi^{\prime}} ({\hat z}) Y_{l m} ({\hat Q}^{\, \prime})
\equiv
\sum\limits_{J \ge 1} C (l 1 J ; 0, \xi^{\prime}, \xi^{\prime})
{\vec Y}_{l 1 J}^{\xi^{\prime}} ({\hat Q}^{\, \prime})  .
\label{vectorY}
\end{equation}

Thus altogether we obtain
\begin{equation}
N_\xi ({\vec Q}) =
\sum\limits_{l m}  Y_{l m}^* ({\hat Q})
\
\sum\limits_{\xi^{\prime}} D_{\xi^{\prime} \xi}^1  ({\hat Q})
\
\int d {\hat Q}^{\, \prime} \,
\sum\limits_{J \ge 1} \,
C (l 1 J ; 0, \xi^{\prime}, \xi^{\prime})
{\vec Y}_{l 1 J}^{\xi^{\prime}} ({\hat Q}^{\, \prime}) \cdot
{\vec I} ( \mid {\vec Q} \mid {\hat Q}^{\, \prime}) .
\label{Nxi2}
\end{equation}
Now we can use~\cite{Rose}
\begin{equation}
Y_{l m}^* ({\hat Q})
= \sqrt{ \frac{ 2 l + 1 }{4 \pi} } \, D_{ m 0}^l ({\hat Q}) ,
\label{Ylmstar}
\end{equation}
as well as
\begin{equation}
\sum\limits_m C (l_1 l_2 L ; m, M-m, M) \, 
D_{m \, m_1}^{l_1} \, D_{M-m \, m_2}^{l_2}
\ = \
C (l_1 l_2 L ; m_1, m_2 , m_1 + m_2) \, D_{M \, m_1+m_2}^{L} ,
\label{sumCG}
\end{equation}
and obtain after rearranging the summation over the magnetic quantum numbers
\begin{equation}
N_\xi ({\vec Q}) =
\sum\limits_{J \ge 1} \,
\sum\limits_{M=-J}^J \,
D_{M \xi}^J  ({\hat Q}) \,
\sum\limits_{l=J,J \pm 1}
\sqrt{ \frac{ 2 l + 1 }{4 \pi} } \,
C (l 1 J ; 0, \xi, \xi) \,
\int d {\hat Q}^{\, \prime} \,
{\vec Y}_{l 1 J}^M ({\hat Q}^{\, \prime}) \cdot
{\vec I} ( \mid {\vec Q} \mid {\hat Q}^{\, \prime}) .
\label{Nxi3}
\end{equation}
This nicely shows the dependence on the photon direction together
with projections of the pure nuclear matrix element into the vector
spherical harmonics. The  latter ones are conventionally called
the electric and magnetic multipole elements. Inserting the
Clebsch-Gordan coefficients one defines
\begin{equation}
T^{el}_{J M} ( \mid {\vec Q} \mid) \equiv
-\frac1{4 \pi} \int d {\hat Q}^{\, \prime} \,
\left\{
\sqrt{\frac{J}{2 J + 1}}
{\vec Y}_{J+1 \, 1 J}^{M} ({\hat Q}^{\, \prime}) +
\sqrt{\frac{J+1}{2 J + 1}}
{\vec Y}_{J-1 \, 1 J}^{M} ({\hat Q}^{\, \prime})
\right\} \cdot
{\vec I} ( \mid {\vec Q} \mid {\hat Q}^{\, \prime}) ,
\label{Tel}
\end{equation}
and
\begin{equation}
T^{mag}_{J M} ( \mid {\vec Q} \mid) \equiv
\frac1{4 \pi} \int d {\hat Q}^{\, \prime} \,
{\vec Y}_{J \, 1 J}^{M} ({\hat Q}^{\, \prime})
\cdot
{\vec I} ( \mid {\vec Q} \mid {\hat Q}^{\, \prime})  .
\label{Tmag}
\end{equation}

This leads then to
\begin{equation}
N_\xi ({\vec Q}) =
-\sqrt{ 2 \pi}
\sum\limits_{J \ge 1} \,
\sqrt{2 J +1}
\sum\limits_{M=-J}^J \,
D_{M \xi}^J  ({\hat Q}) \,
\left\{
\pm T^{mag}_{J M} ( \mid {\vec Q} \mid)  + T^{el}_{J M} ( \mid {\vec Q} \mid)
\right\} ,
\label{Nxi4}
\end{equation}
where $( \pm ) $ refers to $ \xi = \pm 1 $.

Now the identity \cite{Edmonds}
\begin{equation}
{\hat Q} \, Y_{J M} ({\hat Q}) \ = \
- \sqrt{\frac{J+1}{2 J +1}} \, {\vec Y}_{J+1 \, 1 J}^M ({\hat Q})
\ + \
\sqrt{\frac{J}{2 J +1}} \, {\vec Y}_{J-1 \, 1 J}^M ({\hat Q})
\label{identityedmonds}
\end{equation}
applied in (\ref{Tel}) allows us to express
$ {\vec Y}_{J-1 \, 1 J}^M ({\hat Q}) $ in terms of a vector
spherical harmonics with an orbital part larger by 2 and 
 by ${\hat Q} \, Y_{J M} ({\hat Q})$.
This leads to the expression 
$ {\hat Q} \cdot {\vec I} ({\vec Q})$ which occurs in
the continuity equation for the electromagnetic current matrix element:
\begin{equation}
{\vec Q} \cdot {\vec I} ({\vec Q}) =
\langle {\vec P}^{\, \prime} \, \Psi_f^{(-)} \mid
\left[
H, {\hat \rho} (0)
\right] \mid \Psi_i \, {\vec P} \rangle \ = \
(E_f - E_i )
\langle {\vec P}^{\, \prime} \, \Psi_f^{(-)} \mid
{\hat \rho} (0)
\mid \Psi_i \, {\vec P} \rangle .
\label{continuity}
\end{equation}
Here ${\hat \rho} (0) \equiv j^0(0)$ is the density operator.  
Because of energy conservation, $E_f - E_i = \omega$,
the photon energy $\omega = \mid {\vec Q} \mid $.
Thus with
\begin{equation}
\rho ({\vec Q}) \equiv
\langle {\vec P}^{\, \prime} \, \Psi_f^{(-)} \mid
{\hat \rho} (0)
\mid \Psi_i \, {\vec P} \rangle ,
\label{rho}
\end{equation}
we end up with the electric multipole term in the form
\begin{eqnarray}
T^{el}_{J M} ( \mid {\vec Q} \mid) =
-\frac1{4 \pi} \int d {\hat Q}^{\, \prime} \,
\left\{
\sqrt{\frac{2 J +1}{J}}
{\vec Y}_{J+1 \, 1 J}^M ({\hat Q}^{\, \prime})
\cdot
{\vec I} ( \mid {\vec Q} \mid {\hat Q}^{\, \prime})
\right. \nonumber \\
\left. \
+
\sqrt{\frac{J+1}{J}}
Y_{J M} ({\hat Q}^{\, \prime})
\rho ( \mid {\vec Q} \mid {\hat Q}^{\, \prime})
\right\} .
\label{Tel3}
\end{eqnarray}

Summarizing, one finds that by using the identity among vector spherical
harmonics (\ref{identityedmonds})
together with the continuity equation it is possible to replace a 
part of the current matrix element
$ {\vec I} ({\vec Q}) $ by the density matrix element
$ \rho ({\vec Q}) $
and higher multipole contributions. Because the matrix element
$ \rho ({\vec Q}) $
receives two-body contributions only at a higher order in a $p/m$
expansion than the current matrix element \cite{Carlson98}, one
can expect that 
the form (\ref{Tel3}) for the electric multipole contribution
is a better approximation than (\ref{Tel}) even when 
  only the single nucleon density operator is used.
The higher multipole term in (\ref{Tel3}) 
is usually neglected in the literature, but not in our applications. 
We also do not approximate (\ref{Tel3}) in a long wave length limit. 
We take zero as arguments in the nucleon electromagnetic form factors
for all processes with real photons.
In the nonrelativistic framework one should rather take the photon three-momentum
squared. The results based on these two approaches are practically
indistinguishable for low energies but for $\omega$=140 MeV/c  
they lead to differences in the cross section up to about 8\%.
For polarization observables these changes are much smaller (about 1\%).

\subsection{$\pi$ and $\rho$-like meson exchange currents}
\label{ourMEC}

The seminal papers~\cite{Chemtob}
introduced $\pi$- and heavy-meson exchange current (MEC) operators
satisfying the continuity equation with meson-exchange interactions.
But of course they violate the continuity equation in relation
to phenomenological high precision NN forces like AV18, which we employ.
Thus we follow a recipe adapted to phenomenological NN forces.
For the sake of completeness we briefly sketch the derivation
formulated in \cite{Riska1,Riska2,MEC}.
Equivalent  proposals have been given in \cite{MEC2}
and \cite{MEC3}.

High accuracy NN forces like AV18 are not formed
in a pure meson exchange picture but, except for the long range
one-$\pi$ exchange, they contain a phenomenological structure dependent
on a number of parameters. Nevertheless the spin-isospin structure occurring
in a proper one-$\pi$ and one-$\rho$ exchange is present.
For the isovector exchanges
this is the $ {\vec \tau}(1)\cdot {\vec \tau}(2) $ isospin operator.
Also there occur the spin-spin and the  tensor operators.
In addition for the $\rho$-exchange there is a pure central term.
Therefore, this part of a local NN force reads in momentum space
\begin{equation}
V = \sum\limits_{t=0}^1 v_{SS}^t (k) \tilde{\Omega}_{SS} ({\vec k}) \, P_t
  + \sum\limits_{t=0}^1 v_{T}^t (k) \tilde{\Omega}_{T} ({\vec k}) \, P_t \, 
  + \sum\limits_{t=0}^1 v_{C}^t (k) \tilde{\Omega}_{C} \, P_t ,
\label{Vnn}
\end{equation}
with the isospin projection operators
$P_{t=0} = \frac14 ( 1 - {\vec \tau}(1) \cdot {\vec \tau}(2) ) $ and
$P_{t=1} = \frac14 ( 3 + {\vec \tau}(1) \cdot {\vec \tau}(2) ) $, 
and the spin operators
\begin{equation}
\tilde{\Omega}_{SS} ({\vec k}) = {\vec k}^{\, 2} \,  
{\vec \sigma}(1)\cdot {\vec \sigma}(2) ,
\label{Omegass}
\end{equation}
\begin{equation}
\tilde{\Omega}_{T} ({\vec k}) =  {\vec k}^{\, 2} 
\, {\vec \sigma}(1) \cdot {\vec \sigma}(2)  \ - \
3 {\vec \sigma}(1)\cdot {\vec k} \, {\vec \sigma}(2) \cdot {\vec k} , \
\label{Omegat}
\end{equation}
\begin{equation}
\tilde{\Omega}_{C} = 1 .
\label{Omegac}
\end{equation}
In (\ref{Vnn}) $  v_{SS}^t (k)$, $  v_{T}^t (k)$  and 
 the central piece $  v_{C}^t (k)$
are radial functions depending on
$
\mid \vec k \mid \equiv \mid {\vec p}^{\, \prime} -  {\vec p} \mid
$, 
where $ {\vec p}^{\, \prime} $ and $ {\vec p} $ are the final and initial
relative two-nucleon momenta.

Separating the term with $ {\vec \tau}(1)\cdot {\vec \tau}(2) $ one obtains
\begin{equation}
V \rightarrow v_1{\vec \tau}(1)\cdot {\vec \tau}(2) \equiv
 {\vec \tau}(1)\cdot {\vec \tau}(2)
\left(
 v_{SS} (k) \tilde{\Omega}_{SS} ({\vec k})  \ + \
 v_{T} (k) \tilde{\Omega}_{T} ({\vec k}) \ + \
 v_{C} (k) \tilde{\Omega}_{C} 
\right) ,
\label{termt1t2}
\end{equation}
with
\begin{equation}
 v_{SS} (k) \equiv \frac14 \left(  v_{SS}^{t=1} (k) -  v_{SS}^{t=0} (k) 
\right) ,
\label{vss}
\end{equation}
\begin{equation}
 v_{T} (k) \equiv \frac14 \left(  v_{T}^{t=1} (k) -  v_{T}^{t=0} (k) \right) ,
\label{vt}
\end{equation}
\begin{equation}
 v_{C} (k) \equiv \frac14 \left(  v_{C}^{t=1} (k) -  v_{C}^{t=0} (k) \right) .
\label{vc}
\end{equation}

Now one assumes that $v_{SS}^t (k) $  and  $v_{T}^t (k) $  
are built up by the sum
of $\pi$- and $\rho$-like parts
\begin{equation}
 v_{SS}^t \equiv  v_{SS}^{\pi,t} + v_{SS}^{\rho,t} ,
\label{vss2}
\end{equation}
\begin{equation}
 v_{T}^t \equiv  v_{T}^{\pi,t} + v_{T}^{\rho,t} , 
\label{vt2}
\end{equation}
and that these parts obey the same relations which are valid
for the true one-$\pi$ and one-$\rho$ exchange terms
\begin{equation}
v_{SS}^{\pi,t} = - v_{T}^{\pi,t} ,
\label{rel1}
\end{equation}
\begin{equation}
v_{SS}^{\rho,t} = 2 v_{T}^{\rho,t} .
\label{rel2}
\end{equation}
All that taken together allows now to solve for the individual
$\pi$-         and $\rho$-like parts contained in the potential $V$
 in terms of $ v_{SS}^{t}$ and $ v_{T}^{t} $.
According to (\ref{vss})--(\ref{rel2})  one obtains
\begin{eqnarray}
 v_{SS}^{t=1} -  v_{SS}^{t=0} +  v_{T}^{t=1} -  v_{T}^{t=0} \ = \
\underbrace{\left(
 v_{SS}^{\pi,t=1} +  v_{T}^{\pi,t=1}
\right)}_{=0}
\ - \
\underbrace{\left(
 v_{SS}^{\pi,t=0} +  v_{T}^{\pi,t=0}
\right)}_{=0} \nonumber \\
\ + \
\left(
 v_{SS}^{\rho,t=1} +  v_{T}^{\rho,t=1}
\right)
\ - \
\left(
 v_{SS}^{\rho,t=0} +  v_{T}^{\rho,t=0}
\right)
\ = \
\frac32 \,
\left(
 v_{SS}^{\rho,t=1} -  v_{SS}^{\rho,t=0}
\right) .
\label{relation}
\end{eqnarray}
Consequently, the $\rho$-like term of the potential 
in the form (\ref{termt1t2})
is determined via the two functions
$ v_{SS}^{\rho} (k) $ and $ v_{T}^{\rho} (k) $ given by
\begin{eqnarray}
v_{SS}^{\rho} (k) \equiv
\frac14 \, \left(  v_{SS}^{\rho,t=1} -  v_{SS}^{\rho,t=0} \right) \
= \ \frac16 \, \left(  v_{SS}^{t=1} +  v_{T}^{t=1} - v_{SS}^{t=0} 
-  v_{T}^{t=0} \right)
\ = \ 2 \, v_{T}^{\rho} (k) ,
\label{vss3}
\end{eqnarray}
and the $\pi$-like part of the potential in the form  (\ref{termt1t2})
is determined by the two functions $v_{SS}^{\pi}$ and $v_{T}^{\pi}$ given by
\begin{eqnarray}
v_{SS}^{\pi} (k) =
 v_{SS} (k) -  v_{SS}^{\rho} (k) ,
\label{vss4}
\end{eqnarray}
\begin{eqnarray}
v_{T}^{\pi} (k) = v_{T} (k) -  v_{T}^{\rho} (k) .
\label{vss5}
\end{eqnarray}
Also one assumes that $ v_{C} (k) $ is a $\rho$-like object
\begin{equation}
v_{C}^{\rho} (k) = v_{C} (k). 
\label{vc2}
\end{equation}
In this manner the isospin dependent part (\ref{termt1t2})
of the general potential (\ref{Vnn}) is separated into two parts,
a $\pi$-like and $\rho$-like terms.

Now let us regard the continuity equation in the lowest nontrivial
order of a $p/m$ expansion
\begin{equation}
\left[
V , j_{SN}^0 (0)
\right] = \left[ {\vec P} , {\vec j}^{\, exch} (0) \right] .
\label{cntn2}
\end{equation}
Here $V$ is a nonrelativistic NN force, like the lowest order one-$\pi$
or one-$\rho$ exchange potentials, $j_{SN}^0 (0) $ is the single
nucleon density operator taken at the space-time point $0$, 
 ${\vec j}^{\, exch} (0)$ the related 
 exchange current operator, and $\vec P$ the 
total two-nucleon momentum operator. 
 Neglecting in (\ref{sin.55})  
all terms except the first one the matrix element of   $j_{SN}^0 (0) $ is
\begin{equation}
\langle {\vec p}_i^{\, \prime} \mid j_{SN}^0 (0) \mid  {\vec p}_i \rangle \ = \
G_E^p \left( {\vec p}_i^{\, \prime} -  {\vec p}_i \right) \, \Pi^p  \ + \
G_E^n \left( {\vec p}_i^{\, \prime} -  {\vec p}_i \right) \, \Pi^n ,
\label{j0}
\end{equation}
with $G_E^{p,n}$  and $\Pi^{p,n}$ the nucleon Sachs form
factors and the projection
operators  for the proton and neutron, respectively.  
Then the equation (\ref{cntn2})
is easily worked out in momentum space with the result
\begin{eqnarray}
{\vec Q}  &\cdot& \langle \vec p_1~' \vec p_2~' \vert 
  {\vec j}^{~exch} (0)  \vert   \vec p_1 \vec p_2   \rangle
 =   \cr 
  (~  
 &V& ( {\vec p}^{~\prime} , {{\vec Q}\over{2}} + {\vec p} ) -
V ( {\vec p}^{~\prime} -  {{\vec Q}\over{2}} , {\vec p} )  ~) 
~ (~
G_E^p \left( {\vec Q} \right) \, \Pi^p (1)
 + G_E^n \left( {\vec Q} \right) \, \Pi^n (1)
 ~) \cr
 + (~  
&V& ( {\vec p}^{~\prime} , -{{\vec Q}\over{2}} + {\vec p} ) -
V ( {\vec p}^{~\prime} +  {{\vec Q}\over{2}} , {\vec p} ) ~)
 ~ (~
G_E^p ( {\vec Q} ) \, \Pi^p (2)
 + G_E^n ( {\vec Q} ) \, \Pi^n (2)
 ~)  \cr
 +  ~&i&~ 
 (~
 v_1 ( {\vec p}^{~\prime} -{{\vec Q}\over{2}} , {\vec p} ) - 
v_1 ( {\vec p}^{~\prime} +  {{\vec Q}\over{2}} , {\vec p} ) ~) 
 (~ G_E^p( {\vec Q} ) - G_E^n( {\vec Q} ) ~) 
 (~ {\vec \tau}(1) \times {\vec \tau}(2) ~)_3 . 
\label{cnt3}
\end{eqnarray}
For $V$ we assumed the form
\begin{equation}
V (  {\vec p}^{\  \prime} , {\vec p} )
= v_0 (  {\vec p}^{\  \prime} , {\vec p} )
+ v_1 (  {\vec p}^{\  \prime} , {\vec p} ) \,  
{\vec \tau}(1) \cdot {\vec \tau}(2) ,
\label{Vparts}
\end{equation}
in terms of two-nucleon relative momenta. Further,  
 having photon absorption in mind, $ {\vec Q} 
\equiv {\vec P}^{\  \prime} - {\vec P} $
is the photon momentum.

For a purely local potential the first two terms on the right-hand-side of (\ref{cnt3}) vanish. For a pure
one-$\pi$ exchange potential
\begin{equation}
V_\pi = v_\pi (k) {\vec \sigma}(1)\cdot{\vec k} \, 
{\vec \sigma}(2)\cdot{\vec k}
\, {\vec \tau}(1) \cdot {\vec \tau}(2) ,
\label{v1pi}
\end{equation}
with
\begin{equation}
v_\pi (k) = - \frac{f_{\pi N N}^2}{m_\pi^2} \, \frac1{m_\pi^2 + k^2} ,
\label{vpik}
\end{equation}
a simple algebra employing (\ref{cnt3}) leads to the well
known pure pionic exchange current
\cite{Riskaprogress}
\begin{eqnarray}
{\vec j}_\pi^{exch} ( {\vec k}_1 , {\vec k}_2 )
= i \,
\left( G_E^p ( {\vec Q} ) - G_E^n ( {\vec Q} ) \right)
\,
\left( {\vec \tau}(1) \times {\vec \tau}(2) \right)_3  \nonumber \\
\left(
{\vec \sigma}(2) {\vec \sigma}(1) \cdot {\vec k}_1 v_\pi ( k_1 )
\ - \
{\vec \sigma}(1) {\vec \sigma}(2) \cdot {\vec k}_2 v_\pi ( k_2 )  \right.
\nonumber \\
\left.
+ \
\frac{{\vec k}_1 -  {\vec k}_2}{ k_1^2 - k_2^2 } \,
(v_\pi (k_2) - v_\pi (k_1) ) \, {\vec \sigma}(1) \cdot {\vec k}_1 \,
{\vec \sigma}(2) \cdot {\vec k}_2 \right) .
\label{j2pi}
\end{eqnarray}
The momentum ${\vec k}_i \equiv {\vec p}_i^{\, \prime} -  {\vec p}_i $
is the momentum transferred to the nucleon $i$.

Similarly for the pure one-$\rho$ exchange potential
\begin{equation}
V_\rho = v_\rho (k) \left( {\vec \sigma}(1)\times{\vec k} \, \right)\, \cdot \,
\left( {\vec \sigma}(2)\times{\vec k} \, \right) \ + \ v_\rho^S ( k ) ,
\label{vrho}
\end{equation}
with
\begin{equation}
v_\rho (k) = - \left(
\frac{g_{\rho N N}}{2 M}
\right)^2 \, \frac{(1 + \kappa)^2}{m_\rho^2 + k^2} ,
\label{vrhok}
\end{equation}
and
\begin{equation}
v_\rho^S (k) = g_{\rho N N}^2 \, \frac1{m_\rho^2 + k^2} ,
\label{vrhos}
\end{equation}
one obtains going through the corresponding algebra
\begin{eqnarray}
{\vec j}_\rho^{exch\ \prime} 
= i \,
\left( G_E^p ( {\vec Q} ) - G_E^n ( {\vec Q} ) \right)
\,
\left( {\vec \tau}(1) \times {\vec \tau}(2) \right)_3  \
\left[
\frac{{\vec k}_1 -  {\vec k}_2}{ k_1^2 - k_2^2 } \,
( v_\rho^S (k_2) - v_\rho^S (k_1) ) \right. \nonumber \\
-\left(
v_\rho (k_2) {\vec \sigma}(1) \times ( {\vec \sigma}(2) \times {\vec k}_2 )
\ - \
v_\rho (k_1) {\vec \sigma}(2) \times ( {\vec \sigma}(1) \times {\vec k}_1 )
\right) \nonumber \\
\left.
-\frac{v_\rho (k_2) - v_\rho (k_1)}{k_1^2 - k_2^2} \,
(
{\vec \sigma}(1) \times {\vec k}_1
)
\cdot
(
{\vec \sigma}(2) \times {\vec k}_2
)
\
( {\vec k}_1 -  {\vec k}_2)
\right] .
\label{j2rhoprime}
\end{eqnarray}
That extraction of the two-body currents 
$ {\vec j}^{exch}$ from the continuity equation
obviously can determine only the longitudinal (direction ${\hat Q}$)
part of $ {\vec j}^{exch}$ and in the case of the $\rho$-exchange 
in fact one piece required by the underlying Lagrangian 
is missing. 
The full expression is
\begin{eqnarray}
{\vec j}_\rho^{exch} \ = \
{\vec j}_\rho^{exch\ \prime}
\ - \ i \, \left( G_E^p ( {\vec Q} ) - G_E^n ( {\vec Q} ) \right)
\,
\left( {\vec \tau}(1) \times {\vec \tau}(2) \right)_3 \nonumber \\
\frac{v_\rho (k_2) - v_\rho (k_1)}{k_1^2 - k_2^2} \,
\left(
{\vec \sigma}(2) \cdot ( {\vec k}_1 \times {\vec k}_2 ) \ ( {\vec \sigma}(1) \times {\vec k}_1 )
\ + \
{\vec \sigma}(1) \cdot ( {\vec k}_1 
\times {\vec k}_2 ) \ ( {\vec \sigma}(2) \times {\vec k}_2 )
\right) .
\label{j2rho}
\end{eqnarray}
In (\ref{vrhok})--(\ref{vrhos})
$g_{\rho N N}$ and $\kappa$ are the vector and tensor coupling
constants and $m_\rho$ is the $\rho$-mass.
This agrees with the expressions given in \cite{Riska2,MEC}.

Putting now the $\pi$-like part of the phenomenological potential $v_1$
in (\ref{termt1t2}) and (\ref{vss4})--(\ref{vss5})
into the form (\ref{v1pi}) of the pure pion exchange one finds the
correspondence
\begin{equation}
- 3 v_T^\pi ( k ) {\hat =} v_\pi ( k ) .
\label{correspondence}
\end{equation}
This leads to the idea proposed 
in \cite{Riska1,Riska2,MEC} to replace $ v_\pi (k) $
in (\ref{j2pi}) by $ - 3 v_T^\pi ( k ) $ determined
via (\ref{vss4})--(\ref{vss5}) from the phenomenological
potential (\ref{Vnn}).
In this manner one arrives at the $\pi$-like exchange
current which together with
the $\pi$-like part of the 
force fulfills the continuity equation by construction.

Similarly putting the $\rho$-like part of the phenomenological potential
in (\ref{termt1t2}) and (\ref{vss3}) into the form (\ref{vrho})
of the pure $\rho$-exchange leads to the correspondence
\begin{equation}
 v_T^\rho ( k ) {\hat =} v_\rho ( k ) ,
\label{correspondence2}
\end{equation}
and
\begin{equation}
 v_C^\rho ( k ) {\hat =} v_\rho^S ( k ) .
\label{correspondence3}
\end{equation}
Therefore one replaces $v_\rho (k)$ in (\ref{j2rhoprime})--(\ref{j2rho})
by $ v_T^\rho (k)$ given in (\ref{vss3})
and $v_\rho^S ( k )$ in (\ref{j2rhoprime})
by $v_C^\rho ( k ) $ given in (\ref{vc2}).
This leads to the $\rho$-like exchange current which again together with
the $\rho$-like part of the force  fulfills the continuity equation exactly.

It remains to provide
the forms of $v_{SS}^t (k)$, $v_{T}^t (k)$, and $ v_{C}^t (k)$
belonging to the local NN force
\begin{eqnarray}
V = \sum\limits_{t=0}^1 v_{SS}^t (r) 
{\vec \sigma}_1 \cdot {\vec \sigma}_2 \, P_t
\ + \
\sum\limits_{t=0}^1  v_{T}^t (r) \left(
3   {\vec \sigma}_1 \cdot {\vec r} \, {\vec \sigma}_2 \cdot {\vec r} \ - \
{\vec \sigma}_1 \cdot {\vec \sigma}_2
\right) \, P_t
\ + \
\sum\limits_{t=0}^1  v_{C}^t (r)\, P_t .
\label{Vforrho}
\end{eqnarray}
The Fourier transform of (\ref{Vforrho}) results in
\begin{equation}
v_T^t (k) \ = \ \frac{4 \pi}{k^2} \, 
\int\limits_0^\infty dr r^2 j_2 (k r ) v_T^t (r) ,
\label{fourier1}
\end{equation}
\begin{equation}
v_{SS}^t (k) \ = \ \frac{4 \pi}{k^2} \, \int\limits_0^\infty dr r^2
\left[ j_0 (k r ) - 1 \right] \, v_{SS}^t (r) ,
\label{fourier2}
\end{equation}
\begin{equation}
v_C^t (k) \ = \ {4 \pi} \, \int\limits_0^\infty dr r^2 j_0 (k r ) v_C^t (r) .
\label{fourier3}
\end{equation}
The bracket $\left[ j_0 (k r ) - 1 \right]$  
in (\ref{fourier2}) guarantees that the volume integral related
to $ v_{SS}^t (k) $ vanishes, like for the pure one-$\pi$ exchange.

Clearly there are additional two-body currents related
to spin-orbit NN interactions and other momentum dependences
in the NN force AV18. 
They have been considered
in ~\cite{spinorbitMEC} and  ~\cite{spinorbitMEC2},
and appear in general to be of less quantitative importance.
The purely transverse currents are of course
not constrained by the continuity equation.
Among them $\rho \pi \gamma$-, $\omega \pi \gamma$- and $\Delta$- currents
have been considered and we refer the reader to ~\cite{Carlson98}
and references therein for their discussion.
Again they appear to be less important for low energy physics~\cite{Carlson98}.
 Based on current insights they  are clearly strongly model dependent. 

Quite recently \cite{marc04,Marcucci2005} currents have been constructed, which exactly 
fulfill the current continuity equation with the AV18 NN force in combination 
with the UrbanaIX 3NF. The authors follow the steps using minimal 
substitutions as outlaid in 
 \cite{sachs48,nyman}.  
The first observation on those steps is that
${\vec \tau}_i \cdot {\vec \tau}_j $ can be replaced by
\[
{\vec \tau}_i \cdot {\vec \tau}_j  = -1 -
\left( 1 + {\vec \sigma}_i \cdot {\vec \sigma}_j   \right) \,
P^{\rm \, space}{(i,j)}
\]
when applied to an antisymmetric wave function.
Thereby the operator $P^{\rm \, space}{(i,j)}$  exchanges the positions
of particles $i$ and $j$. The key point is then that
 $P^{\rm \, space}{(i,j)}$ can be expressed in terms of
momentum operators as \cite{Wheeler1936,sachs48}
\begin{equation}
   P^{\rm \, space}{(i,j)} =
e^{{\vec r}_{ji} \cdot {\vec \nabla}_i + {\vec r}_{ij} \cdot {\vec \nabla}_j},
\end{equation}
where the ${\vec \nabla}$-operators do not act on the pair distances
$  {\vec r}_{ij} = -  {\vec r}_{ji}$.
In this form one can perform
a minimal substitution to couple to an electromagnetic field
$\vec A (\vec r \, )$. This leads to an expression of the form
\begin{equation}
   P (\vec r \,) =
e^{{\vec a} \cdot {\vec \nabla} + g ( \vec r \,)},
\end{equation}
which can be expressed as
\begin{equation}
   P (\vec r \,) =
e^{\frac1{a} \, \int\limits_{\vec r}^{\vec r + \vec a}  \,
d {\vec s} \, g \left( \vec s \, \right) }
\,
e^{  {\vec a} \cdot {\vec \nabla} }
\end{equation}
with a line integral along the straight line between the positions
${\vec r}$ and ${\vec r} + {\vec a}$.
For the application to the NN force AV18 and the 3NF UrbanaIX
we refer the reader to the very recent paper
\cite{Marcucci2005}. Here, low energy electronuclear observables, 
nd and pd radiative capture reactions and magnetic form factors of 
$^3$He and $^3$H are calculated. Comparative studies of new and old 
current models are performed. It turns out that three-body currents 
give small but 
significant contributions to some of the very low energy observables. 
For detailed information see \cite{marc04,Marcucci2005}.

The interesting issue of modifications for the absorption mechanism
of a photon on hadrons in nuclear medium is addressed in
\cite{Brown04,Harada03}.

\clearpage

\section{The Observables}
\label{observables}

Knowing the nuclear matrix elements for electron scattering on $^3$He ($^3$H) 
the step to the rich variety
of observables based on the one-photon exchange is standard.
Thereby it is assumed that the initial and final nuclear states
are eigenstates of the hadronic four-momentum operator.
This leads to the $S$-matrix element
\begin{equation}
S_{fi} = i ( 2 \pi ) ^4 \,
\delta ( k^{\, \prime} - k + P_f - P_i ) \,
\frac{e^2}{Q^2} \, L_\mu \, N^\mu ,
\label{obse1}
\end{equation}
where $ k (k^{\, \prime}) , P_i (P_f)$
are the initial (final) electron and nuclear four momenta,
\begin{equation}
Q = k - k^{\, \prime} = P_f - P_i
\label{obse2}
\end{equation}
the photon four momentum, and
\begin{equation}
N^\mu \, \equiv \,
\langle f \mid \frac1e j^\mu_{hadron} (0) \mid i \rangle
\label{obse3}
\end{equation}
\begin{equation}
L^\mu \, \equiv \,
\langle k' s' \mid -\frac1e j^\mu_{electron} (0) \mid k s \rangle
\label{obse4}
\end{equation}
the hadronic and electronic matrix elements. In  a nonrelativistic
treatment which we pursue, this underlying property of the hadronic
states to be eigenstates  of the  hadronic four-momentum operator
${\hat P}^\mu$
\begin{equation}
{\hat P}^\mu \mid i,f \rangle = {P}^\mu_{i,f} \mid i,f \rangle
\label{obse5}
\end{equation}
is of course not fulfilled but in the derivation for the expression
of the observables we nevertheless assume this to be true.

The cross section for the transition 
into the final states spanned by  $ d f$ reads
\begin{equation}
d \sigma \, = \,
( 2 \pi ) ^4 \,
\delta ( k' - k + P_f - P_i ) \, \frac{e^4}{Q^4} \,
\left( L_\mu \, L_\nu^* \right) \,
\left( N^\mu \, N^{\nu\, *} \right) \, df  .
\label{obse6}
\end{equation}
For an initially polarized electron with helicity $h$
one obtains by well known steps
\begin{equation}
 L_\mu \, L_\nu^* \, = \,
\frac1{2 k_0^{\, \prime} \, k_0} \,
\frac1{(2 \pi)^6} \,
\left(
k_\mu k_\nu^{\, \prime}
+ k_\nu k_\mu^{\, \prime}  - g_{\mu \nu} k \cdot k^{\, \prime}
- i h \epsilon_{ \mu \nu \alpha \beta} k^\alpha k^{\, \prime \, \beta}
\right) .
\label{obse7}
\end{equation}
Further we regard ultrarelativistic electrons ($m_e \rightarrow 0$)
and use current conservation in the form
\begin{equation}
Q_0 N^0 - {\vec Q}\cdot {\vec N} = 0 .
\label{obse8}
\end{equation}
Thus we can drop some terms in (\ref{obse7}) and  put
\begin{equation}
 L_\mu \, L_\nu^* \, \rightarrow \,
\frac1{4 k_0^{\, \prime} \, k_0} \,
\frac1{(2 \pi)^6} \,
\left(
K_\mu K_\nu + g_{\mu \nu} Q^2
- 2 i h \epsilon_{ \mu \nu \alpha \beta} k^\alpha k^{\, \prime \, \beta}
\right)
\, \equiv \,
\frac1{(2 \pi)^6} \, l_{\mu \nu } ,
\label{obse9}
\end{equation}
with
\begin{equation}
K \equiv k + k^{\, \prime} .
\label{obse10}
\end{equation}

The contraction with $N^\mu N^{\nu \, *} $ is a quite tedious step.
Starting from
\begin{equation}
l_{\mu \nu } N^\mu N^{\nu \, *} \, = \,
\frac1{4 k_0^{\, \prime} \, k_0} \,
\left(
K\cdot N \, K\cdot N^* \, + \,
Q^2 N\cdot N^*  \, 
- 2 i h \epsilon_{ \mu \nu \alpha \beta} k^\alpha k^{\, \prime \, \beta}
N^\mu N^{\nu \, *}
\right)
\label{obse11}
\end{equation}
one uses spherical unit vectors  ${\hat e}_\mu$
to represent the space part of $N^\mu$ as
\begin{equation}
{\vec N} =
{\hat e}_1^{\, *} N_1 +
{\hat e}_{-1}^{\, *} N_{-1} +
{\hat e}_{0} N_{0} .
\label{obse12}
\end{equation}
The assumed  current conservation (\ref{obse8}),
another property of the hadronic
dynamics which is not always exactly fulfilled in the present day practice,
allows to eliminate the component of ${\vec N}$ along ${\vec Q}$
in favor of $N_0$:
\begin{equation}
{\hat Q} \cdot {\vec N} = \frac{Q_0}{\mid {\vec Q} \mid}\, N_0 .
\label{obse13}
\end{equation}
From now on we shall choose the $z$-direction to be
the direction of ${\vec Q}$.
In rewriting  (\ref{obse11})
in terms of $N_{\pm 1}$   and $N_0$ only the kinematical relation
(\ref{obse2}) is used. Further it is convenient to choose the plane spanned by
${\vec k}$ and ${\vec k}^{\, \prime}$
to coincide with the $x-z$ plane and to choose the positive $x$-direction
such that $(k + k^{\, \prime})_x \ge 0 $.
 Then one obtains \cite{Donnelly86}
\begin{eqnarray}
d \sigma \, = \,
( 2 \pi ) ^4 \,
\delta ( P_f - P_i - Q) \, \frac{e^4}{(Q^2)^2} \,
\cos^2 \frac{\vartheta}2 \, df \nonumber \\
\left[
v_L R_L + v_T R_T + v_{TT} R_{TT} + v_{TL} R_{TL}
\, + \,
h \left( v_{T '} R_{T '} + v_{TL '} R_{TL '} \right)
\right] ,
\label{obse14}
\end{eqnarray}
where the purely kinematical functions $v$ are given in terms of
electron properties only ($\vartheta$ is the laboratory 
electron scattering angle)
\begin{eqnarray}
v_L = \frac{(Q^2)^2}{\left( {\vec Q}^{\ 2} \right)^2} \nonumber \\
v_T = -\frac12 \frac{Q^2}{{\vec Q}^{\ 2}} + \tan^2 \frac{\vartheta}2 \nonumber \\
v_{TT} = \frac12 \frac{Q^2}{{\vec Q}^{\ 2}} \nonumber \\
v_{TL} = \frac1{\sqrt{2}}\, \frac{Q^2}{{\vec Q}^{\ 2}} \,
\sqrt{ - \frac{Q^2}{{\vec Q}^{\ 2}} + \tan^2 \frac{\vartheta}2 } \nonumber \\
v_{T '} = \sqrt{ - \frac{Q^2}{{\vec Q}^{\ 2}} + \tan^2 \frac{\vartheta}2 } \tan \frac{\vartheta}2
\nonumber \\
v_{TL '} = \frac1{\sqrt{2}} \frac{Q^2}{{\vec Q}^{\ 2}} \tan \frac{\vartheta}2  .
\label{obse15}
\end{eqnarray}
The nuclear response functions are
\begin{eqnarray}
R_L = \mid N_0 \mid ^2  \nonumber \\
R_T = \mid N_1 \mid ^2  + \mid N_{-1} \mid ^2  \nonumber \\
R_{TT} = 2 \Re \left( N_1 N_{-1}^{\, *} \right) \nonumber \\
R_{TL} = -2 \Re \left( N_0 ( N_1 - N_{-1} )^{\, *}  \right) \nonumber \\
R_{T '} =  \mid N_1 \mid ^2  - \mid N_{-1} \mid ^2  \nonumber \\
R_{TL '} = -2 \Re \left( N_0 ( N_1 + N_{-1} )^{\, *}  \right)  .
\label{obse16}
\end{eqnarray}
This is a general form for the cross section where the polarization
of the hadronic states can still be chosen at will.

Let us first regard elastic scattering on $^3$He. 
Straightforward calculation of 
the phase space factor
\begin{equation}
\rho \equiv \int \delta ( P'-P -Q ) df =
\int \delta ( P'-P -Q ) d {\vec P}^{\, \prime} \,  d {\vec k}^{\, \prime}
\label{obse17}
\end{equation}
in the lab system leads to 
\begin{equation}
\rho = d {\hat k}^{\, \prime} \, \frac{E_{P '}}{M} \,
\frac{k_0^{\, \prime \, 2}}{1 + \frac{k_0}M ( 1 - \cos \vartheta ) } .
\label{obse18}
\end{equation}
Here $M $ is the $^3$He mass, 
$E_{P '}= \sqrt{M^2 + {\vec P}^{\, \prime \, 2} }$, 
and
\begin{equation}
k_0^{\, \prime} = \frac{k_0}{1 + \frac{k_0}M ( 1 - \cos \vartheta ) } .
\label{obse19}
\end{equation}
Then one introduces the Mott cross section
\begin{equation}
\sigma_{Mott} = \frac{\alpha^2 
\cos^2\frac{\vartheta}2}{4 k_0^2 \sin^4 \frac{\vartheta}2} ,
\label{obse20}
\end{equation}
with $\alpha= \frac{e^2}{4 \pi} \approx \frac{1}{137} $ 
and obtains the differential cross section for
 unpolarized electron scattering on an unpolarized  $^3$He target state
 in the lab system
\begin{equation}
\frac{d \sigma }{ d {\hat k}^{\, \prime} } \, = \,
\sigma_{Mott} \, \frac{1}{1 + \frac{k_0}M ( 1 - \cos \vartheta ) } \,
\left[
\frac{(Q^2)^2}{\left( {\vec Q}^{\ 2} \right)^2} {\cal R}_L
+ \left( -\frac12 \frac{Q^2}{{\vec Q}^{\ 2}} 
+ \tan^2 \frac{\vartheta}2  \right) {\cal R}_T
\right] .
\label{obse21}
\end{equation}
We defined longitudinal ${\cal R}_L$ and transversal ${\cal R}_T$ 
response functions, averaged over initial $(m)$ and summed over final $(m')$ 
spin magnetic quantum numbers 
\begin{equation}
{\cal R}_L = (2 \pi )^6 \, \frac{E_{P '}}{M} \, 
\frac12 \sum\limits_{ m m '} \mid N_0 \mid ^2
\label{obse22}
\end{equation}
\begin{equation}
{\cal R}_T = (2 \pi )^6 \, \frac{E_{P '}}{M} \, \frac12 \sum\limits_{ m m '}
\left( \mid N_1 \mid ^2  + \mid N_{-1} \mid ^2 \right) .
\label{obse23}
\end{equation}
This is usually written in terms of the charge and 
magnetic form factors~\cite{Collard65}
\begin{eqnarray}
\frac{d \sigma }{ d {\hat k}^{\, \prime} } \, = \,
\sigma_{Mott} \, \frac{Z^2}{1 + \frac{k_0}M ( 1 - \cos \vartheta ) } \,
\left[
F_{C}^2  \right. \nonumber \\
\left.
- \frac{Q^2}{4 M^2} F_{m}^2 ( 1 + \kappa )^2 \,
\left(  1 + 2 ( 1 - \frac{Q^2}{4 M^2} ) \tan^2 \frac{\vartheta}2  \right)
\right] \, \frac1{1 - \frac{Q^2}{4 M^2} } .
\label{obse24}
\end{eqnarray}
The form factors are normalized as
\begin{equation}
F_{C} \left( Q^2 = 0 \right) = 1
\label{obse25}
\end{equation}
\begin{equation}
F_{m} \left( Q^2 = 0 \right) = 1 ,
\label{obse26}
\end{equation}
such that $(1 + \kappa) $  is the magnetic moment of $^3$He in 
nuclear magnetons ($ \frac{e}{2 m_N}$)~\cite{Kamada92}.

In contrast to (\ref{obse14}) the response functions $R_{TT}$ and $R_{TL}$
do not show up in (\ref{obse24}).
The partial wave decomposition reveals~\cite{Kamada92} that 
they are zero in this case.

It is known that polarizing the initial
electron or initial $^3$He does not lead to  new independent information
\cite{Donnelly86}.

Now we move on to inclusive scattering on $^3$He. Only the scattered
electron is measured and one has to integrate over all final nucleon
momenta. We choose the lab system and work nonrelativistically.
The phase space factor in the pd channel is then
\begin{equation}
\rho_{pd} \equiv
\int \delta ( {\vec P}^{\, \prime} - {\vec Q} ) \,
\delta \left( E_d + \frac{k_d^2}{4 m_N} + \frac{k_p^2}{2 m_N} 
- Q_0 - E_{{}^3{\rm He}} \right) \,
d {\vec k}_d \,
d {\vec k}_p \,
d {\vec k}^{\, \prime} ,
\label{obse27}
\end{equation}
where $E_d~~ (E_{{}^3{\rm He}})$ is the (negative) deuteron $(^3{\rm He})$ binding
energy and ${\vec k}_d ~~({\vec k}_p)$ is
the final deuteron (proton) momenta. Since the nuclear matrix element is
evaluated in terms of Jacobi momenta  
 it is convenient  to
change ${\vec k}_d$  and ${\vec k}_p$  to
\begin{equation}
\label{obse28}
{\vec q} \equiv \frac23 ({\vec k}_p - \frac12 {\vec k}_d )
\end{equation}
\begin{equation}
{\vec P}^{\, \prime} \equiv {\vec k}_p + {\vec k}_d .
\label{obse29}
\end{equation}
This leads then to
\begin{equation}
\rho_{pd} = d {\hat k}^{\, \prime} \, d {\hat q} \,
\frac{2 m_N}3 \mid {\vec q}_0 \mid ,
\label{obse30}
\end{equation}
with
\begin{equation}
\mid {\vec q}_0 \mid
= \sqrt{ \frac{4 m_N}3 \left( Q_0 + E_{{}^3{\rm He}} 
- \frac{{\vec Q}^{\, 2}}{6 m_N} - E_d \right) } .
\label{obse31}
\end{equation}
In the case of the 3N breakup one introduces Jacobi momenta for both
relative motions
\begin{equation}
{\vec p} \equiv \frac12 ({\vec k}_2 - {\vec k}_3 )
\label{obse32}
\end{equation}
\begin{equation}
{\vec q} \equiv \frac23 \left( {\vec k}_1 - \frac12 ({\vec k}_2 
+ {\vec k}_3 ) \right)
\label{obse33}
\end{equation}
on top of the total momentum and obtains
\begin{eqnarray}
\rho_{ppn} \equiv
\int \delta ( {\vec P}^{\, \prime} - {\vec Q} ) \,
\delta \left( \sum\limits_{i=1}^3 \frac{k_i^2}{2 m_N} - Q_0 
- E_{{}^3{\rm He}} \right) \,
d {\vec k}_1 \,
d {\vec k}_2 \,
d {\vec k}_3 \,
d {\vec k}^{\, \prime}  \nonumber \\
= \ d {\vec k}^{\, \prime} \, \int d {\vec q} \, d {\vec p} \,
\delta \left( \frac{p^2}{m_N} + \frac{3 q^2}{4 m_N}  
+ \frac{{\vec Q}^{\, 2}}{6 m_N} - Q_0 - E_{{}^3{\rm He}} \right)
\ = \ d {\vec k}^{\, \prime} \,  \int  d {\hat q} \, 
d {\vec p} \, \frac{2 m_N }3 \, \mid \vec q \mid ,
\label{obse34}
\end{eqnarray}
with
\begin{equation}
\mid {\vec q} \mid
= \sqrt{ \frac{4 m_N}3 \left( Q_0 + E_{{}^3{\rm He}} 
- \frac{{\vec Q}^{\, 2}}{6 m_N} - \frac{p^2}m_N \right) } ,
\label{obse35}
\end{equation}
and the integration over $\mid {\vec p} \mid$ is between
$0$  and $p_{max}$ 
\begin{equation}
p_{max} =
\sqrt{m_N \left( Q_0 + E_{{}^3{\rm He}} 
- \frac{{\vec Q}^{\, 2}}{6 m_N} \right)} .
\label{obse36}
\end{equation}
Again the partial wave decomposition reveals (see \cite{Ishikawa98})
that in the case when electron and $^3$He are unpolarized 
 only two response
functions survive (this is, of course, known from standard 
symmetry arguments)
\begin{eqnarray}
R_L = (2 \pi )^6 \, \frac12 \sum\limits_{ m_d m_p m}
\frac{2 m_N}{3} \, \mid {\vec q}_0 \mid \,
\int d {\hat q} \mid N_0^{pd} \mid ^2  \nonumber \\
 + \
(2 \pi )^6 \, \frac12 \sum\limits_{ m_1 m_2 m_3 m} \frac{2 m_N}{3} \,
\int\limits_0^{p_{max}} d {\vec p} \, d {\hat q} \, \mid {\vec q} \mid \,
\mid N_0^{ppn} \mid ^2 ,
\label{obse37}
\end{eqnarray}
\begin{eqnarray}
R_T = (2 \pi )^6 \, \frac12 \sum\limits_{ m_d m_p m}
\frac{2 m_N}{3} \, \mid {\vec q}_0 \mid \,
\int d {\hat q} \, \left( \mid N_1^{pd} \mid ^2  
+  \mid N_{-1}^{pd} \mid ^2 \right) \nonumber \\
 + \
(2 \pi )^6 \, \frac12 \sum\limits_{ m_1 m_2 m_3 m} \frac{2 m_N}{3} \,
\int\limits_0^{p_{max}} d {\vec p} \, d {\hat q} \, \mid {\vec q} \mid \,
\, \left( \mid N_1^{ppn} \mid ^2  +  \mid N_{-1}^{ppn} \mid ^2 \right) .
\label{obse38}
\end{eqnarray}
Then the inclusive unpolarized scattering cross section reads
\begin{equation}
\frac{d \sigma }{ d {\hat k}^{\, \prime} \, d {k}_0^{\, \prime}  } \, = \,
\sigma_{Mott} \, \left\{  v_L R_L  + v_T R_T \right\} .
\label{obse39}
\end{equation}
In (\ref{obse37}) and (\ref{obse38}) we added superscripts to the nuclear matrix elements according to the
two types of final channels.

At this point it is adequate to describe another manner for evaluating
the two response functions $R_L$ and $R_T$~\cite{Ishikawa94.2,Ishikawa98}.
Both functions are of the type
\begin{equation}
R = \sum \!\!\!\!\!\!\!\int \, df \delta \left( E_f - E_i - Q_0 \right) \,
\left| \langle \Psi_f \mid {\cal O} \mid \Psi_i \rangle \right| ^2 ,
\label{obse40}
\end{equation}
where ${\cal O} $ is an appropriate operator. Since the final states
$ \mid \Psi_f \rangle $ are eigenstates to the Hamiltonian $H$, one can
use closure to evaluate $R$ as
\begin{eqnarray}
R \, = \, \sum \!\!\!\!\!\!\!\int \, df  \,
\langle \Psi_i \mid {\cal O}^\dagger \, \delta \left( H - E_i - Q_0 \right) \,
\mid \Psi_f \rangle
\langle \Psi_f \mid {\cal O} \mid \Psi_i \rangle \nonumber \\
\, = \,
\langle \Psi_i \mid {\cal O}^\dagger \, \delta \left( H - E_i - Q_0 \right) \,
 \mid {\cal O} \mid \Psi_i \rangle .
\label{obse41}
\end{eqnarray}
The bound state does not contribute since $ Q_0 > 0$. The result is easily
converted into
\begin{equation}
R \, = \, -\frac1{\pi} \, \Im
\langle \Psi_i \mid {\cal O}^\dagger \, \frac1{ Q_0 + E_i - H + i \epsilon } \,
 {\cal O} \mid \Psi_i \rangle  .
\label{obse42}
\end{equation}
The remaining task is to evaluate the auxiliary state
\begin{equation}
\mid \Phi \rangle \equiv
\frac1{ Q_0 + E_i - H + i \epsilon } \,
 {\cal O} \mid \Psi_i \rangle ,
\label{obse43}
\end{equation}
which apparently fulfills the inhomogeneous Schr\"odinger equation
\begin{equation}
( E + i \epsilon - H ) \mid \Phi \rangle = {\cal O} \mid \Psi_i \rangle .
\label{obse42.5}
\end{equation}
We introduced $ E \equiv E_i + Q_0$.
The Faddeev scheme is introduced by converting
(\ref{obse42.5}) into
\begin{equation}
\mid \Phi \rangle = G_0 \sum\limits_{i=1}^3  ( V_i + V^{(i)} ) \mid \Phi \rangle
+ G_0 {\cal O} \mid \Psi_i \rangle .
\label{obse42.6}
\end{equation}
Similarly to the 3NF,
the operator $ {\cal O}$ for three identical nucleons
can always be split into three parts ${\cal O}_i$, symmetrical under 
the exchange of particles $j$ and $k$ 
\begin{equation}
 {\cal O} = \sum\limits_{i=1}^3 {\cal O}_i .
\label{obse42.7}
\end{equation}
Therefore, the right hand side decomposes as
\begin{equation}
\mid \Phi \rangle = G_0 \sum\limits_{i=1}^3 \mid \Phi_i \rangle ,
\label{obse42.8}
\end{equation}
where as before $ \mid \Phi_2 \rangle$ and $ \mid \Phi_3 \rangle$
result by cyclical and anticyclical permutations
out of $\mid \Phi_1 \rangle$.
One obtains using the definition (\ref{P})
\begin{equation}
( 1 - V_1 G_0 ) \mid \Phi_1 \rangle =
V_1 G_0 P \mid \Phi_1 \rangle + V^{(1)} G_0 (1 + P ) \mid \Phi_1 \rangle
+ {\cal O}_1  \mid \Psi_i \rangle ,
\label{obse42.9}
\end{equation}
or
\begin{equation}
\mid \Phi_1 \rangle  =
( 1 + t G_0 )  {\cal O}_1  \mid \Psi_i \rangle
+
( t_1 G_0 P + ( 1 + t_1 G_0 ) V^{(1)} G_0 (1 + P ) ) \mid \Phi_1 \rangle  .
\label{obse42.10}
\end{equation}
This is a Faddeev-like integral equation with the same kernel
as in (\ref{T.2}). The response function is then given as
\begin{equation}
R = - \frac3{\pi} \, \Im \langle \Psi_i \mid
{\cal O}_1^\dagger ( 1 + P ) G_0 \mid \Phi_1 \rangle .
\label{obse42.11}
\end{equation}
The factor $3$ arises since we kept only ${\cal O}_1^\dagger$.

The remaining cross section observables are for semi-exclusive
and exclusive reactions on $^3$He. In the case of the processes
${}^3{\rm He}(e,e'p)d $  and ${}^3{\rm He}(e,e' N) NN$,
where only one nucleon is measured in coincidence with
the scattered electron, the plane spanned by the photon momentum and
the detected nucleon momentum (hadronic plane) is in a general case
rotated by an angle $\phi$ with respect to the plane spanned by
the electron momenta (electronic plane). The dependence of the cross
section on $\phi$ can be made explicit by introducing instead of
the  spherical unit vectors ${\hat e}_{\pm 1}$ used up to now and which are
perpendicular to ${\hat Q}$ (chosen in ${\hat z}$ direction) 
two other perpendicular
unit vectors~\cite{Forest83}
\begin{equation}
{\hat e}_\perp \equiv \widehat{{\vec Q} \times {\vec p}}  
= -{\hat x} \sin\phi + {\hat y} \cos\phi
\label{obse44}
\end{equation}
\begin{equation}
{\hat e}_\parallel \equiv \widehat{{\hat e}_\perp \times {\vec Q}}  
= {\hat x} \cos\phi + {\hat y} \sin\phi .
\label{obse45}
\end{equation}
The unit vector ${\hat e}_\perp $ is perpendicular to the hadronic plane
and ${\hat e}_\parallel $ lies in that plane.
 The two pairs of unit vectors are connected by
\begin{equation}
{\hat e}_{\pm 1} \ = \ \mp \frac{e^{\pm i \phi}}{\sqrt{2}} \,
\left(
\pm i {\hat e}_\perp  + {\hat e}_\parallel
\right) .
\label{obse46}
\end{equation}
Introducing the components
\begin{equation}
N_\perp \equiv {\hat e}_\perp \cdot {\vec N} ,
\label{obse47}
\end{equation}
\begin{equation}
N_\parallel \equiv {\hat e}_\parallel \cdot {\vec N} ,
\label{obse48}
\end{equation}
of the nuclear matrix elements one finds the connections
to the components used up to now
\begin{equation}
N_{\pm 1} \ = \ \mp \frac{e^{\pm i \phi}}{\sqrt{2}} \,
\left(
\pm i N_\perp  + N_\parallel
\right) .
\label{obse49}
\end{equation}
The advantage of using $N_\perp$  and $N_\parallel$
in the cross section formula
is to make the $\phi$-dependence explicit.

After a simple algebra, using  (\ref{obse49})
and the definitions (\ref{obse15})
of the $v$'s, one obtains
\begin{eqnarray}
v_L \mid N_0 \mid ^2  \, + \,
v_T \, \left( \mid N_1 \mid ^2  + \mid N_{-1} \mid ^2 \right) \, + \,  \nonumber \\
v_{TT} \, 2 \Re \left( N_1 N_{-1}^{\, *} \right) \, + \,
v_{TL} \, (-2) \Re \left( N_0 ( N_1 - N_{-1} )^{\, *}  \right) \nonumber \\
= \
v_L \mid N_0 \mid ^2  \, + \,
v_T \, \left( \mid N_\perp \mid ^2  
+ \mid N_{\parallel} \mid ^2 \right) \, + \,  \nonumber \\
2 \sqrt{2} \, v_{TL} \, \cos\phi \, 
\Re \left( N_0 N_{\parallel}^{\, *} \right) \, + \,
v_{TT} \, \cos( 2\phi) \left( \mid N_\perp \mid ^2  
- \mid N_{\parallel} \mid ^2 \right)  .
\label{obse50}
\end{eqnarray}
Thereby two terms proportional to $\Re \, \left( N_0 N_{\perp}^{\, *} \right) $
and $\Re \, \left( N_\perp N_{\parallel}^{\, *} \right) $
have been dropped since they vanish here as is seen in a partial 
wave decomposition~\cite{Ishikawa94}.

As is seen from (\ref{obse14}) and (\ref{obse16}) these 
sums generate the cross sections
except for an overall factor. 
The rotational invariance around the $z$-axis (chosen in $\hat Q$ direction) 
guarantees that none of the quantities $N_0$, $N_{\parallel}$ and $N_\perp$ depends on $\phi$. 

It follows that the pd breakup cross section can be written in two forms
\begin{eqnarray}
\frac{d^5 \sigma }{ d {\hat k}^{\, \prime} \, d {k}_0^{\, \prime}
d {\hat q}_0 } \, = \,
\sigma_{Mott} \,
\left[
v_L \mid N_0 \mid ^2  \, + \,
v_T \, \left( \mid N_1 \mid ^2  + \mid N_{-1} \mid ^2 \right) \, 
+ \,  \right.  \nonumber \\
\left.
v_{TT} \, 2 \Re \left( N_1 N_{-1}^{\, *} \right) \, + \,
v_{TL} \, (-2) \Re \left( N_0 ( N_1 - N_{-1} )^{\, *}  \right) 
\right] \, \rho_{pd}
\nonumber \\
= \ \sigma_{Mott} \,
\left[
v_L \mid N_0 \mid ^2  \, + \,
v_T \, \left( \mid N_\perp \mid ^2  + \mid N_{\parallel} 
\mid ^2 \right) \, + \,  \right. \nonumber \\
\left.
2 \sqrt{2} \, v_{TL} \, \cos\phi \, 
\Re \left( N_0 N_{\parallel}^{\, *} \right) \, + \,
v_{TT} \, \cos( 2\phi) \left( \mid N_\perp \mid ^2  
- \mid N_{\parallel} \mid ^2 \right)
\right] \, \rho_{pd} ,
\label{obse51}
\end{eqnarray}
where the second one shows the $\phi$ dependence explicitly.

In the case of the semi-exclusive ${}^3{\rm He}(e,e' p ) p n $
or  ${}^3{\rm He}(e,e' n ) p p $  reactions
one has to integrate over the internal (relative) momentum of
the undetected pair of nucleons and one obtains for the cross section
\begin{eqnarray}
\frac{d^6 \sigma }{ d {\hat k}^{\, \prime} \, d {k}_0^{\, \prime}  d {\vec q} }
\, = \,
\sigma_{Mott} \, {\cal C} \, \frac12 \, m_N \, p \, \int \, d {\hat p} \,
\left[
v_L \mid N_0 \mid ^2  \, + \,
v_T \, \left( \mid N_\perp \mid ^2  + \mid N_{\parallel} \mid ^2 \right) \, 
+ \,  \right. \nonumber \\
\left.
2 \sqrt{2} \, v_{TL} \, \cos\phi \, \Re \left( N_0 N_{\parallel}^{\, *} 
\right) \, + \,
v_{TT} \, \cos( 2\phi) \left( \mid N_\perp \mid ^2  
- \mid N_{\parallel} \mid ^2 \right)
\right] ,
\label{obse52}
\end{eqnarray}
with
\begin{equation}
p \equiv \mid {\vec p} \mid
= \sqrt{m_N \left( Q_0 + E_{{}^3{\rm He}} 
- \frac{{\vec Q}^{\, 2}}{6 m_N} - \frac{3 {\vec q}^{\, 2}}{4 m_N} \right) } .
\label{obse52.1}
\end{equation}
To avoid double-counting the 
${\cal C}$ factor is equal $1/2$ in case when the two undetected nucleons 
are identical (protons). Otherwise ${\cal C}=1$.

Lastly there is the complete breakup of $^3$He measured exclusively
by detecting two nucleons in coincidence with the scattered electron.
This leads to an eightfold differential
cross section. The phase space factor is
\begin{eqnarray}
\rho =
\int \delta ( {\vec P}^{\, \prime} - {\vec Q} ) \,
\delta \left( \sum\limits_{i=1}^3 \frac{k_i^2}{2 m_N} 
- Q_0 - E_{{}^3{\rm He}} \right) \,
d {\vec k}_1 \,
d {\vec k}_2 \,
d {\vec k}_3 \,
d {\vec k}^{\, \prime} \, \nonumber \\
=
d {\vec k}^{\, \prime} \,
d {\hat k}_1 \,
d {\hat k}_2 \,
d E_1 \,
\frac{ m^2 \mid {\vec k}_1 \mid \mid {\vec k}_2 \mid}
{\left| 1- \frac{{\vec k}_2 \cdot {\vec k}_3}{k_2^2} \right|}  ,
\label{obse53}
\end{eqnarray}
where ${\vec k}_3 = {\vec Q} - {\vec k}_1 -{\vec k}_2$
and $ \mid {\vec k}_2 \mid$ is determined kinematically from the 
given energy of the first nucleon $E_1$ and the directions of the first 
(${\hat k}_1$) and the second (${\hat k}_2$) nucleons detected in coincidence.  
At some values of the momenta 
the denominator in (\ref{obse53}) can vanish.  One avoids that singularity by
representing the breakup cross section along the kinematically
allowed locus in the  $E_1-E_2$ plane and parameterizing it by the arc-length
$S$ along that locus. This is a well known and usual device for the
treatment of the Nd breakup process \cite{Ohlsen}.
It leads to
\begin{eqnarray}
\rho =
d {\vec k}^{\, \prime} \,
d {\hat k}_1 \,
d {\hat k}_2 \,
d S \,
\frac{ m^2 \mid {\vec k}_1 \mid \mid {\vec k}_2 \mid}
{
\sqrt{
{\left( 1- \frac{{\vec k}_2 \cdot {\vec k}_3}{k_2^2} \right)^2} \ + \
{\left( 1- \frac{{\vec k}_1 \cdot {\vec k}_3}{k_1^2} \right)^2}
}
} ,
\label{obse54}
\end{eqnarray}
and one obtains the following  form for the complete breakup cross section
expressed in terms of  four  response functions
\begin{equation}
\frac{d^8 \sigma }{ d {\hat k}^{\, \prime} \, d {k}_0^{\, 
\prime} \,  d {\hat k}_1 \, d {\hat k}_2 \, dS } \, = \,
\sigma_{Mott} \,
\left[
v_L R_L + v_T R_T + v_{TT} R_{TT} + v_{TL} R_{TL}
\right] \, \rho .
\label{obse55}
\end{equation}
One can also relate the nuclear matrix elements to the 
unit vectors  ${\hat e}_\perp$
and ${\hat e}_\parallel$ which leads to six structure functions
\cite{Golak95}.

\begin{figure}[!ht]
\begin{center}
\epsfig{file=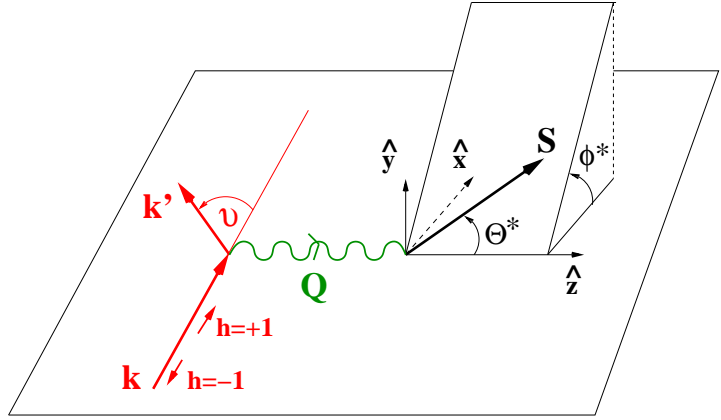,height=6cm}
\caption{
\label{figasymetry}
Definition of the two angles $(\theta^\star , \phi^\star )$
used to specify the initial $^3$He polarization.
}
\end{center}
\end{figure}

The inclusion of polarizations opens a wide field. We only mention
cases which have already been studied experimentally in the 3N system 
 or which are on the list of our predictions (see section \ref{predictions}).  
In inclusive scattering, when only the outgoing electron is detected, 
 two more response functions beyond the ones
in (\ref{obse39}) occur in the
cross section. They go with the helicity $h$ of the initial electron state 
as given in (\ref{obse14}) 
(see \cite{Donnelly86}). This leads naturally to an asymmetry defined by
\begin{equation}
A \ \equiv \
\frac{\sigma (h=+1) - \sigma (h=-1)}{\sigma (h=+1) + \sigma (h=-1)} \ = \
\frac{v_{T '} R_{T '} + v_{TL '} R_{TL '}}{v_{L} R_{L} + v_{T} R_{T}} .
\label{obse56}
\end{equation}
This quantity has been investigated especially  for the case of additionally 
polarized $^3$He.  If the  quantization axis of $^3$He points in the
direction given by the two polar angles $(\theta^\star , \phi^\star )$
(see Fig.~\ref{figasymetry}), the $^3$He state can be written as
\begin{equation}
\mid \Psi_{^3{\rm He}} m \rangle_{(\theta^\star , \phi^\star )} =
\sum\limits_{m '} \mid \Psi_{^3{\rm He}}  m ' \rangle \,
D_{ m ' m } ^{(\frac12)} ( \phi^\star , \theta^\star , 0 ) ,
\label{obse57}
\end{equation}
where $ \mid \Psi_{^3{\rm He}}  m ' \rangle $ is quantized in  $z$-direction.
We refer to \cite{Ishikawa98} where the dependence
on $(\theta^\star , \phi^\star )$ of the two new
response functions has been worked out. There also the generalization
necessary for evaluating the new response functions with the help
of the closure relation is detailed. One obtains the explicit
$(\theta^\star , \phi^\star )$ dependence as \cite{Donnelly86}
\begin{equation}
R_{T '} \equiv -{\tilde R}_{T '} \, \cos\theta^\star
\label{obse58}
\end{equation}
\begin{equation}
R_{TL '} \equiv -2 {\tilde R}_{TL '} \, \sin\theta^\star \, \cos\phi^\star .
\label{obse59}
\end{equation}
Consequently the asymmetry reads now
\begin{equation}
A \ = \
-\, \frac{v_{T '} {\tilde R}_{T '} \,  \cos\theta^\star \ + \
2v_{TL '} {\tilde R}_{TL '} \, \sin\theta^\star \, \cos\phi^\star}
{v_{L} R_{L} + v_{T} R_{T}} .
\label{obse60}
\end{equation}
Polarizing  the $^3$He target spin
($m=1/2$) along the virtual photon direction ${\hat Q}$ 
($\theta^\star=0^\circ$) 
 one selects the transverse asymmetry $A_{T '}$ (proportional to  ${\tilde R}_{T '}$),
 whereas taking $\theta^\star=90^\circ$ one gets the transverse-longitudinal
asymmetry $A_{TL '}$ (proportional to ${\tilde R}_{TL '}$).

In the case of the semiexclusive 
$\overrightarrow{^3{\rm He}}(\vec{e},e'p)pn$
and $\overrightarrow{^3{\rm He}}(\vec{e},e'n)pp$ processes, the asymmetry is
defined analogously to (\ref{obse56}). However, an 
additional integration
over the direction of the relative momentum  of the two
undetected nucleons is needed and, according to (\ref{obse14}), 
two additional response functions, $R_{TT}$ and $R_{TL}$, occur in the 
denominator. One obtains
\begin{equation}
A \ \equiv \
\frac{\int \, d{\hat p} \, \left( v_{T '} R_{T '} + v_{TL '} R_{TL '} \right)}
{\int \, d{\hat p} \, \left( v_{L} R_{L} + v_{T} R_{T} 
 + v_{TT} R_{TT} + v_{TL} R_{TL}
\right) } ,
\label{obse61}
\end{equation}
and one can define again
the asymmetries $A_\perp$ and $A_\parallel$
corresponding to two different initial $^3$He spin orientations
with respect to the photon direction.

In the case of the pd breakup process
$\overrightarrow{^3{\rm He}}(\vec{e},e'p)d$ the asymmetries
are defined correspondingly.

The more intricate processes
$\overrightarrow{^3{\rm He}}(\vec{e},e'\vec{p})d$
and
$\overrightarrow{^3{\rm He}}(\vec{e},e'\vec{d})p$,
where also the polarization of the final particles is measured, 
are dealt with in section~\ref{predictions}.

Spin asymmetries in the exclusive 3N  breakup reaction  have not yet
been  measured or investigated theoretically in the 
3N system  to the best of our
knowledge but the formal extensions are straightforward.

The last group of observables we want to address in this review
are photodisintegration of $^3$He and Nd capture processes. Because of
lack of own experience we shall, however, not discuss Compton scattering
on $^3$He nor Bremsstrahlung in the 3N system.

The $S$-matrix element for photodisintegration into the final channel $f$,
which is either pd or ppn fragmentation, is given as
\begin{equation}
S_{fi} = - i (2 \pi )^4 \, \delta ( P ' - P - Q ) \, 
\frac1{\sqrt{2 \mid {\vec Q} \mid}} \,
\frac1{ (2 \pi )^{3/2}} \,
\langle \Psi_f^{(-)} \, {\vec P}^{\, \prime} \mid {\vec 
\epsilon}_\lambda \cdot {\vec j}(0) \mid
\Psi \, {\vec P} \, \rangle ,
\label{obse62}
\end{equation}
with  $\langle \Psi_f^{(-)}\mid$ the 3N scattering state 
with appropriate asymptotic quantum
numbers $f$. Further ${\vec \epsilon}_\lambda $ is the 
polarization vector for the initial
photon whose momentum defines the z-direction. It results in the
differential cross section
\begin{equation}
d \sigma \ = \
(2 \pi )^4 \frac1{2 \mid {\vec Q} \mid} \,
 \delta ( P ' - P - Q ) \, d f \,
\left|
\langle \Psi_f^{(-)} \, {\vec P}^{\, \prime} 
\mid {\vec \epsilon}_\lambda \cdot {\vec j}(0) \mid
\Psi \, {\vec P} \, \rangle
\right|^2 .
\label{obse63}
\end{equation}
Neglecting any polarization the differential cross section for pd
fragmentation in the lab system is
\begin{equation}
\frac{ d \sigma }{d {\hat k}_p} \ = \
(2 \pi )^4 \, \alpha \, \frac1{2 \mid {\vec Q} \mid} \,
\frac{k_p^2}{\left| \frac{k_p}{m_N} - \frac{{\vec k}_d
\cdot {\vec k}_p}{2 m_N \mid {\vec k}_p \mid }    \right|}
\, \frac12 \, \sum\limits_{ m  m_N m_d} \,
\left(
\mid N_1 \mid^2 + \mid N_{-1} \mid^2
\right) .
\label{obse64}
\end{equation}
In the nuclear matrix elements
\begin{equation}
N_{\pm 1} \equiv
\langle \Psi_{\vec q}^{(-)} \, \mid {\vec j}_{\pm 1} (0) \mid \Psi \, \rangle
\label{obse65}
\end{equation}
the final 3N scattering state depends on the asymptotic Jacobi momentum
${\vec q}$ expressed in terms of the final lab  momenta ${\vec k}_p$ and ${\vec k}_d$ as
\begin{equation}
{\vec q} \equiv \frac23 \left( {\vec k}_p - \frac12 {\vec k}_d \right) .
\label{obse66}
\end{equation}

In the case of the 3N breakup the unpolarized cross section is
\begin{equation}
\frac{ d^5 \sigma }{d {\hat k}_1 \, d {\hat k}_2 \, dS} \ = \
\frac{2 \pi^2 \, \alpha}{E_\gamma} \,
\frac12 \, \sum\limits_{ m  m_1 m_2 m_3} \,
\left(
\mid N_1 \mid^2 + \mid N_{-1} \mid^2
\right)
\,
\frac{ m_N^2 \mid {\vec k}_1 \mid \mid {\vec k}_2 \mid}
{
\sqrt{
{\left( 1- \frac{{\vec k}_2 \cdot {\vec k}_3}{k_2^2} \right)^2} \ + \
{\left( 1- \frac{{\vec k}_1 \cdot {\vec k}_3}{k_1^2} \right)^2}
}
} .
\label{obse67}
\end{equation}
Now the nuclear matrix element depends of course on the asymptotic
Jacobi momenta ${\vec p}$    and  ${\vec q}$ related to the lab momenta
as given before in (\ref{jacobip}) and (\ref{jacobiq}).

The cross sections for the  semi-exclusive processes
${}^3{\rm He}(\gamma,p)$ and ${}^3{\rm He}(\gamma,n)$
in the 3N breakup are
\begin{equation}
\frac{ d^3 \sigma }{d {\hat k}_1 \, d E_1} \ = \
\frac{2 \pi^2 \alpha }{E_\gamma} \, m_N^2 \, \frac12 \,
\mid {\vec k}_1 \mid \,
\mid {\vec p} \mid \, {\cal C} \,
\int d {\hat p} \,
\frac12 \, \sum\limits_{ m  m_1 m_2 m_3} \,
\left(
\mid N_1 \mid^2 + \mid N_{-1} \mid^2
\right) ,
\label{obse68}
\end{equation}
where $\mid {\vec p} \mid $ and ${\hat p}$ are the 
magnitude and direction of the relative
momentum between the undetected nucleons 2 and 3 and  ${\cal C}$ as given 
before. 

The availability of high intensity polarized $\gamma$ sources~\cite{tunl}  
made it possible to measure  semiexclusive reactions 
 with a linearly polarized incoming 
$\gamma$ beam and even  with a polarized $^3$He target. 
 Due to the 
polarization of the incoming $\gamma$'s and/or of the $^3$He target, 
the energy spectrum of the outgoing nucleon taken  at a particular  
polar lab angle $\theta$  depends on the 
azimuthal angle $\phi$, leading 
 to an asymmetry of the measured cross sections. 
 Assuming  that the incoming $\gamma$'s are 
linearly polarized along the x-axis 
with a  nonzero component $P_0^{\gamma}$  and that the $^3$He target  
nucleus  is polarized along the y-axis with  polarization $P_0^{^3He}$, 
one obtains for the cross section measured with a nucleon detector 
 placed at angles ($\theta,\phi$): 
\begin{eqnarray}
\sigma^{pol}_{\gamma,^3He} &=&  \sigma^{unpol}_{\gamma,^3He} [ 1 + 
P_0^{\gamma} ~ \cos (2\phi) ~ A_x^{\gamma} ~ + ~ 
 P_0^{^3He} ~ \cos (\phi) ~ A_y^{^3He} ~ + ~ \cr
&~&P_0^{\gamma}~ \cos (2\phi) ~ P_0^{^3He} ~ \cos (\phi) ~C_{x,y}^{\gamma,^3He} 
~ + ~  
P_0^{\gamma} ~ \sin (2\phi) ~ P_0^{^3He} ~ \sin (\phi) ~C_{y,x}^{\gamma,^3He}
 ].
\label{eq1}
\end{eqnarray}

The analyzing powers $A_x^{\gamma}(\theta)$, $A_{y}^{^3He}(\theta)$ 
and spin correlation coefficients $C_{x,y(y,x)}^{\gamma,^3He}(\theta)$  
are expressed through the nuclear matrix element 
$N_{m_1 m_2 m_3, \lambda  m} \equiv N_{m_i, \lambda  m}$  by: 
\begin{eqnarray}
A_x^{\gamma}(\theta)&\equiv& {{ \sum_{m_i m} (2\Re\lbrace N_{m_i, -1 m}
N_{m_i, +1 m}^*\rbrace) }
\over{\sum_{m_i m}
({\vert N_{m_i, +1 m} \vert }^2 + {\vert N_{m_i, -1 m} \vert }^2 ) }} \cr
A_y^{^3He}(\theta)&\equiv& 
{ { \sum_{m_i} (-2\Im\lbrace N_{m_i, -1  -{1\over{2}} }
N_{m_i, -1 {1\over{2}} }^*\rbrace - 2\Im\lbrace N_{m_i, +1 -{1\over{2}} }
N_{m_i, +1 {1\over{2}} }^*\rbrace )   }
\over{\sum_{m_i m}
({\vert N_{m_i, +1 m} \vert }^2 + {\vert N_{m_i, -1 m} \vert }^2 ) }} \cr
C_{x,y}^{\gamma,^3He}(\theta)&\equiv& 
{ { \sum_{m_i} (-2\Im\lbrace N_{m_i, -1 -{1\over{2}} }
N_{m_i, +1 {1\over{2}} }^*\rbrace + 2\Im\lbrace N_{m_i, -1 {1\over{2}} }
N_{m_i, +1 -{1\over{2}} }^*\rbrace )   }
\over{\sum_{m_i m}
({\vert N_{m_i, +1 m} \vert }^2 + {\vert N_{m_i, -1 m} \vert }^2 ) }} \cr
C_{y,x}^{\gamma,^3He}(\theta)&\equiv& 
{ { \sum_{m_i} (2\Im\lbrace N_{m_i, -1 -{1\over{2}} }
N_{m_i, +1 {1\over{2}} }^*\rbrace + 2\Im\lbrace N_{m_i, -1 {1\over{2}} }
N_{m_i, +1 -{1\over{2}} }^*\rbrace )   }
\over{\sum_{m_i m}
({\vert N_{m_i, +1 m} \vert }^2 + {\vert N_{m_i, -1 m} \vert }^2 ) }} ,
\label{eq2}
\end{eqnarray}
where $m$ is the spin 
projection of the $^3$He target and $m_i$ are the spin 
projections of the outgoing nucleons.

Finally we come to the Nd capture process. The angular distribution
of the photon in the system of total momentum zero neglecting
any polarization is
\begin{equation}
\frac{ d\sigma }{d {\hat Q}} \ = \
(2 \pi)^2 \alpha \frac16 \,
\sum\limits_{ m  m_N m_d} \,
\left(
\mid N_1 \mid^2 + \mid N_{-1} \mid^2
\right) \,
\frac{2 m_N Q}{3 \mid {\vec q}_0 \mid} ,
\label{69}
\end{equation}
where $ {\vec q}_0$ is the relative nucleon-deuteron 
momentum in the initial state,
which also defines the $z$-direction.

For this  reaction vector and tensor analyzing powers have been
measured. This comprises the cases that the initial proton is
polarized perpendicular to the scattering plane or the initial
deuteron is vector or tensor polarized. Now a 
more detailed notation for the nuclear matrix element is needed, namely 
\begin{equation}
N_{ \lambda m , m_N m_d } \, \equiv \,
\langle
\Psi_{^3{\rm He}} m {\vec Q} \mid {\vec \epsilon}_\lambda \cdot {\vec j} (0)
\mid \Psi_{{\vec q}_0} \, m_N m_d \, \rangle
\label{70}
\end{equation}
showing explicitly the dependence on the spin magnetic quantum
numbers $m$, $m_N$, $m_d$ and $\lambda$ of $^3$He, the nucleon, the deuteron
and the photon, respectively.
Then according to the standard formalism \cite{ourreport} one obtains the
nucleon analyzing power $A_y$ as
\begin{equation}
A_y =
i \sqrt{2} \, \frac{\sum\limits_{m_N, m_N^\prime , m_d , \lambda, m}
\sqrt{2} \, (-1)^{\frac12 - m_N} \, C\left(  \frac12 , 
\frac12 , 1 ; m_N^\prime , -m_N , 1 \right) \,
N_{\lambda , m , m_N , m_d } \, N^\star_{\lambda , m , m_N^\prime , m_d }
}
{\sum\limits_{m_N, m_d , m} \left( \mid N_{+1} \mid^2 
+ \mid N_{-1} \mid^2 \right) } ,
\label{obse71}
\end{equation}
the  deuteron vector analyzing power $i T_{11}$
\begin{equation}
 i T_{11} =
i \, \frac{\sum\limits_{m_N, m_d, m_d^\prime , \lambda, m}
\sqrt{3} \, (-1)^{1 - m_d} \, C\left(  1 , 1 , 1 ; 
m_d^\prime , -m_d , 1 \right) \,
N_{\lambda , m , m_N , m_d } \, N^\star_{\lambda , m , m_N , m_d^\prime }
}
{\sum\limits_{m_N, m_d , m} \left( \mid N_{+1} \mid^2 + 
\mid N_{-1} \mid^2 \right) } ,
\label{obse72}
\end{equation}
and the  deuteron tensor analyzing powers $ T_{j k}$
\begin{equation}
 T_{j k} =
\frac{\sum\limits_{m_N, m_d, m_d^\prime , \lambda, m}
\sqrt{3} \, (-1)^{1 - m_d} \, C\left(  1 , 1 , j ; m_d^\prime , 
-m_d , k \right) \,
N_{\lambda , m , m_N , m_d } \, N^\star_{\lambda , m , m_N , m_d^\prime }
}
{\sum\limits_{m_N, m_d , m} \left( \mid N_{+1} \mid^2 
+ \mid N_{-1} \mid^2 \right) } .
\label{obse73}
\end{equation}
Similarly to the unpolarized cross section these observables can be 
 parametrized by e.g. 
 the c.m. scattering angle of the outgoing photon against the direction
of the incoming deuteron.

\section{The Performance}
\label{performance}

This is the most central but also likely less pleasant part of
this review. Here we would like to indicate the way we evaluate the
nuclear matrix elements, which requires the solution of various
types of Faddeev-like equations.

Let us regard the electron induced 
pd breakup matrix element (\ref{Npd}),
where the state $\mid \tilde{U} \rangle $
obeys the Faddeev like equation (\ref{Utilde.2}).
Neglecting all rescattering $\mid \tilde{U} \rangle $
reduces to the driving term in (\ref{Utilde.2}).
This leads to the nuclear matrix element in 
the symmetrized plane wave approximation, denoted as PWIAS,
\begin{equation}
N^{pd, {\rm PWIAS}} = \frac12 \,
\langle \phi_{q'} \mid P (1 + P) {\cal O} \mid \Psi \rangle
= \langle \phi_{q'} \mid (1 + P) {\cal O} \mid \Psi \rangle  .
\label{per1}
\end{equation}
The second equality is due to the identity (\ref{identity1}).
Now we have to insert explicit choices for the operator ${\cal O}$.
In the case of $N_0$ the density operator appears while $N_{\pm 1}$ is driven
by the transversal pieces of the vector current. We start with
the single nucleon contributions for the density and the vector current.
Since ${\cal O}$ is fully symmetrical and
$ \langle \phi_{q'} \mid (1 + P) $ as well as $ \mid \Psi \rangle $ are fully
antisymmetrical, it is sufficient to choose the operators acting
on one nucleon, say nucleon 1,  and multiply that matrix element
by the factor 3. Thus still not specifying the component of $j^{\mu}_{SN}$, one
has in the very first step
\begin{equation}
N^{pd, {\rm PWIAS}} \ \equiv \
3 \, \langle \phi_{q'} \mid (1 + P) j_{SN} (1) \mid \Psi \rangle .
\label{per2}
\end{equation}
The nonrelativistic 3N states are conventionally expressed in terms
of Jacobi momenta $ {\vec p}$ and $ {\vec q}$ as defined in (\ref{jacobip})
and (\ref{jacobiq}) for one choice of the two-nucleon subsystem.
Thus inserting completeness relations one obtains
\begin{equation}
N^{pd, {\rm PWIAS}} \ = \
3 \, \int \langle \phi_{q'} \mid (1 + P) \mid 
{\vec p}^{\, \prime} \, {\vec q}^{\, \prime} \, \rangle
\langle  {\vec p}^{\, \prime} \, {\vec q}^{\, \prime} \mid j_{SN} ( 1 )
 \mid {\vec p} \, {\vec q} \, \rangle
\langle  {\vec p} \, {\vec q} \mid \Psi \rangle ,
\label{per3}
\end{equation}
where of course integration  over $ {\vec p}$ ,
$ {\vec q}$,
${\vec p}^{\, \prime}$,
$ {\vec q}^{\, \prime}$ is assumed.
The free states  $ \mid {\vec p} \, {\vec q} \, \rangle $
      also include spin and isospin magnetic quantum
 numbers for the three nucleons but for the sake of a simpler notation
 we dropped that information and the accompanying discrete sums.
Now because of the overall $\delta$-function
$\delta ( {\vec P}^{\, \prime} - {\vec P} - {\vec Q} ) $  
which is taken care of in evaluating the observables,
the matrix element of the single particle operator
in the space of the 3N Jacobi momenta is
\begin{equation}
\langle  {\vec p}^{\, \prime} \, {\vec q}^{\, \prime} \mid j_{SN} ( 1 )
\mid {\vec p} \, {\vec q}\, \rangle
= J( {\vec Q}, {\vec q} ) \, \delta ( {\vec p}^{\, \prime} - {\vec p} ) \,
\delta \left(  {\vec q}^{\, \prime} - {\vec q} - \frac23 {\vec Q} \right)  ,
\label{per4}
\end{equation}
and it results in
\begin{equation}
N^{pd, {\rm PWIAS}} \ = \
3 \, \int \langle \phi_{q'} \mid (1 + P) \mid {\vec p} \, {\vec q} \, \rangle
\, J( {\vec Q} , {\vec q} \, ) \,
\langle  {\vec p} \, {\vec q} - \frac23  {\vec Q} \mid \Psi \rangle .
\label{per5}
\end{equation}
Three components of $j$ occur in the response functions,
$j^0$ and $ j_{\pm 1}$. According to (\ref{sin.55}) 
the function $J^0 ( {\vec Q} ,{\vec q})$
related to the density operator is in lowest order 
\begin{equation}
J^0 ( {\vec Q} , {\vec q} \, ) = G_E^p ( {\vec Q} ) \, \Pi^p \ 
+ \ G_E^n ( {\vec Q} ) \, \Pi^n ,
\label{per6}
\end{equation}
and according to (\ref{sin.6}) the functions related to 
the spherical components
of the spin current are
\begin{equation}
J_{\pm 1}^{spin} ( {\vec Q} , {\vec q} \, ) = \frac{i}{2 m} \,
\left( G_M^p ( {\vec Q} ) \, \Pi^p \ + \ G_M^n ( {\vec Q} ) \, \Pi^n \right) \,
\left( {\vec \sigma} (1) \times {\vec Q} \right)_{\pm 1} .
\label{per7}
\end{equation}
In the case of the convection current (see (\ref{sin.6}) ) one 
expresses the individual nucleon momenta by the Jacobi momentum $\vec q$ 
\begin{equation}
\frac{{\vec k}_1 + {\vec k}^{\, \prime}_1 }{2 m_N} =
\frac{\frac23 {\vec P} + {\vec Q} + 2 {\vec q} }{2 m_N} .
\label{per8}
\end{equation}
Then in the lab system and choosing the ${\hat z}$ and ${\hat Q}$
directions to coincide
only ${\vec q}$ survives for the spherical components and the corresponding
functions for the convection current are 
\begin{equation}
J_{\pm 1}^{convect} ( {\vec Q} , {\vec q} \, ) = \frac{q_{\pm 1}}{m} \,
\left( G_E^p ( {\vec Q} ) \, \Pi^p \ + \ G_E^n ( {\vec Q} ) \, 
\Pi^n \right) .
\label{per9}
\end{equation}

The bra state in (\ref{per1})-(\ref{per3})  and (\ref{per5}) 
composed of a deuteron and a state of free relative motion of nucleon 1
 and  the deuteron is
\begin{equation}
\langle \phi_{\vec q'} \mid {\vec p} \, {\vec q} \rangle =
\langle \varphi_d \mid {\vec p} \, \rangle \, \delta \left(
{\vec q'} - {\vec q}
\right) .
\label{per10}
\end{equation}

The permutations $P$ are most conveniently evaluated as described in 
 \cite{WGbook} (here we drop the spin-isospin parts for simplicity)
\begin{equation}
P \mid  {\vec p} \, {\vec q} \, \rangle _1 =
 \mid  {\vec p} \, {\vec q} \, \rangle _2
\ + \
 \mid  {\vec p} \, {\vec q} \, \rangle _3
\ = \
\left| \, -\frac12 {\vec p} -\frac34 {\vec q} ,  {\vec p} 
-\frac12 {\vec q} \, \right\rangle _1
\ + \
\left| \, -\frac12 {\vec p} +\frac34 {\vec q} , -{\vec p} 
-\frac12 {\vec q} \, \right\rangle _1 .
\label{per11}
\end{equation}
In (\ref{per11}) we added subscripts. The subscript "1" 
indicates that ${\vec p}$
refers to the subsystem (23) and ${\vec q}$ is the relative 
momentum of particle 1
in relation to the pair (23). This choice appears in (\ref{per5}). Now
$\mid  {\vec p} \, {\vec q} \rangle _2 \ (\mid  
{\vec p} \, {\vec q} \rangle _3)$
signifies that the momenta did not change but they refer to different 
two-body subsystems. The particles
are cyclically (anticyclically) permuted, thus "2" points to the
subsystem (31) and "3" to the subsystem (12). In the second equality
the Jacobi momenta  of the type "2" and "3" are re-expressed 
in terms of linear combinations of the Jacobi momenta of the type "1". Therefore, 
one can again use (\ref{per10}) to evaluate
$\langle \phi_{\vec q'} \mid P \mid {\vec p} \, {\vec q} \rangle $ .

The second part of $\mid \tilde{U} \rangle$ in (\ref{Utilde.2}) depends
on the solution $\mid \tilde{U} \rangle$ of that
equation. In the future it might be advisable to
solve that equation directly in vector variables. First steps in
that direction have been already undertaken \cite{Liu04}.
We still work using partial wave decomposition and would like
to indicate some formal structures. A complete set of basis states
for three nucleons is
\begin{equation}
\mid p \, q \, \alpha \, \rangle
\ \equiv \
\mid  p \, q \, (l s ) j \, (\lambda \frac12 ) I \, 
( j I ) J m ; ( t \frac12 ) T m_T \, \rangle ,
\label{per12}
\end{equation}
where $p$   and $q$ are the magnitudes of Jacobi momenta,
$l$, $s$, and $j$ orbital, spin,  and total angular momentum quantum numbers
of the two-body subsystem, $\lambda$, $\frac12$, and $I$ orbital, spin, 
and total angular momentum quantum numbers of the third particle.
Then $j$ and $I$ are coupled to the total 3N angular momentum $J$.
Finally the two-body subsystem isospin $t$ is coupled with the
one of the third particle to the total isospin $T$.

Because of the identity of the nucleons not all quantum numbers
are allowed and one has the condition
\begin{equation}
(-1)^{ l + s + t } \ = \ -1 .
\label{per13}
\end{equation}
That set of basis states is complete
\begin{equation}
\sum\limits_\alpha \, \int dp p^2 \, \int dq q^2 \,
\mid p \, q \, \alpha \, \rangle
\langle p \, q \, \alpha \mid \, = \, 1 .
\label{per14}
\end{equation}
An equation like (\ref{Utilde.2}) is now projected onto those states
\begin{equation}
\langle p \, q \, \alpha \mid \tilde{U} \rangle \ = \
\langle p \, q \, \alpha \mid {\tilde{U}}^{\, 0} \rangle \ + \
\langle p \, q \, \alpha \mid t G_0 P + \dots \mid \tilde{U} \rangle .
\label{per15}
\end{equation}
We abbreviated the driving term by $\mid {\tilde{U}}^{\, 0} \rangle $
and the dots stand
for the second  part of the integral kernel.
Now $ t G_0 P $ is exactly the kernel which occurs in our standard
Faddeev like integral equation for 3N scattering 
\cite{ourreport,Ubad,ME3}.
In \cite{ourreport,ME3} that
partial wave decomposition has been displayed in all detail,
namely the evaluation of the permutation operator in the chosen basis 
$\vert p q \alpha >$, 
the solution of the Lippmann-Schwinger two-body equation leading
to the representation of $t$ in that basis, and the treatment
of the logarithmic singularities arising from the free propagator $G_0$.
Therefore, we shall not repeat all that here. Clearly one ends up with
a set of coupled integral equations in two variables for each
total angular momentum $J$, total isospin $T$, and 
parity $\pi = (-1)^{l+\lambda}$. 
The second part of the kernel, which involves the three nucleon force $V^{(1)}$
in interference with the NN $t$-operator, appears to be more complex but in fact it is easier 
for numerical treatment.
We refer for its representation to \cite{Huber94}.

Let us now come back to the driving term and regard its projection 
 on the basis states (\ref{per12}) 
in the case of a single nucleon operator  ${\cal O}$:
\begin{eqnarray}
\langle p \, q \, \alpha \mid ( 1 + P ) {\cal O} \mid \Psi \rangle \ = \
3 \, \langle p \, q \, \alpha \mid ( 1 + P ) j_{SN} (1) \mid \Psi \rangle \ =
\nonumber \\
3 \, \langle p \, q \, \alpha \mid j_{SN} (1) \mid \Psi \rangle \ + \
3 \, \langle p \, q \, \alpha \mid P j_{SN} (1) \mid \Psi \rangle .
\label{per16}
\end{eqnarray}
We could extract again the factor 3 since
$\langle p \, q \, \alpha \mid ( 1 + P ) $ is fully
antisymmetrical due to the condition (\ref{per13}). 
 As an example we now show
 for the simplest case
of the density operator  the partial
wave decomposition of
$\langle p \, q \, \alpha \mid j_{SN}^0 (1) \mid \Psi \rangle $.
Comparing (\ref{per4}) and (\ref{per5}) we see that
\begin{equation}
\langle {\vec p} \, {\vec q} \mid  j_{SN}^0 (1) \mid \Psi \rangle \ = \
J^0 ( {\vec Q}) \, \langle {\vec p} \, , \, {\vec q} 
- \frac23 {\vec Q} \mid \Psi \rangle
\label{per17} .
\end{equation}
Consequently
\begin{eqnarray}
\langle p \, q \, \alpha \mid  j_{SN}^0 (1) \mid \Psi \rangle \ = \
J^0 ( {\vec Q}) \, \int \,
\langle p \, q \, \alpha \mid {\vec p}^{\, \prime} 
\, {\vec q}^{\, \prime} \, \rangle
\langle {\vec p}^{\, \prime} \, {\vec q}^{\, \prime} 
-\frac23 {\vec Q}  \mid \Psi \rangle
\nonumber \\
= \ J^0 ( {\vec Q}) \,\int \,
\langle p \, q \, \alpha \mid {\vec p}^{\, \prime} \, 
{\vec q}^{\, \prime} \, \rangle
\sum \!\!\!\!\!\!\!\int \,
\langle {\vec p}^{\, \prime} \, {\vec q}^{\, \prime} -\frac23 {\vec Q} \mid
p '' \, q'' \, \alpha'' \, \rangle \Psi_{\alpha ''} ( p '' , q'' )  .
\label{per18}
\end{eqnarray}
In the second equality we inserted the partial wave decomposition
of the 3N bound state. The wave function components
$ \Psi_{\alpha ''} ( p '' , q'' ) \equiv \langle p'' \, q'' 
\, \alpha '' \mid \Psi \rangle $
result from solving the 3N bound state Faddeev equation \cite{Nogga2003}.
The rather tedious but known steps to evaluate the overlaps between momentum
vector states with shifted vector arguments  and our partial
wave projected basis states as well as the six fold integration 
 can be carried through analytically with the techniques presented in 
\cite{WGbook}. Results for various partial wave projected 
matrix elements can be found in \cite{Kamada92,Golak95.2,Ishikawa98,ourMEC}. 
 As an example the expression in (\ref{per18}) 
results for an arbitrary direction of $\vec Q$ in
\begin{eqnarray}
\langle p \, q \,  (l s ) j \, (\lambda \frac12 ) I \, 
( j I ) J m  ; (t \frac12) T m_T \mid
j_{SN}^0 (1) \mid \Psi \, m '' \, \rangle = \nonumber \\
I(t,T,M_T) \, \sqrt{\pi} \, (-1)^j \, \sqrt{ \hat{J} } \, 
\sqrt{ \hat{I} } \nonumber \\
\sum\limits_{\alpha ''} \,
\delta_{ l '' l} \,  \delta_{ s '' s} \,  
\delta_{ j '' j} \,  \delta_{ t '' t} \,
\sqrt{ \hat{\lambda ''} } \, \sqrt{ \hat{I ''} } \,
\sqrt{ ( 2 \lambda '' + 1 ) ! } \, (-1)^{\lambda ''} \nonumber \\
\sum\limits_{\lambda_1 '' + \lambda_2 '' = \lambda ''} \,
(q)^{\lambda_1 ''} \, \left( 
\frac23 Q \right)^{\lambda_2 ''} \, (-1)^{\lambda_2 ''} \,
\frac1{\sqrt{ ( 2 \lambda_1 '')! \,  ( 2 \lambda_2 '')!}} \nonumber \\
\sum\limits_{k} \, \hat{k} \, (-1)^k \, 
\left( k \lambda_1 '' \lambda ; 0 0 0  \right) \, g_k  \nonumber \\
\sum\limits_{g} \, \left( k \lambda_2 '' g ; 0 0 0  \right)
\,
\left\{ \begin{array}{ccc}
g & \lambda & \lambda '' \\ 
\lambda_1 '' & \lambda_2 '' & k
\end{array}\right\}
\,
\left\{ \begin{array}{ccc}
I & \lambda & \frac12 \\ 
\lambda '' & I '' & g
\end{array}\right\}
\,
\left\{ \begin{array}{ccc}
g & I & I '' \\ 
j & \frac12 & J
\end{array}\right\}
\nonumber \\
\left( J g \frac12 ; m, m '' - m, m '' \right) \,
Y_{g , m '' - m } \left ( {\hat Q} \right ) ,
\label{per19}
\end{eqnarray}
where
\begin{equation}
g_k \equiv g_k \left( p , q , \mid {\vec Q} \mid \, ; \alpha '' \right) \, = \,
\int\limits_{-1}^1 \, d x \, P_k (x ) \,
\frac{\Psi_{\alpha ''} ( p , \tilde{q}) }{ {\tilde{q}}^{\lambda ''}} ,
\label{per20}
\end{equation}
with
\begin{equation}
\tilde{q} = \sqrt{ q^2 + \frac49 \mid {\vec Q} \mid^2 
- \frac43 \mid {\vec Q} \mid q x } ~~.
\label{per21}
\end{equation}
Note that we abbreviate $ {\hat a} \equiv 2 a + 1$.
The isospin factor $I(t,T,M_T)$
arising from the isospin matrix element
\begin{eqnarray}
\left. \left\langle \left( t \frac12 \right) T M_T \right| \,
G_E^p \, \frac12 \left( 1 + {\hat \tau}_z (1) \right)
\ + \
G_E^n \, \frac12 \left( 1 - {\hat \tau}_z (1) \right) \,
\left| \left( t '' \frac12 \right) T '' M_T ''  \, 
\right\rangle \right|_{T''=\frac12}
\label{per22}
\end{eqnarray}
is given as
\begin{eqnarray}
I(t,T,M_T) =
\delta_{ M_T '' M_T} \,
\delta_{ t '' t} \,
\left[
\left( G_E^p + G_E^n \right) \, \frac12 \, \delta_{ T \frac12}
\, \right.
\nonumber \\
\, - \, \left.
\left( G_E^p - G_E^n \right) \, \sqrt{3} \,
\left( 1 \frac12 T ;  0 M_T M_T \right) \, (-1)^t \,
\left\{ \begin{array}{ccc}
t & \frac12 & \frac12 \\ 
1 & T & \frac12
\end{array}\right\}
\right] .
\label{per23}
\end{eqnarray}

The corresponding expressions for the convection and spin currents are
\begin{eqnarray}
\langle p \, q \, \alpha \, J M  \, ; \, T M_T \mid
J_{\tau}^{convect} (1) \mid \Psi \, M '' \, \rangle = \nonumber \\
(-1)^\tau \, \sqrt{\pi} \, \frac{q}{m_N} \, I(t,T,M_T) \,
(-1)^{j + \lambda + 1} \, \sqrt{ \hat{J} } \, \sqrt{ \hat{I} }
\, \sqrt{ \hat{\lambda} }
\nonumber \\
\sum\limits_{\alpha ''} \,
\delta_{ l '' l} \,  \delta_{ s '' s} \,  \delta_{ j '' j} \,
\delta_{ t '' t} \,
\sqrt{ \hat{\lambda ''} } \, \sqrt{ \hat{I ''} } \,
\sqrt{ ( 2 \lambda '' + 1 ) ! } \nonumber \\
\sum\limits_{\lambda_1 '' + \lambda_2 '' = \lambda ''} \,
(q)^{\lambda_1 ''} \, \left( \frac23 Q \right)^{\lambda_2 ''} \,
\frac1{\sqrt{ ( 2 \lambda_1 '')! \,  ( 2 \lambda_2 '')!}} \nonumber \\
\sum\limits_{g_1} \, \left( \lambda 1 g_1 ; 0 0 0  \right) \
\sum\limits_{k} \, \hat{k} \, \left( k \lambda_1 '' g_1 ; 0 0 0
\right) \, g_k  \nonumber \\
\sum\limits_{g_2} \, \left( k \lambda_2 '' g_2 ; 0 0 0  \right)
\,
\left\{ \begin{array}{ccc}
g_2 & g_1 & \lambda ''\\ 
\lambda_1 '' & \lambda_2 ''& k
\end{array}\right\}
\,
Y_{g_2 , M '' - M + \tau} \left ( {\hat Q} \right )  \nonumber \\
\sum\limits_{h} \,  \sqrt{ \hat{h} } \,
\left\{ \begin{array}{ccc}
\lambda & 1 & g_1 \\ 
g_2     & \lambda ''& h
\end{array}\right\}
\,
\left\{ \begin{array}{ccc}
I & \lambda & \frac12 \\ 
\lambda '' & I '' & h
\end{array}\right\}
\,
\left\{ \begin{array}{ccc}
h & I & I '' \\ 
j & \frac12 & J
\end{array}\right\}
\nonumber \\
\left( J h \frac12 ; M, M '' - M, M '' \right) \,
\left( 1 g_2 h ; -\tau, M '' - M + \tau, M ''- M \right) \,
\label{per24}
\end{eqnarray}          
and
\begin{eqnarray}
\langle p \, q \, \alpha \, J M  \, ; \, T M_T \mid
J_{\tau}^{spin} (1) \mid \Psi \, M '' \, \rangle = \nonumber \\
(-3) \, \sqrt{\pi} \, \frac{Q}{m_N} \, \tilde{I}(t,T,M_T) \,
\sqrt{ \hat{I} } \, \sqrt{ \hat{J} } \,
(-1)^{\frac12 + I} \, (-1)^{J + \frac12} \,
\nonumber \\
\sum\limits_{\alpha ''} \,
\delta_{ l '' l} \,  \delta_{ s '' s} \,  \delta_{ j '' j}
\,  \delta_{ t '' t} \,
\sqrt{ ( 2 \lambda '' + 1 ) ! } \,
\sqrt{ \hat{\lambda ''} } \, \sqrt{ \hat{I ''} } \,
\nonumber \\
\sum\limits_{\lambda_1 '' + \lambda_2 '' = \lambda ''} \,
(q)^{\lambda_1 ''} \, \left( \frac23 Q \right)^{\lambda_2 ''} \,
\frac1{\sqrt{ ( 2 \lambda_1 '')! \,  ( 2 \lambda_2 '')!}} \nonumber \\
\sum\limits_{k} \, \hat{k} \, \left( k \lambda_1 '' \lambda
; 0 0 0  \right) \, g_k  \nonumber \\
\sum\limits_{g} \, \sqrt{ \hat{g} } \, \left( k \lambda_2 '' g ; 0 0 0  \right)  
\,
\left\{ \begin{array}{ccc}
g & \lambda & \lambda ''\\ 
\lambda_1 '' & \lambda_2 ''& k
\end{array}\right\}
\,  (-1)^{g} \nonumber \\
\sum\limits_{f} \, \hat{f}
\,
\left\{ \begin{array}{ccc}
g & \lambda & \lambda ''\\ 
\frac12 & I '' & f
\end{array}\right\}
\,
\left\{ \begin{array}{ccc}
1 & \frac12 & \frac12 \\ 
 \lambda & I & f
\end{array}\right\}
\nonumber \\
\sum\limits_{x} \, \hat{x}
\, (-1)^{f+x}
\left\{ \begin{array}{ccc}
g & f & I ''\\ 
j & \frac12 & x
\end{array}\right\}
\,
\left\{ \begin{array}{ccc}
1 & f & I \\ 
j & J & x
\end{array}\right\}
\nonumber \\
\sum\limits_{w} \, \hat{w}
\,
\left\{ \begin{array}{ccc}
1 & 1 & 1 \\ 
g & h & w
\end{array}\right\}
\,
\left\{ \begin{array}{ccc}
J & 1 & x \\ 
g & \frac12 & w
\end{array}\right\}
\nonumber \\
\sum\limits_{h} \,  \frac1{\sqrt{ \hat{h} }} \,
\left( g 1 h ; 0 0 0  \right) \,
\left( 1 w h ; \tau , M '' - M , M '' - M + \tau \right) \,
\nonumber \\
\left( J w \frac12 ; M , M '' - M , M ''\right) \,
Y_{h , M '' - M + \tau} \left ( {\hat Q} \right ) ,
\label{per25}
\end{eqnarray}                      
where in $ \tilde{I}(t,T,M_T) $ the electric nucleon form factors,
$ G_E^p$ and $G_E^n$ are replaced by
the magnetic nucleon form factors, $ G_M^p$ and $G_M^n$.  

The second piece in (\ref{per16}) including the 
permutation operators $P$ is evaluated as
\begin{equation}
\langle p q \alpha \mid P j_{SN} (1) \mid \Psi \rangle =
\sum \!\!\!\!\!\!\!\int \, \langle p q \alpha \mid P \mid 
p ' q ' \alpha ' \rangle
\langle p' q' \alpha' \mid j_{SN} (1) \mid \Psi \rangle .
\label{per26}
\end{equation}
Now the partial wave representation of $P$ can be chosen in various 
forms \cite{Huber94}.
Beside purely geometrical quantities it is always
an integral of two $\delta$-functions over the cosine of an angle,
where the arguments depend on momenta and that cosine.
The two $\delta$-functions express two of the four momenta in terms of 
the other two.
In the case of (\ref{per26}) one chooses
the form of $P$ where $p'$   and $q'$ are expressed in terms of $p$   and $q$:
\begin{equation}
\langle p q \alpha \mid P \mid p ' q ' \alpha ' \rangle
= \int\limits_{-1}^1 \, dx G_{\alpha \, \alpha'} ( p , q , x ) \,
\frac{ \delta ( p'- \pi_1 (p , q, x) ) }{ {p'}^{\, l' + 2} } \,
\frac{ \delta ( q'- \pi_2 (p , q, x) ) }{ {q'}^{\, \lambda ' + 2} }\, .
\label{per27}
\end{equation}
This then leads to
\begin{equation}
\langle p q \alpha \mid P j_{SN} (1) \mid \Psi \rangle =
\sum\limits_{\alpha '} \, \int\limits_{-1}^1 \, dx 
G_{\alpha \, \alpha'} ( p , q , x ) \,
\frac{ \langle \pi_1 ( p,q,x) \, \pi_2 ( p,q,x) \alpha ' 
\mid j_{SN} (1) \mid \Psi \rangle}
{\pi_1 ( p,q,x)^{\, l'}  \, \pi_2 ( p,q,x)^{\, \lambda '} } ,
\label{per28}
\end{equation}
where the functions $G_{\alpha \, \alpha'} ( p , q , x ), \pi_1 ( p,q,x)$ 
and $\pi_2 ( p,q,x)$ are given in \cite{Huber94}

Because one evaluates $\langle p q \alpha \mid j_{SN} (1) \mid \Psi \rangle $
on certain grids in $p$ and $q$, the evaluation of (\ref{per28}) 
requires interpolation.
We use cubic splines of two types \cite{Glockle82,Huber97}.
In this manner the driving term in (\ref{per15}) is determined on 
grids in $p$ and $q$.

We solve the set of coupled integral equations in the two variables $p$ and $q$
by iteration, generating the multiple scattering series 
for each fixed total angular momentum $J$ and parity.
We neglect the coupling of states with total isospin $T= \frac12$ 
and $T= \frac32$,
which is due to charge independence breaking for np and pp forces 
but keep both isospins $T$. The difference between
pp and np forces is, however, taken into account by applying the 
$``\frac23-\frac13''$ rule~\cite{Witala89,23-13rule}.
For the lower $J$-values (especially for $J =\frac12^+$,
the 3N bound state quantum numbers) that multiple scattering series diverges
or converges only very slowly.
For every $J^\pi$-value we sum up the series by the Pad\'e method~\cite{WGbook}  which is a 
very reliable and accurate method.
Because of the rather high dimension of the discretized integral kernel
an iterative procedure is mandatory.
Typical dimensions for the kernel are $100000 \times 100000 $ 
for each $J^\pi$-value.

Once $\langle p q  \alpha  \mid \tilde{U}\rangle $
has been determined, final integrations are required
to arrive at the nuclear matrix elements $ \langle \phi_q \mid P 
\mid \tilde{U} \rangle $
occurring in (\ref{Npd}).
In this case another form of the permutation operator is used, namely

\begin{equation}
\langle p q \alpha \mid P \mid p ' q ' \alpha ' \rangle
= \int\limits_{-1}^1 \, dx \tilde{G}_{\alpha \, \alpha'} ( q , q' , x ) \,
\frac{ \delta ( p - \tilde{\pi}_1 (q , q', x) ) }{ {p}^{\, l + 2} } \,
\frac{ \delta ( p'- \tilde{\pi}_2 (q , q', x) ) }{ {p'}^{\, l' + 2} }\, .
\label{per29}
\end{equation}
The two $\delta$-functions allow to perform the integrations  over $p$ and $p'$
and one encounters the deuteron wave function 
components $\varphi_l ( \tilde{\pi}_1 )$ $(l=0,2)$
and
$\langle  \tilde{\pi}_2 q'  \alpha ' \mid \tilde{U}\rangle $ which can 
be gained by cubic splines interpolation.
We  refer to \cite{ourreport,Ubad,ME3} and references
therein for the detailed notation.

In the case of the complete breakup one encounters the matrix elements  
 (\ref{N.4}), (\ref{N.5}) or (\ref{newuuuuuuuuuu}).
In the case of
$ \langle \varphi_0 \mid P \mid  \tilde{U}\rangle $
we use (\ref{per11}) and apply the permutation $P$ to the left. Then 
we obtain the structure
\begin{equation}
\langle {\vec p}^{\, \prime} \, {\vec q}^{\, \prime} \mid \tilde{U} \rangle
\, = \,
\sum \!\!\!\!\!\!\!\int \,
\langle {\vec p}^{\, \prime} \, {\vec q}^{\, \prime} \mid
p q \alpha \rangle
\langle p q \alpha \mid \tilde{U} \rangle ,
\label{per30}
\end{equation}
with certain linear combinations ${\vec p}^{\, \prime}$    
and ${\vec q}^{\, \prime}$
of the original final momenta ${\vec p}$   and ${\vec q}$. The overlaps
$ \langle {\vec p}^{\, \prime} \, {\vec q}^{\, \prime} 
\mid p q \alpha \rangle $
are trivially given by the very definition of the 
basis states $ \mid p q \alpha> $
in terms of geometrical quantities and spherical harmonics~\cite{romek.thesis}.

The remaining term
$ \langle \varphi_0 \mid t G_0 P \mid \tilde{U}\rangle $
requires just an  application of 
part of the kernel  in (\ref{Utilde.2})
onto $ \mid \tilde{U}\rangle $ and 
 the additional structures in (\ref{Utilde.2}) and (\ref{N.5})
are treated in a corresponding manner. We refer to 
\cite{ourreport} for more details.

The capture matrix element given in (\ref{Ncapture.2}) consists of two terms.
For the second one we need the quantity $T$, which
is part of the 3N breakup amplitude for Nd scattering \cite{ourreport}
and is determined in form of the set of functions 
$ \langle p q \alpha \mid T \rangle $.
The free propagator $G_0$ delivers a simple pole, which we treat
by subtraction. The remaining part
$ \langle \Psi \mid {\cal O} ( 1 + P ) \mid p q \alpha \rangle $
has been discussed before. The first term in (\ref{Ncapture.2})
is apparently closely related to
the first term in (\ref{Npd2}). With these relatively schematic
and brief remarks we end the description of the
performance related to matrix elements and Faddeev like integral equations.
For practitioners more is needed and we refer for details 
to \cite{ourreport,Ubad,ME3,metody3}.

Up to now we addressed the
$\mid p \, q \, \alpha \rangle $-representation of the single
nucleon current. The representation of the two-body currents is
much more complex.
In the 3N space spanned by the Jacobi momenta  $\vec p$ and    $\vec q$
the two-body current related to particles 2 and 3 has the form
\begin{equation}
\langle
{\vec p}^{\, \prime} \, {\vec q}^{\, \prime}
\mid {\vec j} (2,3) \mid {\vec p} \, {\vec q} \rangle
= \delta \left( {\vec q}^{\, \prime} - {\vec q} - \frac13 {\vec Q} \right) \,
{\vec J} \left( {\vec p}_2 , {\vec p}_3 \right) ,
\label{per101}
\end{equation}
where
\begin{eqnarray}
{\vec p}_2 \equiv {\vec k}_2^{\, \prime} - {\vec k}_2
= \frac12{\vec Q} + {\vec p}^{\, \prime} -  {\vec p} ,
\nonumber \\
{\vec p}_3 \equiv {\vec k}_3^{\, \prime} - {\vec k}_3
= \frac12{\vec Q} - {\vec p}^{\, \prime} +  {\vec p} ,
\label{per102}
\end{eqnarray}
are the momentum transfers to nucleons 2 and 3.
The photon momentum $\vec Q$ occurs due to the overall  momentum
conserving $\delta$-function. The structure (\ref{per101}) shows  that the
$\vec p$ and $\vec q$ dependence is separated, what simplifies the
partial wave decomposition quite substantially. Let us split
the basis states as
\begin{equation}
\mid p \, q \, \alpha \rangle =
\mid p \, q \, (l s ) j (\lambda \frac12 ) I  ( j I ) J m 
\rangle \mid ( t \frac12 ) T m_T \rangle
\equiv \mid p \, q \, \alpha_J \rangle \mid \alpha_T \rangle ,
\label{per103}
\end{equation}
and introduce
\begin{equation}
{\cal Y}_{j \mu} (\hat p ) \equiv \sum\limits_{\mu '} C (l s j ; 
\mu - \mu ', \mu ', \mu )
Y_{l,\mu -\mu '} (\hat p) \mid s \, \mu ' \rangle ,
\label{per104}
\end{equation}
and
\begin{equation}
{\cal Y}_{\lambda \nu} (\hat q ) \equiv 
\sum\limits_{\nu '} C (\lambda \frac12 I ; \nu - \nu ', \nu ', \nu )
Y_{\lambda,\nu -\nu '} (\hat q) \mid \frac12 \, \nu ' \rangle .
\label{per105}
\end{equation}
Then
\begin{eqnarray}
\left\langle  p' q' \alpha '\left| {\vec j} ( 2,3 ) \right|
p q \alpha \right\rangle &&
\nonumber \\
& =&  \int d\vec {p'_1} \int d \vec {q'_1}
 \int d\vec {p_1} \int d\vec {q_1}
\left\langle p' q' \alpha \vert \vec {p'_1} \vec {q'_1 }
\right\rangle
\left\langle \vec {p'_1 } \vec { q'_1} \left| j (2,3 )
\right| \vec {p_1} \vec {q_1} \right\rangle
 \left\langle \vec { p_1 } \vec { q_1 } \vert p q \alpha
\right\rangle
\nonumber \\
& =&  \int d\vec {p'_1} \int d \vec {q'_1}
 \int d\vec {p_1} \int d\vec {q_1}
\sum _{ \mu ' \mu } C(j' I' J' ; \mu ', M'-\mu ', M' )
C(j I J;\mu , M-\mu, M)
\nonumber \\
&\times&
 {\cal Y}_{j' \mu'} ^*
( \hat {p'_1}) { {\delta ( p'_1 - p' ) } \over {p'_1}^2 }
{\cal Y } _{I', M'-\mu '} ^* ( \hat {q'_1} )
{{\delta ( q'_1 -q') } \over { q '_1 } ^2 }
\delta ( \vec {q_1} -\vec {q'_1} - {1 \over 3 } \vec Q )
\nonumber \\
&\times&\left\langle \alpha _{T'} \left|
 {\vec{j} \left( \frac12 \vec{Q}+\vec{p}_1\;'-\vec{p}_1,\:
                    \frac12 \vec{Q}-\vec{p}_1\;'+ \vec{p}_1;\;2,3 \right)}
\right| \alpha_T \right\rangle
\nonumber \\
&\times&
{\cal Y}_{j \mu}
( \hat {p_1})
{\cal Y } _{I, M-\mu} ( \hat {q_1} )
{ {\delta ( p_1 - p ) } \over {p_1}^2 }
{{\delta ( q_1 -q) } \over { q_1 } ^2 } .
\label{per106}
\end{eqnarray}
Here ${\vec p}_1 , {\vec q}_1$ and ${\vec p}_1^{\, \prime} , 
{\vec q}_1^{\, \prime}$
are Jacobi momenta of the type (23), what leads immediately to
\begin{eqnarray}
\left\langle  p' q' \alpha ' \left| \vec j ( 2,3 ) 
\right| p q \alpha \right\rangle &&
\nonumber \\
&=&
\sum _{ \mu ' \mu } C(j' I' J' ; \mu' , M' -\mu', M')
C(j I J;\mu , M-\mu, M)
\nonumber \\
&\times&
\vec I_2 ( p',p,Q; (l's')j' \mu ' \alpha_{T'},(ls)j \mu \alpha_T )
\nonumber \\
&\times&
I _3( q',q,Q ; (\lambda' {1\over2}) I' M'-\mu ' 
,( \lambda {1 \over 2}) I M- \mu ) ,
\label{per107}
\end{eqnarray}
with
\begin{eqnarray}
\vec
I_2 (  p',p,Q; (l's')j' \mu ' \alpha_{T'},(ls)j \mu, \alpha_T )
\nonumber \\
= \int d\hat  {p'}\int d\hat {p}  {\cal Y}_{j' \mu'} ^*
( \hat {p'})  \left\langle \alpha _{T'} \left|
{\vec{j} \left( \frac12 \vec{Q}+\vec{p}\;'-\vec{p},\:
                    \frac12 \vec{Q}-\vec{p}\;'+\vec{p};\;2,3 \right)}
\right| \alpha_T \right\rangle  {\cal Y}_{j \mu}
( \hat {p}) ,
\label{per108}
\end{eqnarray}
and
\begin{eqnarray}
I _3( q',q,Q ; (\lambda' {1\over2}) I' M'-\mu ' ,( \lambda {1 \over
2}) I M- \mu )
\nonumber \\
=
\int d\hat {q'} {\cal Y } _{I', M'-\mu '} ^* ( \hat {q'} )
{{\delta ( q -\vert \vec {q'} + {1 \over 3} \vec Q \vert ) } \over { q } ^2 }
{\cal Y } _{I, M-\mu}  ( \widehat {\vec {q'} + {1 \over 3} \vec Q} ) .
\label{per109}
\end{eqnarray}
In the nuclear matrix elements the current is applied onto the $^3$He state.
Therefore we need the basic building blocks 
$ \langle p q \alpha \mid {\vec j} ( 2,3) \mid \Psi \rangle $.
We obtain
\begin{eqnarray}
 \left\langle p' q'  \alpha ' \left| \vec j ( 2,3 )
\right| \Psi \right\rangle
= \sum _{ \alpha }
\int p^2 d p  q^2 d q
 \left\langle p' q'  \alpha ' \left| \vec j ( 2,3 )
 \right| p q \alpha \right\rangle
\left\langle p q \alpha | \Psi \right\rangle
\nonumber \\
 = \sum _{ \alpha } \int p^2 d p  \sum_{\mu \mu ' }
C(j' I' J' ; \mu' , M' -\mu', M')
C(j I J,\mu ; M-\mu, M )
\nonumber \\
\times
\vec I_2 ( p',p,Q; (l's')j' \mu ' \alpha _{T'},
(ls)j \mu \alpha_T )
\nonumber \\
\times
\tilde I _3( p, q',Q ; (\lambda' {1\over2}) I' M'-\mu ' ,( \lambda {1 \over
2}) I M- \mu ) ,
\label{per110}
\end{eqnarray}
with
\begin{eqnarray}
 \tilde I _3(p, q',Q ; (\lambda' {1\over2}) I' M'-\mu ' ,( \lambda {1 \over
2}) I M- \mu ) =
\nonumber \\
 \int d\hat {q'} {\cal Y } _{I', M'-\mu '} ^* ( \hat {q'} )
{{ \left\langle p, \vert  \vec {q'} + {1 \over 3} \vec Q \vert , \alpha \vert
 \Psi \right\rangle  }  }
{\cal Y } _{I, M-\mu}  ( \widehat {\vec {q'} + {1 \over 3} \vec Q} ){\rm .}
\label{per111}
\end{eqnarray}
The angular integration in $\tilde{I}_3$ can be performed by well established
analytical steps (see \cite{WGbook}). The much harder task is the reliable
evaluation of ${\vec I}_2$. It is convenient to decompose the current as
\begin{equation}
{\vec j} (2,3) = G_E^V \; \sum\limits_{k\kappa}
{\vec O}^{\, k\kappa} \left( {\vec p}_2 , {\vec p}_3 \right) \;
\left\{ \sigma(2) \otimes \sigma(3) \right\}_{k\kappa}
\;
{i \: {\left[ {\vec{\tau}(2)} \times {\vec{\tau}(3)} \right] }_z} .
\label{per112}
\end{equation}
The $\pi$- and $\rho$-like currents given in (\ref{j2pi}) and (\ref{j2rho})
are of that type.
The complex angular momentum algebra is detailed in \cite{ourMEC} and we refer
the reader to that reference. In \cite{ourMEC} we also evaluate those integrals
directly in a numerical manner to check the validity and accuracy
independently. Benchmark studies are displayed there, which we think
are very useful for practitioners, since the momentum space
representation of the two-body currents requires great care.

\clearpage

\section{Comparison with data}
\label{data}

Our theoretical framework is nonrelativistic. This limits the range
of data we can analyze. Unfortunately, in addition, quite a few data
are not well documented in the literature with respect to the necessary angular
and energy averaging. Therefore, a quantitative comparison of such data to
our theory is no longer possible. Under all these limitations
we are aware of only a  restricted data set, which we shall
display now.

In the following, the dynamical input for the theoretical calculations
is always the NN force AV18 alone or together
with the 3N force UrbanaIX~\cite{urIX}. 
Including the 3N force, the 
resulting binding energies for $^3$He and $^3$H 
are 7.746 and 8.476 MeV, respectively, 
which is sufficiently close to the experimental 
values (7.718109$\pm$0.000010) MeV and
(8.481855$\pm$0.000013) MeV \cite{Wapstra85}.
The AV18 potential includes electromagnetic
forces \cite{Nogga2003,AV18}.
They are all kept in our treatment of the two 3N bound states but for the 
3N continuum we keep only the strong forces.

On top of the standard single nucleon current, we employ the $\pi$-
and $\rho$-like two-body currents related to AV18. In the case 
of photodisintegration
we also show examples based on the  Siegert approximation as defined
in section~\ref{ourSiegert}. 
Technically we still rely on a partial wave decomposition
which is always converged within our typical numerical accuracy
of about 1-2 \% in the observables.

\subsection{Elastic electron scattering on 
\boldmath{$^3$}He and \boldmath{$^3$}H}

It has been known for a long time \cite{Kloet74,Barroso75} that the
3N charge and magnetic form factors
require two-body densities
and two-body currents. The two-body density is already a
relativistic correction and therefore strictly spoken already outside
our framework. Nevertheless, we follow \cite{Riskaprogress}
and use the one-$\pi$
and one-$\rho$ exchange process.
Nowadays ~\cite{Carlson98} the radial functions "$v$" are also taken 
from the $\pi$- and $\rho$-like
pieces of AV18. In all calculations the UrbanaIX 3NF is included. 

\begin{figure}[!ht]
\begin{center}
\epsfig{file=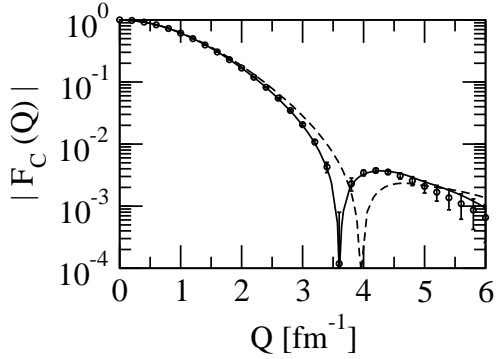,bb=10 515 300 715,clip=true,height=5cm} 
\caption{The charge form factor of $^3$H as a function 
of $Q \equiv \sqrt{Q^2}$ for the single nucleon density
given alone by $G_E$ (dashed curve) and including the Darwin-Foldy and
spin-orbit terms as well as the two-body $\pi$- and $\rho$-like 
densities (solid curve). Data are from \cite{elasticff.data}.}
\label{figFc3H}
\end{center}
\end{figure}

\begin{figure}[!ht]
\begin{center}
\epsfig{file=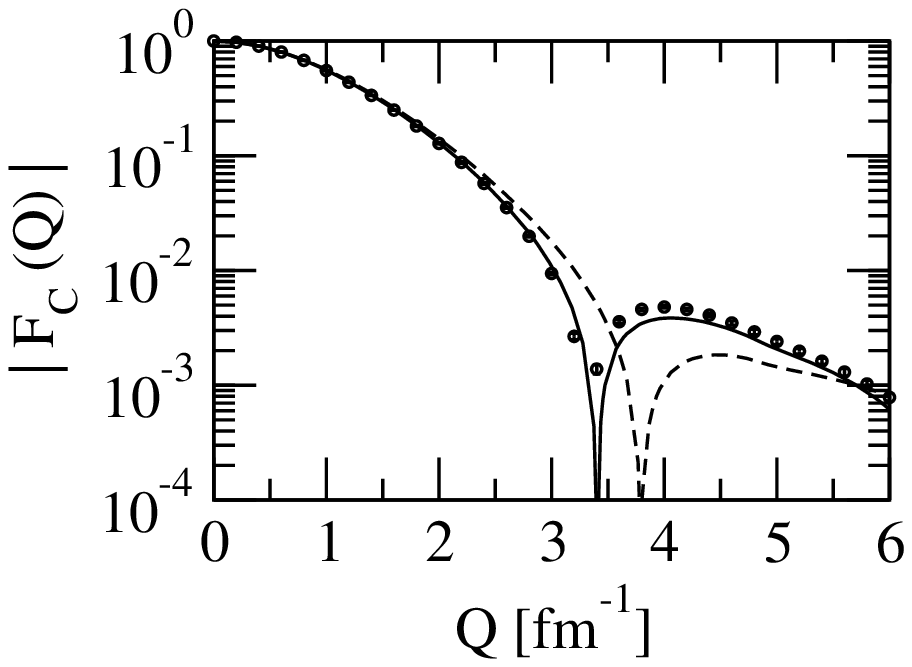,bb=10 515 300 715,clip=true,height=5cm}
\caption{The same as in Fig.~\ref{figFc3H}, now for $^3$He.
Data are from \cite{elasticff.data}.}
\label{figFc3He}
\end{center}
\end{figure}

\begin{figure}[!ht]
\begin{center}
\epsfig{file=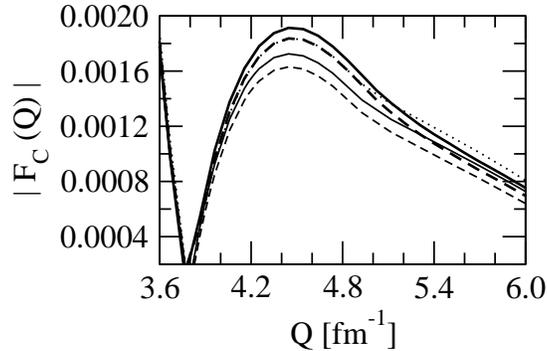,bb=10 515 300 715,clip=true,height=5cm} 
\caption{The effects of relativity in the single nucleon density 
for the charge form factor of $^3$He.
$G_E$ alone (thin dotted),
$G_E$ + Darwin-Foldy (thin dashed),
$G_E$ + Darwin-Foldy + spin-orbit (thin solid),
the first spin independent term in (70) (thick dashed),
the full relativistic density (thick solid).}
\label{fig7}
\end{center}
\end{figure}

\begin{figure}[!ht]
\begin{center}
\epsfig{file=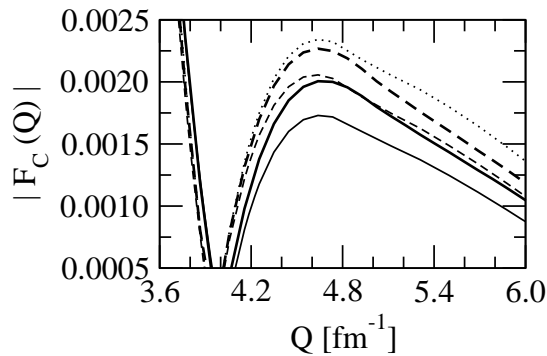,bb=10 515 300 715,clip=true,height=5cm}
\caption{The effects of relativity in the single nucleon density 
for the charge form factor of $^3$H. Curves as in Fig.~\ref{fig7}.}
\label{fig8}
\end{center}
\end{figure}

Our results for the charge form factors of $^3$He and $^3$H
are shown in Figs.~\ref{figFc3H} and \ref{figFc3He}.
The dashed curves
are  based on the single nucleon density, solely given by G$_E$ (not including
the Darwin-Foldy and spin-orbit terms as in (\ref{sin.55})). They  
start to deviate
strongly from the data for momenta above $\approx$2.5 fm$^{-1}$. The
solid curve includes in addition to the Darwin-Foldy and spin-orbit terms the
two-body $\pi-\rho$ densities. All that shifts theory rather
close into the data for $^3$H. This is also true at least up to 
about 3 fm$^{-1}$ for $^3$He.
Since in this review
we concentrate on a regime which can be called 
dominantly nonrelativistic we do not
comment on missing dynamics  responsible for  the  strong
deviations above around 3 fm$^{-1}$
in $^3$He and on the possibly accidental agreement for $^3$H. Nevertheless,
we would like
to illustrate the effects
of relativity in the single nucleon density operator in 
Figs.~\ref{fig7}-\ref{fig8}.
Since for $Q$-values below about 3 fm$^{-1}$ the changes in F$_C$ going beyond
the single density operator G$_E$ caused by relativity stay below about 5 \%,
we show only the effects for the higher $Q$-values. There we 
can choose a linear scale
and display five curves according to different choices of the 
single nucleon density operator.
We see in Fig.~\ref{fig7} for $^3$He that the Darwin-Foldy
term added to the nonrelativistic single nucleon current 
 operator ($G_E$) shifts the theory downwards,
while further adding  the spin-orbit term
reduces that downward shift. We also display the full relativistic result
according to the first term (spin independent) in (\ref{sin.3}).
In the maximum this is identical to the nonrelativistic result. Thus the
terms additional to the Darwin Foldy term cancel its contribution completely
in this case.
Finally, the complete relativistic single
nucleon density operator shifts the theory upwards beyond 
the nonrelativistic result by
about $5 \%$ in the maximum.

In the case of $^3$H both the Darwin-Foldy and the spin-orbit terms shift
theory downwards from the nonrelativistic result and the full
relativistic curve ends up below the nonrelativistic 
one by about 14 \% in the maximum.

\begin{figure}[!ht]
\begin{center}
\epsfig{file=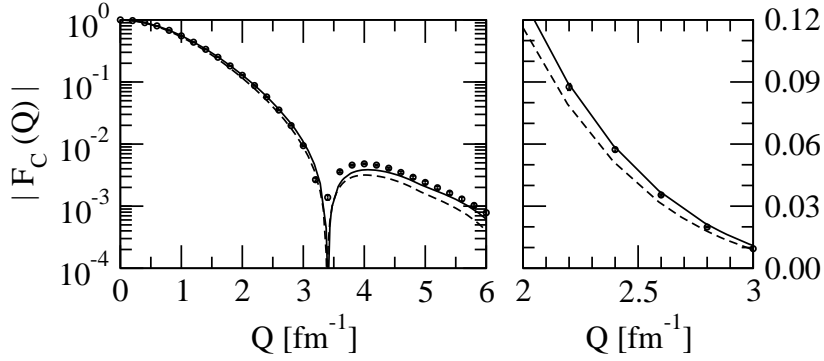,bb=10 515 475 715,clip=true,height=5cm}
\caption{The 3NF effects for the charge form factor of $^3$He. The solid curve 
 is the same as in  Fig.~\ref{figFc3He}.
For the dashed curve only the 3NF has been dropped
in the bound state wave function.
In the right panel the $Q$-range is restricted to $ 2 \le Q \le 3$ 
and the linear scale for $F_C ( Q) $ is used.}
\label{fig9}
\end{center}
\end{figure}

\begin{figure}[!ht]
\begin{center}
\epsfig{file=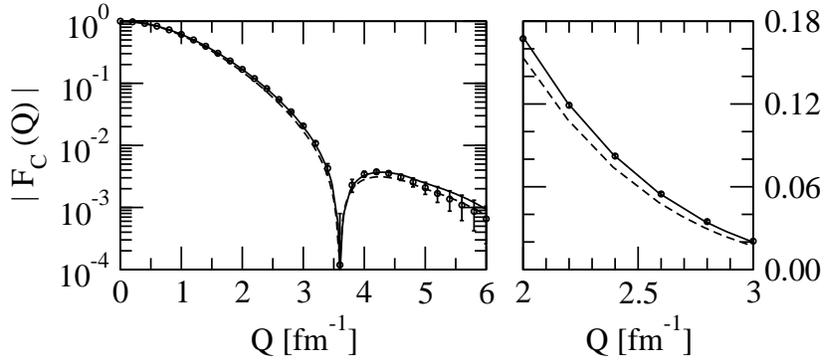,bb=10 515 475 715,clip=true,height=5cm}
\caption{The 3NF effects for the charge form factor of $^3$H.
Curves as in Fig.~\ref{fig9}.}
\label{fig10}
\end{center}
\end{figure}

After this small excursion into relativistic features, we display 
 noticeable  effects
of the 3NF in  Figs.~\ref{fig9}-\ref{fig10}.
For both nuclei, $^3$He and $^3$H, the addition of the 3NF shifts the theory
closer into 
the data for the lower $Q$-range, on which we concentrate.
For $^3$He the effects grow with $Q$ from $0$ to about 12 \% 
at $Q$= 2 fm$^{-1}$ and about 20 \% at $Q$= 3 fm$^{-1}$.
For $^3$H they are slightly smaller (17 \% at $Q$= 3 fm$^{-1}$).
3NF effects on the charge form factor have been investigated
earlier in \cite{Friar87} showing a similar tendency.

\begin{figure}[!ht]
\begin{center}
\epsfig{file=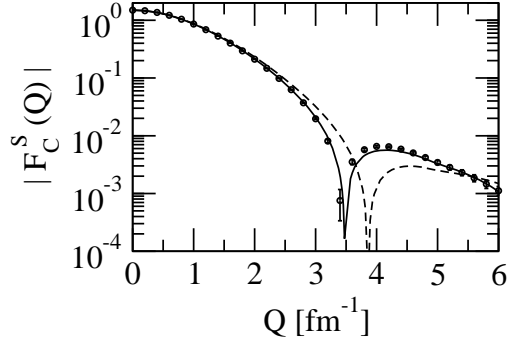,bb=10 515 300 715,clip=true,height=5cm}
\caption{The 3N isoscalar charge form factor.
Curves as in Fig.~\ref{figFc3H}.}
\label{fig11}
\end{center}
\end{figure}

\begin{figure}[!ht]
\begin{center}
\epsfig{file=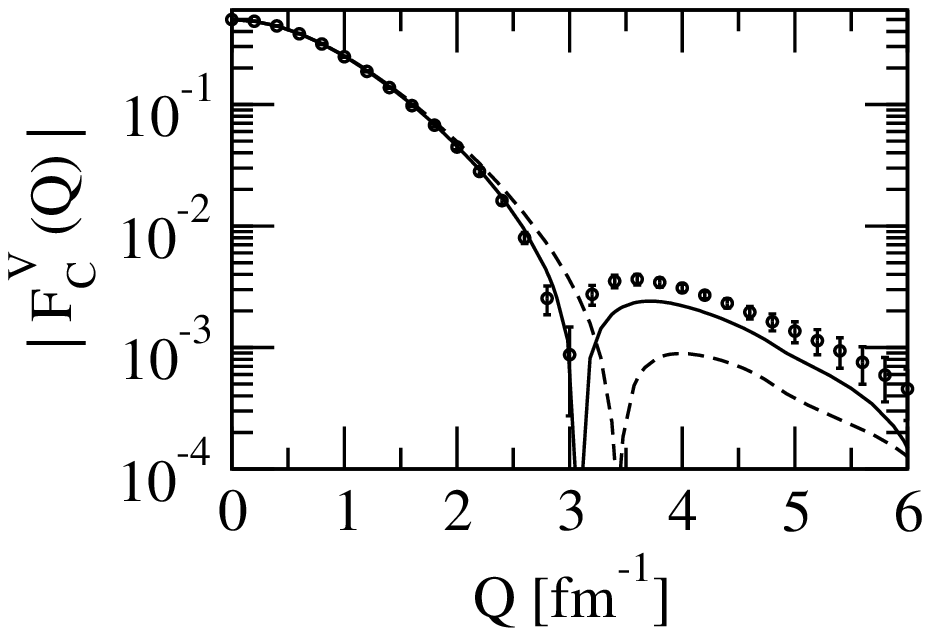,bb=10 515 300 715,clip=true,height=5cm}
\caption{The 3N isovector charge form factor.
Curves as in Fig.~\ref{figFc3H}.}
\label{fig12}
\end{center}
\end{figure}

Since at  higher $Q$-values  the comparison 
 between theory and experiment differs 
in quality for $^3$He and $^3$H, it is
common to look into the isoscalar and isovector charge form factors defined as

\begin{equation}
F_C^{S,V} = \frac12 [2 F_C(^3He) \pm F_C(^3H) ] .
\end{equation}

They are displayed in Figs.~\ref{fig11}-\ref{fig12} together with the data.
For the lower $Q$-values the agreement  with the data 
is in both cases quite good
but in the higher $Q$-range the isovector form factor, which is  sensitive
to our two-body density, underestimates the data significantly.
We refer the reader to \cite{charge2b} and \cite{Carlson98}
for further discussions on that higher $Q$-range and an inclusion of different
components of the charge density operator. 
Including additional parts in the two-body density in
\cite{charge2b} leads to a remarkably good description of the
data. Similarly, the Hanover group could describe the data
very well with a single $\Delta$-isobar admixture 
and including several selected
relativistic corrections \cite{Strueve87,Deltuva2004}.
In \cite{Marcucci98}
 the first 
 time three-nucleon currents related to the $2\pi$-exchange 3NF have been
 included. 
 Also variational Monte Carlo techniques based on realistic NN and 3N forces 
have been successfully applied and similar results for the elastic form 
factors have been achieved~\cite{wiringa91}. 
 We also would like to draw attention to the work in
\cite{Henning95} 
where relations between isoscalar charge form factors of two- and
three-nucleon systems were studied and inconsistencies were found
using the "standard" model of meson exchange currents.

\begin{figure}[!ht]
\begin{center}
\epsfig{file=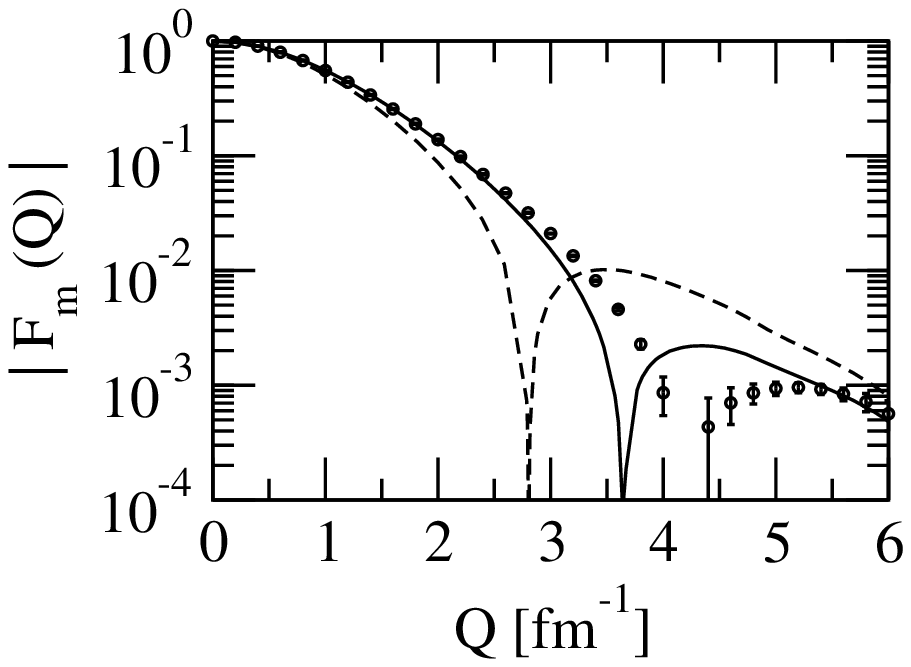,bb=10 515 300 715,clip=true,height=5cm}
\caption{The magnetic form factor of $^3$He.
         The dashed line represents the results
	 obtained with the nonrelativistic single nucleon current operator
	 from (72) (with $F_1$ replaced by $G_E$) and the solid line includes 
the effects of the $\pi$- and $\rho$-like
	 meson exchange currents. Data are from \cite{elasticff.data}.}
\label{fig13}
\end{center}
\end{figure}

\begin{figure}[!ht]
\begin{center}
\epsfig{file=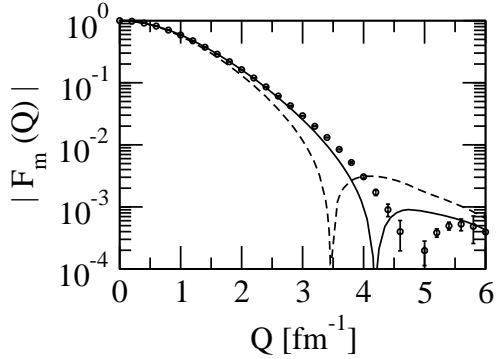,bb=10 515 300 715,clip=true,height=5cm}
\caption{The magnetic form factor of $^3$H. Curves as in Fig.~\ref{fig13}. 
 Data are from \cite{elasticff.data}.}
\label{fig14}
\end{center}
\end{figure}

\begin{figure}[!ht]
\begin{center}
\epsfig{file=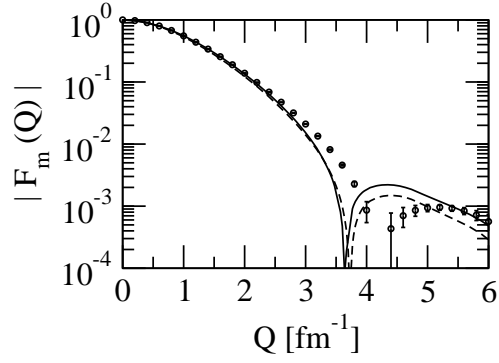,bb=10 515 300 715,clip=true,height=5cm}
\caption{The 3NF effects for the magnetic form factor of $^3$He.
The solid curve is the same as in Fig.~\ref{fig13}.
In the case of the dashed curve only the 3NF has
been dropped in the bound state wave function.}
\label{fig15}
\end{center}
\end{figure}

\begin{figure}[!ht]
\begin{center}
\epsfig{file=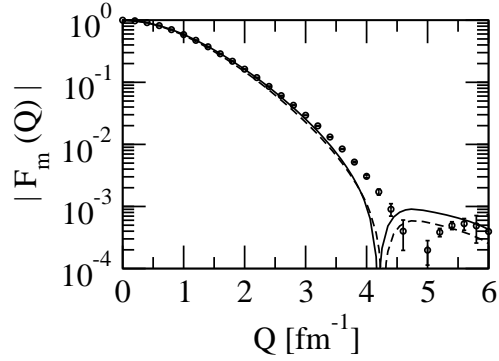,bb=10 515 300 715,clip=true,height=5cm}
\caption{The 3NF effects for the magnetic form factor of $^3$H. Curves
as in Fig.~\ref{fig15}.}
\label{fig16}
\end{center}
\end{figure}

\begin{figure}[!ht]
\begin{center}
\epsfig{file=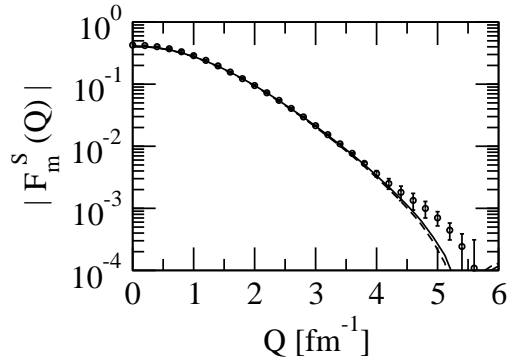,bb=10 515 300 715,clip=true,height=5cm}
\caption{The 3N isoscalar magnetic form factor.
The curves as in Fig.~\ref{fig13}.}
\label{fig17}
\end{center}
\end{figure}

\begin{figure}[!ht]
\begin{center}
\epsfig{file=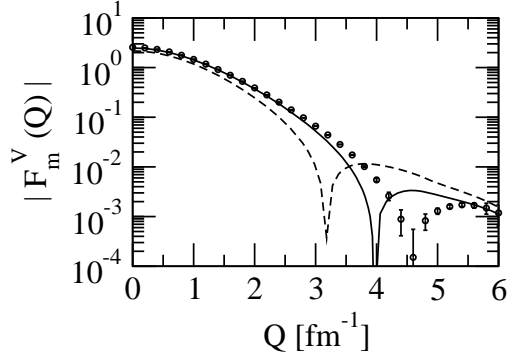,bb=10 515 300 715,clip=true,height=5cm}
\caption{The 3N isovector magnetic form factor.
The curves as in Fig.~\ref{fig13}.}
\label{fig18}
\end{center}
\end{figure}

Now we regard the magnetic form factors of $^3$He and $^3$H
in Figs.~\ref{fig13} and \ref{fig14}.
Here the situation is more demanding in
relation to the choice of the two-body current operators.
Up to about 2.5 fm$^{-1}$ the agreement with the data is quite good but
beyond that it is very insufficient. The effects of the 3NF
slightly improve the agreement in the lower $Q$-range as
displayed in Figs.~\ref{fig15} and \ref{fig16}. In Fig.~\ref{fig17}
we show the isoscalar magnetic 
form factor which is in quite good agreement with
the data up to about 4 fm$^{-1}$, while the two-body current dependent
isovector magnetic form factor is dramatically off the data in the
higher $Q$-range as shown in Fig.~\ref{fig18}.

\begin{figure}[!ht]
\begin{center}
\epsfig{file=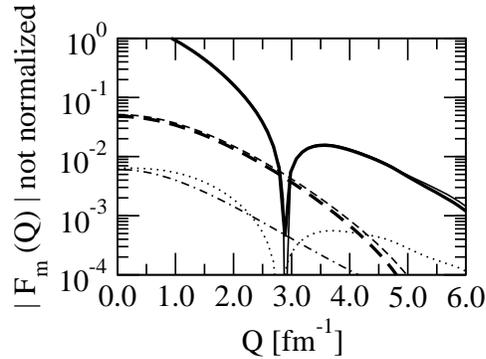,bb=10 515 300 725,clip=true,height=5cm}
\caption{The convection (thin dashed) and the spin part (thin solid)
of the nonrelativistic single nucleon current, 
and the four parts of the relativistic single nucleon current given 
 in (\ref{sin.4new})  
 (first (thick dashed),
second (dot-dashed),
third (dotted),
fourth (thick solid) ) 
for the magnetic form factor of $^3$He.
The thin and thick solid lines practically overlap.}
\label{fig19}
\end{center}
\end{figure}

\begin{figure}[!ht]
\begin{center}
\epsfig{file=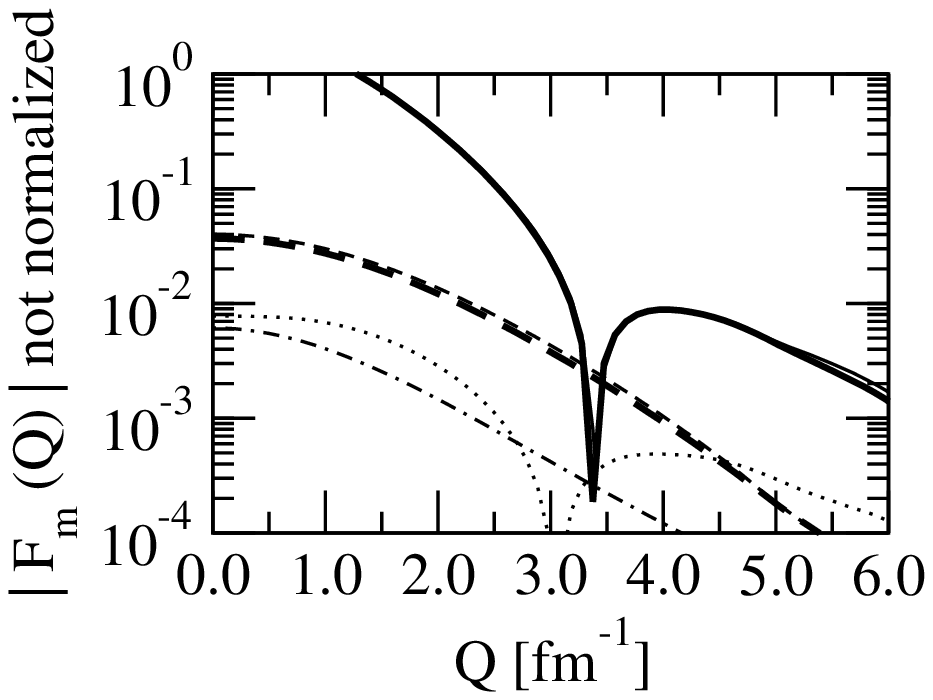,bb=10 515 300 725,clip=true,height=5cm}
\caption{The same as in Fig.~\ref{fig19} but for $^3$H.}
\label{fig20}
\end{center}
\end{figure}

Finally we come to the relativistic effects  in the single nucleon current
operator given in (\ref{sin.4}). As in all our results we choose
the laboratory frame for which the total momentum of the initial $^3$He is zero
and work with the Jacobi momenta defined in (\ref{jacobip}) 
and (\ref{jacobiq}).
Assuming that the photon couples to nucleon 1, the initial ($\vec p$)
and final (${\vec p}^{\ \prime}$) individual momenta of the struck nucleon are
given in terms of the Jacobi momentum $\vec q$ and the 
three-momentum transfer $\vec Q$ as
\begin{equation}
\vec p = \vec q - \frac23 \vec Q
\label{p1vec}
\end{equation}
\begin{equation}
{\vec p}^{\ \prime}  = \vec q + \frac13 \vec Q .
\label{p1primevec}
\end{equation}
Further we put $\vec Q \parallel {\hat z} $, which simplifies the calculation
of the spherical $\tau = \pm 1$ components of the current operator
given in (\ref{sin.4}).  Since
\begin{equation}
\left( {\vec p}^{\ \prime} +  {\vec p} \right)_\tau =  
2 \left(  {\vec q} \, \right)_\tau  ,
\label{pprimeplusp}
\end{equation}

\begin{equation}
\left( \frac{{\vec p}^{\ \prime}}{p_0 + m_N} +  
\frac{{\vec p}}{p_0^\prime + m_N} \right)_\tau =
\left( \frac{1}{p_0 + m_N} +  \frac{1}{p_0^\prime + m_N} 
\right) \, \left(  {\vec q} \, \right)_\tau ,
\label{pprimeplusp2}
\end{equation}
\begin{equation}
\left( {\vec p}  \times {\vec \sigma} \right)_\tau =
\left( {\vec \sigma} \times {\vec q} \right)_\tau -
\frac23 \left(  {\vec Q} \times {\vec \sigma} \right)_\tau ,
\label{ptimessigma}
\end{equation}
and
\begin{equation}
\left( {\vec \sigma} \times {\vec p}^{\ \prime}  \right)_\tau =
\left( {\vec \sigma} \times {\vec q} \right)_\tau +
\frac13 \left( {\vec \sigma} \times  {\vec Q} \right)_\tau ,
\label{pprimetimessigma}
\end{equation}
we rewrite (\ref{sin.4}) as
\begin{eqnarray}
J_\tau =
A \left\{ - 2 F_2 \left(  1 -
 { {{\vec p}^{\, \prime} \cdot {\vec p}} \over { ({p_0} + m_N) ({p_0}' + m_N)} }
 \right)
+ G_M  \left( { 1 \over {{p_0} + m_N} } 
+ { 1 \over {{p_0}' + m_N} } \right) \, \right\} \,  (\vec q \, )_\tau \cr
 + 2 A F_2 { { 1  } \over { ({p_0} + m_N) ({p_0}' + m_N)} } \,
 i {\vec \sigma} \cdot \left({\vec Q} 
\times {\vec q} \, \right) \, (\vec q \, )_\tau \cr
 + A G_M \left( { 1 \over {({p_0} + m_N)} }  
- { 1 \over {({p_0}' + m_N)} } \right)
 \, i \left( {\vec q} \times {\vec \sigma} \, \right)_\tau \cr
 + A G_M \left( \frac23 { 1 \over {({p_0} + m_N)} }  
+ \frac13 { 1 \over {({p_0}' + m_N)} } \right)
 \, i \left( {\vec \sigma} \times {\vec Q} \, \right)_\tau  .
		      \label{sin.4new}
		      \end{eqnarray}
The first and the last parts in (\ref{sin.4new})
correspond in the non-relativistic limit to the convection
and spin current, respectively.
The second and the third parts disappear in the non-relativistic limit
and turn out to be less important.
This is shown in Figs.~\ref{fig19} (\ref{fig20})
for $^3$He ($^3$H). We see that the convection part, nonrelativistically
and relativistically, is unimportant. It is the spin part which provides the
dominant contribution and the relativistic effects are quite insignificant.

The magnetic form factors have been studied
by other groups as well \cite{Strueve87,Marcucci98,Deltuva2004},
where more sophisticated currents
and $\Delta$-admixtures have been included. This shifts theory much closer
to the data, especially at the higher $Q$-values, which are not in the
focus of this review. Therefore, we do not comment further on all
that. Finally, we would like  to draw attention to a first attempt within the
Bethe-Salpeter approach in the Faddeev form
\cite{Rupp92}
which, however, due to severe truncations cannot yet been
conclusively confronted to data.

We end up with showing the charge radii and the magnetic
moments of $^3$He and $^3$H in Table~\ref{tabrcmm}. 
In all cases the inclusion
of the 3NF improves the description of the data. Small discrepancies
remain. The agreement in the case
of the magnetic moments is somewhat better using the enriched dynamics in
\cite{Marcucci98,Deltuva2004}.

\begin{table}
\begin{tabular}{cccccccc}
   &\multicolumn{2}{c}{$^3$He} & & 
\multicolumn{2}{c}{$^3$H}\\ \hline
   &  $r_{ch}$ [fm] & & $\mu$ & & $r_{ch}$ [fm] 
& & $\mu$ \\ \hline
without 3NF & 2.025 & & -2.054 & & 1.788 & & 2.883 \\
with 3NF & 1.932  & & -2.071 & & 1.722 & & 2.891 \\
exp.     & 1.959 $\pm$ 0.030 & & -2.127 & & 1.755 $\pm$ 0.086 & & 2.979 \\
\end{tabular}
\caption{\label{tabrcmm}The theoretical predictions including MEC and experimental values
for the $^3$He and $^3$H charge radii and magnetic moments. Data are
from \cite{Carlson98,elasticff.data} and \cite{tilley}.}
\end{table}

\subsection{Inclusive electron scattering 
on \boldmath{$^3$}He and \boldmath{$^3$}H}
\label{sub6b} 

\begin{figure}[!ht]
\begin{center}
\epsfig{file=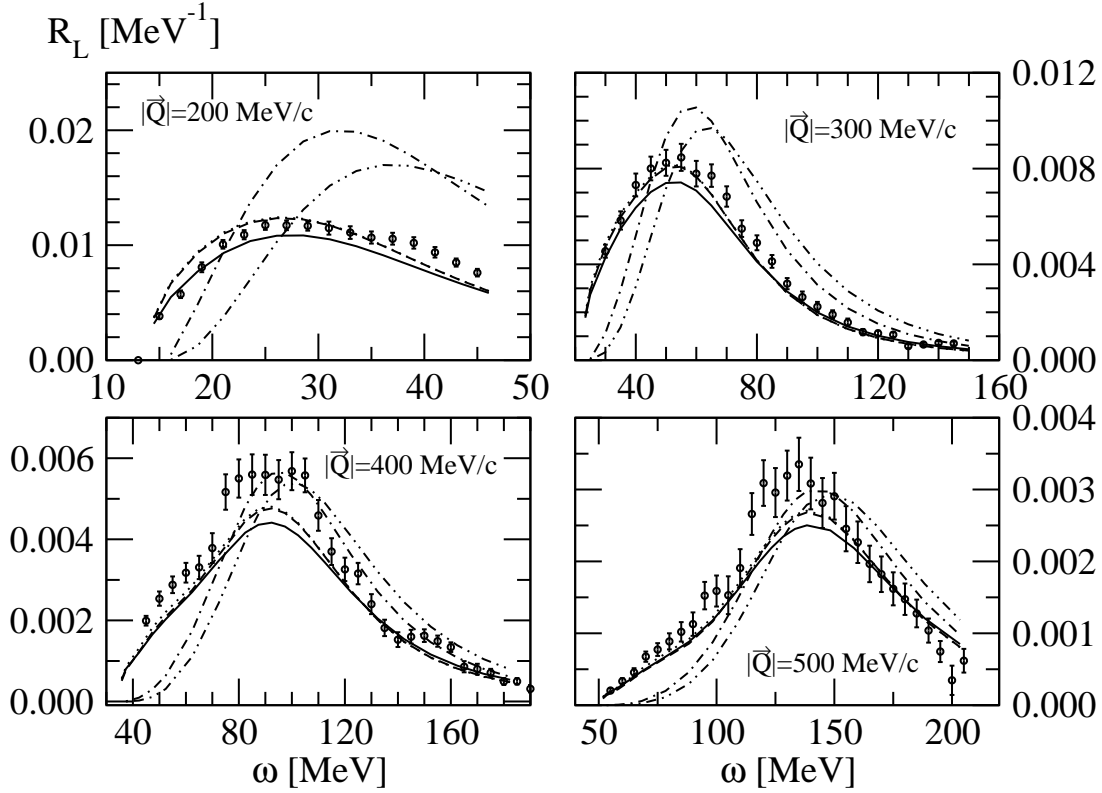,bb=40 360 560 740,clip=true,height=110mm}
\caption{The longitudinal response function of $^3$H
for different magnitudes of the three-momentum transfer.
The double-dot-dashed curve represents 
the prediction based on the extreme PWIA.
The dot-dashed curve was obtained under the assumption that FSI
acts only in one two-nucleon subsystem (the so-called FSI23),
the dotted curve takes the full FSI into account but 
neglects MEC and 3NF effects.
The $\pi$- and $\rho$-like two-body densities are accounted for 
 in the dashed curve
and finally the full dynamics including MEC and the 3NF is given 
by the solid curve.
The dotted and dashed curves practically overlap.
Data are from \cite{Dow88}.}
\label{figRL3H}
\end{center}
\end{figure}

\begin{figure}[!ht]
\begin{center}
\epsfig{file=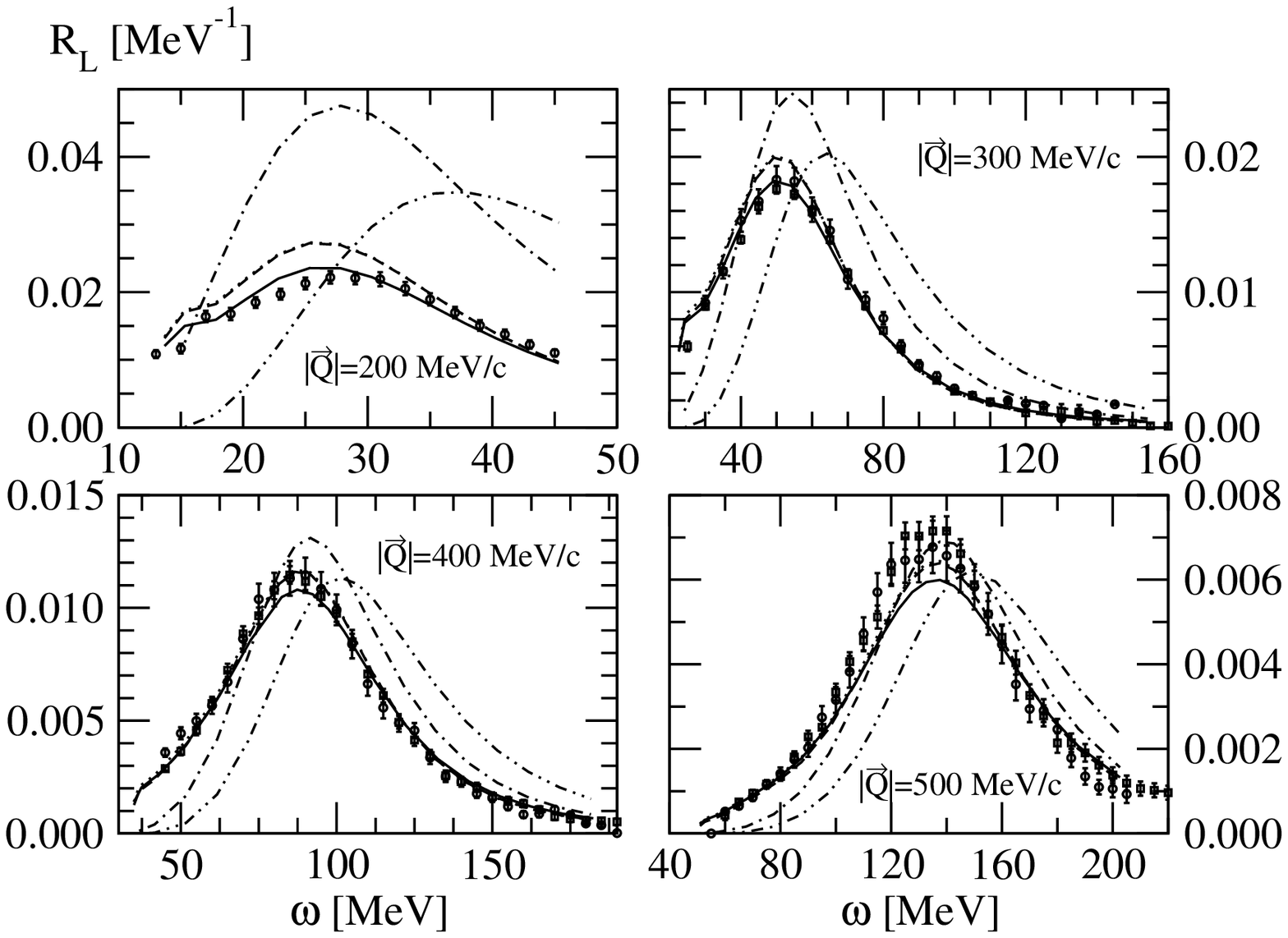,bb=40 360 560 740,clip=true,height=110mm} 
\caption{The same as in Fig.~\ref{figRL3H} for $^3$He.
Data are from \cite{Dow88} (circles) and from \cite{Marchand85} (squares).}
\label{figRL3He}
\end{center}
\end{figure}

Without polarization two response functions $R_L$ and $R_T$ 
defined in (\ref{obse37})-(\ref{obse39}) can be
measured using a Rosenbluth separation method. We 
compare in Figs.~\ref{figRL3H} and \ref{figRL3He}
for $^3$H and $^3$He 
the data to our theory for the longitudinal response function $R_L$ 
depending on the energy transfer $\omega = Q_0$ 
at the $\mid \vec Q \mid$-values
200, 300, 400 and 500 MeV/c.
As can be seen already at $\mid \vec Q \mid$= 500 MeV/c the 
experimental and theoretical
peak positions are slightly different.
This is already the result of our non-relativistic kinematics
and could be cured by improving the kinematics. We have not done that
and will concentrate on the lower $\mid \vec Q \mid$-values.

The two types of plane wave approximations, PWIA and PWIAS (not shown), 
are very much off the data
at 200 and 300 MeV/c. There the inclusion of the rescattering 
in the final state is strongly needed. 
 We would like to point out, that we distinguish between final state 
interaction effects  when the nucleons in the final state 
are interacting only through  NN forces (FSI) and  when both two- and 
three-nucleon forces are acting. In the following, we also present the results 
of the simplified treatment of FSI, where the interaction is restricted only to the spectator 
nucleons 2 and 3 (FSI23). A more detailed explanation of FSI23 is given in 
section \ref{subpredc}.  
 For both nuclei, the  FSI predictions are 
 close to the data at $\vert \vec Q \vert =$ 200 and 300 MeV/c. 
  The effects of the two-body density  are marginal in case of the $^3$H, but 
 noticeable for $^3$He. The 3NF effects 
are clearly  visible. Note that 3NF effects are taken consistently 
into account, in the bound and in the scattering states.
For the PWIA, FSI23, FSI, FSI+MEC results the 3N bound states
obtained without 3NF are used.
In the case of $^3$H the 3N force effects lower theory too 
much and lead to an
underprediction of the data while for $^3$He theory goes right 
away into the data. The
underprediction of theory in the case of $^3$H is clearly visible 
at Q= 400 MeV/c. It is also of
interest to notice the tendency that the nuclear interaction 
effects in the continuum
decrease with increasing $\mid \vec Q \mid$-values. 

\begin{figure}[!ht]
\begin{center}
\epsfig{file=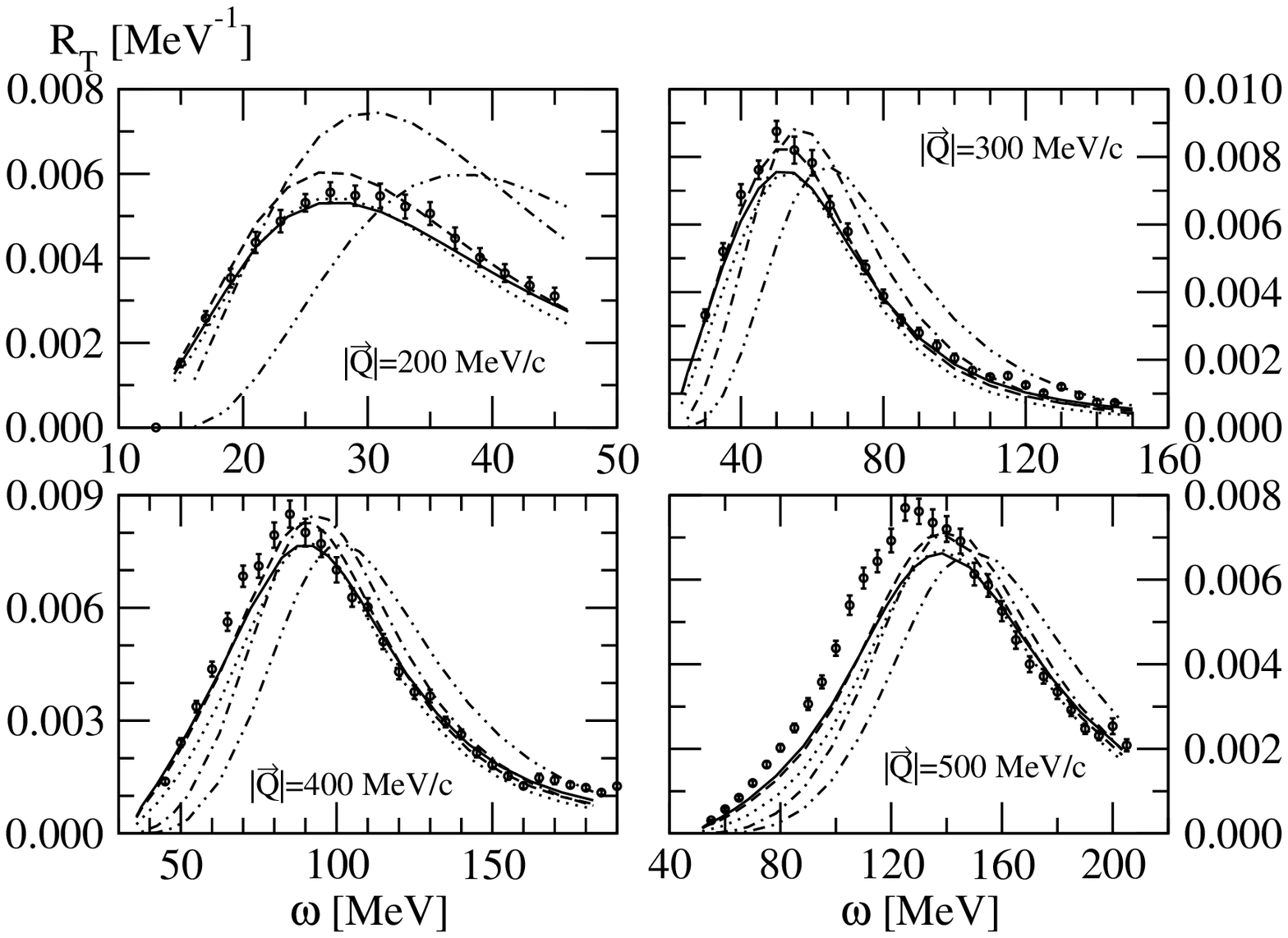,bb=40 360 560 740,clip=true,height=110mm}
\caption{The transverse response functions $R_T$ for  $^3$H.
Curves as in Fig.~\ref{figRL3H}, except that the two-body density
is replaced by the two-body current.
Data are from \cite{Dow88}.}
\label{figRT3H}
\end{center}
\end{figure}

\begin{figure}[!ht]
\begin{center}
\epsfig{file=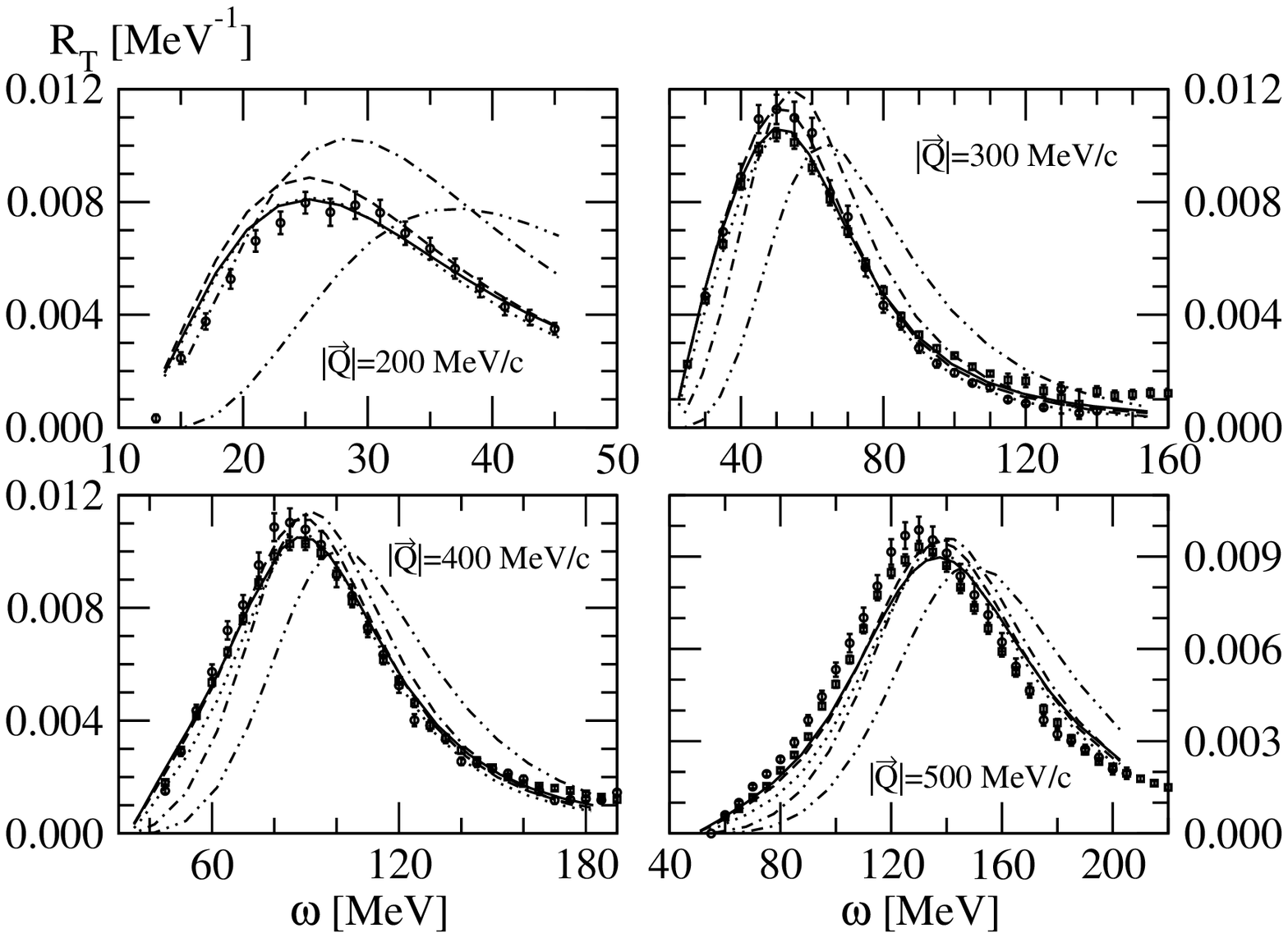,bb=40 360 560 740,clip=true,height=110mm}
\caption{The same as in Fig.~\ref{figRT3H} for $^3$He.
Data are from \cite{Dow88} (circles) and from \cite{Marchand85} (squares).}
\label{figRT3He}
\end{center}
\end{figure}

The situation in the case of the transverse response function $R_T$, shown in 
Figs.~\ref{figRT3H} and \ref{figRT3He}
is different. The tendency that the interaction effects in the
continuum decrease with increasing $\mid \vec Q \mid$-values 
starts earlier than for $R_L$. 
Further,  the MEC
effects are quite strong, as is well known, but are essentially 
compensated by the 3NF
effects in the maxima. In the lower and upper energy wings of the 
peaks the addition of the
3NF has little effect. Overall the agreement of data and theory 
for our complete
prediction (NN and 3N forces plus MEC) is quite good for both nuclei,
$^3$H and $^3$He, at $\mid \vec Q \mid$= 200, 300 and 400 MeV/c. 
 The very interesting interplay of 3NF and MEC effects
would make a renewed, more precise 
 measurement very interesting. Finally, like for $R_L$,
relativistic effects, at least the ones of kinematical origin,
are clearly visible at $\mid \vec Q \mid$= 500 MeV/c.

\begin{figure}[!ht]
\begin{center}
\epsfig{file=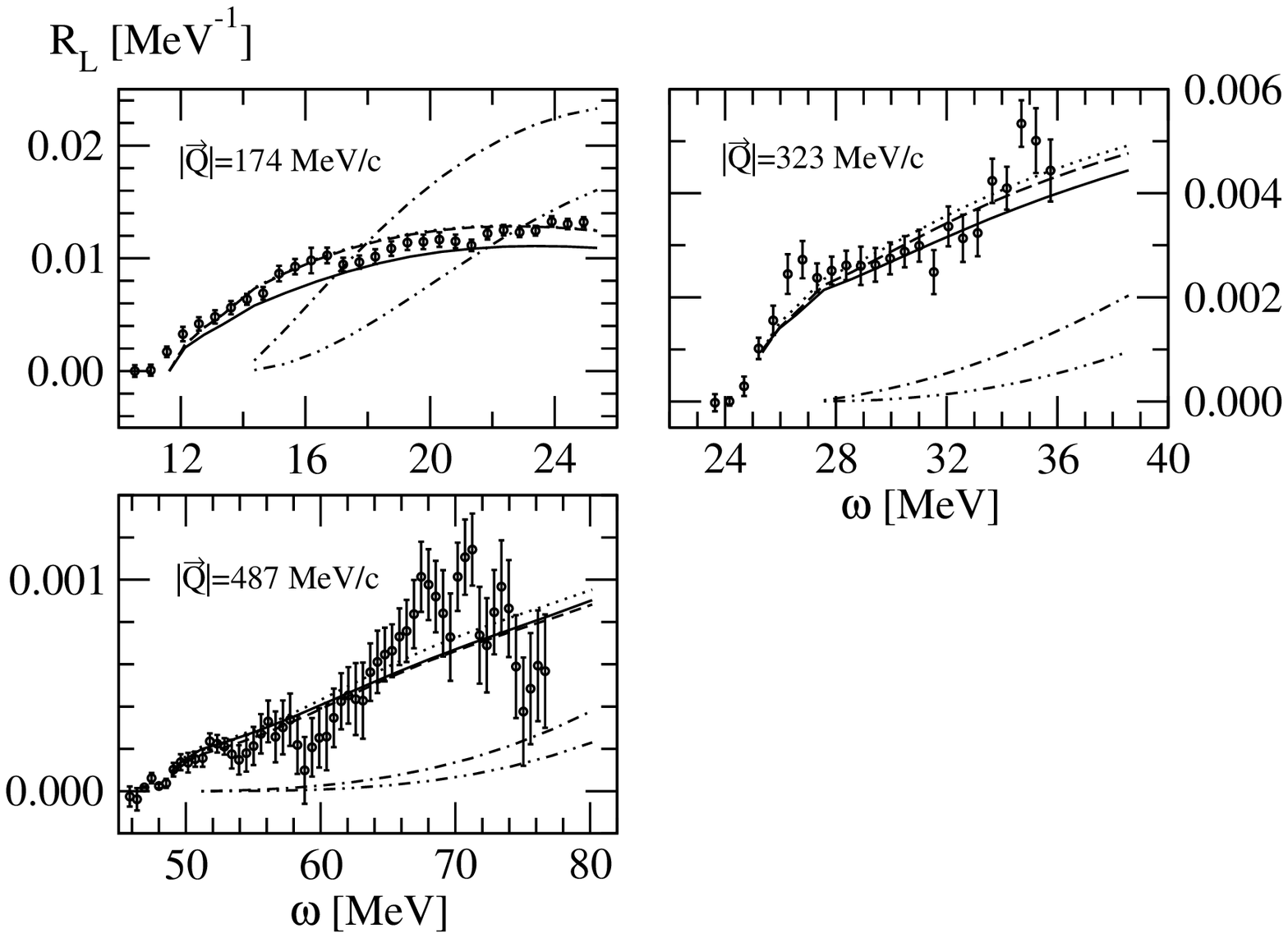,bb=40 360 560 740,clip=true,height=110mm}
\caption{The longitudinal response function $R_L$ for $^3$H.
Curves as in Fig.~\ref{figRL3H}.
Data are from \cite{Retzlaff94}.}
\label{figRetzRL3H}
\end{center}
\end{figure}

\begin{figure}[!ht]
\begin{center}
\epsfig{file=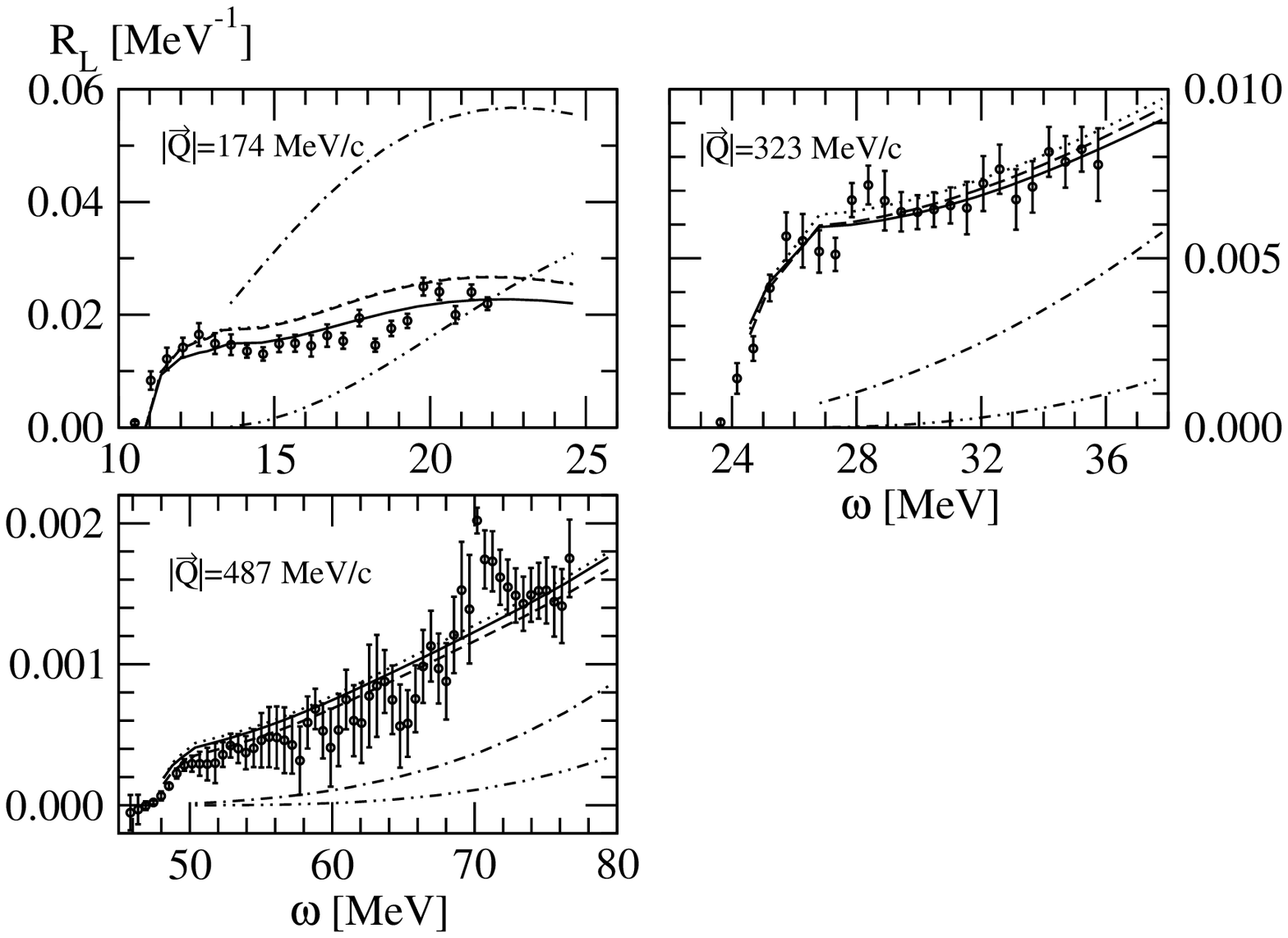,bb=40 360 560 740,clip=true,height=110mm}
\caption{The same as in Fig.~\ref{figRetzRL3H} for $^3$He.
Data are from \cite{Retzlaff94}.}
\label{figRetzRL3He}
\end{center}
\end{figure}

\begin{figure}[!ht]
\begin{center}
\epsfig{file=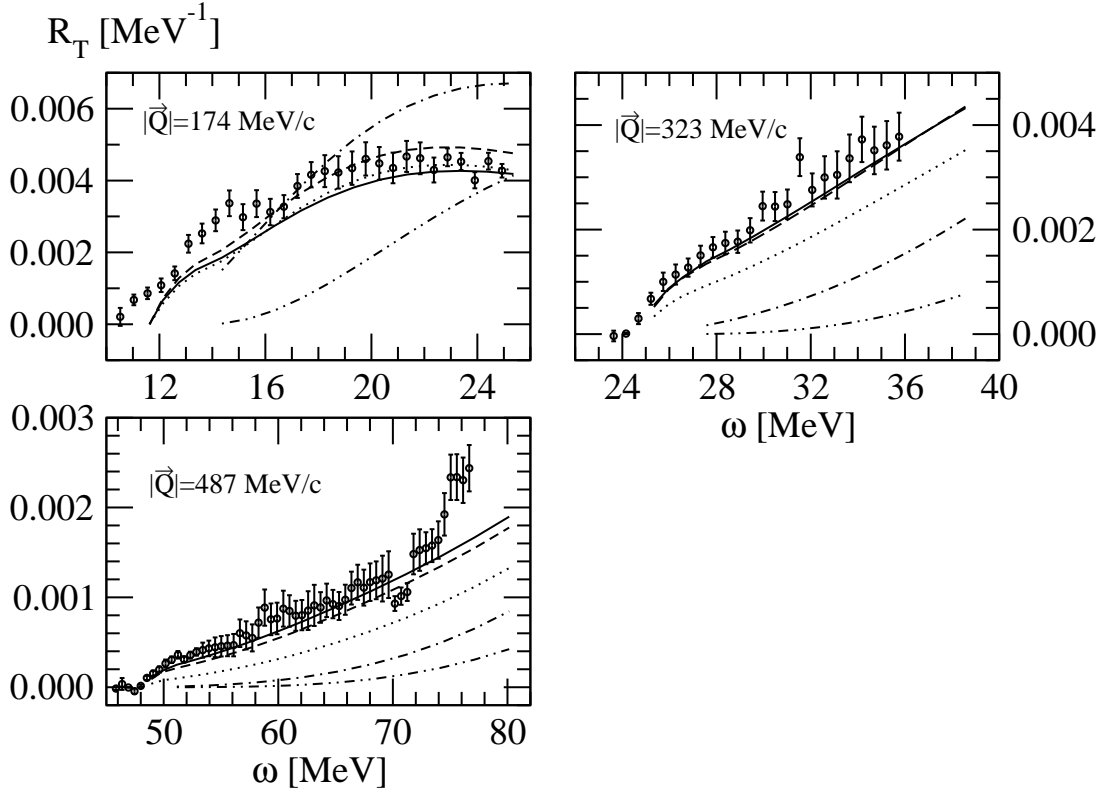,bb=40 360 560 740,clip=true,height=110mm}
\caption{The transverse response function $R_T$ for $^3$H.
Curves as in Fig.~\ref{figRL3H}.
Data are from \cite{Retzlaff94}.}
\label{figRetzRT3H}
\end{center}
\end{figure}

\begin{figure}[!ht]
\begin{center}
\epsfig{file=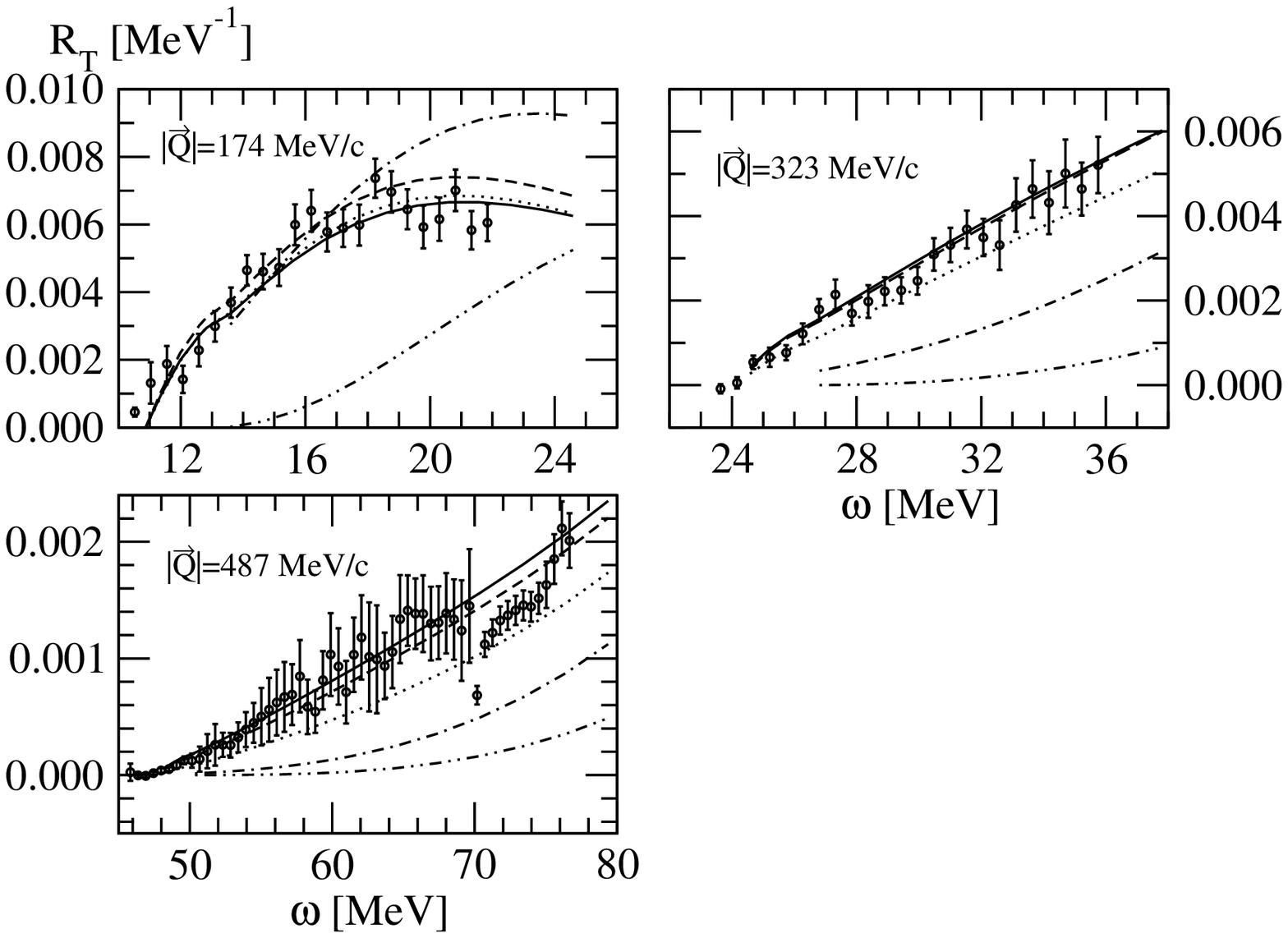,bb=40 360 560 740,clip=true,height=110mm}
\caption{The same as in Fig.~\ref{figRetzRT3H} for $^3$He.
Data are from \cite{Retzlaff94}.}
\label{figRetzRT3He}
\end{center}
\end{figure}

There are also inclusive data~\cite{Retzlaff94} 
for $\mid \vec Q \mid$= 174, 323 and 487  MeV/c,
starting at threshold.
They are plotted as a function of the
energy transfer $\omega$
in Figs.~\ref{figRetzRL3H}-~\ref{figRetzRT3He} in comparison to our theory.
The overall agreement of our complete theory with the data is quite good, 
for both nuclei,
$^3$H and $^3$He. Not including full FSI
would be a disaster
for all $\mid \vec Q \mid$-values: namely predictions based on the two 
simplest approximations,
PWIA and FSI23, are far away from the data.
In the case of $R_L$ 3NF as well as MEC
effects turn out to be small, except at $\mid \vec Q \mid$= 174 MeV/c 
where 3NF effects 
for $^3$He 
shift theory downwards in direction to the data.
For $R_T$ MEC effects are again quite significant, shifting the theory 
upwards.  
The counteractive effect of the 3NF is
only seen at $\mid \vec Q \mid$= 174 MeV/c.

For a smaller $\omega$ region (not explicitly displayed) there are also 
results for $R_L$ of the Trento group~\cite{efros04}. They use the same 
dynamical input but without two-body densities. Taking that into account 
the agreement between ours and the Trento group results 
in the case of $^3$H is good. In case of $^3$He a quantitative 
comparison is not possible, since we do not include the Coulomb force in the 
continuum. Going to the $\omega$-region below the three-body breakup in case of $^3$He
one can compare to the results from the
Pisa group~\cite{Pisa.low}.
Despite the fact that we do not include the 
Coulomb force in continuum we find 
in the case of $R_L$ 
a similar increase of the two-body density effect with increasing 
$\vert \vec Q \vert $-values, namely a shift downwards. 
In the case of $R_T$ 
we find a similar, upwards shift of the 
two-body current effects. Our curves including FSI+MEC+3NF lie a bit 
higher in comparison to the ones in~\cite{Pisa.low} but still rather 
close to the data.

\begin{figure}[!ht]
\begin{center}
\epsfig{file=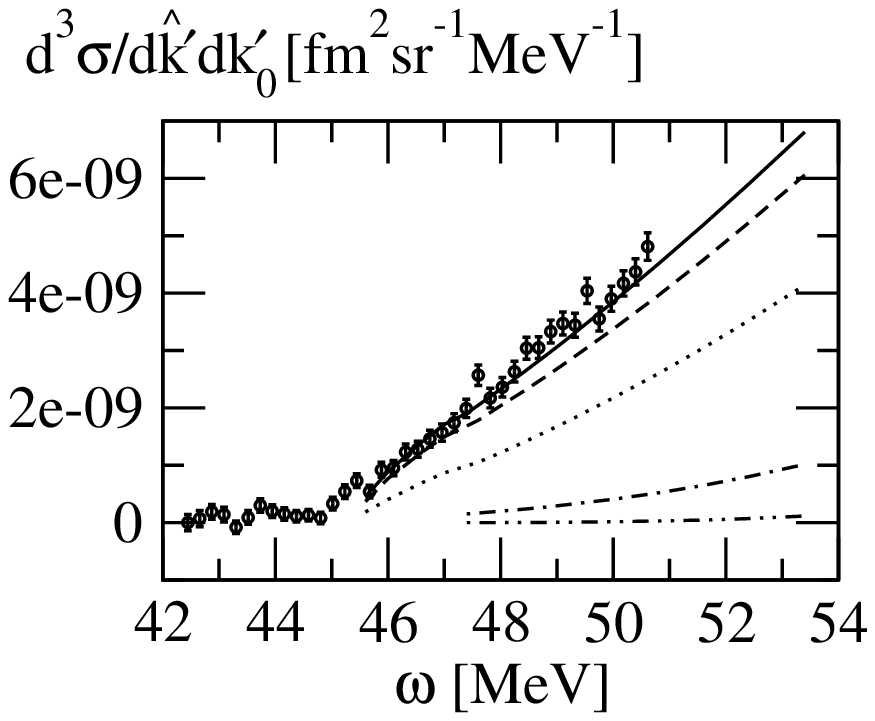,bb=10 515 300 740,clip=true,height=6cm}
\caption{The inclusive differential cross section
$\frac{d^3 \sigma }{ d {\hat k}^{\, \prime} \, d {k}_0^{\, \prime} }$
taken at $\vartheta$= 160$^\circ$ as a function of the energy transfer $\omega$
for the electron beam energy of 263 MeV. Curves as in Fig.~\ref{figRL3H}. 
Data are from  \cite{Hicks}.}
\label{figHicks}
\end{center}
\end{figure}

We would like to present another set of data for 
threshold electrodisintegration of $^3$He
\cite{Hicks}, where the electron scattering angle was 160$^\circ$, 
emphasizing the contribution from R$_T$.
The cross section 
 is shown in Fig.~\ref{figHicks}. Again the absolute need
for interaction in the continuum is obvious, 
but furthermore also significant effects
of MEC and 3NF are visible. The agreement of our theory with 
the data is very good. Further data
displayed in \cite{Hicks} require relativity and are therefore not shown here 
(see, however, \cite{Deltuva2004}, where some selected relativistic corrections 
shift theory in the direction to the data).

\begin{figure}[!ht]
\begin{center}
\epsfig{file=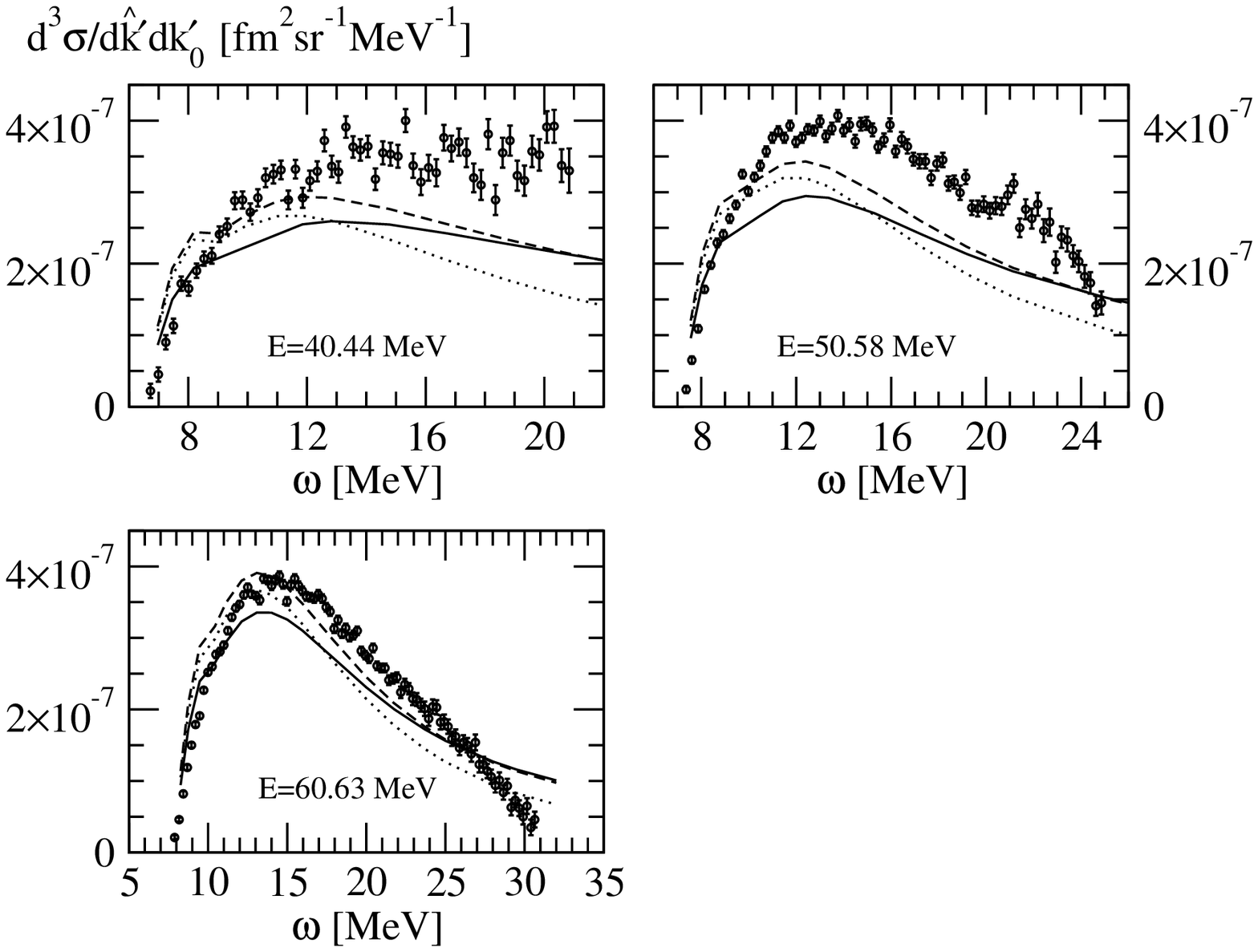,bb=37 340 560 740,clip=true,height=110mm}
\caption{The inclusive differential cross section
$\frac{d^3 \sigma }{ d {\hat k}^{\, \prime} \, d {k}_0^{\, \prime} }$
taken at $\vartheta$= 180$^\circ$ as a function of the energy transfer $\omega$
for three electron beam energies. For the description of the curves see 
Fig.~\ref{figRL3H}. 
Data are from  \cite{Jones79}.}
\label{figFagg}
\end{center}
\end{figure}

Finally, Fig.~\ref{figFagg} displays 180$^\circ$ inelastic electron scattering
cross sections for  $^3$He at rather
low incident electron energies \cite{Jones79}. We see three theoretical
curves, one with AV18 alone in the continuum (the dotted curve), then
using AV18 alone + MEC (the dashed curve) and finally our most
complete calculation with AV18+MEC+3NF (the solid curve). There are 
strong up and 
down effects  against
the pure NN force predictions adding MEC and the 3NF. Though our most complete
theoretical  prediction is close to the data at the strong rise for the lowest
excitation energies, it clearly underpredicts the data at the higher 
excitation energies.

Previous calculations for the inclusive responses aside from the
pioneering one \cite{Meijgaard90-90-92} mentioned in the 
introduction, appeared in \cite{mart95,efros04,Pisa.low}. 
 In \cite{mart95} the longitudinal response was
determined with the LIT method combined with a Faddeev decomposition and
carried through in momentum space. A qualitative agreement with
experimental data was achieved using the Bonn B~\cite{bonnb} 
 NN interaction and the
nonrelativistic single nucleon density operator.
In \cite{efros04} the longitudinal response functions were determined via the
LIT method using correlated sums of products of hyperspherical
functions, hyperspherical harmonics, and spin-isospin factors. The
configuration space Bonn A~\cite{bonnb} and AV18 NN  potentials including the
UrbanaIX and  Tucson-Melbourne~\cite{TM99} 3NF's were used and standard
 relativistic corrections  of lowest order for the density operator
 were included. Quite 
remarkable is the fact that, because the
LIT method requires only bound state-like solutions, it was possible
to include the Coulomb force also in the  final state.The results for
$\vert Q \vert$-values up to 500 MeV/c are quite similar to 
the ones shown above. We
mention the decrease of the peak heights adding a 3NF and the different
effects on $^3$H and $^3$He, namely an underestimation for $^3$H and a
reasonable agreement for $^3$He. Also the $R_L$ results in \cite{efros04} 
for the $^3$H data
in \cite{Retzlaff94} agree quite well with ours shown 
in Fig.~\ref{figRetzRL3H}  except for
$\vert Q \vert=487$~MeV/c, where in \cite{efros04} an 
overestimation  is visible.
The same data of \cite{Retzlaff94} were also 
analyzed in \cite{Pisa.low}, now for the
longitudinal as well as the transversal responses, but staying below the
three-body breakup. The $^3$He and pd scattering state wave functions were
obtained variationally with the pair correlated hyperspherical harmonics
method. Again the AV18 and AV18+UrbanaIX  were used and the 
 Coulomb force was  also fully
included. The currents and densities are as described above, but
additional pieces are added which are not constrained by the current
conservation. This also includes terms related to the $\Delta$-excitation.
The agreement to the data is comparable to the one shown above.
Below the three-body threshold the results for the longitudinal response
for $^3$He agree  well with the ones in \cite{efros04} 
at the two larger
$\vert Q \vert$-values.

Finally the Euclidean longitudinal and transversal responses have been
worked out . As  a Laplace transform in the energy transfer $\omega$ 
the response is mapped onto an imaginary time $\tau$. The technically very
attractive feature is that the Euclidean response can be cast in a
path-integral form which can be naturally evaluated with Monte Carlo
techniques. We refer the reader to the original literature 
\cite{Carlson98,carl94,carl92,carl94a}
 for the interesting  insights into the propagation of charge with
increasing $\tau$ and the comparison with correspondingly transformed
data. 
 In \cite{carlson02} a thorough study on both (longitudinal and transversal) 
Euclidean response functions have been performed and compared to the world 
data. This also includes $^4He$, which sheds light on the access to the 
transverse quasi-elastic strengths. Also sum rules techniques 
 were employed to study the T/L ratios. 
 Nevertheless the Laplace transform of the data looses details and
appears not to be a substitute of evaluating the responses  directly 
for fixed $\vert Q \vert $.

\begin{figure}[!ht]
\begin{center}
\epsfig{file=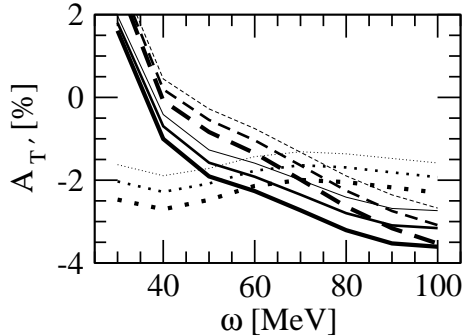,bb=10 515 300 715,clip=true,height=5cm}
\caption{Sensitivity of the asymmetry $A_{T^\prime}$  to the changes
of $G_M^n$ for the four-momentum transfer squared $Q^2$= 0.1 (GeV/c)$^2$. 
Three groups of curves with $0.9 G_M^n$, $1.0 G_M^n$
and $1.1 G_M^n$ are shown for PWIA (dotted), FSI23 (dashed) and
FSI+MEC+3NF (solid). In each case the upper curve is for 0.9, the
middle one  for 1.0 and the lower one for 1.1.}
\label{gmnsensitivity}
\end{center}
\end{figure}

\begin{figure}[!ht]
\begin{center}
\epsfig{file=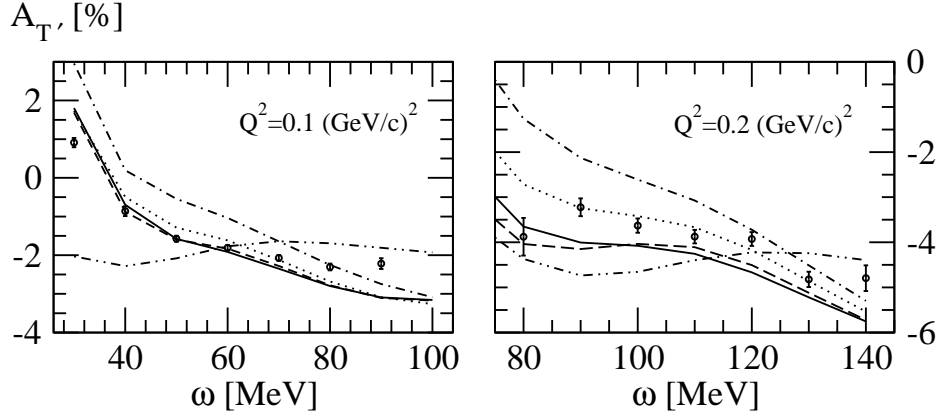,bb=37 520 560 740,clip=true,height=6cm}
\caption{The asymmetry data from \cite{Xu00} against theory
for two different four-momentum transfers 
squared $Q^2$= 0.1 (GeV/c)$^2$ (left) and 0.2 (GeV/c)$^2$ (right).
PWIA (double-dot-dashed),
FSI23 (dot-dashed),
FSI with NN forces alone (dotted),
FSI with NN forces alone + MEC (dashed) and adding in the 3NF (solid).}
\label{figXu}
\end{center}
\end{figure}

In addition, data are available for the cases,
where both initial particles, the electron and $^3$He, are polarized.
This allows to access
two more response functions, $R_{T'}$ and $R_{TL'}$. Data for
$R_{T'}$ and  $R_{TL'}$ alone are not yet taken
to the best of our knowledge,
only asymmetries. In PWIA $R_{T'} \propto (G_M^n)^2 $~\cite{Blankleider84}.
Thus measurements
concentrated on the transversal
asymmetry $A_{T'}$ ($\theta^\star \approx 0^\circ$) what  according 
to (\ref{obse60}) focuses on $R_{T'}$. That sensitivity to 
the magnetic form factor of
the neutron survives despite the fact 
that PWIA is insufficient~\cite{Xu00,withZiemer}.
This is documented in Fig.~\ref{gmnsensitivity}
for $A_{T\prime}$. We show three groups of curves 
 where within each group $G_M^n$ is multiplied by the factors 0.9, 1.0 and 1.1. 
The sensitivities 
to changes of $G_M^n$ values are very
similar whether one uses PWIA, FSI23 or our complete picture
 FSI+MEC+3NF. Therefore the measurement of $A_{T\prime}$ for polarized
$^3$He is a good tool to extract $G_M^n$ because we can 
consider $G_M^n$ the only unknown dynamical input for our calculations.
The dependence of $A_{T'}$ on the electric form factor of the
neutron $G_E^n$, which still has  rather big error bars, is negligible.
Therefore one can use the measured values of $A_{T'}$ and adjust  $G_M^n$. For
the detailed procedure we refer to \cite{Xu00,withZiemer}. 
The theoretical results
against the data are displayed in Fig.~\ref{figXu}
for  $Q^2$= 0.1 and 0.2 $({\rm GeV/c})^2$.
While PWIA has the wrong slope, already the inclusion
of the NN interaction in the spectator pair (FSI23)
leads to the correct shape, though it lies high above the data.
 Complete FSI is important and the NN force prediction alone comes
rather close to the data. On top we show the MEC effects
which are quite noticeable and the somewhat smaller  3NF effects.
The latter ones   lower
the theoretical prediction on top of the shift caused by MEC.
A direct comparison of our new results to the ones
presented in~\cite{Xu00,withZiemer} reveals
some differences. The reasons for  those differences are manifold.
In~\cite{Xu00} we did not use AV18 plus the explicit $\pi$- and $\rho$-like
two-body currents but Bonn B
with the standard $\pi$- and $\rho$-currents augmented
by the strong form factors of Bonn B.
In addition we use now $G_E^p$ in the charge density
and
$G_E^p$ and $G_E^n$ in the convection current, while in the
previous works we employed $F_1^p$ and $F_1^n$.
We also replaced the
H\"ohler models for the electromagnetic form
factors~\cite{holerff} 
by the electromagnetic form factors
\cite{HammerMeissner}
based on a dispersion theoretical analysis. Further
now we also add the  two-body
density. 

At this point we would like to add a more conceptual remark. 
In the spirit of a Hamiltonian approach
the arguments of the nucleon form factors are the difference of the
four-momenta of the nucleons squared, before and after the photon absorption, 
and not the four-momentum squared of the photon, 
which would be required in a manifestly covariant formalism. 
The reason is that in a Hamiltonian formalism, 
where the nucleons are on the mass-shell 
one has only three-momentum conservation at the photon vertex. 
Then since we nearly always neglect
relativistic features we choose as arguments of the electromagnetic
form factors just $(\vec Q)^2$. In the case of real photons and in our momentum 
region, $(\vec Q)^2$ is very small and we put it simply to zero.

In addition to all that we allow now for np and pp (nn) forces using the
``$\frac23-\frac13$'' rule, while in the previous work \cite{withZiemer,Xu00}
we used np forces only.
Finally the deuteron and
the $^3$He wave functions are generated with all the electromagnetic pieces
of the AV18 interaction
and thus especially the pp Coulomb force as the dominant part
is now taken into account in $^3$He.
Based on all that and noting that in ~\cite{Xu00} the theory was averaged over
the spectrometer acceptances using a Monte Carlo
simulation, while in Fig.~\ref{figXu} we show point geometry results,
 some differences to the previous results had  to be expected.
 Therefore one has to accept that a renewed extraction of G$_M^n$
from the data given in~\cite{Xu00} would provide a slightly different result.
We did not perform that study since we have no more
access to the experimental conditions and moreover our 
theory is anyhow only some
intermediate step toward a more basic concept.

\begin{figure}[!ht]
\begin{center}
\epsfig{file=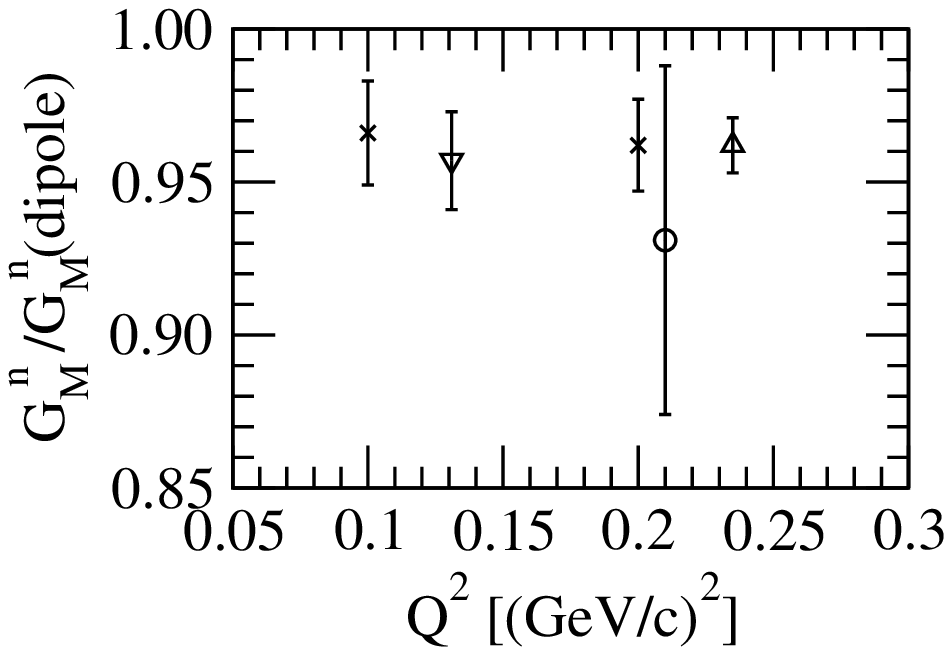,bb=20 515 300 
715,clip=true,height=5cm}
\caption{
$G_M^n$-values extracted from different measurements
on the deuteron
(
\protect\cite{Anklin94} ($\bigtriangledown$),
 \protect\cite{Anklin98} ($\bigtriangleup$))
and on $^3$He
(\protect\cite{Gao94} ($\circ$),
\protect\cite{Xu00} ($\times$)).
For the sake of visibility the two deuteron results
($\bigtriangledown$ and $\bigtriangleup$) are shifted sidewards
but belong to $Q^2$= 0.1 and 0.2 ${\rm (GeV/c)}^2$, respectively.
}
\label{figExtraxtedGmn}
\end{center}
\end{figure}

The resulting values for $G_M^n$ extracted in ~\cite{Xu00}
are shown in Fig.~\ref{figExtraxtedGmn}
together with the values extracted from the deuteron \cite{Anklin94,Anklin98}.
The agreement between the two
totally independent approaches is very good, though one should keep
the above remarks in mind.

The analysis of the $A_{T'}$  data  
 at $Q^2$= 0.3--0.6 $({\rm GeV/c})^2$
 also measured in \cite{Xu00}  is outside the present theoretical
framework and we refer the
reader to \cite{Xu03},  
where $G_M^n$-values 
  were extracted under the assumption of a plane wave impulse
approximation. This work uses the concept of 
the spin dependent spectral function of the three-body system 
and employs realistic forces \cite{kiev97}. 
 The polarized responses  $R_{TL'}$ and $R_{T'}$ were
evaluated with the aim to minimize the model dependence in the
extraction of the neutron electromagnetic form factors. Thereby the
prominent role of the proton contributions got illuminated.

\begin{figure}[!ht]
\begin{center}
\epsfig{file=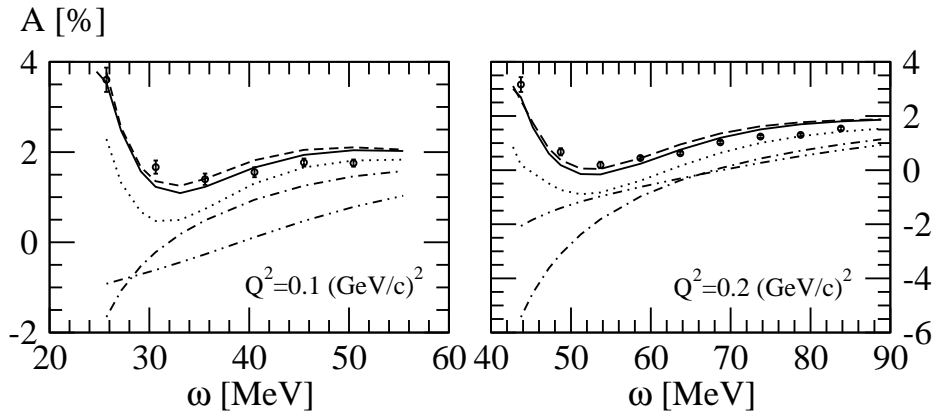,bb=37 520 560 740,clip=true,height=6cm}
\caption{The asymmetry data from \cite{Xiong01} against theory
for two different four-momentum transfers   
squared $Q^2$= 0.1 (GeV/c)$^2$ (left) and 0.2 (GeV/c)$^2$ (right).
Curves as in Fig.~\ref{figXu}.}
\label{figFeng}
\end{center}
\end{figure}

The interplay of both response functions $R_{T'}$ and $R_{TL'}$
in (\ref{obse60}) has been
investigated in \cite{Xiong01} by choosing $\theta^\star $ in a small
range around
135$^\circ$.
The resulting asymmetries $A$ are  shown in comparison  to our
theoretical results for  $Q^2$= 0.1 and 0.2 $({\rm GeV/c})^2$
in Fig.~\ref{figFeng}.
Again MEC's effects are quite important, and they 
are slightly modified by the addition of the 3NF.
The agreement with the data is quite good.

Finally we want to draw attention to the question whether 
signatures of short-range NN correlations can be extracted from
inclusive responses. A nice general introduction with appropriate
references is given in \cite{Carlson98}, thus we shall not repeat it here. The
key-point is to regard the energy integral over the longitudinal response
function (Coulomb sum rule), which can be separated into nucleon 
 form factor parts, the elastic charge form factor of the nucleus, and a third
part, which under the simplest assumption is the Fourier transform of the
proton-proton  correlation function. As nicely shown in 
\cite{schia93}  that third part
 is in addition strongly influenced  by relativistic corrections
 and two-body pieces in the density operator. Unfortunately that third
part carries large experimental error bars due to the strong
 cancellations of the Coulomb sum with
the  first two parts. That third part would be an excellent piece of
information on nuclear dynamics if the data base could be improved, 
especially the high wings of the longitudinal responses. An older
investigation of our collaboration can be found in 
 \cite{jgol95}. 

\clearpage

\subsection{Electron induced pd breakup of \boldmath{$^3$}He}
\label{sub6c}

\begin{figure}[htb]
\begin{center}
\epsfig{file=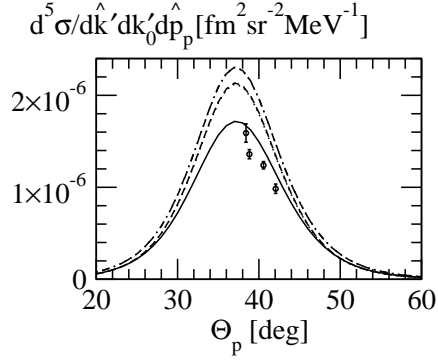,bb=30 515 300 740,clip=true,height=5cm} 
\caption{\label{pdT1}
Proton angular distribution for the T1 configuration from 
Table~\ref{tableNIKHEF}. 
 The double-dot-dashed curve represents 
the prediction based on PWIA.
The dot-dashed curve is obtained under the assumption of PWIAS (which 
overlaps with PWIA),  
the dotted curve takes the full FSI into account but 
neglects MEC and 3NF effects.
The $\pi$- and $\rho$-like two-body densities are accounted for additionally  
 in the dashed curve (which overlaps with FSI), 
and finally, the full dynamics including MEC and the 3NF is given 
by the solid curve.
 Data are from \cite{Keizer86}.
      }
\end{center}
\end{figure}

\begin{figure}[htb]
\begin{center}
\epsfig{file=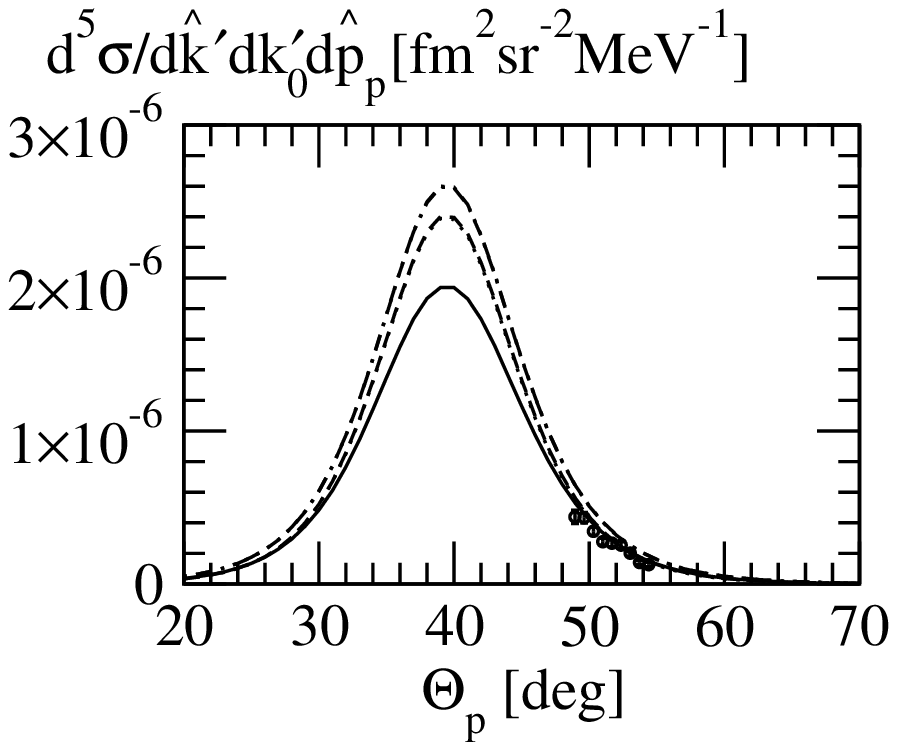,bb=30 515 300 740,clip=true,height=5cm}
\caption{\label{pdC2}
Proton angular distribution for the C2 configuration  from 
Table~\ref{tableNIKHEF}. Curves as in Fig.~\ref{pdT1}. 
 Data are from \cite{Keizer86}.
      }
\end{center}
\end{figure}

\begin{figure}[htb]
\begin{center}
\epsfig{file=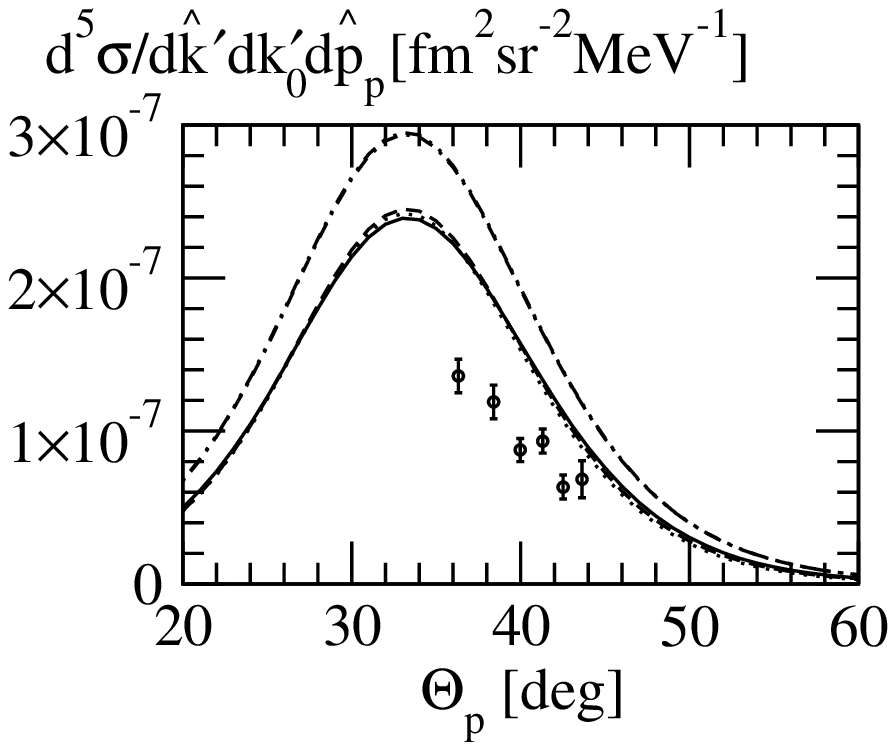,bb=30 515 300 740,clip=true,height=5cm}
\caption{\label{pdT2}
Proton angular distribution for the T2 configuration  from 
Table~\ref{tableNIKHEF}. Curves as in Fig.~\ref{pdT1}. 
 Data are from \cite{Keizer86}.
      }
\end{center}
\end{figure}

\begin{figure}[htb]
\begin{center}
\epsfig{file=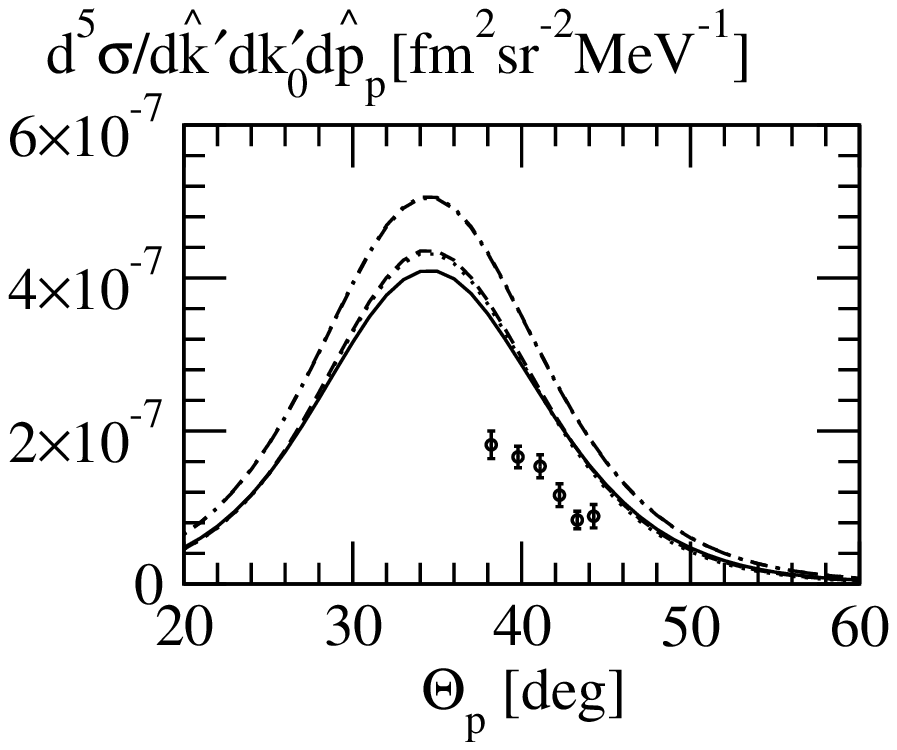,bb=30 515 300 740,clip=true,height=5cm}
\caption{\label{pdC3}
Proton angular distribution for the C3 configuration  from 
Table~\ref{tableNIKHEF}. Curves as in Fig.~\ref{pdT1}. 
 Data are from \cite{Keizer86}.
      }
\end{center}
\end{figure}

\begin{figure}[htb]
\begin{center}
\epsfig{file=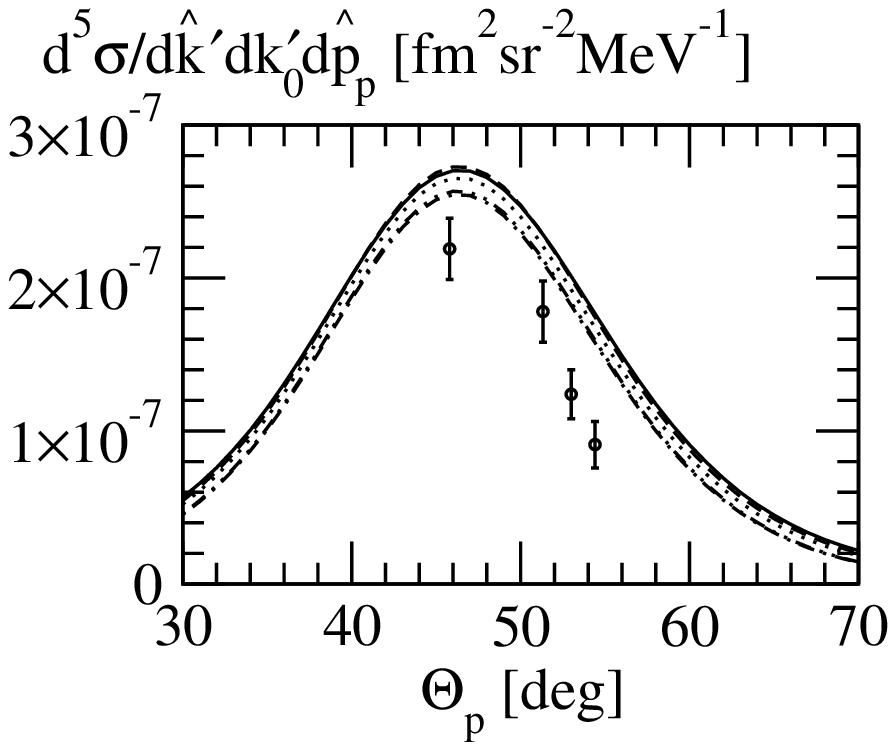,bb=30 515 300 740,clip=true,height=5cm}
\caption{\label{pdC1}
Proton angular distribution for the C1 configuration  from 
Table~\ref{tableNIKHEF}. Curves as in Fig.~\ref{pdT1}. 
 Data are from \cite{Keizer86}.
      }
\end{center}
\end{figure}

There is a big group of data taken at NIKHEF \cite{NIKHEFpd},
presented in \cite{Keizer86} and communicated to us by E. Jans \cite{jans}.
The different kinematical
conditions named as in \cite{Keizer86}
are shown in Table~\ref{tableNIKHEF}. The proton and
deuteron momenta lie in the plane spanned by the electron momenta. For
the configurations T1, T2 and C1 data were taken for proton scattering
angles close to the photon direction, while for C2 and C3 the
proton directions are further off. The quasi free scattering
condition $Q_0 = \frac{ {\vec Q}^{\ 2} } {2 m_N}$ is not covered by the data.
We show in Figs.~\ref{pdT1}-\ref{pdC1} the angular
distributions of the proton against the electron beam direction for
those five configurations.
Since one is close to the quasi free nucleon knockout peak the photon
is absorbed mostly by one nucleon and in plane wave impulse
approximation the antisymmetrization
plays no role, in other words the PWIA result is very close to the PWIAS
result. Also in all cases  except C1 PWIA is totally insufficient. The
MEC
effects are  insignificant. For the similar kinematics T1 and C2 
 the 3NF effects are quite strong and together with the NN force
move the theory quite close into the data.
Going to higher energy
transfers the situation changes and the 3NF effects are unimportant.
This is seen for the kinematics C3 and T2. Finally, in the case of
C1, with a relatively small energy transfer and for the high three
momentum transfer like in the other cases the FSI, MEC and 3NF effects
are all small and all curves overshoot the data somewhat.

\begin{table}[hbt]
\begin{center}
\begin{tabular}{ccccc}
\hline
      & $k_0$ & $\theta_e$ &  $\omega$ &  $Q$     \\
      & (MeV) &    (deg)   &  (MeV)    & (MeV/c)  \\
\hline
T1    & 367.1 &  85.0      & 107.1     &  431.0   \\
T2    & 367.1 &  85.0      & 143.8     &  412.7   \\
C1    & 390.0 &  74.4      &  66.1     &  434.8   \\
C2    & 390.0 &  79.0      & 110.4     &  434.4   \\
C3    & 390.0 &  83.0      & 145.1     &  434.5   \\
HR    & 390.0 &  39.7      & 113.0     &  250.2   \\
\hline
\end{tabular}
\end{center}
\caption[]
{
The NIKHEF electron kinematics specified by different
kinematical quantities.}
\label{tableNIKHEF}
\end{table}

\begin{figure}[htb]
\begin{center}
\epsfig{file=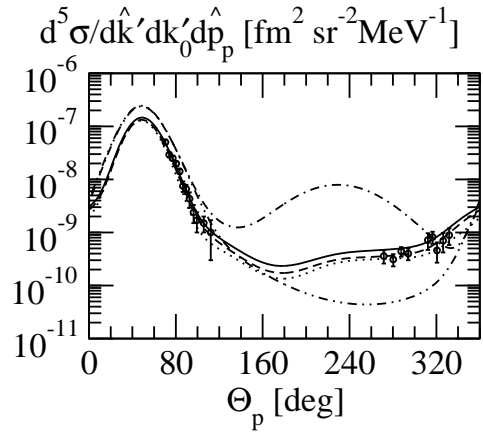,bb=30 515 300 740,clip=true,height=6cm}
\caption{\label{pdHR}
Proton angular distribution for the HR configuration  from 
Table~\ref{tableNIKHEF}. Curves as in Fig.~\ref{pdT1}. 
 Data are from \cite{Keizer86}.
      }
\end{center}
\end{figure}

Another set of data under the $HR$ kinematics from Table~\ref{tableNIKHEF}
is shown in Fig.~\ref{pdHR}.
The data are on the slopes of the proton and deuteron
knockout peaks.
The deuteron knockout peak lies around $\theta_p$ = 240$^\circ$.
The figure shows nicely how in plane wave impulse approximation
the symmetrized version PWIAS deviates
around 90$^\circ$ from the unsymmetrized version PWIA and the absorption
of the photon by the other two nucleons takes over and leads to a
second peak, the deuteron knockout peak. But the nuclear force effects
in the final continuum are extremely important there and shift theory
downwards by about one order of magnitude. Also in the slope of the
proton knockout peak the final state interactions in the continuum are
important. In both cases the agreement with the data is quite good.

\begin{figure}[htb]
\begin{center}
\epsfig{file=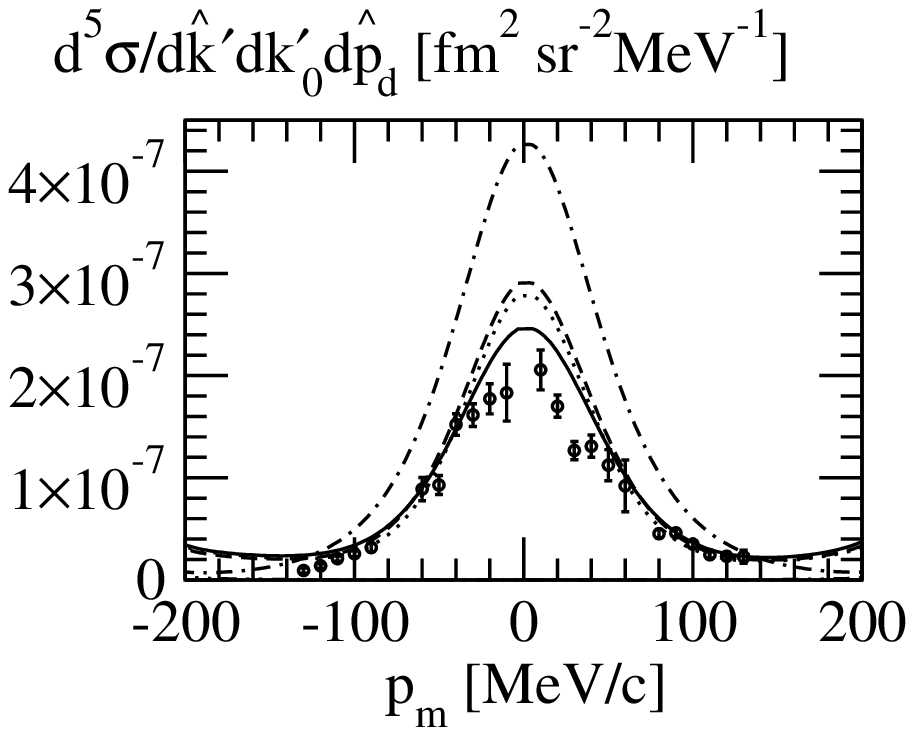,bb=10 515 300 737,clip=true,height=5cm}
\caption{\label{pdBL4}
Deuteron knockout cross section as a function of the missing
(i.e. proton) momentum $p_m$ for the following electron kinematics:
$k_0$= 370 MeV, $\omega$= 50 MeV, $\mid \vec Q \mid$= 412 MeV/c. 
 PWIAS (dot-dashed line),
FSI (dotted line), FSI+MEC (dashed line) and FSI+MEC+3NF (solid line)
results are compared to experimental data from \cite{Spaltro.qomega}.
 Note that the PWIA result is very small and therefore not displayed.
      }
\end{center}
\end{figure}

\begin{figure}[htb]
\begin{center}
\epsfig{file=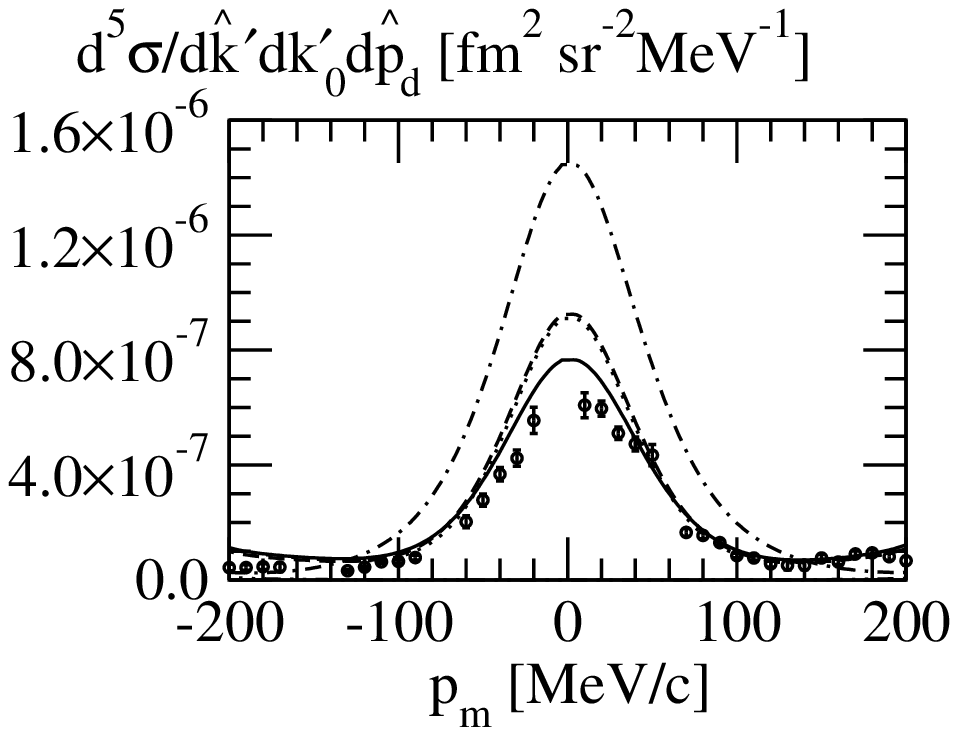,bb=10 515 300 737,clip=true,height=5cm}
\caption{\label{pdBL1}
The same as in Fig.~\ref{pdBL4}
for $k_0$= 576 MeV, $\omega$= 50 MeV, $\mid \vec Q \mid$= 412 MeV/c.
Data are from \cite{Spaltro.qomega}.
      }
\end{center}
\end{figure}

\begin{figure}[htb]
\begin{center}
\epsfig{file=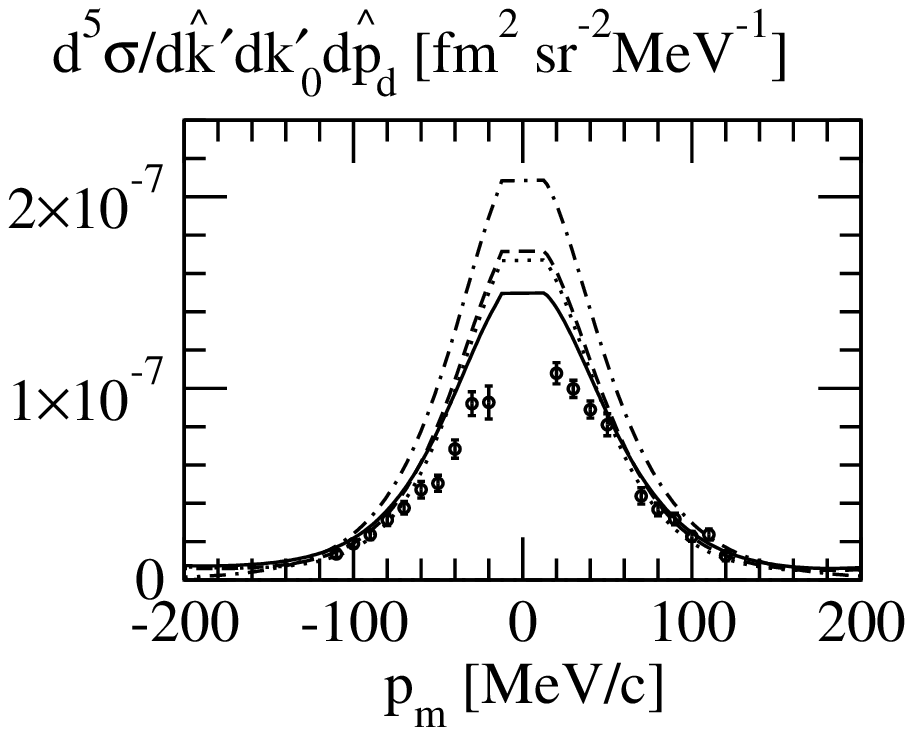,bb=10 515 300 737,clip=true,height=5cm}
\caption{\label{pdBL2}
The same as in Fig.~\ref{pdBL4}
for $k_0$= 576 MeV, $\omega$= 70 MeV, $\mid \vec Q \mid$= 504 MeV/c.
Data are from \cite{Spaltro.qomega}.
      }
\end{center}
\end{figure}

\begin{figure}[htb]
\begin{center}
\epsfig{file=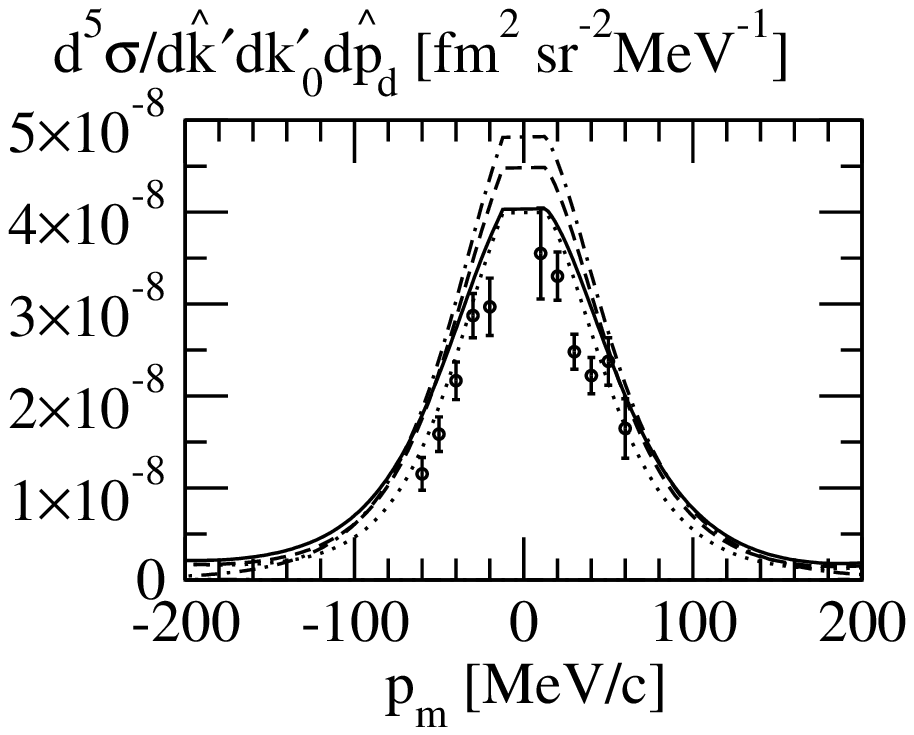,bb=10 515 300 737,clip=true,height=5cm}
\caption{\label{pdBL5}
The same as in Fig.~\ref{pdBL4}
for $k_0$= 370 MeV, $\omega$= 70 MeV, $\mid \vec Q \mid$= 504 MeV/c.
Data are from \cite{Spaltro.qomega}.
      }
\end{center}
\end{figure}

\begin{figure}[htb]
\begin{center}
\epsfig{file=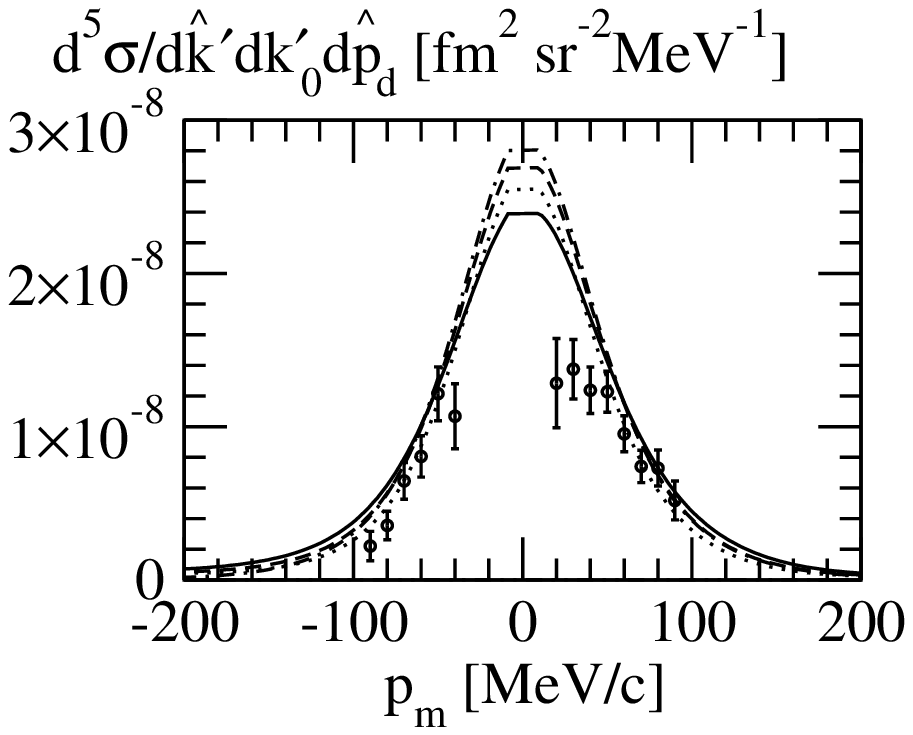,bb=10 515 300 737,clip=true,height=5cm}
\caption{\label{pdBL3}
The same as in Fig.~\ref{pdBL4}
for $k_0$= 576 MeV, $\omega$= 100 MeV, $\mid \vec Q \mid$= 604 MeV/c.
Data are from \cite{Spaltro.qomega}.
      }
\end{center}
\end{figure}

Now we concentrate on the deuteron knockout peak and compare
the data from \cite{Spaltro.qomega} and
 the theory in Figs.~\ref{pdBL4}-\ref{pdBL3}. The PWIA result is extremely 
small and not displayed. 
In all cases shown the 3NF effects on top of the NN
force contributions in the continuum are quite important and move
theory  close to the data. Note that for $\omega$= 50 MeV,
$\mid \vec Q \mid$= 412 MeV/c and $\omega$= 70 MeV, $\mid \vec Q \mid$= 
504 MeV/c
the nuclear matrix elements are similar 
 but the electron kinematics are quite different, which weights the
different response functions differently. Thus in the case of
$\omega$= 70 MeV, $\mid \vec Q \mid$= 504 MeV/c the MEC effects are significant,
while in the other case they are insignificant.
In all deuteron knockout peaks the theory  clearly 
overestimates the data. 
 Thus a renewed measurement concentrating on the
missing momentum $p_m$=0 would be very desirable.

\begin{figure}[htb]
\begin{center}
\epsfig{file=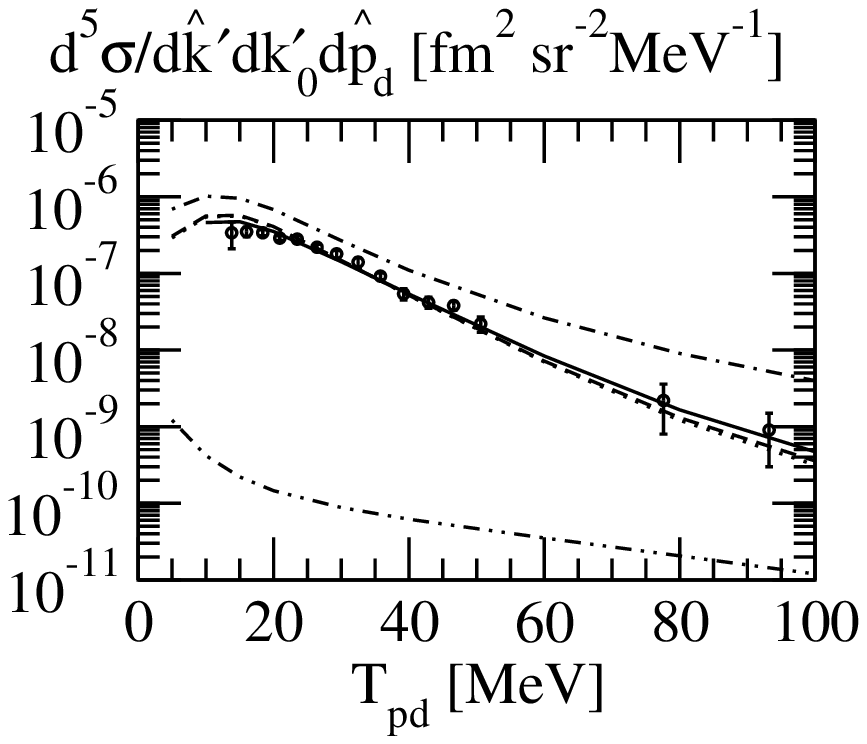,bb=10 515 300 737,clip=true,height=5cm}
\caption{\label{Dknockout}
Deuteron knockout cross section as a function of the relative 
 proton-deuteron  energy $T_{pd}$
for the parallel kinematics with $E_e$= 390 MeV and $\mid \vec Q \mid$= 
380 MeV/c. 
Curves as in Fig.~\ref{pdBL4}. Data are from \cite{Keizer86}.
      }
\end{center}
\end{figure}

Further kinematical configurations in search for the deuteron knockout
are related to the "D-kinematics" in \cite{Keizer86}. In this 
case the direction
of the deuteron has been chosen parallel to the photon direction
and the data were taken for $k_0$= 390 MeV, $\mid \vec Q \mid$= 380 MeV/c
and are displayed in Fig.~\ref{Dknockout} as a function
of the relative kinetic energy  $T_{pd}$ of the proton and the deuteron.
 The agreement is quite good and the  effects of the nucleon 
interactions in the 
continuum are decisive. None of those data points correspond exactly to the
quasi free peak position, where we experienced the discrepancies in
Figs.~\ref{pdBL4}-\ref{pdBL3}.

\begin{figure}[htb]
\begin{center}
\epsfig{file=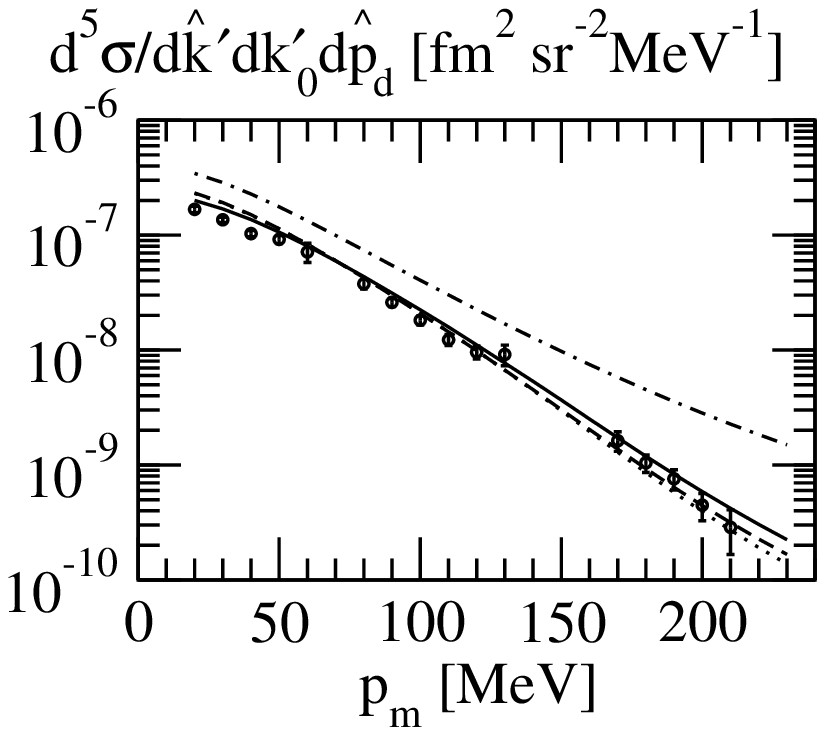,bb=10 515 300 737,clip=true,height=5cm}
\caption{\label{Spaltro.parallel.e370.q412}
Deuteron knockout cross section as a function of the missing momentum $p_m$
for the parallel kinematics with $E_e$= 370 MeV and $\mid \vec Q \mid$= 412 MeV/c. 
 Curves as in Fig.~\ref{pdBL4} but the PWIA results are not displayed. 
Data are from \cite{Spaltro.parallel}.
}
\end{center}
\end{figure}

\begin{figure}[htb]
\begin{center}
\epsfig{file=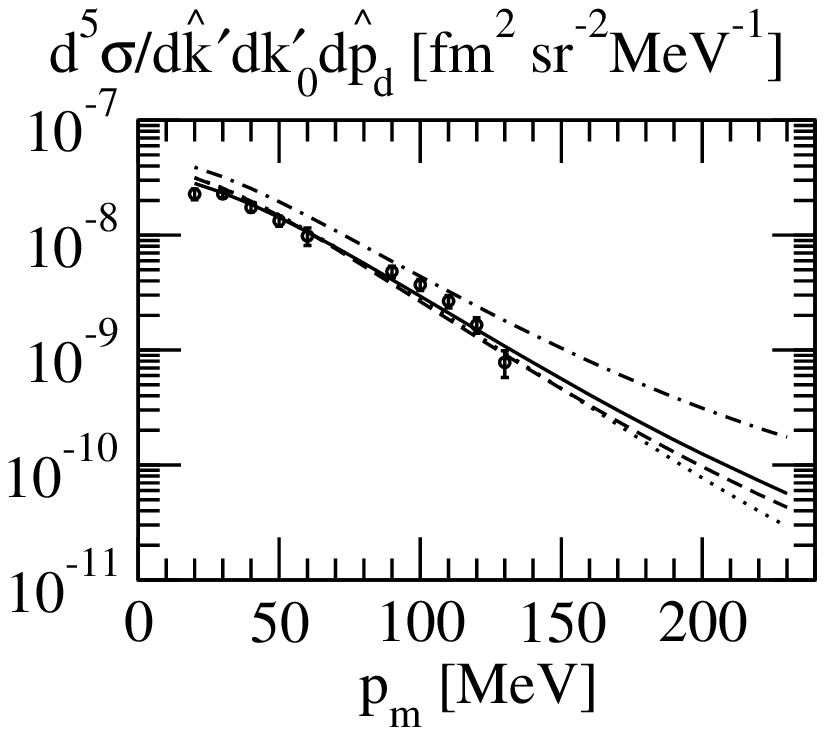,bb=10 515 300 737,clip=true,height=5cm} 
\caption{\label{Spaltro.parallel.e370.q504}
The same as in Fig.~\ref{Spaltro.parallel.e370.q412}
 for $E_e$= 370 MeV and $\mid \vec Q \mid$= 504 MeV/c.
}
\end{center}
\end{figure}

\begin{figure}[htb]
\begin{center}
\epsfig{file=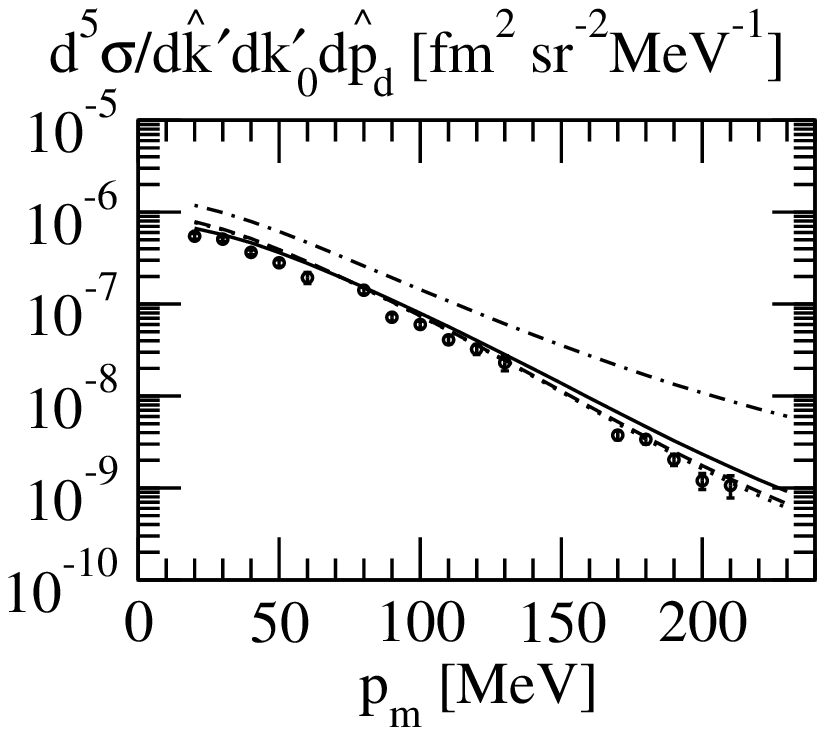,bb=10 515 300 737,clip=true,height=5cm}
\caption{\label{Spaltro.parallel.e576.q412}
The same as in Fig.~\ref{Spaltro.parallel.e370.q412}
 for $E_e$= 576 MeV and $\mid \vec Q \mid$= 412 MeV/c.
}
\end{center}
\end{figure}

\begin{figure}[htb]
\begin{center}
\epsfig{file=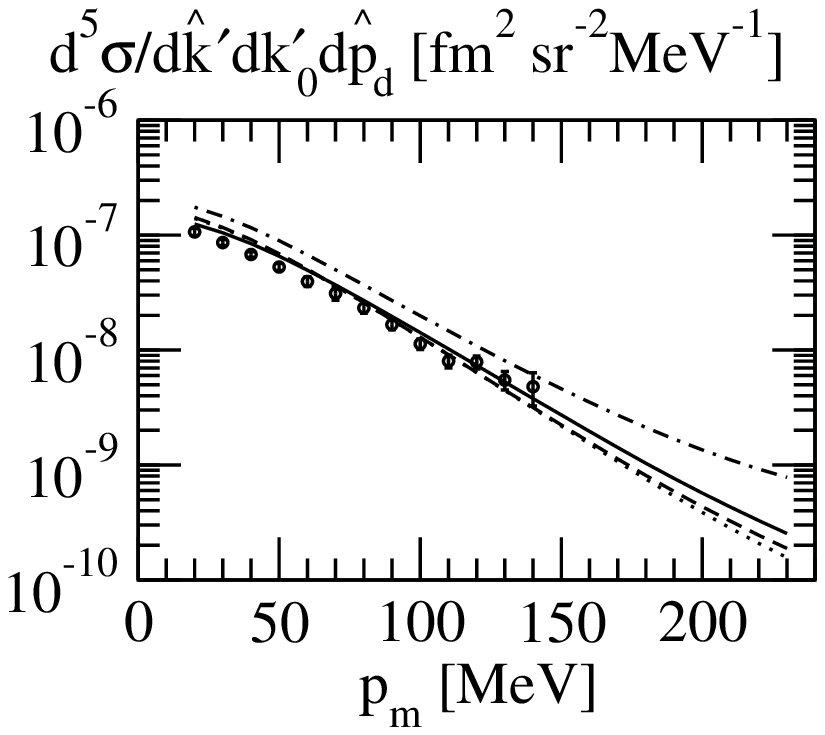,bb=10 515 300 737,clip=true,height=5cm}
\caption{\label{Spaltro.parallel.e576.q504}
The same as in Fig.~\ref{Spaltro.parallel.e370.q412}
 for $E_e$= 576 MeV and $\mid \vec Q \mid$= 504 MeV/c.
}
\end{center}
\end{figure}

\begin{figure}[htb]
\begin{center}
\epsfig{file=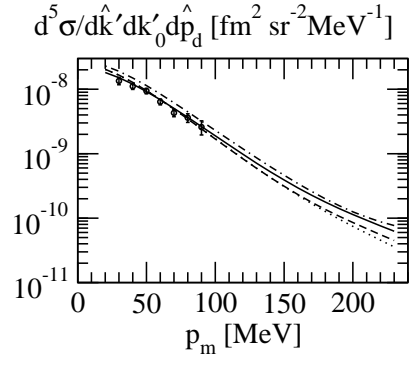,bb=10 515 300 737,clip=true,height=5cm}
\caption{\label{Spaltro.parallel.e576.q604}
The same as in Fig.~\ref{Spaltro.parallel.e370.q412}
 for $E_e$= 576 MeV and $\mid \vec Q \mid$= 604 MeV/c.
}
\end{center}
\end{figure}

Another set of data \cite{Spaltro.parallel} in parallel deuteron knockout
kinematics is compared to our theory 
in Figs.~\ref{Spaltro.parallel.e370.q412}--\ref{Spaltro.parallel.e576.q604}.
Data were taken at three different $\mid \vec Q \mid$-values
($\mid \vec Q \mid$= 412, 504 and 604 MeV/c) and at two electron beam energies
($E_e$= 370 and 576 MeV).
In all cases FSI is
quite important, whereas the addition of MEC's and/or 3NF's yields only
marginal shifts, at least in the range of $p_m$ values, which were covered
by the data.
 The agreement with the data is reasonably good.

Recently these data have been reanalyzed in \cite{Yuan02.1} including 
a single $\Delta$-isobar excitation. The results, agreements and 
disagreements, are very similar to 
 ours shown in Figs.~\ref{pdT1}-\ref{Spaltro.parallel.e576.q604}. 


\clearpage

\subsection{Nd radiative capture and the time reversed 
Nd photodisintegration of 3N bound states.}
\label{subsec6d}

Photon angular distributions for pd capture have been measured
for a wide range of energies.
We show in Fig.~\ref{fig6d1} the cross section data in comparison to our theory
with different dynamical ingredients.
We rely either on the explicit MEC's for $\pi-$ and $\rho$-like
exchanges  or on  the Siegert approach as described in section~\ref{ourSiegert}
and show results either for the NN force AV18 alone
or together with the UrbanaIX 3N force.
We see an overall  good  agreement in the 
AV18 + UrbanaIX model together with explicit MEC's. Also the Siegert 
predictions for that
choice of the interactions are similar. Since our MEC currents are not 
fully consistent
to the forces, one cannot expect equality of these two approaches. 
At the higher 
energies the higher
multipoles play a role. All the multipoles are kept in
our Siegert approach only on the level of the single  nucleon current. 
Nevertheless for the cross
sections that Siegert approach does reasonably well in conjunction 
with the 3N force.
This is not the case for AV18 with Siegert, while AV18 together with MEC's is 
much closer to the
data.
We also would like to point out that the addition of the 3N force decreases
 the cross section
at the lower energies below $\approx 30$~MeV and 
increases it at the higher ones.  
In ~\cite{2bphot} we argued
that this is not only a scaling effect with the 3N binding energy 
as often claimed in the
literature  but at the higher energies it is also caused  by
the action of the 3N force in the continuum.

In Fig.~\ref{fig6d2} we show photon angular distributions for 
nd capture around 10 MeV neutron lab
energy. The situation is very similar to the case of pd capture.
\begin{figure}[!ht]
\begin{center}
\caption{
The c.m. pd capture cross sections  at various deuteron lab
energies and four different dynamical inputs: MEC + AV18 (dashed line), 
Siegert + AV18 (dot-dashed line),
MEC + AV18 + UrbanaIX (solid line), 
Siegert + AV18 + UrbanaIX (dotted line).
Data at 10 MeV are from \cite{Goeckner}, at 19.8 and  
29.6 MeV from \cite{Belt},
at 95 MeV from \cite{Pitts}, at 200 MeV circles from \cite{Pickar} and 
x-es from \cite{Yagita2003}, and at 400 MeV from \cite{Pickar}.
}
\label{fig6d1}
\end{center}
\end{figure}

\begin{figure}[!ht]
\begin{center}
\epsfig{file=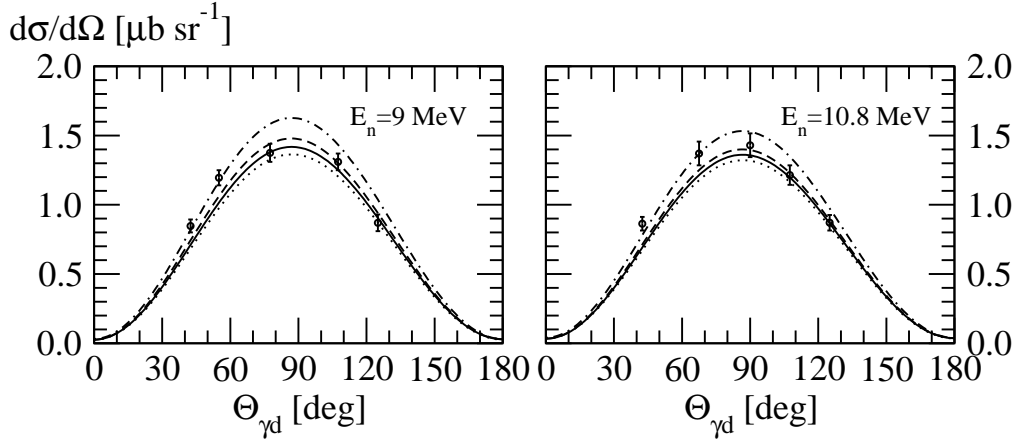,bb=20 520 550 735,clip=true,height=6cm}
\caption{
\label{fig6d2}
The c.m. nd capture  cross sections at 9.0 and 10.8 MeV
neutron lab energies. Curves  as in Fig.~\ref{fig6d1}. Data 
at 9.0 MeV and 10.8 MeV are from~\cite{Mitev}.
}
\end{center}
\end{figure}

\begin{figure}[!ht]
\begin{center}
\epsfig{file=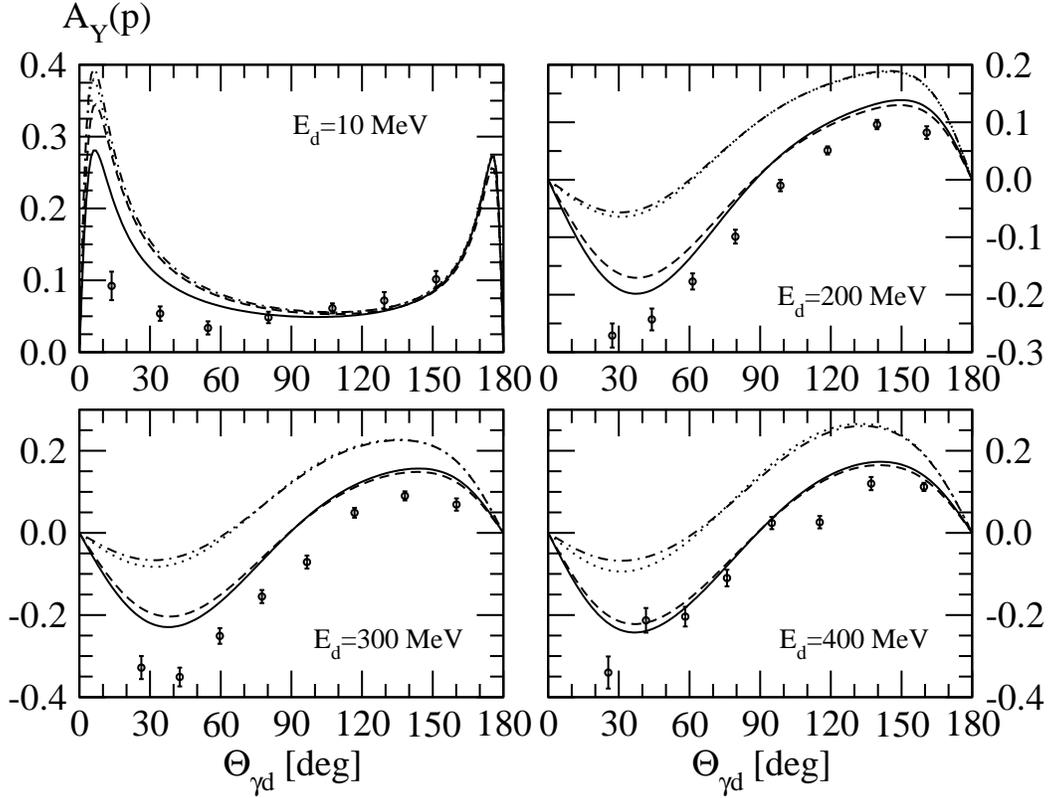,bb=40 360 560 740,clip=true,height=110mm} 
\caption{
\label{fig6d3}
The c.m. angular distributions for the proton analyzing powers $A_y (p)$ in pd
    capture at various deuteron lab energies. Curves 
as in Fig.~\ref{fig6d1}. Data at 10 MeV from \cite{Goeckner}, at
    200, 300 and 400 MeV from \cite{Pickar}.
}
\end{center}
\end{figure}

Then there is a rich set of polarization observables in pd capture.
Proton analyzing powers $A_y (p)$ at $E_d$= 10, 200, 300 and 400 MeV 
are shown in Fig.~\ref{fig6d3}.
At the deuteron lab energy of 10 MeV the choice MEC+AV18+UrbanaIX 
comes closest to the data
but nevertheless fails significantly at the smaller angles. For the 
three much higher
energies the two curves with explicit MEC's come significantly 
closer to the data than
the Siegert predictions but are too high at $E_d$ = 200 and 300 MeV 
in relation to the data.
Maybe it is accidental that there is a good agreement for the MEC 
predictions at $E_d$ = 400
MeV.

Like for the proton analyzing power $A_y (p)$ we face a serious 
discrepancy for the deuteron
vector analyzing power iT$_{11}$ (in the spherical notation
iT$_{11}$ = $\frac{\sqrt{3}}{2}$ A$_y$(d) ).
 This
is shown in Fig. \ref{fig6d4}. Again the MEC+AV18+UrbanaIX model
comes closest to the data for the
deuteron lab energies E$_d$ = 10, 17.5, 29, 45, and 200~MeV. 
At E$_d$ = 95 MeV all our predictions show a
strong slope not seen in the data. In view of 
the strong discrepancy and the
relatively  large experimental error bars a renewed, more precise 
measurement at this
energy would be very
useful to challenge  improved theoretical approaches in the future.

\begin{figure}[!ht]
\begin{center}
\epsfig{file=fig54_rep.Ed17p5_AYD_iT11.ps,bb=50 170 560
740,clip=true,height=15cm}
\caption{
\label{fig6d4}
 The c.m.  angular distributions 
for the deuteron vector analyzing power iT$_{11}$ in
pd capture at various deuteron lab energies. 
Curves as in Fig.~\ref{fig6d1}. Data at 10 MeV 
from \cite{Goeckner},
at 17.5 MeV from \cite{Sagara}, at 29 and 45 MeV 
from \cite{Klechneva}, at 95 MeV from \cite{Pitts}, and 
at 200 MeV \cite{Yagita2003}.
}
\end{center}
\end{figure}

Finally we look into the group of tensor analyzing powers. The spherical 
and cartesian notations are connected as
\begin{eqnarray}
A_{xx} &=& \sqrt{3} T_{22} - \frac{\sqrt{2}}{2} T_{20} \\
A_{yy} &=& -\sqrt{3} T_{22} - \frac{\sqrt{2}}{2} T_{20} \\
A_{xz} &=& -\sqrt{3} T_{21} .
\end{eqnarray}
The observables T$_{20}$, T$_{21}$ and T$_{22}$ are displayed
in Fig.~\ref{fig6d5}. Overall there is a good agreement but  
the accuracy of the data does not allow a
clear distinction  among the four theoretical predictions. T$_{21}$,  
and to an even  larger extent
T$_{22}$, turn out to be quite independent to the dynamical  input. In 
the case of T$_{20}$ the
explicit MEC picture reproduces the data at small  angles better than 
the Siegert approach.

\begin{figure}[!ht]
\begin{center}
\caption{
\label{fig6d5}
The c.m. angular distributions for the tensor  
analyzing powers T$_{20}$, T$_{21}$ and T$_{22}$
for pd capture at low energies. Curves as in Fig.~\ref{fig6d1}. Data 
at 10 MeV from \cite{Goeckner}, and at
19.8 MeV from \cite{Vetterli}(circles) and form \cite{Schmid00} (squares).
}
\end{center}
\end{figure}

Next we show A$_{yy}$ in Fig.~\ref{fig6d6}. The data at 
the two lowest energies are fairly well described by the
explicit MEC choice. In the case of 45 MeV theory is somewhat 
too low and especially at very
backward angles one misses the few data points totally. At 95 MeV 
all our predictions are also too
low. Finally, Fig.~\ref{fig6d7}  shows A$_{xx}$ and A$_{zz}$ 
which agree fairly well with the explicit MEC approach. 

\begin{figure}[!ht]
\begin{center}
\epsfig{file=fig57_rep.Ed17p5_AYY.ps,bb=40 360 560 740,clip=true,height=110mm}
\caption{
\label{fig6d6}
The c.m. angular distributions for the 
tensor analyzing power A$_{yy}$  for
pd capture at various energies. Curves as in Fig.~\ref{fig6d1}.  
Data at 17.5 MeV from \cite{Sagara}, at 29 MeV
from \cite{Jourdan86} (square) and \cite{Klechneva} (circles), at 45 MeV 
from \cite{Klechneva} (circles) and \cite{Anklin} (squares) and at 95 MeV from \cite{Pitts}.
}
\end{center}
\end{figure}

In \cite{Yuan02.2} the two-body photodisintegration of the 3N bound
state as well as the time reversed process have also been studied
including a $\Delta$-isobar excitation. The selected results shown there
are very similar to the ones displayed above. The difference in 
 Fig.~\ref{fig6d5} 
to Fig.~11 of \cite{Yuan02.2} is due to a wrong choice of angles in 
\cite{Yuan02.2}. If replotted
the outcome in \cite{Yuan02.2} is quite similar to the one shown above.  
 
The  pd and nd captures at  very low energies ( 0-100 keV c.m. energies) have
considerable astrophysical relevance for studies of stellar structure
and evolution and of big-bang nucleosynthesis. Since single nucleon
currents are insufficient to connect the dominant S-state components 
of the two- and three-body bound states, small components of the wave
 functions acquire importance and even more the additional many-body
currents. Therefore these reactions deserve a careful study. 
We refer to \cite{Carlson98} for an 
introduction to these very low energy processes.
 In a series of
papers \cite{vivia97,ma97,wulf99,Pisa.low,vivia03,marc04,Marcucci2005} 
 these processes were investigated, experimentally and theoretically. In
the most recent papers the two-body currents have been supplemented such,
that they fulfill exactly the continuity equation related to the NN
force AV18 and even three-nucleon currents have been added. We show in
 Fig.~\ref{obskijevsky} the cross section and 
spin observables for pd radiative capture
at $E_{c.m.} = 3.33$~MeV obtained in \cite{marc04,Marcucci2005}  
 with the AV18+UrbanaIX Hamiltonian model. 
These results document an important stride forwards, since the current 
used is fully consistent to the force model in the sense, that 
the continuity equation is exactly fulfilled. Some discrepancies in $A_y$ 
and $iT_{11}$ remain.

\begin{figure}[!ht]
\begin{center}
\epsfig{file=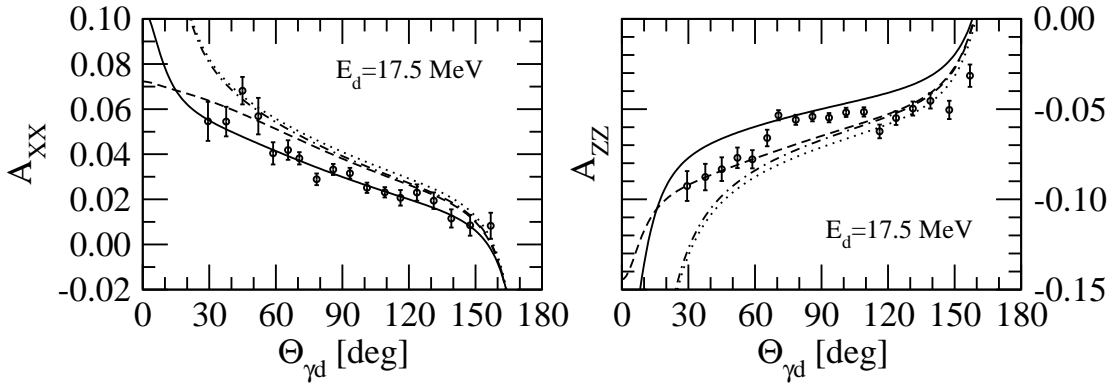,bb=20 520 570 740,clip=true,height=6cm}
\caption{
\label{fig6d7}
 The c.m.  angular distributions for the tensor 
analyzing powers A$_{xx}$
and
 A$_{zz}$  for
pd capture at 17.5 MeV. Curves as in Fig.~\ref{fig6d1}. Data 
from \cite{Sagara}.
}
\end{center}
\end{figure}

Finally, we address photodisintegration of $^3$He. In 
Figs.~\ref{fig6d8} and \ref{fig6d9}
we display the cross section 
of the two-body (pd) breakup in the c.m. system at two fixed angles, 
one at $\theta_d^{c.m.} = 90^\circ$ 
and one at $\theta_d^{lab} = 103.05^\circ$. The first one is 
shown for lower photon energies
and the second one for higher ones, which beyond about 
E$_\gamma$ = 150 MeV are strictly
spoken outside the region where our theoretical framework is 
adequate. For the low energy
region we display the predictions of Siegert and explicit MEC's for 
NN and NN + 3NF,
respectively, while for the higher energy region  only the explicit 
MEC predictions are
shown.  In both energy regions the MEC+AV18 +UrbanaIX predictions 
are in reasonably good
agreement with the data except in the peak area around E$_\gamma$  = 10 MeV. 
This photon energy 
corresponds to E$_d^{lab} = 13.47$~MeV in the time reversed pd 
capture reaction. 
As seen in Fig.~\ref{fig6d1}, in that case there is a good agreement 
with the  data at $\theta_{\gamma d} = 90^\circ$. Thus
we have to conclude that the data in Figs.~\ref{fig6d1} 
and the lower ones in Fig.~\ref{fig6d8} are 
inconsistent. This
calls for an experimental  clarification. 

In the total pd breakup cross section given as a function of $E_{\gamma}$ in 
Fig.~\ref{fig63ver2.2b} there is a big spread in the experimental 
data, which makes any definite conclusion impossible.

\begin{figure}[htb]
\begin{center}
\epsfig{file=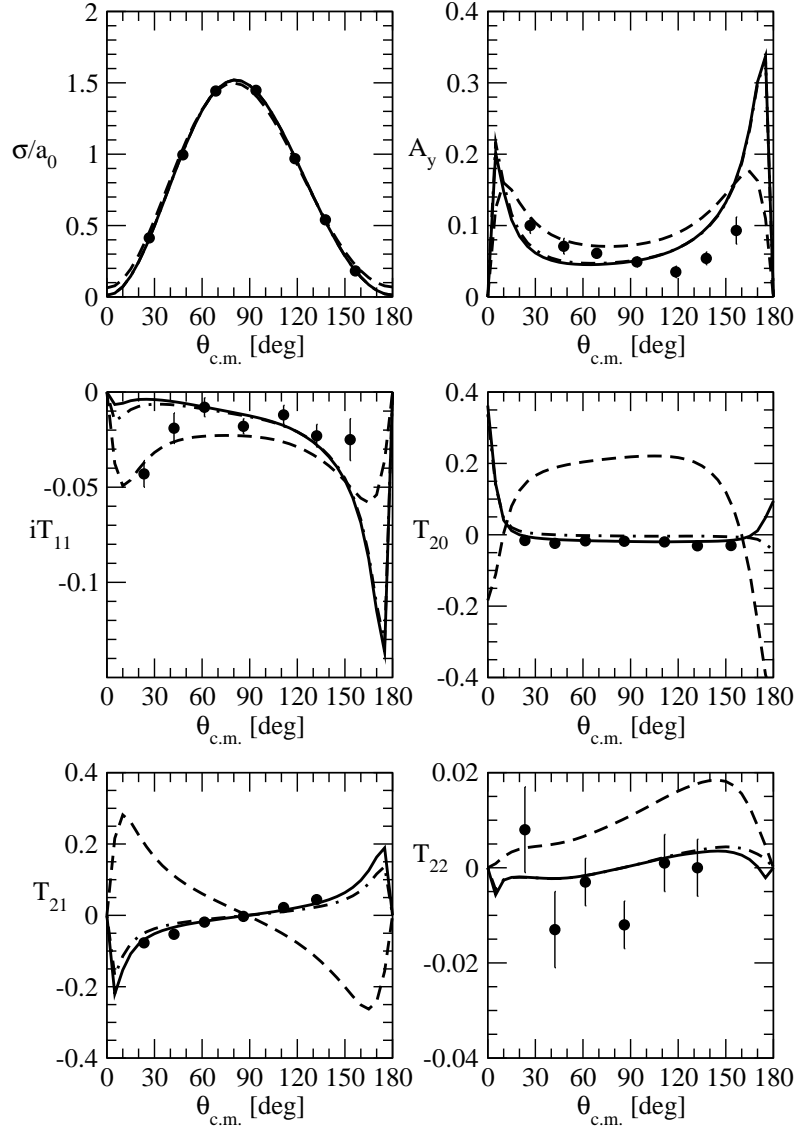,height=15cm}
\caption{\label{obskijevsky}
The pd capture cross section and spin observables at $E_{c.m.}=3.33$~MeV  
obtained with the AV18+UrbanaIX Hamiltonian model~\cite{marc04,Marcucci2005}. 
 The dashed, dot-dashed and solid curves correspond to the calculation 
with  one-body only, with one- and two-body, and with one-, two- and 
three-body  currents. For details see~\cite{marc04,Marcucci2005}. 
 The data are from~\cite{Goeckner}.
      }
      \end{center}
      \end{figure}

\begin{figure}[!ht]
\begin{center}
\epsfig{file=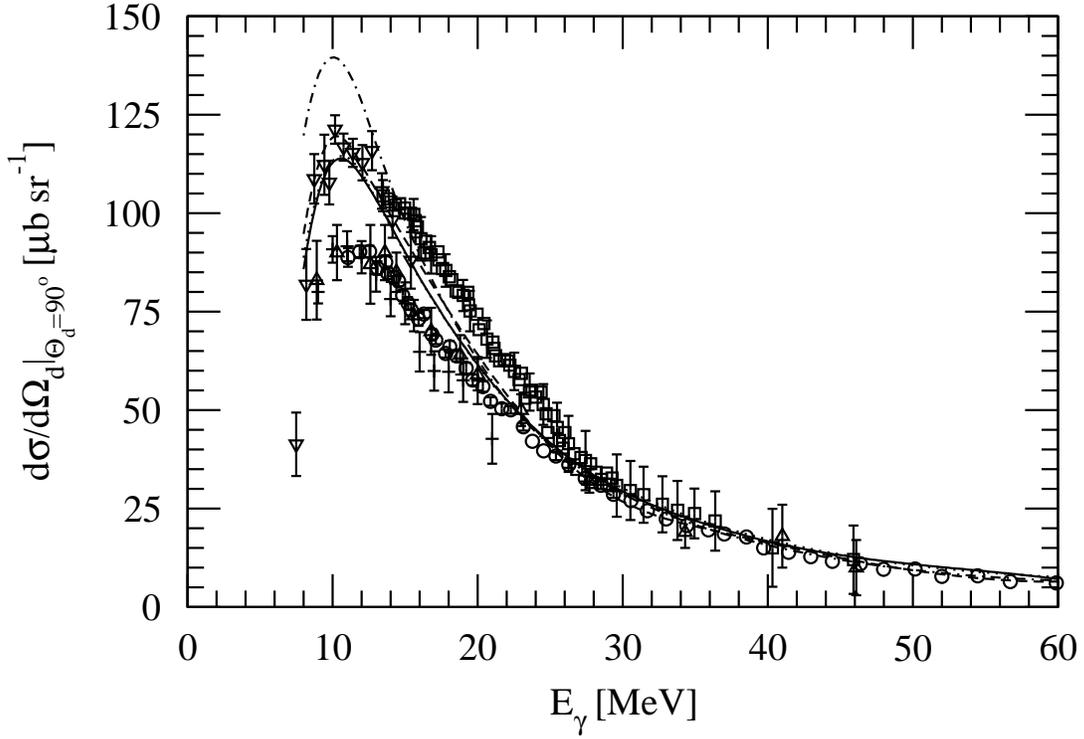,bb=50 530 300 715,clip=true,height=11cm}
\caption{
\label{fig6d8}
Deuteron angular distribution for the process
$^3{\rm He}(\gamma,d)p $ at $\theta_d^{c.m.}=90^\circ$  
as a function of the photon lab energy $E_\gamma$. 
Curves as in Fig.~\ref{fig6d1}.
 The dotted curve practically overlaps with the solid one.  
Since the kinematical shift from the laboratory
to the c.m. system is not significant, we combine the data
for the 90$^\circ$ laboratory angle (up triangles~\cite{Stewart})
with the ones for the 90$^\circ$ c.m. angle (circles \cite{Ticcioni}). 
The squares are the data from \cite{Kundu71}, pluses from \cite{Bermannew}, 
and 
down triangles from \cite{Skopik19}.
}
\end{center}
\end{figure}

\begin{figure}[!ht]
\begin{center}
\epsfig{file=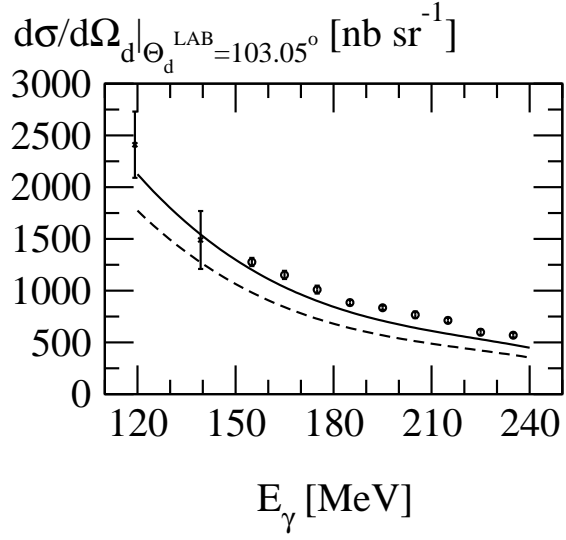,bb=40 500 300 740,clip=true,height=7.5cm}
\caption{
\label{fig6d9}
Deuteron angular distribution for the process
$^3{\rm He}(\gamma,d)p $ at $\theta_d^{lab} = 103.05^\circ$  
 as a function of the photon energy $E_\gamma$.
Curves show results of calculations with the AV18 + UrbanaIX (solid) 
 and with the AV18 alone (dashed). 
Explicit $\pi$- and $\rho$-like MEC's are included in the current operator.
Data are from \cite{Sober} (x-es) and \cite{Ofallon} (circles).
}
\end{center}
\end{figure}

\begin{figure}[!ht]
\begin{center}
\epsfig{file=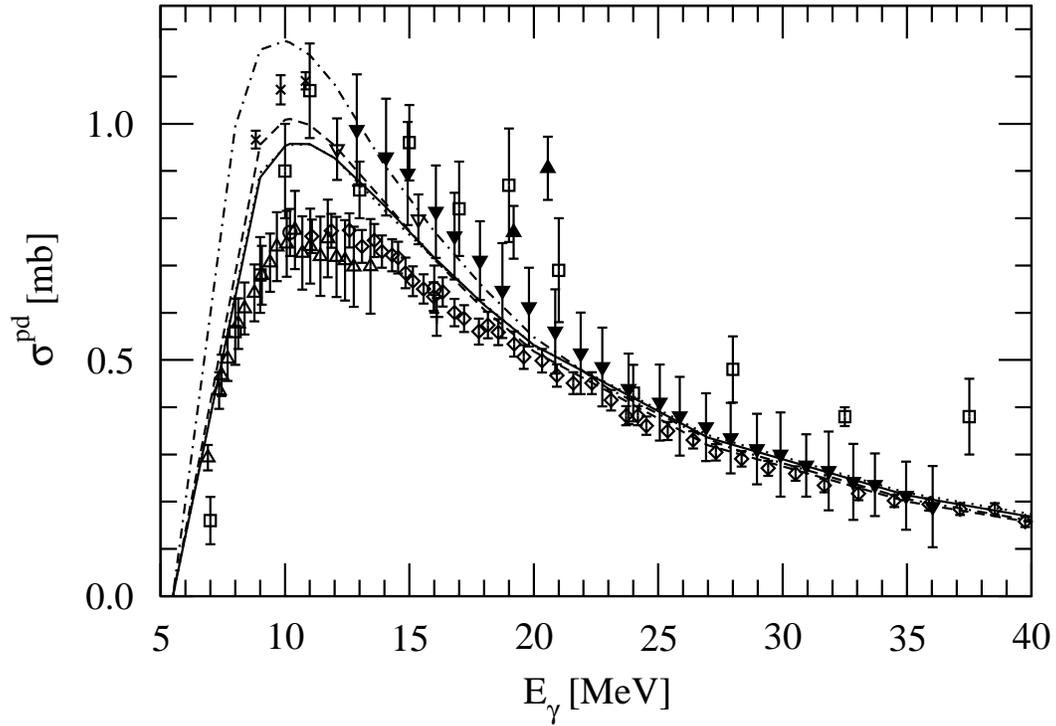,bb=50 530 300
715,clip=true,height=11cm}
\caption{
\label{fig63ver2.2b}
Total $^3$He($\gamma$,p)d cross sections. Curves as in Fig.~\ref{fig6d1}. 
The predictions with UrbanaIX both for Siegert and MEC 
(dotted and solid, respectively) are practically overlapping. 
Data are from: 
 \cite{Fetisov65} (squares), 
 \cite{wolfli66} (up triangles), 
 \cite{Belt} (down triangles), 
  \cite{Wounde71} (full up  triangles), 
 \cite{Ticcioni} (diamonds), 
  \cite{Matthews74} (crosses), 
 \cite{Skopik19} (x-es), 
 \cite{Kundu71} (full down), 
 \cite{Nagai} (circles).
}
\end{center}
\end{figure}

\clearpage

\subsection{Three-nucleon photodisintegration of the $^3$He}

\begin{figure}[!ht]
\begin{center}
\epsfig{file=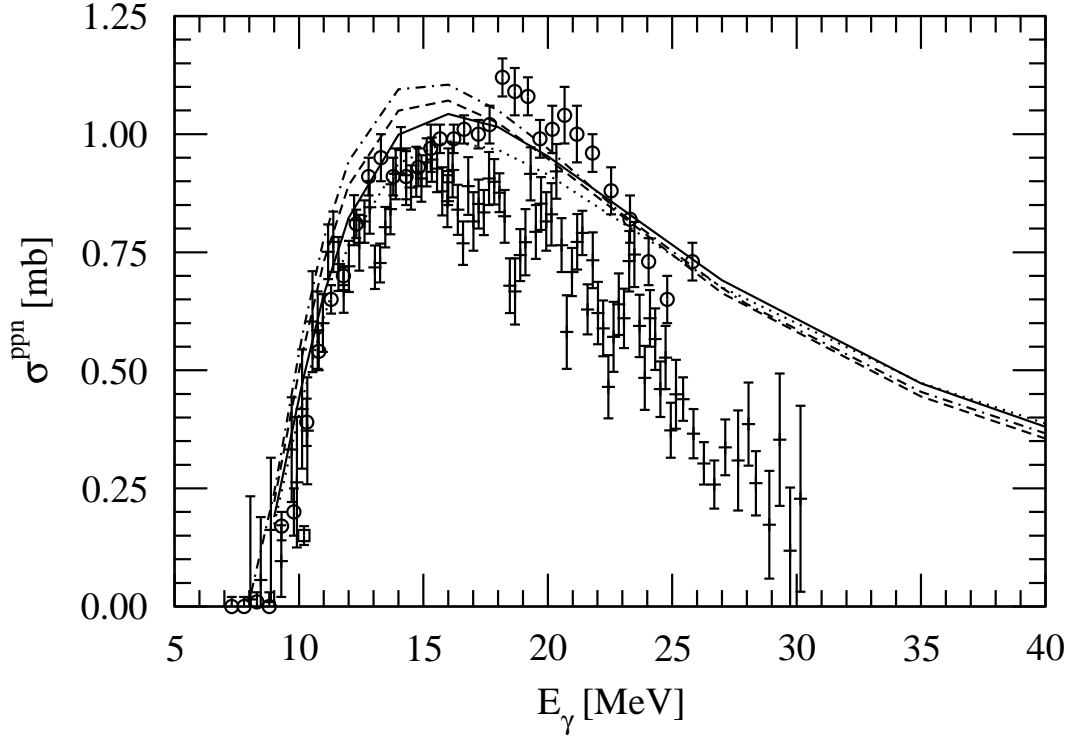,bb=50 530 300
715,clip=true,height=11cm}
\caption{
\label{fig63ver2.3b}
Total $^3$He($\gamma$,pp)n cross sections. 
Curves as in Fig.~\ref{fig6d1}.
Data are from: \cite{Faul} (crosses), \cite{Berman} (circles), 
  and \cite{Nagai} (squares).
         }
\end{center}
\end{figure}

We display in Fig.~\ref{fig63ver2.3b} 
  the total 3N breakup cross section as a
function of the photon energy in the lab system. There is again a 
big spread in the experimental
 data
which precludes any definite conclusion. Especially 
the quick decline of one group of 3N
breakup data in comparison to our theoretical predictions 
is challenging, both for
experiment and theory.



\begin{figure}[!ht]
\begin{center}
\epsfig{file=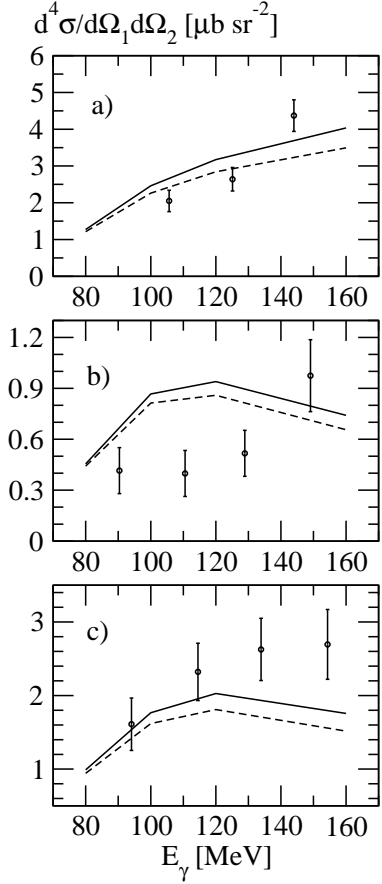,bb=50 200 310 740,clip=true,height=12cm}
\caption{
\label{fig6d11} 
The four-fold differential cross sections
${ {d^{\, 4} \sigma} \over { d \Omega_1 \, d \Omega_2 } }$
for the $^3$He($\gamma$,pp)n process
as a function of $E_\gamma$ in comparison to data from~\cite{Sarty} 
for the angular configurations LR-RL 
$(\Theta_1=81.0^\circ, \Phi_1=0.0^\circ, 
\Theta_2=80.3^\circ, \Phi_2=180.0^\circ)$ (a) ,
LL-RR 
$(\Theta_1=92.2^\circ, \Phi_1=0.0^\circ, 
\Theta_2=91.4^\circ, \Phi_2=180.0^\circ)$ (b) 
and LL-RL+LR-RR
$(\Theta_1=91.7^\circ, \Phi_1=0.0^\circ,
\Theta_2=80.9^\circ, \Phi_2=180.0^\circ)$ and
$(\Theta_1=81.5^\circ, \Phi_1=0.0^\circ,
\Theta_2=90.8^\circ, \Phi_2=180.0^\circ)$
(c). The solid curve
is for AV18+UrbanaIX+MEC, the dashed curve for AV18+MEC.
}
\end{center}
\end{figure}

\begin{figure}[!ht]
\begin{center}
\epsfig{file=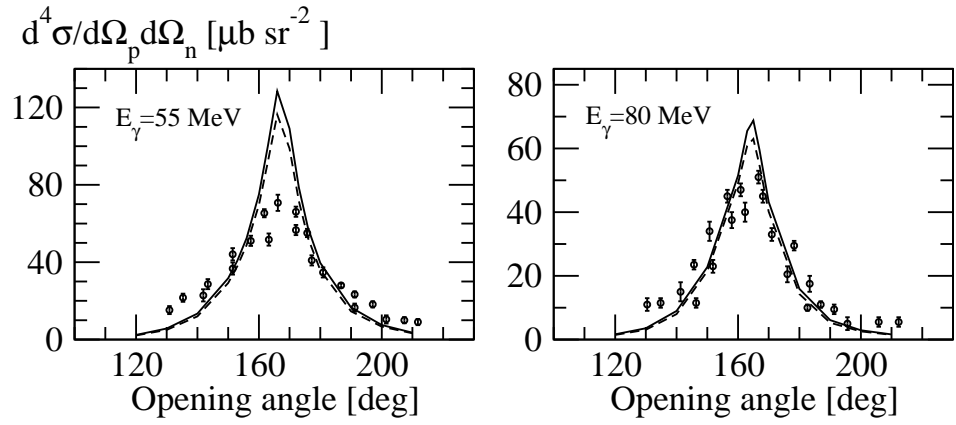,bb=37 520 560 740,clip=true,height=6cm}
\caption{
\label{fig6d12}
The four-fold differential cross sections
${ {d^{\, 4} \sigma} \over { d \Omega_1 \, d \Omega_2 } }$
against the opening angle at $E_\gamma$= 55 (a) and 80 MeV (b)
for the $^3$He($\gamma$,pn)p process
in comparison to data from~\cite{Kolb}. The data in (b) were
taken at $E_\gamma$= 85 MeV. Curves as in Fig.\ref{fig6d11}.
}
\end{center}
\end{figure}

In case of exclusive data we are only aware of two measurements.
In Figs.~\ref{fig6d11} and \ref{fig6d12} we show the four-fold 
differential 3N breakup
cross sections $\frac{d^4\sigma}{d\Omega_1 d\Omega_2}$
for the detection of two nucleons in coincidence. In Fig.~\ref{fig6d11} 
 the dependence on the incoming photon energy 
of the angular configurations, called LR-RL, LL-RR and LL-RL + LR - RR 
in \cite{Sarty}, are
investigated and compared to two of our predictions. Unfortunately we 
lack the
information about angle acceptances of the detectors and therefore the 
comparison could  be
only a rough and qualitative one.

In Fig.~\ref{fig6d12} 
the four-fold differential  cross sections are
displayed as a function of the opening  angle between the 
outgoing neutron and a proton.
Again the exact experimental  conditions were not accessible to us 
and therefore the comparison of our point geometry 
theory and data has to be taken with care. These two experiments 
clearly demonstrate that
data of those types are accessible. Renewed measurements with 
experimentalists and
theoreticians working closely together would be very valuable 
to test the complex
interplay of the dynamical ingredients. 
Recently, in \cite{Deltuva04} the three-nucleon photonuclear reactions with 
$\Delta$-isobar excitation have been analyzed with similar 
results to the ones shown above. 

To close this section, we would like to draw attention 
to a benchmark calculation of the
three-nucleon photodisintegration \cite{gol02},
where the LIT method has been compared to our momentum space Faddeev
treatment. The agreement was quite good. We think that due to the  very
complex dynamics and the numerical challenges 
such benchmarks are necessary to make sure
that the theoretical  predictions really reflect exactly the
dynamical  input and justify the strong efforts of experimental groups.

\clearpage

\section{Predictions}
\label{predictions}

The nuclear forces, on which we base our predictions in this review,
AV18 and UrbanaIX, describe the whole wealth of NN data and reproduce the 
$^3$H and the $^4$He binding energies with high accuracy.
This makes them a very often used tool for predictions 
in nuclear systems. It is certainly 
the "state-of-the-art" of the traditional approach to nuclear physics.
We also employ the $\pi$- and $\rho$-like
two-body currents, which are linked to AV18 using the continuity equation
and in this sense consistent.
These currents are considered the dominant ones.
These dynamical ingredients should already describe 
a wide range of processes. Obviously it is important 
to challenge this scenario and to find its limitations. 
In this section we go beyond the comparison to existing data
and propose additional observables that will probe 
the dynamics more stringently. 
This is, of course, an incomplete and a subjective list
but we hope that it can nevertheless guide future experimental efforts.

\subsection{Inclusive electron scattering on \boldmath{$^3$}He}

\begin{figure}[!ht]
\begin{center}
\epsfig{file=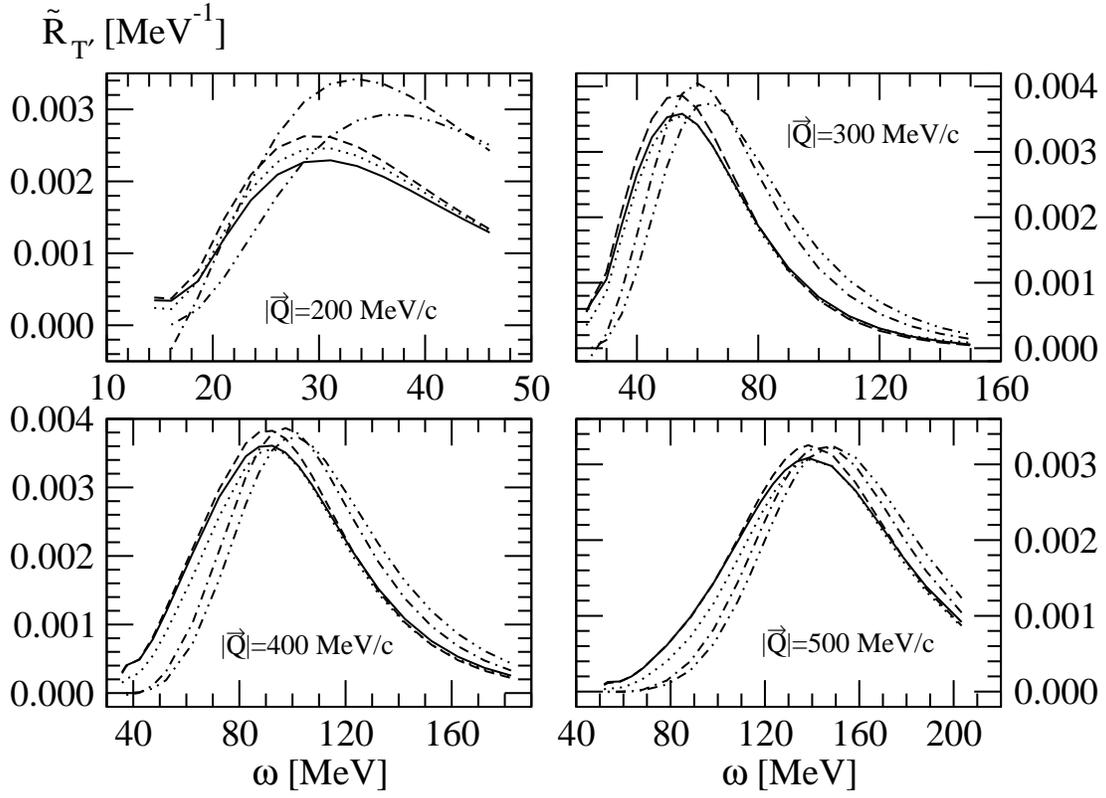,bb=40 360 560 740,clip=true,height=110mm}
\caption{
The response function $\tilde{R}_{T'}$ of $^3$H. Curves as in Fig.~\ref{figRL3H}.
}
\label{figRTP3H}
\end{center}
\end{figure}

\begin{figure}[!ht]
\begin{center}
\epsfig{file=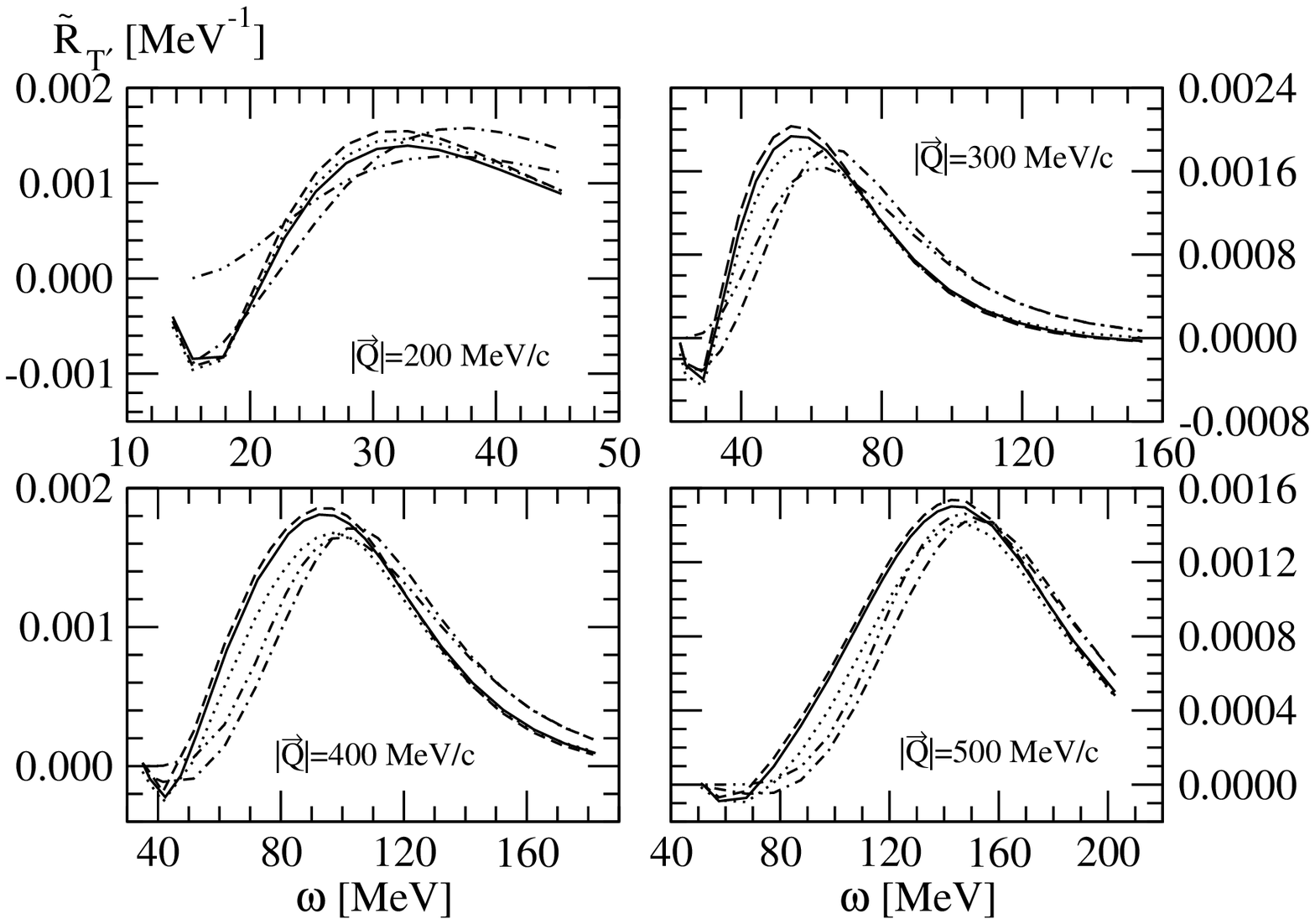,bb=40 360 560 740,clip=true,height=110mm}
\caption{
The response function $\tilde{R}_{T'}$ of $^3$He. Curves as in Fig.~\ref{figRL3H}.
}
\label{figRTP3He}
\end{center}
\end{figure}

\begin{figure}[!ht]
\begin{center}
\epsfig{file=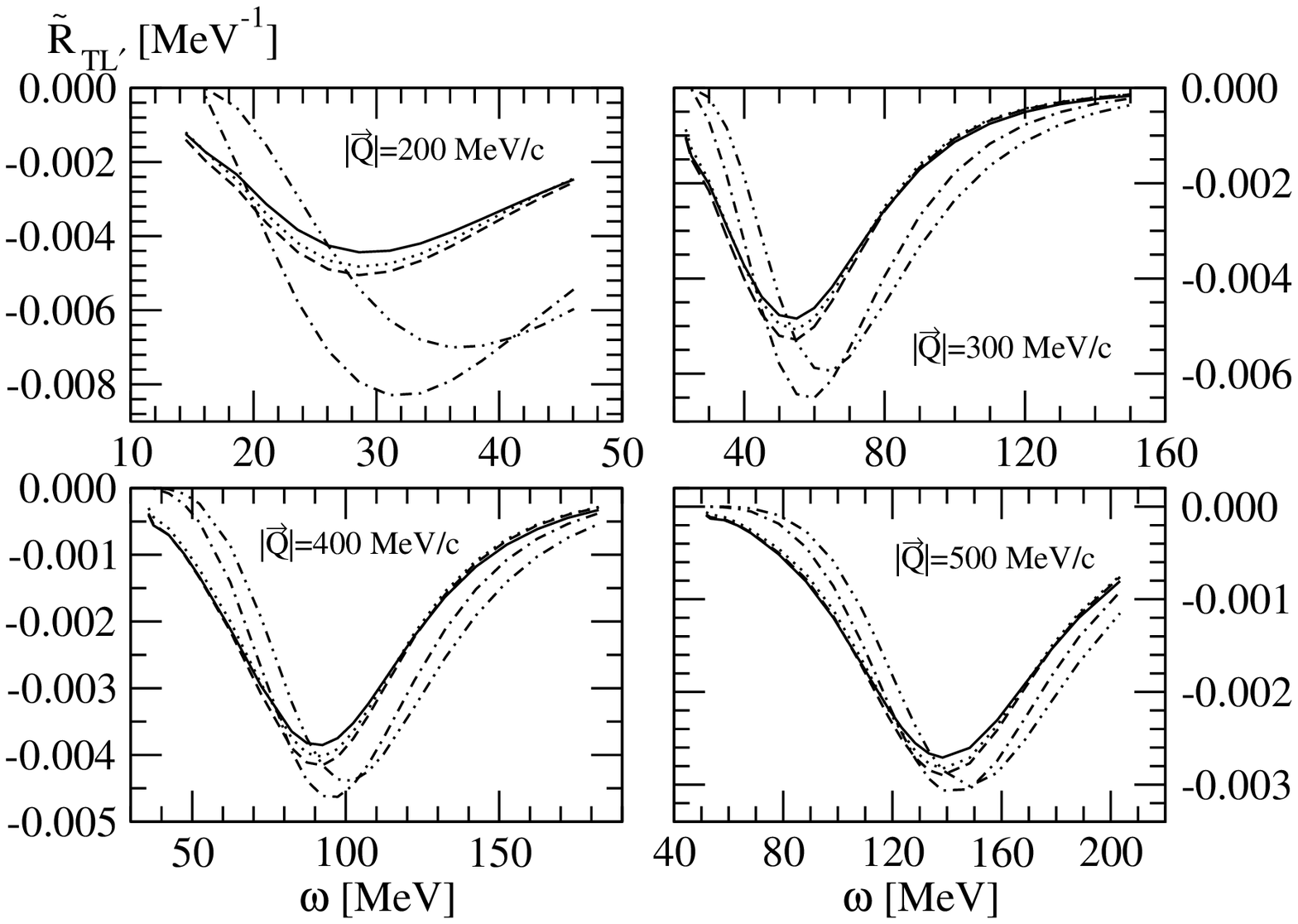,bb=40 360 560 740,clip=true,height=110mm}
\caption{
The response function $\tilde{R}_{TL'}$ of $^3$H. Curves as in Fig.~\ref{figRL3H}.
}
\label{figRTLP3H}
\end{center}
\end{figure}

\begin{figure}[!ht]
\begin{center}
\epsfig{file=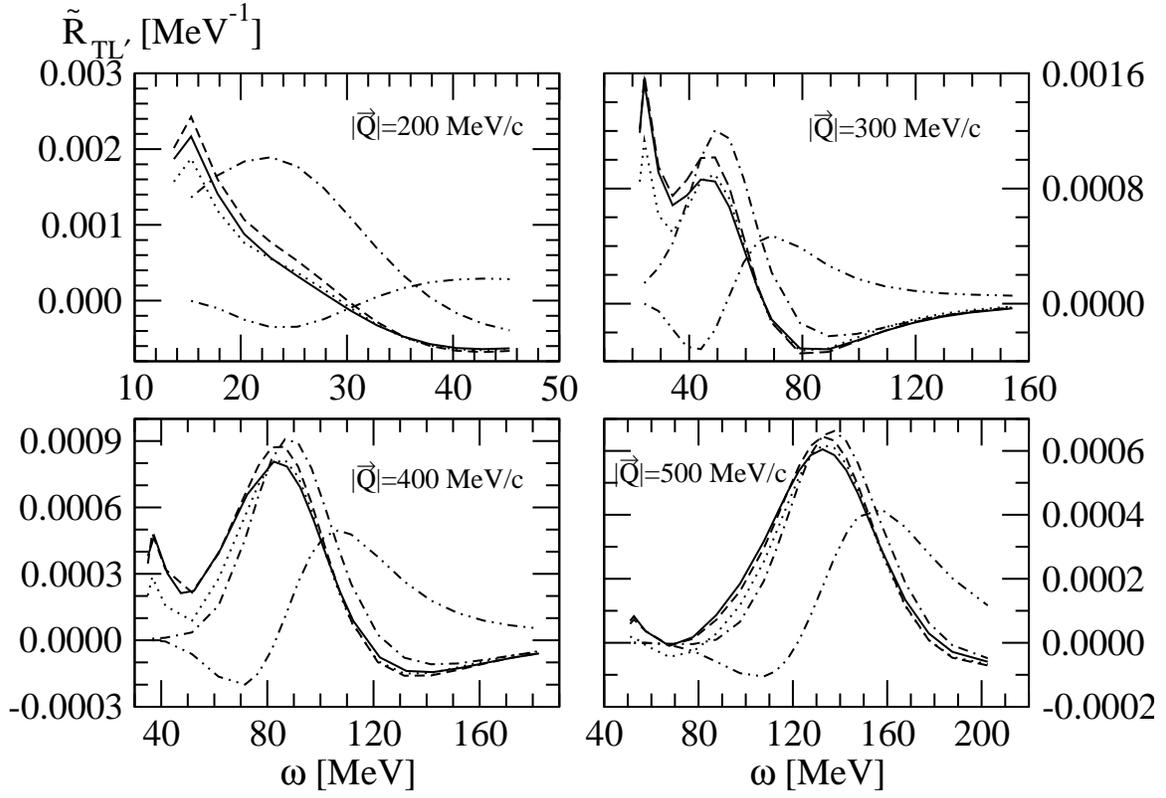,bb=30 360 560 740,clip=true,height=110mm}
\caption{
The response function $\tilde{R}_{TL'}$ of $^3$He. Curves as in Fig.~\ref{figRL3H}.
}
\label{figRTLP3He}
\end{center}
\end{figure}

In section~\ref{sub6b} we showed data for the helicity asymmetries. 
They depend on the initial $^3$He spin direction and on the 
two response functions  $\tilde{R}_{T'}$ and  $\tilde{R}_{TL'}$.
Their measurement by itself
appears to be interesting, since they show a great sensitivity
to the dynamical input as is illustrated in 
Figs.~\ref{figRTP3H}-~\ref{figRTLP3He}. 
 Especially interesting 
appears $\tilde{R}_{TL'}$ for $^3$He, which in addition exhibits
 a strong variation in shape from $\mid \vec Q \mid$= 200 over 300 and 400 to 500 MeV/c. 

\clearpage

\subsection{Electron induced pd breakup of \boldmath{$^3$}He}

\begin{figure}[!ht]
\begin{center}
\epsfig{file=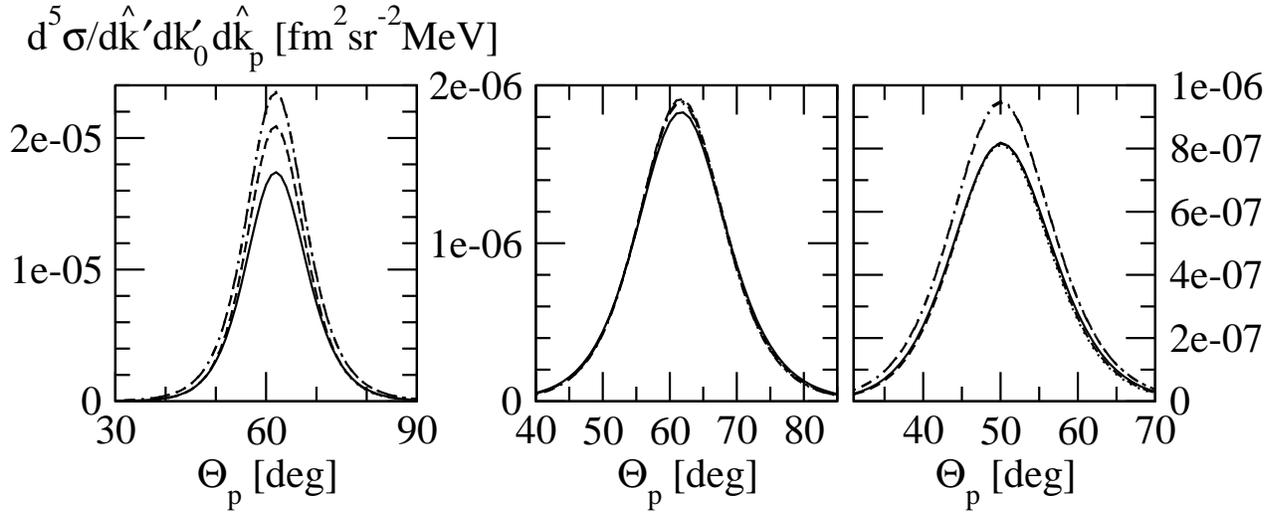,bb=40 520 570 740,clip=true,height=70mm}
\caption{
Proton angular distribution in the vicinity of the proton knockout peak
as a function of the lab 
proton angle $\theta_p$ (measured with respect to the electron beam) 
for three selected electron configurations:
$k_0$= 854.5 MeV, $\vartheta$= 27.9$^\circ$, $k_0^\prime$= 750.9 MeV (left),
$k_0$= 854.5 MeV, $\vartheta$= 35.5$^\circ$, $k_0^\prime$= 754.5 MeV (center)
and
$k_0$= 854.5 MeV, $\vartheta$= 35.7$^\circ$, $k_0^\prime$= 652.3 MeV (right).
The different curves are PWIA (double-dot-dashed line),
PWIAS (dot-dashed line - overlaps with PWIA),
FSI (dotted line),
FSI+MEC (dashed line)
and FSI+MEC+3NF (solid line)
predictions.
}
\label{fig69a_COREL02.3.2}
\end{center}
\end{figure}

\begin{figure}[!ht]
\begin{center}
\epsfig{file=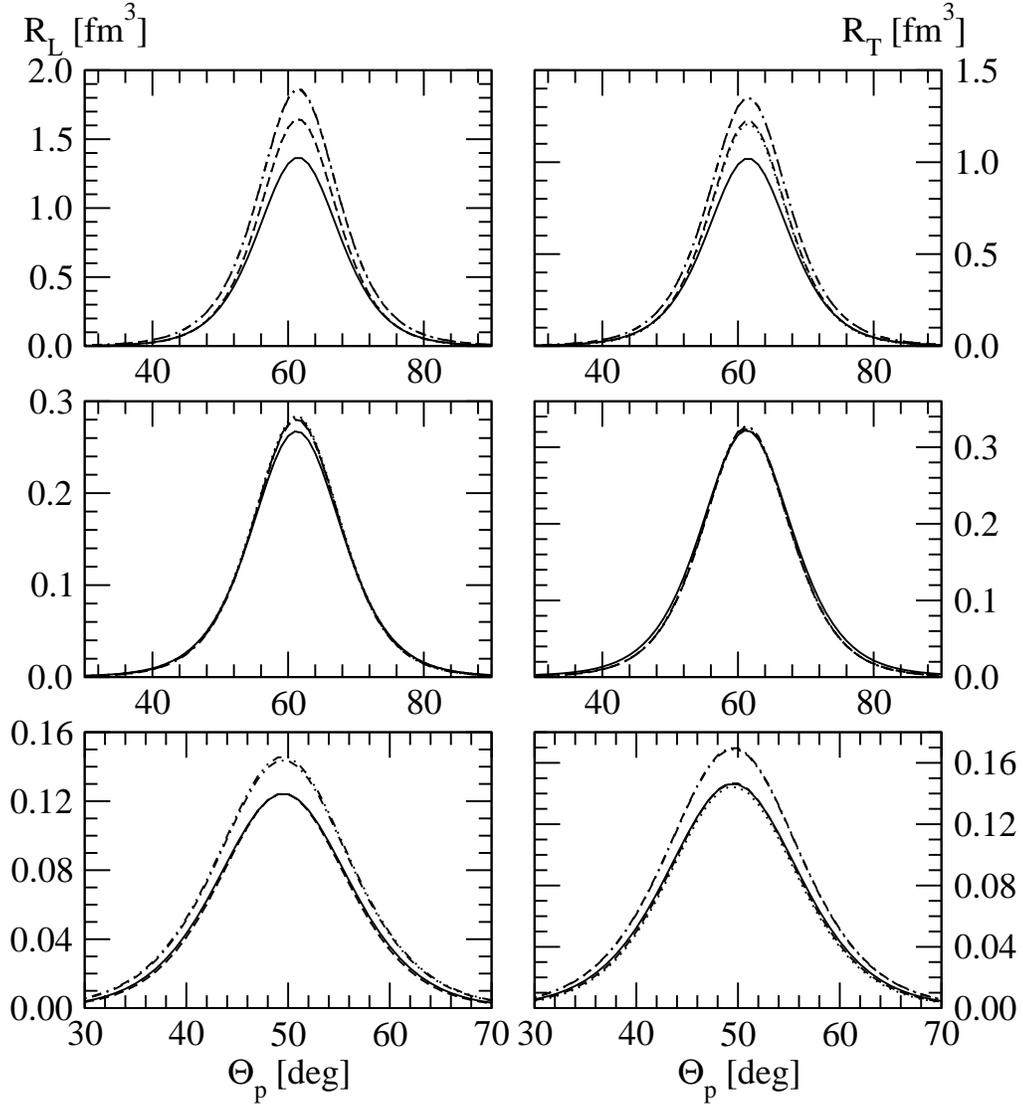,bb=25 200 550 740,clip=true,height=15cm}
\caption{
The longitudinal $R_L$ (left) and transversal $R_T$ (right) responses 
 in the vicinity of the proton knockout peak
as a function of the lab proton  angle $\theta_p$
for the same three electron 
configurations as in Fig.~\ref{fig69a_COREL02.3.2}: 
$k_0$= 854.5 MeV, $\vartheta$= 27.9$^\circ$, $k_0^\prime$= 750.9 MeV (upper
row),
$k_0$= 854.5 MeV, $\vartheta$= 35.5$^\circ$, $k_0^\prime$= 754.5 MeV (middle
row)
and
$k_0$= 854.5 MeV, $\vartheta$= 35.7$^\circ$, $k_0^\prime$= 652.3 MeV (bottom
row).
Curves as in in Fig.~\ref{fig69a_COREL02.3.2}.
}
\label{fig69b_CORREL02.p1.2}
\end{center}
\end{figure}

\begin{figure}[!ht]
\begin{center}
\epsfig{file=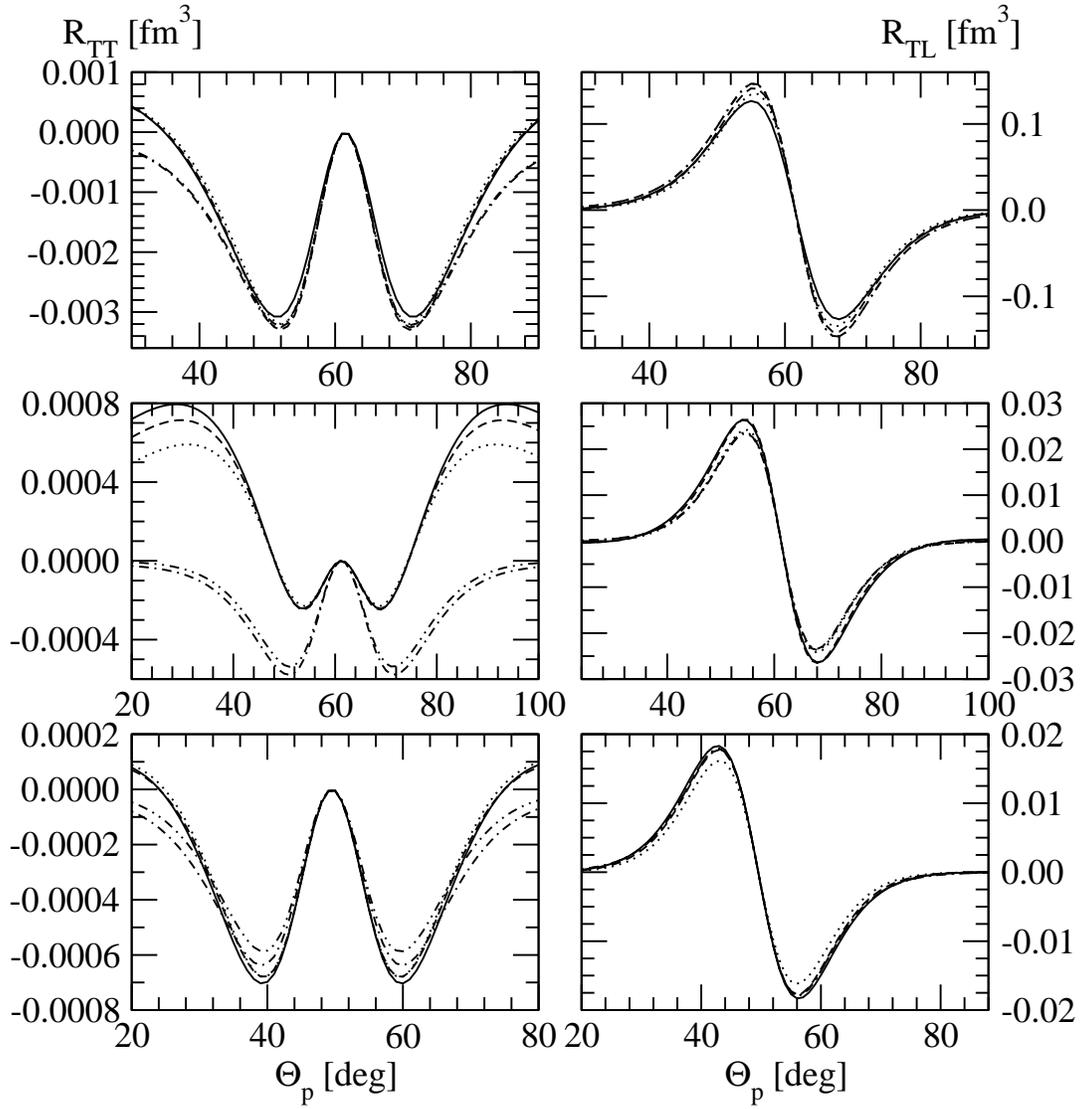,bb=25 200 550 740,clip=true,height=15cm}
\caption{
The same as in Fig.~\ref{fig69b_CORREL02.p1.2}
 but for the $R_{TT}$ (left) and $R_{TL}$ (right) responses. 
}
\label{fig69c_CORREL02.p3.2}
\end{center}
\end{figure}

In view of the discrepancies between theory and data displayed
in section~\ref{sub6c}  it appears advisable to repeat measurements
and to study that process more systematically. In the case of
proton knockout there are cases where  FSI is
negligible and the MEC's we use do not contribute either.
Therefore the angular distribution of the proton around quasi-elastic
kinematics is determined by the simplest ingredients: the $^3$He wave
function, the single nucleon current and the deuteron wave function.
It appears very natural to us that this most simple scenario
should be tested in the first place. To the best of our knowledge
this has not been done up to now. Further, in other cases 
 FSI and/or 3NF effects show up. This is illustrated in 
Figs.~\ref{fig69a_COREL02.3.2}--\ref{fig69c_CORREL02.p3.2}.
 The cross sections shown in Fig.~\ref{fig69a_COREL02.3.2} displays three  
electron configurations (see figure caption). In the left one the FSI effect 
alone is insignificant and is then strongly modified by the inclusion 
of MEC and the 3NF. In the middle one 
PWIA is essentially sufficient and in the 
right one FSI is significant but the addition of the 3NF has no 
further effect. This quite different behavior is of course present in the 
two dominant response functions $R_L$ and $R_T$ 
displayed in Fig.~\ref{fig69b_CORREL02.p1.2}. Finally for the sake of 
completeness the two very small responses $R_{TT}$ and $R_{TL}$ are 
shown in Fig.~\ref{fig69c_CORREL02.p3.2}. 

\begin{figure}[!ht]
\begin{center}
\epsfig{file=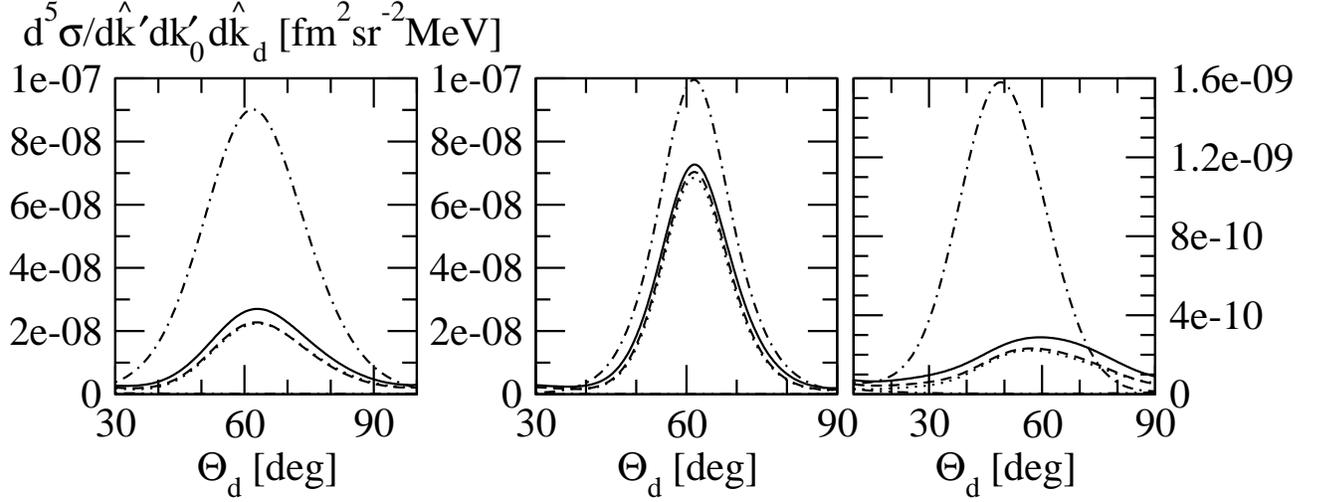,bb=30 520 590 740,clip=true,height=70mm}
\caption{
Deuteron angular distribution in the vicinity of the deuteron knockout peak
as a function of the lab 
deuteron  angle $\theta_d$ 
for the same electron configurations as in Fig.~\ref{fig69a_COREL02.3.2} :
$k_0$= 854.5 MeV, $\vartheta$= 27.9$^\circ$, $k_0^\prime$= 750.9 MeV (left),
$k_0$= 854.5 MeV, $\vartheta$= 35.5$^\circ$, $k_0^\prime$= 754.5 MeV (center)
and
$k_0$= 854.5 MeV, $\vartheta$= 35.7$^\circ$, $k_0^\prime$= 652.3 MeV (right). 
Curves as in in Fig.~\ref{fig69a_COREL02.3.2} with the exception of PWIA, 
which is too small to be visible.
}
\label{fig69d_CORREL02.3.d}
\end{center}
\end{figure}

The situation is different in the deuteron peak area corresponding
to proton angles around 240~$^\circ$. For our $\mid \vec Q \mid$-values
below about 500 MeV/c our theory tells that it is not possible to knock
out the deuteron without FSI. Though the effects of FSI and MEC's
decrease going to higher $\mid \vec Q \mid$-values, sizable effects remain.
This is illustrated in 
 Figs.~\ref{fig69d_CORREL02.3.d}--\ref{fig69f_CORREL02.p3.2.d}. 
   The cross section shown in Fig.~\ref{fig69d_CORREL02.3.d}  
 exhibits very strong shifts
from the PWIAS predictions to the full results generated by FSI and 3NF.
The detailed view into the underlying response functions $R_L$ and $R_T$ in
 Fig.~\ref{fig69e_CORREL02.p1.2.d} 
 show that the MEC contributions in $R_T$ are different in the three
configurations. Interesting is also the shift of the peak position in
the third configuration for the full against the PWIAS result. Note also
that in the cases shown in Fig.~\ref{fig69d_CORREL02.3.d} the effect of the 3NF
moves theory upwards while in Figs.~\ref{pdBL4}--\ref{pdBL3} the 3NF 
effects cause 
a shift downwards. Apparently, there is an intricate dependence on the
kinematical conditions. Again for the sake of completeness the two
smallest response functions are displayed 
in Fig.~\ref{fig69f_CORREL02.p3.2.d}. 
  Precise  new measurements would
be very helpful to test existing and future  dynamical inputs. 
Having the proton and the deuteron peak areas under control one
would have covered essentially the full angular range.

\begin{figure}[!ht]
\begin{center}
\caption{
The longitudinal $R_L$ (left) and transversal $R_T$ (right) responses 
 in the vicinity of the deuteron knockout peak 
as a function of the lab deuteron  angle $\theta_d$ 
for the same  three electron 
configurations as in Fig.~\ref{fig69d_CORREL02.3.d}: 
$k_0$= 854.5 MeV, $\vartheta$= 27.9$^\circ$, $k_0^\prime$= 750.9 MeV (upper
row),
$k_0$= 854.5 MeV, $\vartheta$= 35.5$^\circ$, $k_0^\prime$= 754.5 MeV (middle
row)
and
$k_0$= 854.5 MeV, $\vartheta$= 35.7$^\circ$, $k_0^\prime$= 652.3 MeV (bottom
row).
Curves as in in Fig.~\ref{fig69a_COREL02.3.2}.
}
\label{fig69e_CORREL02.p1.2.d}
\end{center}
\end{figure}

\begin{figure}[!ht]
\begin{center}
\caption{
The same as in Fig.~\ref{fig69e_CORREL02.p1.2.d}
but for the $R_{TT}$ (left) and $R_{TL}$ (right) responses. 
}
\label{fig69f_CORREL02.p3.2.d}
\end{center}
\end{figure}

\clearpage

\subsection{Semiexclusive nucleon knockout processes}
\label{subpredc}

\begin{figure}[htb]
\begin{center}
\epsfig{file=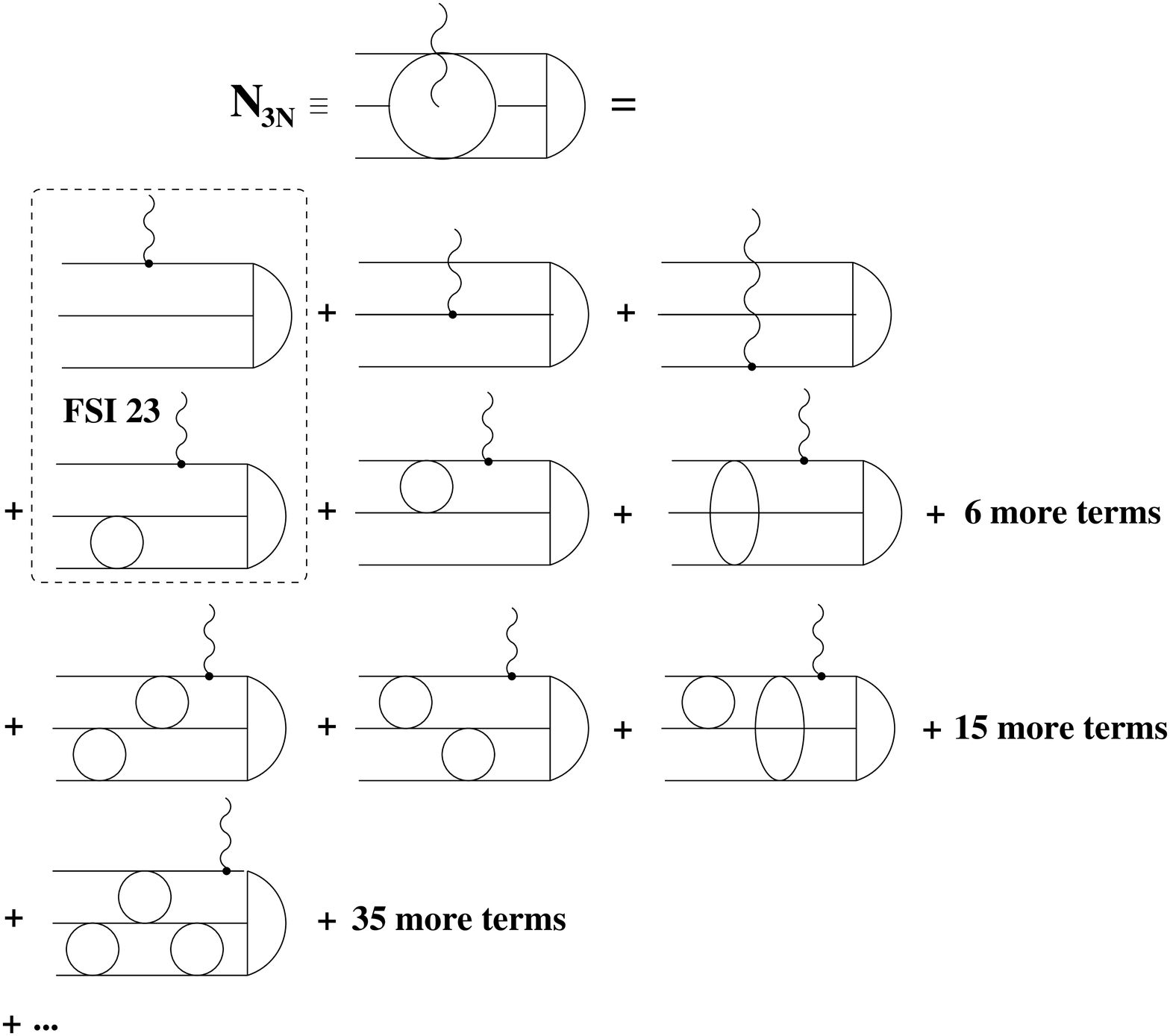,height=10cm}
\caption{\label{diagram3N.2}
Diagrammatic representation of the nuclear matrix element
for the three-body electrodisintegration of $^3$He.
The open circles and ovals represent the two-body $t$-matrices.
Three horizontal lines between photon absorption and forces, and between
forces describe free propagation. The half-moon symbol on the very right
stands for $^3$He.
      }
      \end{center}
      \end{figure}

The analysis of the process
$^3{\rm He}(e,e'p)pn$ has often been done 
approximately using the concept of the spectral function.
The underlying picture is simple. The photon is assumed to be absorbed
by the knocked out proton and the remaining two  nucleons are not
involved in the photon absorption process nor do they interact
with the knocked out proton. The only FSI kept is between the
spectator neutron and proton. This  is technically very easy to calculate,
since beside
the single nucleon current 
 only the $^3$He wave function and the NN $t$-matrix enter. 
Then only the two processes
inside the dashed box of Fig.~\ref{diagram3N.2} are kept. 
This leads to the definition of
 the spectral function
\begin{eqnarray}
S(k,E) = \frac{ m_N \, p}{2} \, \frac12 \sum\limits_m \,
\sum\limits_{m_1, m_2, m_3} \,
\int d{\hat p} \,
\left|
\sqrt{6} \, \langle \nu_1 \nu_2 \nu_3 \mid
\langle m_1 m_2 m_3 \mid  \, \langle {\vec p} \, {\vec k} \mid
( 1 + t G_0  ) \mid \Psi_i m \,
\rangle
\right|^2 .
\label{eq6a}
\end{eqnarray}
The arguments of $S$ are the magnitude $k$ of the missing momentum
\begin{equation}
k \equiv \mid {\vec Q} - {\vec k}_p \mid
\label{eq7}
\end{equation}
and the excitation energy $E$ (missing energy) of the undetected np pair.
 Nonrelativistically
\begin{equation}
E \equiv \frac{p^2}{m_N} ,
\label{eq8}
\end{equation}
where $p$ is the relative momentum of the undetected nucleons.
In addition we completed the notation by adding the isospin quantum numbers
$\nu_i$.
 That strongly reduced treatment of FSI restricted only to 
the spectator nucleons 2 and 3  has already been introduced and denoted 
as FSI23 ($t\equiv t_{23}$). 
 One finds the relations~\cite{ourspectral}
\begin{eqnarray}
   S(k,E) = \frac12 \, m_N \, p \, \frac1{(G_E)^2}
\int d {\hat p} R_L (FSI23) \nonumber \\
          = \frac12 \, m_N \, p \, \frac{2 m_N^2}{{\mid \vec Q \mid}^2 (G_M)^2}
\int d {\hat p} R_T (FSI23) .
\label{eq9}
\end{eqnarray}
This form is convenient to compare to the treatment including
the complete FSI and we define the quantities
\begin{eqnarray}
   S_L (Full) = \frac12 \, m_N \, p \, \frac1{(G_E)^2}
\int d {\hat p} R_L (Full) \nonumber \\
   S_T (Full) = \frac12 \, m_N \, p \, \frac{2 m_N^2}{{\mid \vec Q \mid}^2 (G_M)^2}
\int d {\hat p} R_T (Full) ,
\label{eq9.2}
\end{eqnarray}
which enter directly into the semiexclusive cross section.

Using that simple picture of FSI23 the cross section
factorizes into a kinematical factor, the electron proton cross section, 
and into the spectral function as shown below in (\ref{xs_spectral}).

\begin{figure}[!ht]
\begin{center}
\epsfig{file=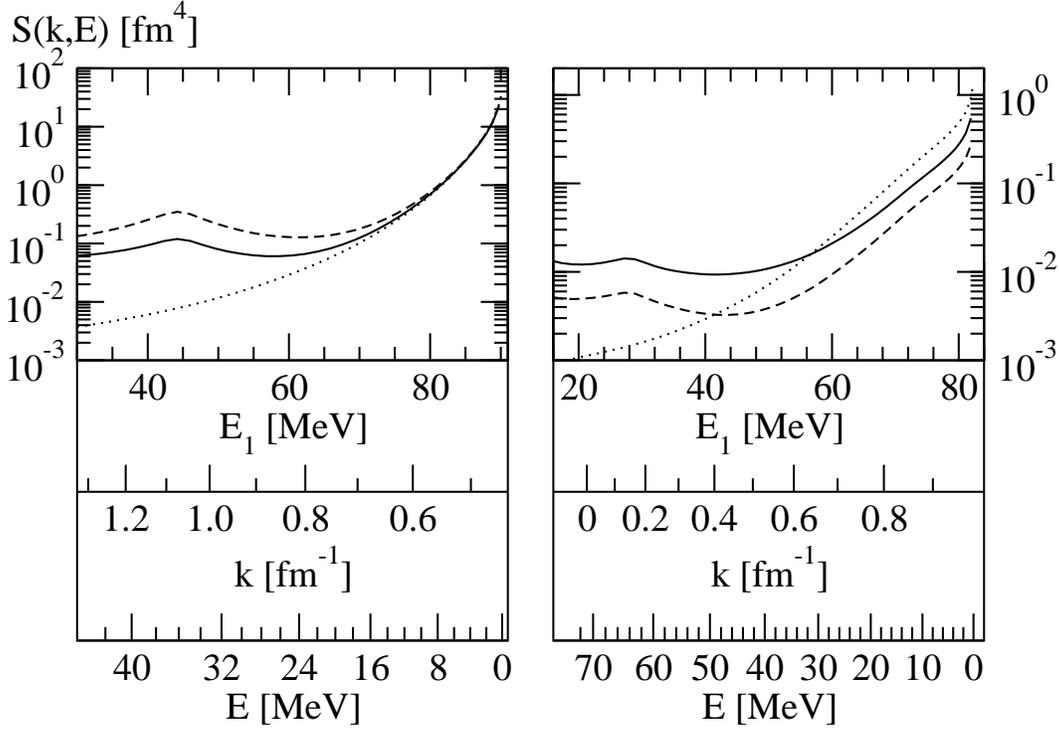,bb=50 395 550 735,clip=true,height=10cm}
\caption{
The spectral function $S(k,E)$ for the proton knockout (dotted line) 
 for two fixed ($\omega-\mid \vec Q \mid$) pairs:  
$\omega$= 100 MeV, $\mid \vec Q \mid$= 500 MeV/c
(left)  and  $\omega$= 100 MeV, $\mid \vec Q \mid$= 200 MeV/c (right)
as a function of the ejected proton energy $E_1$ for the parallel
kinematics ${\vec p}_1 \parallel {\vec Q}$.
The corresponding values of $k$ and $E$ are also indicated. 
 The dashed line is the result based on the full treatment of FSI 
 but neglecting MEC and
3NF effects in the form of Eq.~(10) in ~\cite{ourspectral}
for the response functions $R_L$ 
and the solid line is the 
 corresponding result for the response functions $R_T$.
}
\label{fig70a_ppn-I25.v2}
\end{center}
\end{figure}

We performed very recently a thorough
investigation~\cite{ourspectral} on the validity of that 
approximation in the domain
of nonrelativistic kinematics. We assumed the most favorable condition
of parallel kinematics (${\vec k}_p \parallel {\vec Q}$).
The result was that only for very small
missing momenta $k \equiv p_m$ and missing energies $E$
the use of the spectral function is quantitatively justified. To each
$(k,E)$ pair under parallel condition, $\omega \equiv Q_0$ and $\mid {\vec Q} \mid$
are connected by a quadratic
equation. Then we found that there is a domain of $(k,E)$-values  
where, at least with increasing $\mid {\vec Q} \mid$, 
  $S_L (Full)$ and $S_T (Full)$
approach $S$.
 But unexpectedly for our present insight even for quite small
$k$-values but increasing $E$ values that simplified picture is
invalid. We refer the interested reader for details to \cite{ourspectral}.
In any case again there are clear cut cases, where the concept of the spectral
function is valid and they should be tested against precise data. Like
in the quasi elastic proton peak for pd breakup, also in this case the theoretical
ingredients are quite simple: just the $^3$He wave function, the NN $t$-matrix,
and the  single  nucleon current. The quantitative validity of this
simple picture should be tested in the first place. But then it is also
very interesting to probe the FSI, MEC and the 3N force effects if
one takes other $(k,E)$ pairs.
 We illustrate two of various cases from \cite{ourspectral}  
 in Fig.~\ref{fig70a_ppn-I25.v2}. The quantities (\ref{eq9.2}) 
  together with the spectral function are plotted as a function of
the ejected proton energy $E_1$ for parallel kinematics 
 ${\vec k}_1 \parallel  \vec Q$. We see in
 the left part of the Fig.~\ref{fig70a_ppn-I25.v2} 
that at the upper end of $E_1$ the three curves approach each
other, thus the spectral function concept works very well. The
corresponding decreasing values of k and E for increasing $E_1$ are also
indicated. A counter example is shown in the right part of the 
 Fig.~\ref{fig70a_ppn-I25.v2},  where the use of the
spectral function would be a very poor approximation.

In the case of the process
$^3{\rm He}(e,e'n)pp$ the concept of the spectral function
is useful only for $R_T$ but not  for $R_L$. In the case of $R_L$ 
it is a totally insufficient approximation. The reason is of course the
smallness of $G_E^n$ and the strong interference of the photoabsorption 
on the protons.

If the approximation leading to the spectral function
was valid, it could be quite reliably used to extract
electromagnetic nucleon form factors.
The cross section factorizes as~\cite{ourspectral}
\begin{eqnarray}
\frac{d ^6 \sigma}{ d k_0^\prime d \hat{k}^\prime d \hat{k}_1 d E_1 } =
\sigma_{\rm Mott} \,
\left[  v_L (G_E)^2 + v_T 
\frac{{\mid \vec Q \mid}^2 (G_M)^2}{2 m_N^2} \right] \, S (k, E) \, m_N \, k_1  .
\label{xs_spectral}
\end{eqnarray}
This should be experimentally tested 
for $G_E^p$ and $G_M^p$ since the proton form factors are known and then
be applied to $G_M^n$.

Since our investigation in \cite{ourspectral} was restricted
to the nonrelativistic domain, it does not provide information
for the relativistic region.
In \cite{kiev97,ciofi} the spectral function concept has been studied
at higher $\omega$ and $\mid \vec Q \mid$-values.
The verification of such an assumption
requires a full-fledged relativistic framework
including on top of FSI 3N forces and MEC's.

\begin{figure}[!ht]
\begin{center}
\epsfig{file=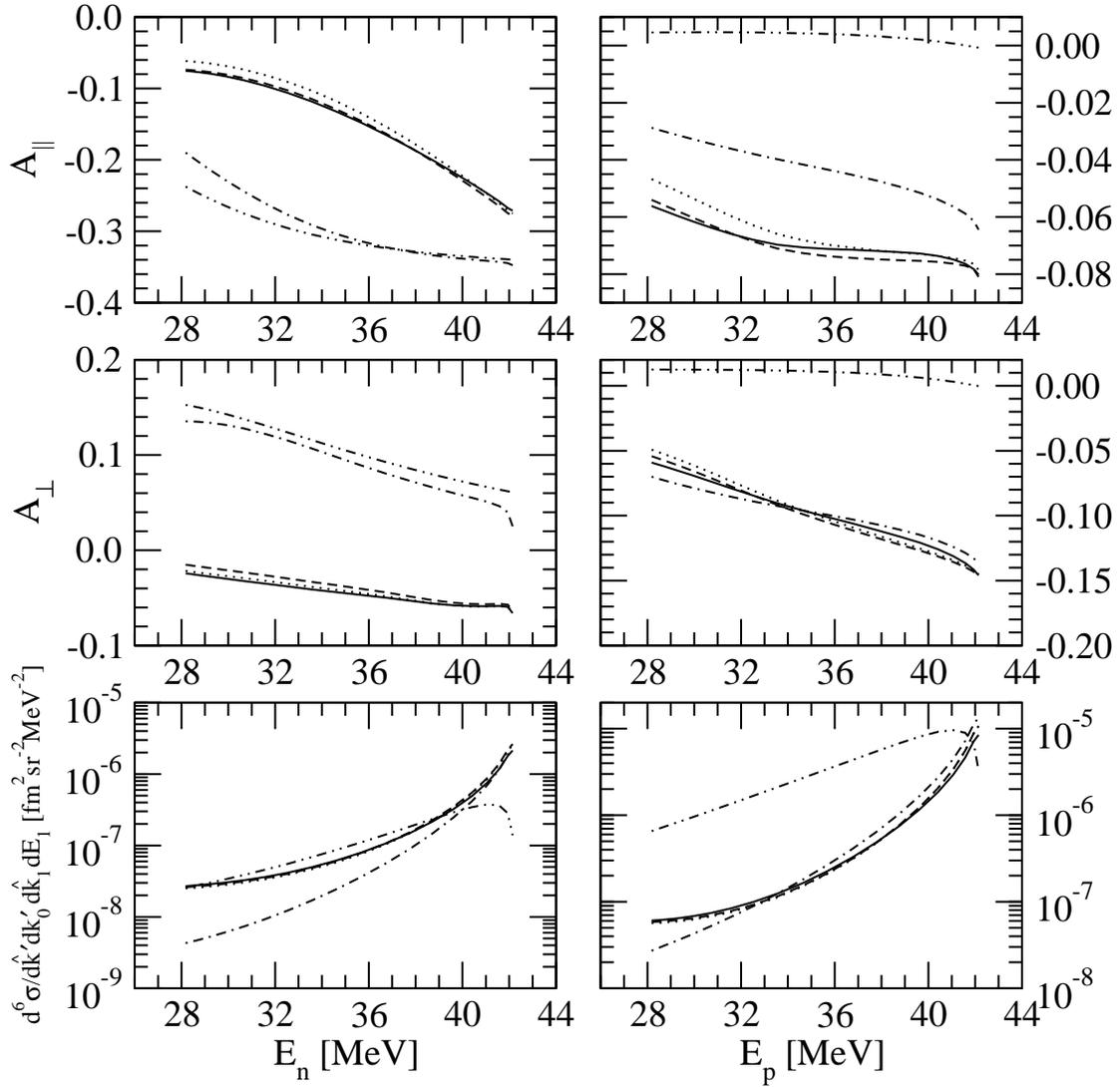,bb=25 200 550 720,clip=true,height=15cm}
\caption{
\label{asymmetry1.e1.1}
The dependence of $A_{\parallel}$,  $A_{\perp}$, and the cross section 
on the energy 
of the outgoing neutron (left column) and proton (right column) in the 
$ \vec{^3He} ({\vec e},e' N)NN$ reaction for
$E_e$ = 854.5 MeV, $\omega$ = 50 MeV, $\mid \vec Q \mid$= 300 MeV/c. 
 The double-dot-dashed, dot-dashed, dotted, dashed, and solid curves are 
based on PWIA, FSI23, FSI, FSI+MEC, and FSI+MEC+3NF, respectively. }
\end{center}
\end{figure}

\begin{figure}[!ht]
\begin{center}
\epsfig{file=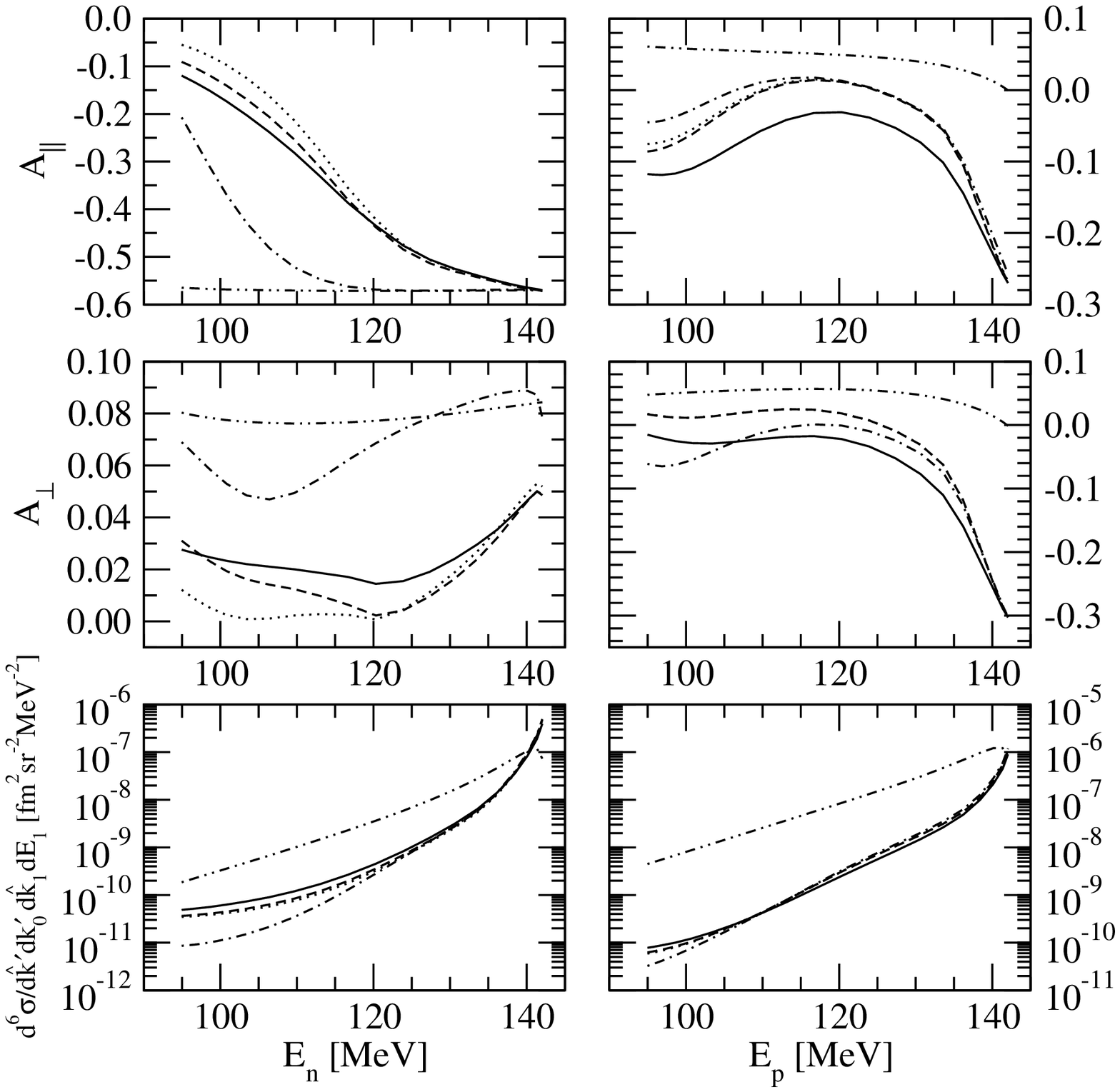,bb=25 200 550 720,clip=true,height=15cm}
\caption{
The same as in Fig.~\ref{asymmetry1.e1.1}  but for 
$E_e$ = 854.5 MeV, $\omega$ = 150 MeV, $\mid \vec Q \mid$= 500 MeV/c. 
}
\label{asymmetry1.e1.2}
\end{center}
\end{figure}

Now let us regard the semiexclusive process
$ \overrightarrow{^3{\rm He}} ({\vec e},e' n)pp$
for an initially polarized $^3$He and polarized 
electron. The asymmetries (199) for
parallel, $A_\parallel$, and perpendicular, $A_\perp$, orientation
of the $^3$He spin in
relation to the photon direction are proportional to $(G_M^n)^2$ and
$G_E^n \, G_M^n$, 
respectively,  under the simplifying assumptions of PWIA and
the restriction of the $^3$He state to the principal S state 
\cite{withZiemer,Blankleider84,friar1990}. 
  If that sensitivity survives for the full
dynamics one can extract the neutron form factors.
In Figs.~\ref{asymmetry1.e1.1} and \ref{asymmetry1.e1.2} we provide
$A_\parallel$ and $A_\perp$
as a function of the ejected neutron energy for two
kinematical  conditions 
using
different dynamical assumptions. We choose the
most favorite  configuration, where the neutron is ejected parallel to
the photon. In both figures five curves are shown,
PWIA, FSI23, FSI, FSI+MEC and FSI+MEC+3NF. For $\omega$ = 50 MeV, 
$\mid \vec Q \mid$= 300
MeV/c, $Q^2 = $ 0.087 $({\rm GeV/c})^2$ shown in Fig.~\ref{asymmetry1.e1.1}, 
$A_\parallel$ for PWIA and FSI23
stays far off the results gained under FSI and with the further
ingredients MEC and 3NF. Thus the extraction of $(G_M^n)^2$ under the
simplifying assumptions of PWIA or FSI23 would require big corrections.
In the case of $A_\perp$ that is also the case but $A_\perp$ is anyhow very
small. Only out of curiosity we add the corresponding results for the
proton ejection. Since the polarization of the proton inside polarized
$^3$He is very small, PWIA is of course far away from the other
results. Then, in Fig.~\ref{asymmetry1.e1.2},
for $\omega = 150$~MeV, $\mid \vec Q \mid$= 500 MeV/c, $Q^2 =$ 0.228 $({\rm GeV/c})^2$
the situation is quite different. All curves for $A_\parallel$
coincide at the upper end of the neutron energy. This should
allow one to extract $(G_M^n)^2$
without big corrections. However, for $A_\perp$ large 
corrections remain. In the case of the proton ejection the
energy
dependence of both asymmetries is totally different from the neutron ones what 
would be interesting to  check 
experimentally. Also proton asymmetries reveal sizable 3NF effects. 
For the sake of completeness
and orientation about the magnitudes of the cross sections we also
include their values in the Figs.~\ref{asymmetry1.e1.1}--\ref{asymmetry1.e1.2}.
These two examples just illustrate that both processes, neutron  as
well as proton  emission,  provide interesting tests of the dynamical
inputs if  accurate data can be gained.


In \cite{withZiemer}
the process $ \overrightarrow{^3{\rm He}} ({\vec e},e' n)pp$
has been applied to extract $G_E^n$ for $Q^2 \approx$ 0.35 $({\rm GeV/c})^2$.
As expected, it turned out that the full FSI was required.
The resulting $G_E^n$-value
was quite different from the one extracted under the assumption
that polarized $^3$He is just a polarized neutron target~\cite{Becker}.
Despite the inclusion of FSI the theoretical analysis in \cite{withZiemer}
was a bit overstretched since we relied on a nonrelativistic framework.
The extracted value for $G_E^n$ might have differed
a bit if relativity and MEC's had been included.

At higher $Q^2$-values the situation appears to be more favorable for the
application of 
the theoretically simple approach offered by FSI23 
as argued in \cite{Rome,Bermuth03}.

The cross section for the semi-exclusive
$ {\vec e} \left( {\overrightarrow{^3{\rm He}}},e' N \right) $ reaction
can also be cast in the following general form \cite{Laget91-92}
\begin{equation}
\sigma (h, {\vec A} )  =  
\sigma_0 \left[ 1 + {\vec S} \cdot {\vec A}^{\ 0}  
+  h \left( A_e + {\vec S} \cdot {\vec A}^{\ '} \right) \right] ,
\end{equation}
where $\sigma_0$ is the unpolarized cross section,
$A_e$ is the electron analyzing power,
${\vec A}^{\ 0}$ the $^3$He target analyzing power
and ${\vec A}^{\ '}  $ are the spin correlation parameters.
The target analyzing power is accessible in experiments where unpolarized
electrons are scattered on the polarized $^3$He.
Due to the symmetry properties only the component of ${\vec A}^{\ 0}$
perpendicular to the electron plane (usually denoted as ${A}^{\ 0}_y$)
is different from zero. This observable 
provides direct information on the importance of FSI 
because it vanishes for calculations
neglecting totally (PWIA and PWIAS) or partly (FSI23)
the final state interactions
among the three outgoing nucleons. 

${A}^{\ 0}_y$ was measured
at MAMI \cite{Bermuth03} and 
this experiment supplied very interesting insight into
the reaction mechanism, even
though the experimental conditions required a lot of integrations over
the relevant parts of the phase space.
It turned out that at $Q^2$= 0.37 (GeV/c)$^2$
the analyzing power ${A}^{\ 0}_{y\, (e,e'n)}$ results
from a coupling of the virtual photon followed by
proton-neutron rescattering. Also a different sensitivity to MEC
for ${A}^{\ 0}_{y\, (e,e'n)}$ and ${A}^{\ 0}_{y\, (e,e'p)}$
was confirmed in \cite{Bermuth03}.

\clearpage
\subsection{The electron induced complete 3N breakup process}

The process $^3$He(e,e'pp)n has been measured 
in the NIKHEF facility~\cite{Groep}.
Unfortunately the kinematical conditions were outside of the 
nonrelativistic domain and the comparison  to our
theory  was generally unsuccessful. Discrepancies
up to factors of 4-5 showed up. Possibly the neglecting of the
$\Delta$-degrees of freedom was the strongest theoretical defect 
(see \cite{Deltuva2004}). Such
a theoretical analysis requires a close interaction  of theory
and experiment since the data are taken in a regime far off from the point geometry, 
where  the two protons are detected
at fixed angles and fixed energies
and this in coincidence with the electron, also detected point-wise.
Due to the smallness of the cross
section, quite  large portions of the  phase
space have to be covered with  large energy and angular bins
to arrive at breakup observables with reasonably small error bars.
 Nevertheless, we would like to show 
 in Figs.~\ref{ppn_examples.1}--\ref{ppn_examples.4} 
 some examples of eightfold
differential cross sections along the kinematical locus for selected
breakup configurations.  
   In Fig.~\ref{ppn_examples.1}
  three final state interaction peaks are shown where
the individual contributions of FSI, MEC and 3NF differ quite
strongly from one peak to the other. Quasi free scattering with
 one final nucleon momentum zero is
shown in Fig.~\ref{ppn_examples.2}. Again the 
individual contributions of the three
dynamical ingredients, FSI, MEC and 3NF among each other and against the
PWIAS prediction differ significantly. The space star configuration is
shown in Fig.~\ref{ppn_examples.3} for two electron kinematics.  
 Very strong dynamical effects
beyond PWIA(S) and FSI23 are seen. Finally cross sections for two
electron kinematics are shown in Fig.~\ref{ppn_examples.4} 
where two nucleons emerge 
back to
back collinear with the photon momentum $\vec Q$.

A second exclusive $^3$He(e,e'pn)p experiment was performed at MAMI 
and is presently
analyzed~\cite{MAMIexclusive}. But again the kinematics is outside of our
nonrelativistic domain.

Insight into the NN correlations in a nucleus is an old issue.
In a recent measurement~\cite{Weinstein} an idea proposed also 
in \cite{actapolonica}
has been realized.
The idea is that the photon is assumed to be absorbed by one nucleon alone,
which is knocked out in the  direction of the photon. The other two spectator 
nucleons leave $^3$He back to back and are assumed not to interact
with the knocked out nucleon. This is the same picture as the one
underlying the spectral function. But now one regards the fully
exclusive process and aims at  the relative  momentum distribution of
the two spectator nucleons. If they did not interact also with
each other, one would see directly the relative momentum distribution
of the two nucleons inside $^3$He. In our notation using Jacobi momenta this
quantity is
\begin{equation}
C (p) = 3 \, \sum\limits_{m} \, \sum\limits_{m_1,m_2,m_3} \,
\left| \Psi ( {\vec p}, {\vec q}= 0) \right|^2 .
\label{Cp}
\end{equation}

We investigated that scenario allowing for the complete FSI, for the
interaction just among the two spectator nucleons (FSI23), and for 
 the case of no FSI at all and no antisymmetrization
in the final state (PWIA). It is easy to see \cite{actapolonica}
that the only two response functions surviving for parallel
kinematics are related to $ C(p)$
in PWIA as
%
\begin{eqnarray}
C ( p ) =
\frac12 \, \sum\limits_{m} \, \sum\limits_{m_1,m_2,m_3} \,
R_L^{PWIA} / G_E ^2
\nonumber \\[4pt]
C ( p ) =
\frac12 \, \sum\limits_{m} \, \sum\limits_{m_1,m_2,m_3} \,
2 m_N^2 \, R_T^{PWIA} / \left( {\vec Q}^2 \, G_M^2 \, \right)
\label{eq:15}
\end{eqnarray}
%
Therefore we investigated
$ \frac12 \, \sum\limits_{m} \, \sum\limits_{m_1,m_2,m_3} \, R_L/(G_E)^2$
and
$ \frac12 \, \sum\limits_{m} \, \sum\limits_{m_1,m_2,m_3} \,
2 m_N^2 \, R_T / \left( {\vec Q}^2 \, G_M^2 \, \right)$
as a function of $p$ for different $\mid \vec Q \mid$-values
and for a fixed sequence of the isospin magnetic quantum numbers.
In PWIA this is just $C (p)$ and the question is
whether, at least with increasing $\mid \vec Q \mid$, FSI looses importance 
for this geometry and $C (p)$ can be extracted.
 It turned out that this does not happen. 
Interestingly, with increasing $\mid \vec Q \mid$-values 
and for proton knockout  
the FSI23 and FSI predictions  approach each other.
However, there still remains a noticeable 
shift toward the result which in addition includes the 3NF's. 
  Thus one has no direct access to $C (p)$. 
 If one is satisfied, however, with a less quantitative result and does 
not pay attention to the shift caused by that additional 3NF effect, 
one has access to a modified $C (p)$ quantity, where the two spectator
nucleons, while leaving $^3$He, interact strongly by the NN t-matrix.
Therefore since the $t$-matrix is rather well 
 under control one can at least approach 
the momentum distribution inside $^3$He modified only by that additional
final state interaction. This is illustrated in 
Fig.~\ref{fig72a_ppn.n0.new.500}.   
Note that this final state interaction leads to a reduction by a factor
10 and more.
The curves in Fig.~\ref{fig72a_ppn.n0.new.500} 
refer to a fixed angle of 90$^\circ$ 
between ${\vec p}$
and $ {\vec Q}$, but for other angles qualitatively the situation is unchanged.
It would be very interesting if these configurations could be measured
in our nonrelativistic domain.

In the case when a neutron is knocked out  
$ \frac12 \, \sum\limits_{m} \, \sum\limits_{m_1,m_2,m_3} \, R_L/(G_E^n)^2$ 
 behaves 
differently and the
FSI23 approximation is unjustified. It is only for
 $ \frac12 \, \sum\limits_{m} \, \sum\limits_{m_1,m_2,m_3} \, 
R_T/( \frac{{\vec Q}^{\, 2}}{2 m_N^2} \, (G_M^n)^2 )$    that the
situation is as favorable as for the proton knockout~\cite{actapolonica}. 
This is displayed in Fig.~\ref{fig72b_npp.n0.new.500}. For 
larger $\mid \vec Q \mid$-values ($\mid \vec Q \mid$ = 600~MeV/c), however, 
 we found that also for $R_L/(G_E^n)^2$ the situation resembles 
the one for the proton. 


\begin{figure}[!ht]
\begin{center}
\caption{
The eightfold  full breakup cross section
$d^8\sigma/(d{\hat k}^{\,\prime} d k_0^\prime d{\hat k}_1 d{\hat k}_2 dS)$
along the arc-length $S$ of the
kinematically allowed locus in the $E_1$-$E_2$ plane
for three different electron configurations:
$k_0$= 854.5 MeV, $\vartheta$= 27.9$^\circ$, $k_0^\prime$= 750.9 MeV (left),
$k_0$= 854.5 MeV, $\vartheta$= 35.5$^\circ$, $k_0^\prime$= 754.5 MeV (center)
and
$k_0$= 854.5 MeV, $\vartheta$= 35.7$^\circ$, $k_0^\prime$= 652.3 MeV (right).
PWIA (double-dash-dotted line),
PWIAS (double-dot-dashed line -  overlaps with PWIA),
FSI23 (dot-dashed line),
FSI (dotted line),
FSI+MEC (dashed line)
and FSI+MEC+3NF (solid line)
predictions are shown.
Particles 1 and 2 are protons.
The angles of the outgoing nucleons in the system
where ${\vec Q} \parallel {\hat z}$
($\theta_1$=60.0$^\circ$, $\phi_1$=0.0$^\circ$, $\theta_2$= 51.0$^\circ$,
$\phi_1$= 180.0 $^\circ$ (left),
 $\theta_1$=60.0$^\circ$, $\phi_1$=0.0$^\circ$, $\theta_2$= 34.0$^\circ$,
$\phi_1$= 180.0 $^\circ$ (center),
 $\theta_1$=60.0$^\circ$, $\phi_1$=0.0$^\circ$, $\theta_2$= 59.0$^\circ$,
$\phi_1$= 180.0 $^\circ$ (right))
are chosen in such a way that
the peaks correspond to the kinematical condition ${\vec k}_2 = {\vec k}_3$ 
 (the final state interaction condition).
}
\label{ppn_examples.1}
\end{center}
\end{figure}
\begin{figure}[!ht]
\begin{center}
\caption{
The eightfold  full breakup cross section
$d^8\sigma/(d{\hat k}^{\,\prime} d k_0^\prime d{\hat k}_1 d{\hat k}_2 dS)$
along the arc-length $S$ of the
kinematically allowed locus in the $E_1$-$E_2$ plane
for three different electron configurations:
$k_0$= 854.5 MeV, $\vartheta$= 27.9$^\circ$, $k_0^\prime$= 750.9 MeV (upper
row),
$k_0$= 854.5 MeV, $\vartheta$= 35.5$^\circ$, $k_0^\prime$= 754.5 MeV (middle
row)
and
$k_0$= 854.5 MeV, $\vartheta$= 35.7$^\circ$, $k_0^\prime$= 652.3 MeV (lower
row).
Curves as in Fig.~\ref{ppn_examples.1}.
In the left panel particles 1 and 2 are protons
while in the right panel particles 1 and 2 are neutron and proton,
respectively.
The angles of the outgoing nucleons in the system
where ${\vec Q} \parallel {\hat z}$
($\theta_1$=60.0$^\circ$, $\phi_1$=0.0$^\circ$, $\theta_2$= 37.0$^\circ$,
$\phi_1$= 180.0 $^\circ$ (upper row),
 $\theta_1$=30.0$^\circ$, $\phi_1$=0.0$^\circ$, $\theta_2$= 34.0$^\circ$,
$\phi_1$= 180.0 $^\circ$ (middle row),
 $\theta_1$=60.0$^\circ$, $\phi_1$=0.0$^\circ$, $\theta_2$= 49.0$^\circ$,
$\phi_1$= 180.0 $^\circ$ (lower row))
are chosen in such a way that
the quasi free kinematical condition ${\vec k}_3 = 0$ is fulfilled
for one central point on the locus.
}
\label{ppn_examples.2}
\end{center}
\end{figure}

\begin{figure}[!ht]
\begin{center}
\caption{
The eightfold  full breakup cross section
$d^8\sigma/(d{\hat k}^{\,\prime} d k_0^\prime d{\hat k}_1 d{\hat k}_2 dS)$
along the arc-length $S$ of the
kinematically allowed locus in the $E_1$-$E_2$ plane
for two different electron configurations:
$k_0$= 854.5 MeV, $\vartheta$= 27.9$^\circ$, $k_0^\prime$= 750.9 MeV (left),
and
$k_0$= 854.5 MeV, $\vartheta$= 35.7$^\circ$, $k_0^\prime$= 652.3 MeV (right).
Curves as in Fig.~\ref{ppn_examples.1}.
Particles 1 and 2 are protons.
The angles of the outgoing nucleons in the system
where ${\vec Q} \parallel {\hat z}$
($\theta_1$=57.0$^\circ$, $\phi_1$=0.0$^\circ$, $\theta_2$= 57.0$^\circ$,
$\phi_1$= 120.0 $^\circ$ (left),
 $\theta_1$=61.5$^\circ$, $\phi_1$=0.0$^\circ$, $\theta_2$= 61.5$^\circ$,
$\phi_1$= 120.0 $^\circ$ (right))
are chosen in such a way that
in the c.m. system all particles momenta are equal and form the so called
``Mercedes star'' in a plane perpendicular to ${\vec Q}$
for one point on the locus (the space-star kinematical condition).
}
\label{ppn_examples.3}
\end{center}
\end{figure}

\begin{figure}[!ht]
\begin{center}
\caption{
The eightfold  full breakup cross section
$d^8\sigma/(d{\hat k}^{\,\prime} d k_0^\prime d{\hat k}_1 d{\hat k}_2 dS)$
along the arc-length $S$ of the
kinematically allowed locus in the $E_1$-$E_2$ plane
for two different electron configurations:
$k_0$= 854.5 MeV, $\vartheta$= 27.9$^\circ$, $k_0^\prime$= 750.9 MeV (left),
and
$k_0$= 854.5 MeV, $\vartheta$= 35.5$^\circ$, $k_0^\prime$= 754.5 MeV (right).
Curves as in Fig.~\ref{ppn_examples.1}.
Particles 1 and 2 are protons.
The momentum of particle 1 is parallel to ${\vec Q}$
and the momentum of particle 2 is anti-parallel to ${\vec Q}$.
}
\label{ppn_examples.4}
\end{center}
\end{figure}


\begin{figure}[!ht]
\begin{center}
\epsfig{file=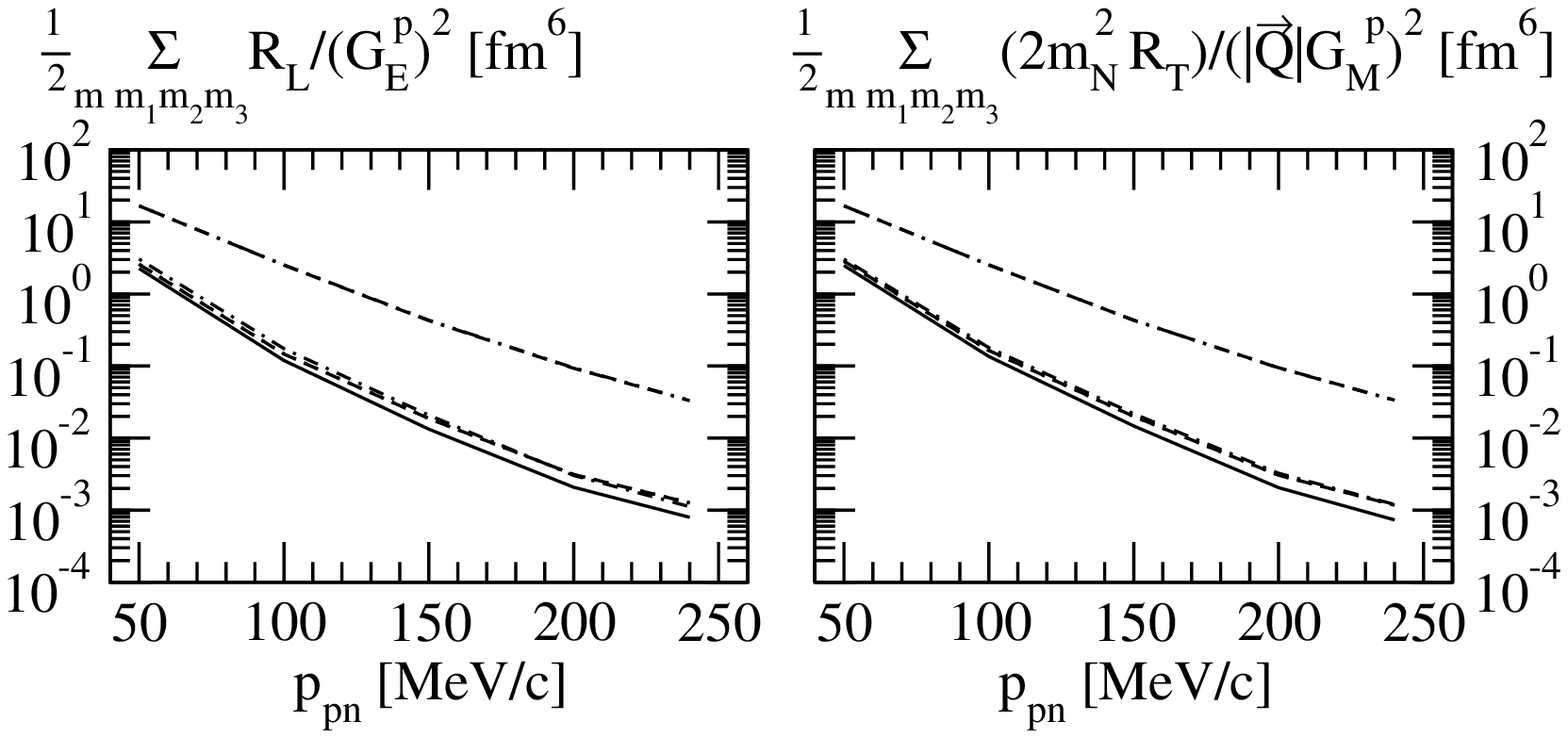,bb=50 515 540 750,clip=true,height=7cm}
\caption{
The quantities from (\ref{eq:15}) in the case of proton knockout 
 for different dynamical assumptions 
for $\vert \vec Q \vert $ = 500 MeV/c
as functions of the relative momentum $p_{pn}$ 
in the spectator proton-neutron subsystem. 
 For the description of the curves see Fig.~\ref{asymmetry1.e1.1}. 
}
\label{fig72a_ppn.n0.new.500}
\end{center}
\end{figure}

\begin{figure}[!ht]
\begin{center}
\epsfig{file=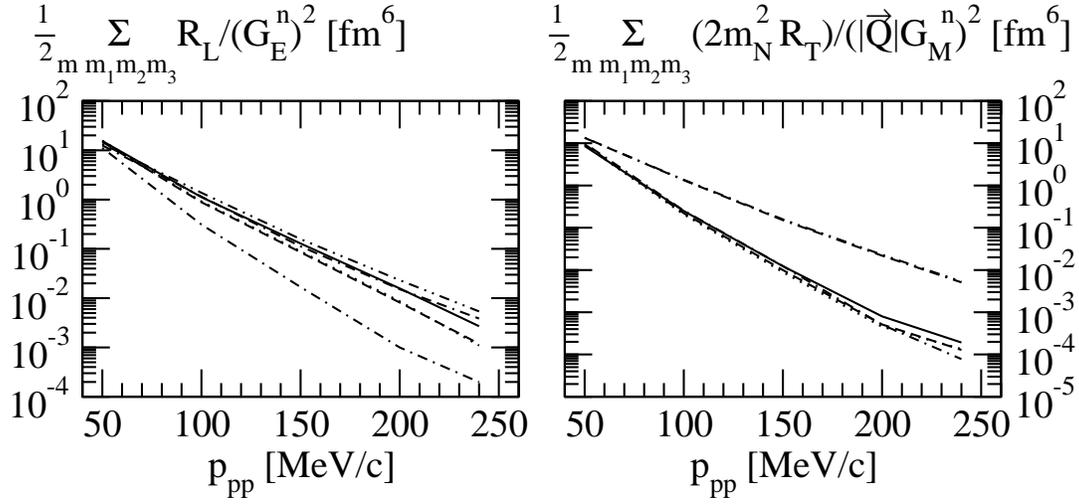,bb=50 515 540 750,clip=true,height=7cm}
\caption{
The quantities from (\ref{eq:15}) in the case of neutron knockout 
for different dynamical assumptions 
for $\vert \vec Q \vert $ = 500 MeV/c
as functions of the relative momentum $p_{pp}$ 
in the spectator proton-proton subsystem. 
Curves as in Fig.~\ref{asymmetry1.e1.1}. 
}
\label{fig72b_npp.n0.new.500}
\end{center}
\end{figure}

\clearpage

\subsection{Spin dependent momentum distributions of polarized proton-deuteron
 clusters in polarized \boldmath{$^3$}He}

We address the question whether momentum distributions of polarized
${\vec d}{\vec p}$ clusters in spin oriented $^3$He are accessible through the
$ \overrightarrow{^3{\rm He}} (e,e'{\vec p})d$
or
$ \overrightarrow{^3{\rm He}} (e,e'{\vec d})p$
processes. Optimal kinematical conditions are that
the polarization of $^3$He and the polarization 
 of the knocked out proton (deuteron)
 together with the momenta ${\vec k}_p$ and  ${\vec k}_d$ of 
the final proton and 
deuteron are collinear
to the photon momentum. The spin dependent momentum distribution
of proton-deuteron clusters inside $^3$He is defined as
\begin{equation}
{\cal Y} ( m, m_d, m_p ; {\vec q}_0 ) \ \equiv \
\left\langle \Psi m \left|
| \phi_d m_d  \rangle
| {\vec q}_0 \, \frac12 m_p  \rangle
\langle {\vec q}_0 \,  \frac12 m_p |
\langle \phi_d m_d |
\right| \Psi m \right\rangle ,
\label{pd.eq1}
\end{equation}
where $ {\vec q}_0 $ is the c.m. proton momentum (the deuteron 
momentum is $-{\vec q}_0 $) and
$m_p$, $m_d$, and $m$ are spin magnetic quantum 
numbers for the proton, deuteron, 
and $^3$He. This  can be expressed as \cite{pdclust}
\begin{eqnarray}
{\cal Y} ( m, m_d, m_p ; {\vec q}_0 ) \, = \,
\left|
\sum_{\lambda = 0,2 } \,
Y_{ \lambda , m - m_d - m_p } ( {\hat q}_0 ) \,
C( 1 I_\lambda \frac12 ; m_d , m - m_d , m ) \right. \cr
\left. C( \lambda \frac12 I_\lambda ; m - m_d - m_p , m_p, m - m_d ) \
H_\lambda (q_0)\right| ^2,
\label{pd.eq3.3}
\end{eqnarray}
in terms of the auxiliary quantity $ H_\lambda (q_0) $
\begin{eqnarray}
H_\lambda (q_0) \,\equiv \,
\sum_{l = 0,2 } \,
\int_0^\infty d p \, p^2 \,  \phi_l (p) \,
\langle p q_0 \alpha_{l \lambda} | \Psi \rangle \ \ , \ \lambda = 0, 2.
\label{p.eq3.4}
\end{eqnarray}
Here  $\lambda$ is the relative orbital angular momentum of the proton
with respect to the deuteron inside $^3$He. $\phi_l (p)$  and
$\langle p q \alpha \mid \Psi \rangle $ are wave
function components of the deuteron and $^3$He, respectively. Thus the
dependence on the direction ${\hat q}_0 $ and the magnetic quantum numbers
is nicely separated. 

\begin{figure}[!ht]
\begin{center}
\epsfig{file=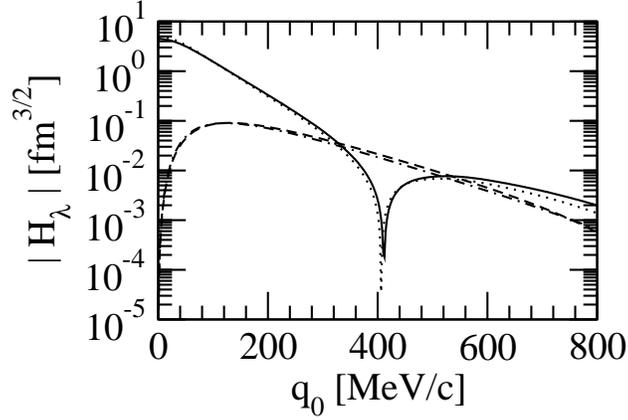,bb=10 515 305 715,clip=true,height=6cm}
\caption{
Absolute value of $ H_\lambda (q_0) $ defined in (\ref{p.eq3.4})
for $\lambda=0$ (solid) and $\lambda=2$ (dashed) calculated with the $^3$He
bound state including the UrbanaIX 3N force.
Corresponding curves neglecting 3N force effects
(dotted for $\lambda=0$ and dot-dashed for $\lambda=2$)
are also shown.
Note $ H_0 (q_0) < 0 $ for $ q_0 > $ 400 MeV/c, while
$ H_2 (q_0) $ remains always positive for the shown
$q_0$-values.
}
\label{fig72c_spindep.fig1}
\end{center}
\end{figure}

\begin{figure}[!ht]
\begin{center}
\epsfig{file=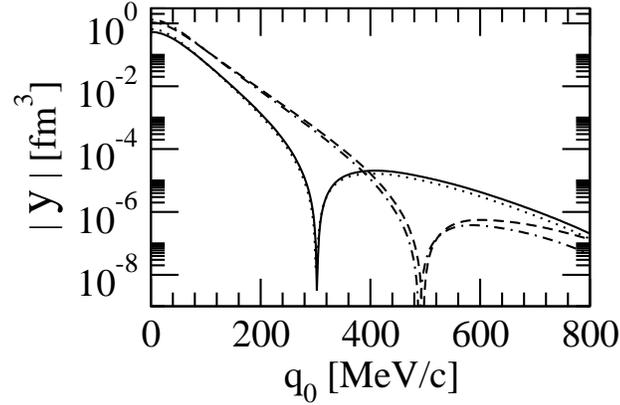,bb=10 515 300 715,clip=true,height=6cm}
\caption{
Spin-dependent momentum distributions
${\cal Y} ( m= \frac12, m_d= 0, m_p= \frac12 ; | {\vec q}_0 | {\hat z} )$
(solid line)
and
${\cal Y} ( m= \frac12, m_d= 1, m_p=-\frac12 ; | {\vec q}_0 | {\hat z} )$
(dashed line)
defined in (\ref{pd.eq3.3})
for ${\vec p} {\vec d}$ clusters in $^3$He when UrbanaIX 3NF is included. 
Corresponding curves neglecting 3N force effects
 are: dotted for $m_p= \frac12$ and dot-dashed for $m_p=-\frac12$).
}
\label{fig72d_spindep.fig2}
\end{center}
\end{figure}

\begin{figure}[!ht]
\begin{center}
\epsfig{file=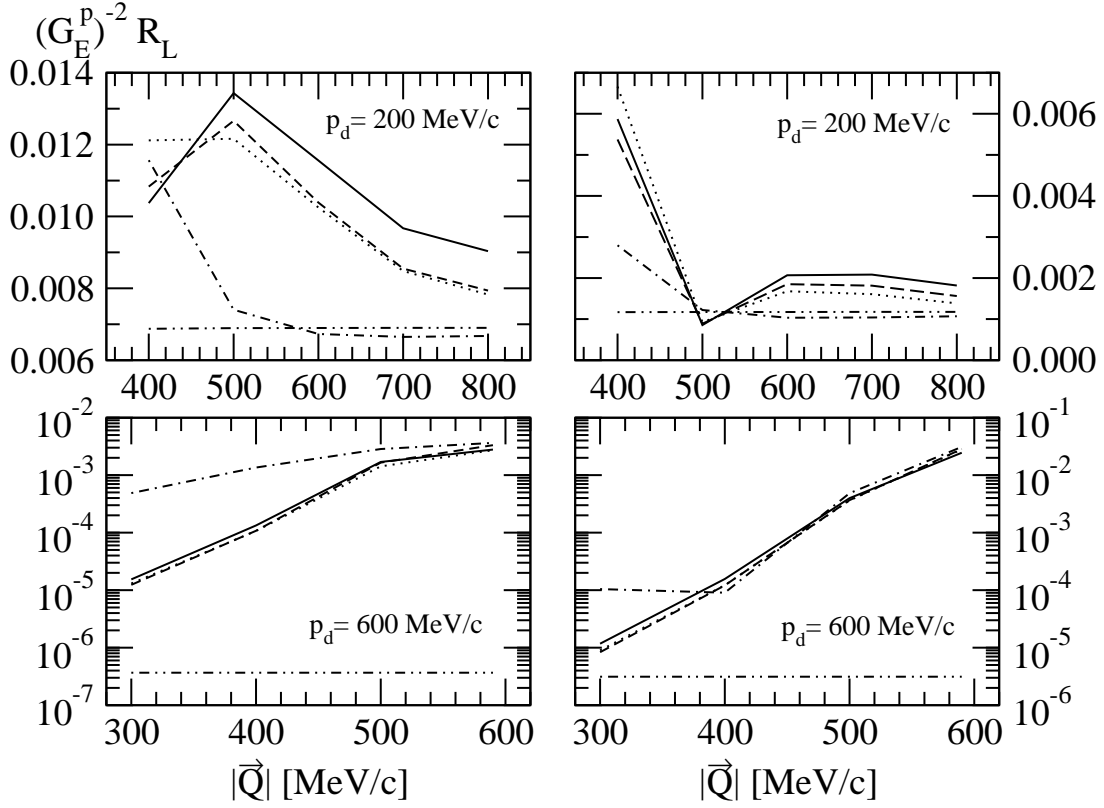,bb=40 360 560 740,clip=true,height=110mm}
\caption{
$\frac1{ (G_E^p)^2 } \, R_L$
as a function of the three-momentum
transfer $\mid \vec Q \mid$ for $p_d$= 200 MeV/c (upper row) and $p_d$= 600 MeV/c (lower row).
Two left panel figures are for the $m = \frac12, m_d = 1, m_p = -\frac12$
and two right panel figures for the $m = \frac12, m_d = 0, m_p = \frac12$
combination of the spin magnetic quantum numbers.
The curves correspond to the PWIA (double-dot-dashed), PWIAS
(dot-dashed),
FSI (dotted), FSI+MEC (dashed) and FSI+MEC+3NF (solid) results.
}
\label{fig72e_wlpd200.1}
\end{center}
\end{figure}

\begin{figure}[!ht]
\begin{center}
\epsfig{file=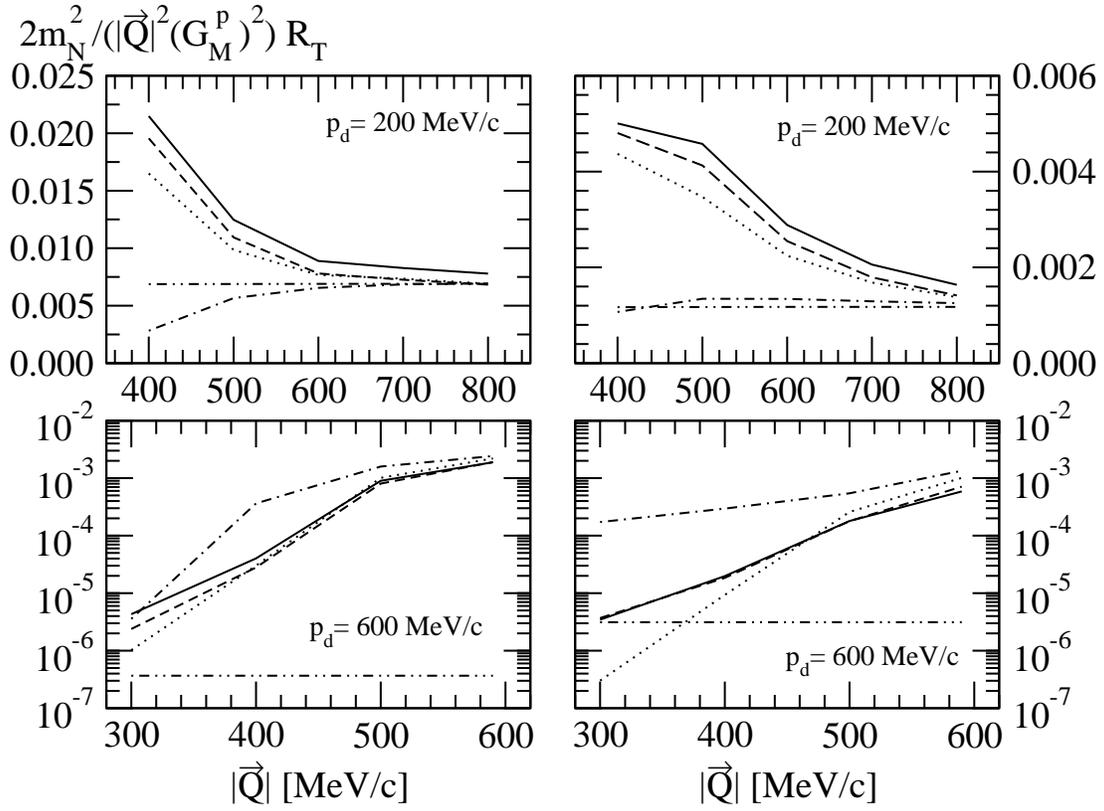,bb=40 360 560 740,clip=true,height=110mm}
\caption{
The same as in Fig.~\ref{fig72e_wlpd200.1} but for
$\frac{2 m_N^2}{{\mid \vec Q \mid}^2 (G_M^p)^2 } \, R_T$.
}
\label{fig72f_wtpd200.1}
\end{center}
\end{figure}

We display $H_\lambda (q_0) $ in 
 Fig.~\ref{fig72c_spindep.fig1}.
This shows that $\lambda$=0 dominates the momentum distribution   ${\cal Y} $
for small relative angular momenta and $H_0$ has a 
node around $q_0$= 400 MeV/c.
Near and above that value  the $s$- and $d$-wave contributions are comparable.
The momentum distribution itself is shown in Fig.~\ref{fig72d_spindep.fig2} 
 for the case that
${\hat q}_0$ points into the direction of the spin 
quantization axis and $^3$He
is polarized with $m = \frac12$. The polarizations of 
the proton and the deuteron
are chosen as $m_d = 0$, $m_p = \frac12$  and  $m_d = 1$, $m_p = -\frac12$, 
respectively.
We see an interesting shift in the minima from $q_0=300$ to $q_0=500$~MeV/c, 
if the polarization of the proton (deuteron) switches from a 
parallel (perpendicular)
to an anti-parallel (parallel) orientation in relation to the
spin direction of $^3$He.  It is easily worked out \cite{pdclust}
that the two momentum distributions shown
in Fig.~\ref{fig72d_spindep.fig2}
coincide in PWIA with the functions 
${\hat R}_L \equiv R_L/(G_E^p)^2  $
and
${\hat R}_T \equiv R_T/( \frac{{\vec Q}^{\, 2}}{2 m_N^2} \, (G_M^p)^2 ) $,
when one fixes the spin projections corresponding 
to the two combinations  and chooses the deuteron lab
momentum $p_d = q_0$.
In~\cite{pdclust} we investigated these two quantities allowing for the
complete FSI (without and with 3NF's), for  
 antisymmetrization, and for  the inclusion of MEC's
as a function of increasing $\mid \vec Q \mid$. The question is whether they
approach the two momentum distributions. The results are quite
intricate within the range of $\mid \vec Q \mid$-values we took into account
($\mid \vec Q \mid \le$ 800 MeV/c).
We show in Fig.~\ref{fig72e_wlpd200.1} $ {\hat R}_L$ and 
 in Fig.~\ref{fig72f_wtpd200.1}   $ {\hat R}_T$ in comparison 
to the PWIA results,
which are directly  the momentum distributions for the two
magnetic quantum number combinations. This illustration
refers to two deuteron momenta $p_d = 200$ and $600$~MeV/c. 
 For $p_d$= 200 MeV/c  $ {\hat R}_L$ and $ {\hat R}_T$ 
 have a tendency to approach
 ${\cal Y}  ( m, m_d, m_p ; {\vec q}_0 ) $  
within our momentum range $\vert \vec Q \vert$, 
but the 3NF effects are quite noticeable. 
 For  $p_d = 600$~MeV/c, however,  this is not the case.
It turned out when looking into several $p_d$-values that the two momentum
distributions could be accessed in our restricted  $\mid \vec Q \mid$-range
at its upper end only for very small deuteron momenta. For the higher 
deuteron momenta 
the FSI and MEC effects precluded that approach. We refer to 
 \cite{pdclust} for a more detailed discussion. The measurement
of that polarized setup would be very interesting since all the
dynamics comes into the play.

\clearpage

\subsection{3N Photodisintegration of \boldmath{$^3$}He}

The semiexclusive $^3{\rm He}(\gamma,N)NN$ reaction
where only one nucleon is detected appears to be rather easily 
accessible. We show 
in Figs.~\ref{fig6d13}--\ref{fig6d15} the
 energy spectra of the outgoing nucleon at several  nucleon emission lab angles.
The structures for proton and neutron emissions  are quite
different. While the structures for E$_\gamma$=12 and 40 MeV 
are similar there is a quite noticeable change
in shapes when going to E$_\gamma$=120 MeV. The 3NF effects in 
the predictions with explicit MEC's
are relatively small. Due to the semiexclusive character they 
are washed out in relation to rather
significant effects in the exclusive processes discussed below.
Also the Siegert approach including the 3NF is shown 
 and it deviates, especially
at 120 MeV,
from the explicit MEC predictions. 
 We refer to \cite{3bphot} for discussions and insights into the
complex underlying interplays. 
 In any case MEC effects are very strong at the 
two higher energies
and measurements would be very rewarding to test the theoretical
predictions.

\begin{figure}[!ht]
\begin{center}
\caption{
\label{fig6d13}
The semiexclusive $^3$He($\gamma$,N)NN processes for neutron or 
 proton emissions at various lab angles and E$_\gamma$=12 MeV. 
The solid curve corresponds to MEC+AV18+UrbanaIX dynamics, the 
dashed curve to MEC+AV18, the dotted curve to 
Siegert+AV18+UrbanaIX and the dot-dashed curve to AV18 
with the single nucleon current only.
}
\end{center}
\end{figure}

\begin{figure}[!ht]
\begin{center}
\caption{
\label{fig6d14}
The same as in Fig.~\ref{fig6d13} but at  E$_\gamma$=40 MeV.
}
\end{center}
\end{figure}

\begin{figure}[!ht]
\begin{center}
\caption{
\label{fig6d15}
The same as in Fig.~\ref{fig6d13} but at  E$_\gamma$=120 MeV.
}
\end{center}
\end{figure}

For the case of semiexclusive reactions with polarized $\gamma$ 
 and/or polarized $\overrightarrow {^3He}$ we calculated  the spin observables 
of (\ref{eq2}) in a range of  outgoing nucleon  lab angles from  
$\theta = 10^o$ to $170^o$. In Figs.~\ref{fig1v2}  and 
 \ref{fig2v2} we show our  predictions 
 for the nucleon outgoing angle 
$\theta=90^o$ at $E_{\gamma}=12$~MeV and  $E_{\gamma}=40$~MeV, respectively. 
 The $A_x^{\gamma}$  analyzing power  
is  large and its magnitude  
approaches one at the higher neutron energies. 
 It is 
practically insensitive to the inclusion of the 3NF both for outgoing neutron 
 and proton. 
 In contrast, the $A_y^{^3He}$ analyzing 
power and spin correlation coefficients $C_{x,y(y,x)}$ 
exhibit rather large 
sensitivity to the 3NF at $E_{\gamma} = 12$~MeV when the outgoing 
neutron is measured.  The 3NF effects modify the magnitude of the three 
observables in a very similar way. 
 Both spin 
correlation coefficients are quite similar to each other and to $A_y^{^3He}$.  
They are approximately of the same magnitude but 
of opposite sign to $A_y^{^3He}$. 
The effects of the 3NF extend over a large energy and angular range 
  of the outgoing neutron  
and in some cases are as large as  $\approx 20 \%$. 
Similar statements are true when the outgoing proton is measured. In this case, however, 
 the largest 3NF effects appear in the region of high energies of the 
 outgoing proton. 
 At $E_{\gamma} = 40$~MeV the 3NF effects are drastically reduced. 
 These  results show that it would be very interesting to measure such 
spin observables.

\begin{figure}[!ht]
\begin{center}
\caption{
\label{fig1v2}
The analyzing powers and spin correlation 
coefficients as a function of the outgoing 
neutron (left) or proton (right) lab energy 
for the $\stackrel{\longrightarrow}{^3He}(\vec{\gamma},n)pp$ 
($\stackrel{\longrightarrow}{^3{\rm He}}(\vec{\gamma},p)pn$) 
reaction at $E_{\gamma} = 12$~MeV and $\theta_{lab}=90^o$. 
The dashed curve is the prediction based on MEC+AV18 and the  solid on 
MEC+AV18+UrbanaIX. 
}
\end{center}
\end{figure}

\begin{figure}[!ht]
\begin{center}
\caption{
\label{fig2v2}
The same as in Fig.~\ref{fig1v2} but 
 at $E_{\gamma} = 40$~MeV.
}
\end{center}
\end{figure}

\clearpage

Then we come to the most informative process, the exclusive
$^3{\rm He}(\gamma,pp)n$ reaction.
We scanned the full phase space and searched for 3N force effects
by switching on and off the 3N force. To have a quantitative
measure we defined
\begin{equation}
\Delta(\Omega_1, \Omega_2, S) \ \equiv \
\mid   {d^{\, 5} \sigma}^{\rm NN+3NF} - {d^{\, 5} \sigma}^{\rm NN} \mid  /
    {d^{\, 5} \sigma}^{\rm NN} \times 100\%\;  ,
\label{Delta1}
\end{equation}
where $\Omega_1$, $\Omega_2$  are the directions of the two 
 outgoing protons and $S$ is 
 the position on the kinematical locus. 
In this manner we can associate $\Delta$-values to all regions in
phase space.
In order to locate phase space regions
uniquely, we show three two-dimensional plots.
The first one is the $\Theta_1-\Theta_2$
plane for the two polar angles of the proton detectors. The second one is the
$\Theta_1-\Phi_{12}$ plane,
where $\Phi_{12} \equiv \mid \Phi_1 - \Phi_2 \mid$ is the
relative azimuthal angle of these two detectors. Finally, the third one
is the E$_1$-E$_2$ plane
for the correlated energies of the two detected protons.
To fill the three planes we proceed as follows. The whole phase-space
is filled with discrete points corresponding to certain grids
in $\Theta_1,\Theta_2,\Phi_1,\Phi_2$, and E$_1$. For $\Theta_1$ and $\Theta_2$
fixed we search for the maximal value of $\Delta$ in the 3-dimensional
subspace spanned by $\Phi_1,\Phi_2$, and E$_1$. Then we combine
those maximal $\Delta$-values into three groups and associate certain
grey tones to those group values.
Next we choose a fixed $\Theta_1$ and $\Phi_{12}=\mid \Phi_2 \mid$ (one can put
$\Phi_1=0^{\circ}$) and search again for the maximal values of $\Delta$
in the 2-dimensional subspace spanned by  $\Theta_2$ and E$_1$.
The same grey tones and groupings are then applied. Finally, in the
E$_1$-E$_2$ plane we search for the maximal $\Delta$-values in the three
dimensional subspace spanned by $\Theta_1,\Theta_2,\Phi_{12}$ and repeat the
procedure. For a larger number of groups see~\cite{romek.thesis}.
This procedure has been  applied and the results are shown in 
Figs.~\ref{mapka.3NFeffects}-\ref{mapka.ZBIORCZY}.
We performed this investigation for 
three photon lab energies E$_\gamma$= 12, 40 and 120 MeV.
The results presented in  Fig.~\ref{mapka.3NFeffects} are based
on AV18+UrbanaIX and the explicit MECs.
The choice of the border values for the three groups is of course arbitrary.
The group with the largest effects according to those choices appear
in dark tone
and the group with the smallest effects appear in light tone.
The remaining group with 3NF effects in between is located in the white areas.

\begin{figure}[!ht]
\begin{center}
\caption{
\label{mapka.3NFeffects}
 The 3NF effects spread over the full 3N breakup phase-space. It is mapped 
into the $\Theta_1 - \Theta_2$, $\Theta_1 - \Phi_{12}$ and the 
$E_1 - E_2$ planes. The three rows refer to the three photon lab energies 
 $E_{\gamma} = 12, 40$ and $120$~MeV. Regions where the effects are largest 
are shown in dark and regions with smallest effects in light tone. 
In the white regions the effects lie between the two border values given to the 
right of each row. These results are based on AV18+MEC and AV18+UrbanaIX+MEC 
 predictions.
}
\end{center}
\end{figure}

\begin{figure}[!ht]
\begin{center}
\caption{
\label{mapka.CROSS}
The distribution of the magnitudes of the cross sections over the full 
3N breakup  phase-space for the three $\gamma$ energies as in 
Fig.~\ref{mapka.3NFeffects}. 
Now the white areas belong to the smallest cross section values and 
the light and dark tone regions to the cross section values 
as indicated on the right for each row. 
}
\end{center}
\end{figure}
\begin{figure}[!ht]
\begin{center}
\caption{
\label{mapka.ZBIORCZY}
The regions in the 3N phase-space where the breakup cross sections for 
$E_{\gamma}= 120$~MeV are larger than $0.1~~ \mu$b~sr$^{-2}$~MeV$^{-1}$, 
 the 3NF effects are larger than $20 \%$, and the choice of the 
 two-body current 
between MEC and Siegert causes effects not greater than $10 \%$. 
}
\end{center}
\end{figure}

As an example let us regard $E_{\gamma}=120$~MeV. Large 3NF effects 
are predicted for instance for the detector angles $\theta_1, \theta_2 \le 
60^{\circ}$ and all relative azimuthal angles $\phi_{12}$. The energies 
$E_1$, $E_2$ lie on a kinematical locus and the 3NF effects are largest 
as displayed by the dark spots. In addition there are smaller 
regions like $\theta_1$ as before but $\theta_2 \approx 180^{\circ}$.

In order to plan experiments in the future the absolute values of the fivefold
differential cross sections are important. Therefore we show those values in
Fig.~\ref{mapka.CROSS} again arranged in three groups. Here the 
white area refers to the
smallest cross section values. It is easily seen investigating the kinematics,
that the configurations corresponding to
 the darkest group are of the type of FSI or close to it.
We show two examples in Figs.~\ref{konfig7}-\ref{konfig11}. 

Finally, we locate the regions in phase space 
 for $E_{\gamma} = 120$~MeV 
 where the cross sections are measurable (larger than $0.1~~ \mu$b~sr$^{-2}$~MeV$^{-1}$) 
 and the 3NF effects are larger than $20 \%$. 
 Despite the fact that our Siegert approach is less suited 
for $E_{\gamma}=120$~MeV we also performed calculations 
and added the further condition, that the two choices of currents, 
explicit MEC or Siegert,  deviate at most by $10 \%$. This selects 
configurations which are dominated by 3NF effects and to a smaller extent by 
the choice of the current (among the ones we had at our 
 disposal).  
 Those small groups of configurations in phase space are
displayed in Fig.\ref{mapka.ZBIORCZY}.

\begin{figure}[!hb]
\begin{center}
\epsfig{file=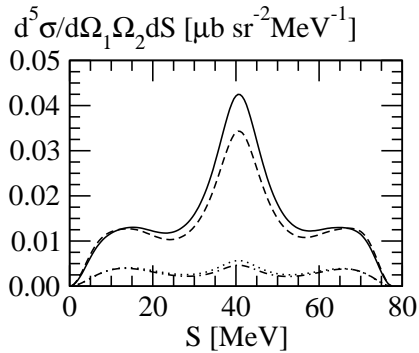,bb=45 520 300 740,clip=true,height=5cm}
\caption{
Fivefold differential cross sections for the angular
con\-fi\-gu\-ra\-tion
$\Theta_1=88^\circ$, $\Phi_1=0^\circ$,
$\Theta_2=100^\circ$, $\Phi_2=11^\circ$, at
photon lab energy E$_\gamma$=120 MeV. The AV18 predictions with single-nucleon
current and with single nucleon current + MEC are given by
 the dot-dashed and dashed curves, respectively.
The corresponding AV18+MEC+UrbanaIX predictions
are given by the dotted and solid curves, respectively.
}
\label{konfig9}
\end{center}
\end{figure}

\begin{figure}[!htb]
\begin{center}
\epsfig{file=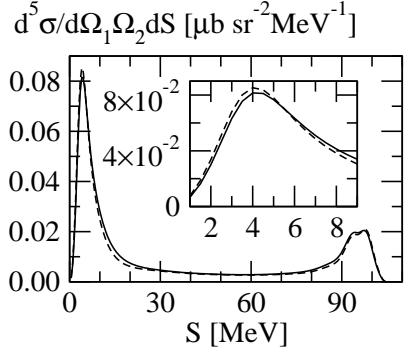,bb=45 520 300 740,clip=true,height=5cm}
\caption{
Fivefold differential cross sections for the angular
con\-fi\-gu\-ra\-tion $\Theta_1=30^\circ$, $\Phi_1=0^\circ$,
$\Theta_2=145^\circ$, $\Phi_2=77^\circ$, at photon lab energy
E$_\gamma$=120 MeV. The AV18+MEC predictions  are given by
 the dashed curve and the corresponding AV18+UrbanaIX predictions
are given by the solid curve.
}
\label{konfig10}

\end{center}
\end{figure}

\begin{figure}[!htb]
\begin{center}
\epsfig{file=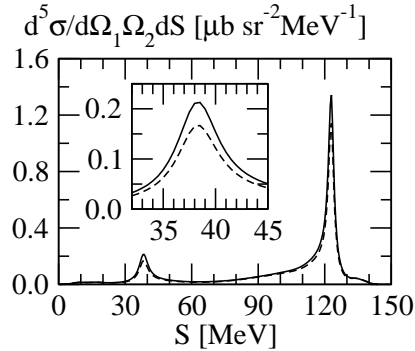,bb=45 520 300 740,clip=true,height=5cm}
\caption{The same as in Fig.~\ref{konfig10} but for
$\Theta_1=142^\circ, \Phi_1=0^\circ,
\Theta_2=27^\circ, \Phi_2=180^\circ$.
}
\label{konfig7}
\end{center}
\end{figure}

\begin{figure}[!htb]
\begin{center}
\epsfig{file=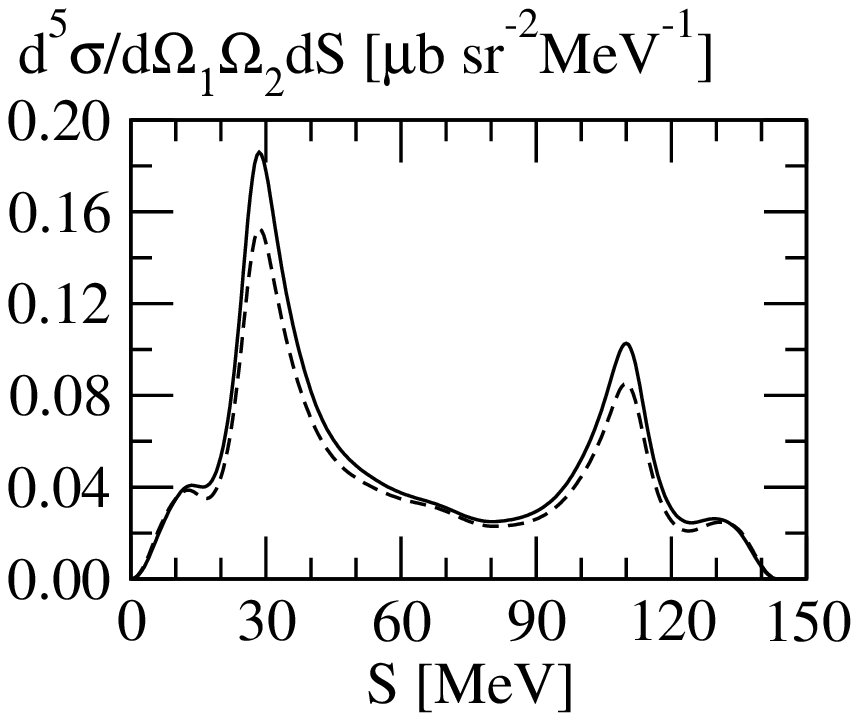,bb=45 520 300 740,clip=true,height=5cm}
\caption{
The same as in Fig.~\ref{konfig10} but for
$\Theta_1=70^\circ, \Phi_1=0^\circ,
\Theta_2=100^\circ, \Phi_2=180^\circ$.}
\label{konfig11}
\end{center}
\end{figure}

 Now we show some configurations for fixed angles along the S-curve
displaying different situations. In Fig.~\ref{konfig9} we see large 
two-body current
and some 3NF effects. In contrast in Fig.~\ref{konfig10} only very small
3NF effects appear. Finally in Figs.~\ref{konfig7} and~\ref{konfig11} 
large 3NF effects show up
in FSI peaks. That variety of current and 3NF effects 
 would be a fruitful and detailed source of information on the dynamics 
and an experimental investigation appears very worthwhile. 

In actual experiments one is far away from our point geometry results 
and a certain
amount of integration over angular regions and energy intervals has to be
accepted.
As an illustration we regard the two peaks in Fig.~\ref{konfig11},
which in point geometry exhibit 3NF effects of $\approx 20~\%$ 
($\approx 23~\%$) for 
the left (right) peak. Will they survive if 
 the cross sections will
be summed up over certain angular and energy regions? To that aim
we integrated the cross sections over all four 
 angles and single nucleon energy, allowing for deviations up to 
 $5^\circ$ around the central values for the  angles and $5$~MeV 
 in one  of the single nucleon energies, E$_1$, where the S-curve 
in Fig.~\ref{konfig11}
is related to a kinematical locus in the E$_1$-E$_2$ plane. This summation
is repeated replacing  $5^\circ$ in angles and $5$~MeV in energy by 
$10^\circ$ in angles and $10$~MeV in energy.
The resulting cross section values are displayed in Table~\ref{tableright1}  
  without and with
3NF. Their ratios   around $1.20$ show still a 
significant effect.
 From those
cross section values as well as from the magnitude of the effects an
experimental realization appears feasible~\cite{Adlerpriv}.

\begin{table}[htb]
\begin{center}
\begin{tabular}{l | c | c | c | c | c }
\hline
         S=110 MeV             & choice~I      & choice~II 
 &         S=30 MeV         & choice~I      & choice~II   \\  
\hline
without 3NF           & 0.683E-07   & 0.138E-05    
 &
without 3NF           & 0.234E-06   & 0.386E-05    \\
with    3NF           & 0.824E-07   & 0.166E-05     
 & with    3NF           & 0.280E-06   & 0.451E-05     \\ 
\hline
ratio                 &  1.21         & 1.20        
 & ratio                  & 1.20         & 1.17        
\end{tabular}
\end{center}
\caption[]
{Integrated cross sections $\Delta \sigma $ (in fm$^2$) 
 at $E_\gamma$= 120 MeV  
without and with 3NF for the two
choices of integration ranges (see text).
This refers to the two peaks 
in Fig.~\ref{konfig11} around $S=110$ and $30$~MeV. 
The ratios are practically as large as for point geometry.
}
\label{tableright1}
\end{table}

Note that in~\cite{romek.thesis} the NN interaction was taken in the
form of the $np$-interaction only, while in the present work and 
in \cite{3bphot} we include
$pp$ and $nn$ interactions by the "$\frac{2}{3}+\frac{1}{3}$" 
rule~\cite{Witala89,23-13rule}.
We refer to \cite{3bphot} and \cite{romek.thesis}
where several additional investigations are displayed.

Often in the literature photodisintegration is treated by keeping only the
lowest multipole
E1. This extreme low energy assumption would be quite insufficient
for nearly all phase
space regions and for all three photon energies studied in this paper.
This can again be quantified and we find, that even at
12 MeV there are plenty of breakup
configurations where the electric multipole E1 alone would deviate 
 by more than 20\% from the result when all multipoles are included.
Again for  detailed plots see~\cite{3bphot,romek.thesis}.

\clearpage

\section{Addendum}
\label{addendum}

We would like to add brief remarks on 
 several issues also relevant in the 3N system and 
 which have not been addressed in
this review: relativistic approaches, y-scaling and weak processes. 
 These remarks will  mostly serve to provide recent
references.

The {\bf covariant spectator theory} includes relativity in a manifestly
covariant way. It restricts  all but one of the particles to their mass
shell, which leads to the technically welcome property that all loop
integrations are three-dimensional. Also the manifest covariance goes
with the property that all boosts are kinematic and the off-shell
particle has negative energy components. Cluster separability holds
which in a Hamiltonian approach has been formally solved in \cite{polcoe} 
 but presents  a
 big challenge in the practical application. The spectator equations have
been applied to the NN system including electromagnetic processes as well
as to the 3N bound state. Most recently a complete Feynman diagram
expansion for the electromagnetic form factors and the three-body photo-
and electro-disintegration of the three-body bound state has been
derived. For the long list of references see the most recent ones 
~\cite{gross04,stadler04}

Another approach including relativity is the {\bf relativistic Hamiltonian
dynamics}. The seminal paper  is by E.P. Wigner~\cite{wigner39}. 
 It lays the ground for the physical 
requirements of special relativity in quantum mechanics leading to
the necessary and sufficient conditions for the 
 existence of an unitary (ray) representation
of the inhomogeneous Lorentz group (Poincare group) in the quantum
mechanical Hilbert space. Further seminal papers are  by  
P.A.M.  Dirac~\cite{dirac49} who 
  introduced in the Hamiltonian formulation  the "point","instant" and "front" 
 forms of dynamics. B. Bakamjian and
L.H. Thomas~\cite{bakam53} 
  constructed the first
relativistic quantum mechanical model of two interacting particles in
Dirac's "instant" form of dynamics. L. Foldy~\cite{foldy61} 
 pointed to the importance of macroscopic locality (cluster separability).
 F. Coester~\cite{coester65} 
 extended the work by Bakamjian and Thomas to systems of three particles with a
scattering operator consistent with the principle of macroscopic
locality. Finally S.N. Sokolov~\cite{sokolov77} 
 generalized relativistic Hamiltonian dynamics to N particles under the
 condition of macroscopic locality. Motivated by Sokolov's work F. Coester
and W.N. Polyzou~\cite{polcoe} 
 treated cluster properties for any
fixed number of particles in the instant-, front- and point-forms of the
 dynamics. The review of relativistic Hamiltonian dynamics by B.D. Keister and
W.N. Polyzou~\cite{keister91}
 includes in addition to basic concepts the material specific for the three-body
problem, on how to treat spin, and on matrix elements of tensor and spin
operators (currents).
 The most general  treatment of the  two-body problem
in relativistic dynamics appeared in \cite{polyzu89}. 
 It is not limited to Diracs forms of dynamics. They 
  are replaced by representations of Poincare Clebsch
 Gordon coefficients. Special choices lead to Diracs forms. This approach was
 generalized in \cite{polyzu2002} 
 to many bodies. 
Particle production was included  in \cite{polyzu2003}.  
 For relativistic variational Monte Carlo calculations of the  N-body
bound states the paper \cite{klink1996} 
 is suited. 
The Balian-Brezin method for treating angular momentum reduction in
the Faddev equation \cite{balian69} 
 has been generalized to the relativistic case in \cite{jean1994}.  
 A very basic investigation \cite{polyzu1985} 
 shows, that given a relativistic Hamiltonian dynamics it is possible to
construct a conserved covariant current operator that satisfies cluster
 properties and which will produce any kind of experimental form factors.
 In other words, it shows, that Poincare invariance, current covariance and
 cluster properties do not constrain form factors.

The above  citations refer to basic formalisms and we refrain to list 
 the  various applications of Hamiltonian dynamics to
 electron scattering, which are anyhow mostly carried through 
 for hadron form factors. This 
 is outside the scope of this review. We restrict ourselves only to 
 a few recent ones,  
which provide references to earlier work and to studies by J. Carbonell and 
collaborators:  
\cite{sengbush,coester040,carbon98,carbon99}. 
All that work briefly addressed  opens the doors to generalize what has
 been presented in this review into a relativistic Hamiltonian  scheme.

The issue of {\bf y-scaling} has been nicely discussed in~\cite{day90} 
 including a rich list of references, among them the seminal work in  
 \cite{west75} by G. West and the theoretical investigations
 based on plane wave impulse approximation by the Rome and Rehovot groups
 \cite{ciofi89a,gurwitz87}, 
 to mention just those two. Under PWIA it can be shown that the cross
section in inclusive electron scattering, which depends on 
$ \vert \vec Q \vert $ 
and $\omega$,  
at high momentum transfers,  
 after the cross section has been divided by an appropriately chosen single
 nucleon cross section,  
is a function of a single variable y. 
 This y-scaling variable is itself a function of 
 $ \vert \vec Q \vert $ and $\omega$. 
Of course the question is, whether the underlying
assumptions
are realized in nature and especially, whether the interaction of the knocked
out nucleon with the recoiling system can be neglected or sufficiently well
taken  into account. That issue has been critically studied in two model
investigations, one in a nonrelativistic two-nucleon model \cite{hub90a},  
 and one in a light
front formalism of relativistic quantum mechanics \cite{poly96},   
 where a conserved
model  hadronic current operator has been used which is covariant
with respect to a unitary representation of the Poincare group. 

{\bf Weak processes} have been discussed 
in \cite{Carlson98} where also references to
previous work can be found. A more recent study \cite{vivia03} 
evaluated the  decay rate for the  process $\mu^-~+~^3He~\to~^3H~+~\nu_{\mu}$ 
 including angular correlation parameters. The total rate agreed nicely
with experiment and showed only a weak model dependence. The two-body
currents, which turned out to be significant, could be
constrained in the tritium beta decay. This paper also provides some
clues on the induced pseudoscalar form factor $G_{ps}$. The process 
 $\mu^-~+~^3He~\to~d~+~n~+~\nu_{\mu}$   has
been studied  in \cite{skibmu} using  a Faddeev treatment for bound
and
continuum states. Only the single nucleon current has been employed.
 Very large effects of the final state interaction have been found,
 which brought  theory into the vicinity of the experimental decay rate 
 $d\Gamma / dE_d$~\cite{cummings,kuhn}.

\clearpage

\section{Summary and Outlook}
\label{summary}

This review has been devoted to electron and photon induced processes in the
3N system, restricted  to a mostly nonrelativistic kinematical regime.
We focused on  the Faddeev scheme which for the various
processes has been outlaid in some detail. This guarantees rigorous
solutions of the 3N bound and scattering states for any type of NN and 3N
forces.
 Naturally the electromagnetic currents play a central role, too.
Since this issue of current has been dealt with at many places in the
literature we were relatively brief and just described 
 the two-body currents which were  used in
our calculations on top of the 
standard single nucleon current. 
 These are the dominant $\pi$- and $\rho$-like currents
related to the NN force AV18. Then we provided expressions
for the rich set of observables and explained in some detail how the
different algebraic elements in the formalism are prepared in an angular
momentum decomposition for the numerical implementation.

The bulk of
the review has been devoted to a comparison of theory and experiment
and to theoretical predictions. The latter ones, if confronted with the data 
in the future, would challenge the dynamical assumptions even 
more stringently and systematically than what has been achieved up to now.
Our theoretical results which are compared to data are based on the
AV18/UrbanaIX Hamiltonian model and one- and the dominant two-body
currents related to AV18. The rich set of data comprises elastic
electron scattering on $^3$He and $^3$H, inclusive electron scattering on
$^3$He and $^3$H, nucleon-deuteron radiative capture and the time reversed
process of pd photodisintegration of $^3$He  and finally the 3N
photodisintegration of $^3$He. We tried to include as many as possible of  the
data situated in our nonrelativistic regime, which we qualitatively
defined by $\vert \vec Q  \vert \le  500$~MeV/c for 
the virtual photon and the three-nucleon
c.m. energy below the pion threshold. Clearly also in that kinematical
domain some effects of relativity will be  visible but they are not dealt
with in this review except for a small study for the elastic
electron scattering process on $^3$He. 

The elastic form factors of $^3$He and $^3$H are rather well described in the
low momentum region $q \le 3$~fm$^{-1}$. The presence of the 3NF is noticeable
and its effect goes in the right direction toward the
data. Our results are very similar to the ones achieved by the
Hanover group, which rely on a single $\Delta$-isobar admixture model 
instead of an explicit 3NF. They are also 
similar to predictions  of the Pisa group and
collaborators, who apply the same model Hamiltonian as used in this review,
but include additional currents. These currents 
 when applied  in the higher $\vert \vec Q  \vert$-domain not studied
in this review significantly improve the agreement with the data.
 
The two inclusive response functions, $R_L$ and $R_T$, in inclusive
unpolarized electron scattering on $^3$He and $^3$H show overall a good
agreement between theory and experiment with a slight underestimation,
however, of $R_L$ in case of $^3$H. Interesting is the interplay of 3NF's and
the two-body currents for $R_T$, which have a tendency to cancel each
other under our (restricted) dynamical assumptions. 
 When a comparison was possible 
 the results by the 
other groups are very similar to ours. The Pisa
and Trento groups are able to include the pp Coulomb force which is an
important step forward. It will be very interesting to see its
quantitative effect in detailed future studies, especially in the low
momentum regime.

If one allows for polarization of the incoming electron and the $^3$He
target two more response functions, $R_{T'}$ and $R_{TL'}$, in inclusive 
electron scattering are accessible with
related asymmetries. We showed that the sensitivity to the magnetic form
factor of the neutron survives in the transversal asymmetry $A_{T'}$
despite the fact that all dynamical ingredients, FSI, 
3NF effects and MEC's play an important role in the low
momentum  region. This has been used to extract $G_M^n$ for $Q^2 = 0.1$ and
 $0.2$~(GeV/c)$^2$, which are 
  in good agreement with the $G_M^n$-values extracted using
a deuteron target. 

We also draw attention to the Coulomb  sum rule which
in principle is an excellent source of information on two-body
correlations modified by two-body density and relativistic
effects. Unfortunately, due to strong cancellations the part of the Coulomb
sum which carries that information has large error bars, thus an
improved  set of data would be very informative.

In case of the pd electrodisintegration  of $^3$He we faced both, agreement
and disagreement, around the quasi elastic proton knockout peak. This is
a quite unsatisfactory situation, especially since a renewed
theoretical analysis by the Hanover group with a $\Delta$-isobar admixture
and therefore with a different  dynamics found very similar results.
This deserves further theoretical studies. In case of the deuteron knock
out peak we also face disagreement, namely a severe overestimation  of the data
 in the neighborhood of  missing momentum $p_m = 0$ despite the fact, 
 that the 3NF for the measured configurations 
moves theory significantly in the direction of the data. 
 For another  set of data in parallel deuteron knockout kinematics
 the agreement with the data looks better    but does not include 
the  situation with $p_m = 0$.  In relation to both peak areas 
 we think that the pd
electrodisintegration of $^3$He requires further efforts both in experiment 
and theory, as will be also addressed below.

In radiative Nd capture  the cross  section data are 
rather well described  over a wide range
 of energies. This is not the
case for the spin observables $A_y(p)$, $i T_{11}$ and the 
 tensor analyzing powers
$T_{ij}$. There remains much room for improvements in the dynamical
inputs. An important step forward in that direction  has been done very 
recently by the Pisa group with collaborators. 
 They  completed the
 current  related to the AV18/UrbanaIX model Hamiltonian
 what indeed improved  the
 agreement between the theory  and experiment in the very low energy
 regime. But also there  some discrepancies remained in the two vector 
 analyzing powers. 
 Since similar discrepancies 
 are also present in pure Nd scattering~\cite{ourreport}  
 they might have a common origin, presumably missing spin structures in
the 3NF.

The experimental situation in pd photodisintegration of $^3$He 
 is quite controversial as has been displayed for the energy dependence of the 
cross section at a fixed angle and for the integrated cross section.

The photon induced 3N breakup of $^3$He is still a rather unsettled issue.
 The total 
breakup cross section   data are severely controversial which
  precludes a conclusion about
the validity of the theory for that process. The very few more exclusive
data for that complete breakup unfortunately could not be analyzed properly 
by us since the experimental  conditions about angular and energy 
 acceptances were not
sufficiently well documented in the literature. In any case, our point
geometry results are at least in the neighborhood of those data given in 
 the form 
 $d^4\sigma/d\Omega_1 d\Omega_2$.

In view of the existing data we think that 
 more systematic measurements with  possibly improved 
accuracy are needed
to get better insight into the validity of the  dynamical
assumptions. For that aim we provided a few 
theoretical predictions, some of which at least
will hopefully be addressed in future experiments. 

The two response
functions $\tilde{R}_{T'}$ and $\tilde{R}_{TL'}$ appearing with the helicity of the incoming
electron show a great sensitivity to the dynamical input and especially
 $\tilde{R}_{TL'}$ for $^3$He shows a strong variation in shape as a function of 
$\vert \vec Q \vert $. 

The electron induced pd breakup of $^3$He poses  questions.
For the proton knockout peak region we have shown three quite different
cases. One is affected separately by FSI and by the 3NF, another one is
 predominantly  just given by PWIA alone, and  a third one just by FSI
 with no
effect of the 3NF. The second one would be especially important to be
verified by experiment, since only the simplest ingredients enter, the
 $^3$He state, the deuteron state, and the single nucleon current. In all
three cases the effects of MEC's are negligible.

In case of the deuteron knockout peak we also have selected three different 
situations in relation to the strength of the FSI effect and the MEC 
contributions. Since the deuteron is composite the mechanism of knockout 
is more complex than for the nucleon knockout. 

Despite the fact that the concept of the spectral function has been 
widely used in the literature we think that a more systematic 
approach to the situations where it is predicted to be 
useful and where it fails  would be adequate. We displayed two examples 
out of many described before in~\cite{ourspectral}. 

The semiexclusive process $\overrightarrow {^3He} (\vec e,e'N)NN$, where 
both initial particles
are polarized, would be also an interesting source of information about
the interplay of dynamical  ingredients. We showed two kinematical
conditions. In one  the asymmetry $A_{\parallel}$  for the upper 
 end of the
knocked out neutron energy spectrum 
 would be suitable to extract $G_M^n$, since all
curves, PWIA, FSI23, FSI, FSI + MEC and finally FSI+MEC+3NF coincide
there. In the other case the PWIA result differs strongly  from the others and
 large  corrections are necessary. The asymmetry $A_{\perp}$,  which in PWIA
is proportional to $G_E^n \cdot G_M^n$, requires in both chosen kinematical 
configurations always
strong corrections from FSI. Since $^3$He carries little proton polarization
the corresponding asymmetries are strongly influenced by final state
interactions. For one kinematical condition we found that FSI23 alone
would be quite misleading for  $A_{\parallel}$  but completely 
sufficient for $A_{\perp}$, 
while for the other kinematical condition 3NF effects are significant
for both asymmetries  except at the upper end of the proton energy spectrum, 
where all curves (except PWIA) coincide. We think that also these
different scenarios deserve a systematic experimental  study.

We also investigated the question, whether  the two-nucleon relative
momentum  distribution inside $^3$He could be approached experimentally.
We showed that in our kinematical regime this is not possible but at
least for  proton knockout under parallel kinematics the FSI23 dynamics
is sufficient. Thus the relative momentum distribution
folded with the NN t-matrix would be accessible, except for 
 an additional small shift caused by the action of the
3NF.
In the case of the neutron knockout only the transversal response
function exhibits that feature. For high $\vert  \vec Q \vert $-values, 
 however, also
 $R_L$ can be expected to behave similarly. 

Finally, in the field of 
electrodisintegration we investigated the spin dependent momentum
distribution for polarized proton-deuteron clusters in polarized $^3$He. For
the processes $\overrightarrow {^3He}(e,e'\vec p)d$ and 
 $\overrightarrow {^3He}(e,e'\vec d)p$ under 
fully collinear condition it turned out, that only
for rather low $p_d$ momenta  we found a tendency that the two
responses $R_L$ and $R_T$ properly divided by the electromagnetic nucleon
form factors approach the sought-for momentum distributions for
increasing $\vert \vec Q \vert $-values; otherwise 
FSI and 3NF effects  preclude that. Nevertheless a measurement 
of that polarized setup would be
quite interesting since all the dynamics comes into the play. 

3N photodisintegration of $^3$He comprises a lot of detailed dynamical
information. We found that the semiexclusive reactions 
 $^3He(\gamma,p)pn$ and $^3He(\gamma,n)pp$   show 
quite a different dependence on the emitted nucleon energy and the emission
angles. In all cases the 3NF effects are mostly washed out due to the
integration over part of the phase space. To the best of our knowledge no
data are available, but they would be very informative. 

If one allows for polarization for the incoming photon
and/or $^3$He, analyzing powers and 
spin correlation  coefficients can be
measured in the semiexclusive processes. We found that at $E_{\gamma} = 12$~MeV,
  especially for neutron emission, 3N force effects are quite
significant in $A_y^{^3He}$  and in the spin correlation coefficients, 
while  $A_x^{\gamma}$    has
no noticeable 3NF dependence. At $E_\gamma = 40$~MeV the 3NF effects have
essentially disappeared. No data are available to the best of our
knowledge.

Our last predictions in this review are for the most informative process,
the exclusive $^3He(\gamma,pp)n$ reaction. We scanned the full phase
space for 3NF effects and located the regions where they are as large
as $20 \%$ and above. Even after averaging over certain angular and energy
intervals carried out in two  examples, the magnitudes of these effects
survived. Precise and well documented data (for future analysis and 
possibly  new
dynamics) would be very important.

The comparison of data and theory in this review clearly demonstrated
that the chosen dynamics, forces and currents, is more or less adequate.
In most cases we encountered fair to good agreement with the data  but
 also in some
cases  clear discrepancies. Since for pure hadronic processes in
few-nucleon systems, especially in the well investigated 3N continuum,
the AV18/UrbanaIX Hamiltonian  model leads to similar 
agreements and disagreements,  the
reason for certain discrepancies in the electromagnetically induced processes
 cannot be searched alone in the additional ingredient, the electromagnetic
current operator, but also in the deficiencies of that Hamiltonian  model.
 Certainly additional spin structures in the 3NF model are required. This
has been already noticed  in pure 3N scattering 
 \cite{ourreport,sekigu,ermisch,kuros02,witala2001} but also 
in the description of spectra of light nuclei \cite{Carlson98,wiringa2000a}. 
 Additional 3NF models
introduced recently \cite{pieper2001a} improved the theoretical spectra. 
 Therefore, proceeding in this manner and 
  allowing for corresponding additional currents and
 relativistic features might be one way to go to achieve more
quantitative results. 

Another approach emerged in recent years based on effective field
theory,
either in the pion-less form or explicitly including the pion degrees of
freedom in a form constrained by spontaneously broken chiral symmetry
and including explicitly broken parts. This is a systematic
approach which is controlled in   the low momentum region by a
smallness parameter. Therefore the predictions can 
  be improved systematically and
theoretical errors can be estimated. This new approach to low energy
nuclear physics is very promising. It relies on effective Langragians, 
which allow  for well defined couplings to electroweak fields,  provides
internal connections between NN and many-nucleon forces, and generates
systematically relativistic corrections. Of course this approach is
restricted to generic external momenta below a certain mass scale.

We refer the reader to several reviews \cite{chiralrev} on these kind of
approaches and cite only a short subjective list of papers out of very
many, which we think are very relevant to investigate few-nucleon 
 systems without and
with electroweak probes. More references can be found there. The approach
to nuclear forces based on effective field theory constrained by chiral
symmetry goes back to S. Weinberg~\cite{wein91}.  
First applications were pioneered in \cite{ordon}. This was 
followed up in an extended and improved manner in 
 \cite{kaiser2001b,epel98,entem03} 
 pushing NN forces to next-to-next-to-next-to leading order (N$^3$LO) in
the chiral expansion. Thereby it has to be emphasized that the
3NF's and beyond are consistent to the NN forces. Various
applications \cite{epelrev,Evgeni} clearly demonstrated the success 
of that approach. In the
pion-less form, restricted to a lower momentum regime, also convincing
successful strides have been performed \cite{bedaq}. 
 Coupling to electroweak fields
has been investigated without and with explicit 
pions \cite{Rho91,kubis,kaplan}.  We
expect that these approaches will put low energy nuclear physics
including electroweak processes on a firm ground and will enable
 well founded applications like for astrophysical issues.

This review has been closed in January 2005 . We would like to
apologize to the authors whose work has not been sufficiently well
presented or whose work has not been cited at all.

\acknowledgments
This work was supported by the Polish Committee for Scientific Research
under grant no. 2P03B00825, by the NATO grant no. PST.CLG.978943, 
and by DOE under grants nos. DE-FG03-00ER41132 and DE-FC02-01ER41187. 
One of us (W.G.) would like to thank the Foundation for Polish Science 
for the financial support during his stay in Krak\'ow.  
The numerical calculations have been performed on the Cray SV1 and T3E and 
on the IBM Regatta p690+ of the NIC in J\"ulich, Germany.

\end{document}